**Kurdistan Region Government**

**Ministry of Higher Education and Scientific Research**

**University of Sulaimani**

**College of Agricultural Engineering Sciences**

# GENETIC DIVERSITY OF BARLEY ACCESSIONS AND THEIR RESPONSE UNDER ABIOTIC STRESSES USING DIFFERENT APPROACHES

**A Dissertation**

**Submitted to the Council of the College of Agricultural Engineering Sciences at the University of Sulaimani in Partial Fulfillment of the Requirements for the Degree of Doctor of Philosophy**

**in**

**Crop Science**

**Plant Biotechnology**

By

**Djshwar Dhahir Lateef**

B.Sc., Crop Science (2006), College of Agriculture, University of Sulaimani

M.Sc., Plant Genetics and Crop Improvement (2013), School of Biological Science, University of East Anglia

Supervisors

**Dr. Nawroz Abdul-razzak Tahir**
Professor

**Dr. Kamil Mahmood Mustafa**
Assistant Professor

**2022 A. D.** **2722 K.**

# Summary

In this investigation, five separate experiments were carried out. The first experiments were examined the molecular characteristics of 59 barley accessions collected from different regions in Iraq using three different molecular markers (ISSR, CDDP, and SCoT). A total of 391 amplified polymorphic bands were generated using forty-four ISSR, nine CDDP, and twelve SCoT primers, which they totally observed 255, 35, and 101 polymorphic bands respectively. The mean values of PIC for ISSR, CDDP, and SCoT markers were 0.74, 0.63, and 0.80, respectively, indicating the efficiency of the underlying markers in detecting polymorphic status among the studied barley accessions. Based on the respective markers, the barley accessions were classified and clustered into two main groups using the UPGMA and population structure analysis. Results of clustal analyses showed that the variation patterns corresponded with the geographical distribution of barley accessions.

Under present of stress conditions, barley's agronomic production is highly affected. Abundant germplasm resources, and reliable evaluation platforms are used to identify and select drought and heavy metal tolerant barley accessions. In this context regarding the second experimnets, 59 barley accessions were subjected to drought and cadmium stresses at the seedling stage. Polyethylene glycol (PEG 6000)-simulated drought with two concentrations of 10.25% and 20.50% was performed, while for cadmium stress, three conc. (125, 250, and 500 Mm) were used against the barley accessions. Results showed significant changes of barley accessions in their morphological characteristics such as: (Germination percentage, root and shoot length, and seed vigor) and biochemical profiles including: (Proline content, total proline content, antioxidant (DPPH), soulable suger content, and guaiacol peroxidase).

To confirm dependability of the screening criteria for drought and cadmium tolerance in barley accessions, the third experiments were carried out by selecting four barley accessions. MORA (AC36) and ABN (AC37) as drought resistant, Black-Garmiyan (AC47) and Bujayl 2-Shaqlawa (AC53) as a drought susceptible. The four accessions were subjected to eight drought treatments including control and intraction conditions at three different growth stages; tillering (S1), flowering (S2), and anthesis (S3) under plastic house condition. Similarly, based on early screening responses by all barley accessions, six accessions, namely Acsad-14 (AC29), ABN (AC37), and Arabi aswad (AC38) were used as cadmium resistance, and Black-Garmiyan (AC47), Black-Chiman (AC48), and Bujayl 1-Shaqlawa (AC52) were used as cadmium susceptible. The six accessions were subjected to cadmium stress with the concentration of 500 μM at tillering (S1) and flowering (S2) and double cadmium stress conditions at both (S1+S2). The selected drought and cadmium tolerant barley accessions showed less reduction in total yield and related traits in comparison with sensitive accessions under both stress conditions. The activities of biochemical traits that involve increasing scavenging free radicals against the availability of stress conditions were dramatically increased by considering drought-resistant barley accessions. However, these accessions maintained the lipid peroxidation even under the presence of severe drought stress conditions. The significant results obtained here biochemically and morphologically for all studied traits were positively correlated with the results of early screening of all barley accessions, confirming the dependability of the screening for drought and cadmium resistance accessions at the seedling stage.

The fourth experiments regarding the influence of using plant residues as biosorbents against cadmium under plastic house conditions at the seedling stage were conducted. Accessions AC29, AC37, AC48, and AC52 were exposed to 500 μM cadmium stress with and without extract applications. Each accession received four treatments, including a negative control (only water), a positive control (cadmium+water), oakleaf residues+cadmium, and guandela plant parts residue+cadmium. The modified residues with NaOH improved the overall performance of barley accessions, including the two considered susceptible barley accessions, AC48 and AC52, morphologically and biochemically, especially under the presence of Oakleaf and followed by guandela. Given the potential benefits, they are recommended to apply as a practical application when cadmium is available in the soil profile.

The final experiments were conducted to study the effects of Moringa Plant Extract (MPE) on the biomass and growth yield productivity for a collection of barley accessions. They were treated with a foliar application of MPE (1:30 v/v, MPE/Water). The field experimental design was a randomized complete block design (RCBD) with three replications. Results revealed that all studied traits were associated with barley growth and yield productivities. Both parameters were significantly increased as consequence of MPE application in comparison with the control condition, suggesting that the foliar application of MPE can be effectively used as a natural growth enhancer to improve the yield productivity and growth of barley.

# CHAPTER ONE

# INTRODUCTION

Barley (*Hordeum vulgare* L.) is considered to be the oldest and one of the main cereal crops cultivated by humans, and it is the fourth largest cereal plant after corn, wheat, and rice worldwide in terms of global production (Weiss, Zohary and Hopf, 2012). In general, barley is self-pollinating, diploid with a genome that contains 7 chromosomes. As a human food source, farmers have been used the wild type (*H. vulgare ssp. spontaneum*) of domesticated barley before the domestication of barley (*Hordeum vulgare* ssp) (Weiss, Zohary and Hopf, 2012). In archaeological sites in the Fertile Crescent dating back about 10,000 years, domesticated and wild barley were found, which is assumed to be the origin of barley domestication. The kernels of wild barley have their own indicators as they are fragile and have smooth rachis, which can be distinguished from those of cultivated barley (Weiss, Zohary and Hopf, 2012). In the world, there are over (32) species of the Hordeum genus, including polyploid, diploid, perennial, and annual types. During the process of selection and domestication, six-rowed and two-rowed barleys have emerged (Kilian *et al.*, 2006). Six-rowed barley is believed to be deriven from two-rowed wild barley as a result of selection and mutations which morphologically have different spikes (Komatsuda *et al.*, 2007).

World Food Security announced serious concerned about the growing world population. According to recent estimates, the world population will reach 9.8 billion or more by 2050 (DESA, 2019), and meeting the needs of this estimate of the global population is likely to require a doubling in food production. So, it is critical to improve the productivity not only food crops but also crops used for livestock fodder to meet the food demands of the near future (Qaim, 2020). For this purpose, inclusive genome and population structure explorations of genetic diversity in different plant species have received great attentions. While plants can adapt to different environmental conditions, such as fluctuating of climate conditions, the genetic diversity in a given species allows the plants to overcome these obstacles in a good manner (Raza *et al.*, 2019). The genetic variation in a population of a plant species, containing landraces, cultivars, as well as wild individuals, is an essential resource for the development of sustainable agricultural practices and increasing food production (Latutrie, Gourcilleau and Pujol, 2019). For that reason, to discover useful genes in crops, assessment of genetic diversity in the available genetic resources is the main step to conducting the plant research project. Nowadays, plants face multiple harsh environmental conditions such as salinity, drought, extreme low and high temperatures, the toxicity of heavy metals and nutrient deficiencies, collectively characterized as abiotic stresses (Vats, 2018). Among all of these environmental stresses, drought is the most severe issue, which globally affects crop production as a result of decreasing rainfall and global warming (Raza *et al.*, 2019). Barley is one of the most essential crops in many developing countries including Iraq. Barley is considered to be the only possible rain-fed crop that farmers can grow it in most of those countries. During the dry season, it is often subjected to extreme water limitations (Sadok and Tamang, 2019). In general, it is possible to define drought stress as a shortage of water in which dramatic morphological, physiological, biochemical and molecular changes can be observed in plants. All these changes reduce crop production and plant growth (Anjum *et al.*, 2011). Although barley can be grown in a variety of agro-climatic environments, one of the major challenges to barley productivity in these environments is drought stress (Xie *et al.*, 2018). Depending on the local environment, the stress of drought could take place at any stage of plant growth. For that reason, genotypes should be tested for their ability to tolerate drought in different growth stages, as some genotypes may tolerate drought stress at an early stage (seedling, germination) but are sensitive at later stages such as floweringand vice versa.

From both natural and anthropogenic sources, heavy metals are unstoppably released into the environment. The former usually does not harm plants, which includes volcanic activity and weathering of the bedrock (Schutzendubel and Polle, 2002) while the latter has strong direct effects on plants it comprises the left-over chemical products resulting from industrial activities and agricultural wastes such as pesticides, fertilizers, and herbicides. These wastes have involved in increasing the accumulation of heavy metals in soil. One of the most severe anthropogenic environmental stressors is soil pollution with heavy metals, causing plant growth inhibition and effects negatively on its productivity. In addition, heavy metals can cause long-term toxic effects on the health of the ecosystem as they are non-degradable (Ganeshamurthy, Varalakshmi and Sumangala, 2016; Shiyu *et al.*, 2020).

Chemical fertilizers are considered an essential factor that affects the outcome of barley products (Alemu *et al.*, 2017). Regrettably, in the case of using this type of fertilizer, despite its cost, it causes the degradation of soil and pollution of the environment. Lately, finding different and natural sources of bio-stimulants has received a great attention for improving the productivity of crops and accomplishing sustainable agriculture (Parađiković *et al.*, 2019; Nephali *et al.*, 2020). There are many sources of bio-stimulants that are commonly used in agriculture. For instance, humic acid (Li *et al.*, 2019), chitosan and chitin derivatives (Sharif *et al.*, 2018), seaweed extracts (Nabti, Jha and Hartmann, 2017), and plant extracts (Drobek, Frąc and Cybulska, 2019). Besides the use of these extracts as bio stimulators, there are many other advantages, such as chelating heavy metals, which will be addressed in literature sections.

The objectives of this study were to identify:

1- Genetic diversity among 59 barley accessions collected across Iraq using three different molecular markers, namely: ISSR, CDDP, and SCoT.
2- Morphological and phytochemical responses of these barley accessions under the presence of different drought and cadmium heavy metal stress conditions at seedling stage.
3- Physiological and biochemical activities of selected barley accessions were evaluated under plastic house conditions under both dought and cadmium heavy metal stress conditions.
4- The morphological and biochemical effects of oak leaf and kanger plant residues on barley accessions exposed to cadmium stress conditions in a plastic house at an early growth stage.
5- The influences of moringa plant water extracts on growth and yield contributed traits on collected barley accessions under natural filed conditions.

# CHAPTER TWO
# LITERATURE REVIEW

## 2.1 The Role of Genetic Diversity

Genetic variability describes genetic differences among individuals of the same species, and it is an important tool for conservation and plant breeding programs (Singh, 2015). In the past, morphological characterization was the only parameter for the evaluation of genetic variation. With this method, only a reduced set of *loci* is estimated and the bias derived from environmental variability was not be able to be measured (Singh and Singh, 2015). For more precise evaluation, DNA molecular markers have been used to determine the molecular variability that is not affected by the environment, and they can be applied to DNA derived from each growth stage. Molecular markers are sequences or genes localized on chromosomes that encode a specific trait, and they are applied to define and understand the genetic variability in several organisms (Li *et al.*, 2016).

For successful breeding and producing new cultivars, genetic diversity is an essential tool for this purpose. The study of genetic and phenotypic diversity to identify groups with similar genotypes is important for evaluating and using genetic resources. Saroei, Cheghamirza and Zarei (2017) demonstrate that knowledge of the genetic diversity of barley germplasm is necessary for finding diverse parental combinations and creating segregating progeny with high genetic variability for selection.

To determine genetic variation among plant species, morphological markers are widely conducted by farmers and breeders. This includes morphological classification of diverse entries grown in the field. The morphological characteristics are the strongest determination factor of the agronomic value and plants' taxonomic classification (Luo, Xia and Lu, 2019). Morphological evaluations are inexpensive, easy to perform, and do not require expensive technology. However, the requirements for large areas of land and human workers over a period of time make it costly (Andersen and Lübberstedt, 2003).

When compared to other marker systems, the morphological traits suffer from the limitations of environmental fluctuations and subjective characterization (Choudhary, Choudhary and Shekhawat, 2008). Different sets of morphological parameters are taken into consideration for assessing the variations among tested plant materials, which include both farmer-preferred parameters such as (high yield, large seed, and so on) and breeder-preferred traits such as (disease, drought, and other stressors). In a wheat breeding program using this type of marker, a successive linkage was found between male sterility and a dwarf phenotype. In this respect, the sterility of males can be indirectly selected by the straight selection of dwarf plants in the segregation population (Yang *et al.*, 2009).

Another marker system for assessing genetic diversity was used, which is recognized as biochemical markers (isozyme markers). The mechanism of this type of marker depends on the difference in specific banding forms of the proteins separated on SDS PAGE (Farooq and Azam, 2002). These markers have the feature of near genetic neutrality, which can be used for linkage map construction and genetic variability studies (Kumar, 1999). To study genetic diversity and construct genetic linkage maps, biochemical markers are considered the first group of molecular markers that were commonly used in the 1990s. Since the number of polymorphic *loci* is insufficient, the use of isozymes is limited for assessing genetic diversity in plant breeding programs. To detect quantitative genetic analysis in barley (Kleinhofs *et al.*, 1993), wheat (Moghaddam, Ehdaie and Waines, 2000), and rice (Fuentes *et al.*, 1999), isozymes are used. Molecular markers can offer genomic information for plant evaluation, which is critical for successful breeding in any studied plant species before entering the next cycle of selection, and also support tracking polymorphisms with no clear phenotypic trait parameters (Paux *et al.*, 2010). In general, molecular markers are fragments of DNA showing (mutations and /or variations that can be used to detect polymorphism between alleles of a gene for a specific sequence of DNA or different varieties. Such fragments are linked to exact locations within the genome and may be detected by using certain molecular technologies. Molecular characterization can play a role in uncovering the history and estimating the diversity, distinctiveness, and population structure. It can also serve as an aid in the genetic management of small populations, to avoid excessive inbreeding. Several investigations have been described within and between-population diversity (Ramesh *et al.*, 2020). As a consequence, the information about genetic relationships among barley genotypes and their morphological characteristics will be a very useful tool in barley cross-breeding programs. In the present review, based on the importance of molecular marker techniques in which they are used to examine genetic diversity for barley genotypes, an attempt has been made to demonstrate overall concepts in the area of barley genetic diversity, especially PCR-based markers, which could help extend the knowledge of genetic diversity in barley crops.

### 2.1.2 Random amplified polymorphic DNA (RAPD)

This method is quite handy and cheap to use by the researchers, and no previous information on the studied genome is required. It is based on the amplification of random DNA fragments by PCR with primers of random nucleotide sequences. Single RAPD primers can hybridize to several hundred sites within the target DNA, and then the amplified DNA products can be visualized by gel electrophoresis (Agarwal, Shrivastava and Padh, 2008).

For assessment of genetic differences in populations and species, fingerprinting, and study of phylogenetic relationships among species and subspecies, these markers have been commonly used in diverse plant species and are still conducted by many researchers (Nybom and Bartish, 2000; Sirijan *et al.*, 2020). However, these markers, like other techniques, have limitations, and they are not ideal for genome mapping due to the length of the primer sequence, which is short and randomly amplified in the genome. In addition, allelic differences in heterozygotes cannot be detected using RAPD markers, which can only provide dominant markers (Edwards and McCouch, 2007).

Regarding barley investigations in terms of diversity using this molecular marker, a group of researchers (Guasmi *et al.*, 2012) conducted their study on 80 barley varieties from South Tunisia to determine genetic differences among tested verities using two molecular approaches ISSR and RAPD markes. Regarding RAPD markers, out of 10 primers, 3 of them were amplified and yielded 17 polymorphic bands. In addition, two clusters were estimated for the studied barley varieties using both molecular set techniques. Moreover, a few attempts were carried out by other researchers using this molecular method as they conducted their study using only five barley cultivars and four RAPD primers (Mylonas *et al.*, 2014), 34 fragments were amplified, but 8 of them were polymorphic, and the polymorphic information content (PIC) ranged from 0.32 to 0.48. On the other hand, extensive experiments were carried out for 31 barley landraces collected from three countries: Algeria, Tunisia, and Egypt, and their primary goal was to assess the genetic differences between these landraces as collected from different geographical regions (Allel *et al.*, 2017). The dendrogram in their study clustered all the 31 barley landraces into two major groups. In addition, out of 132 amplified bands, 112 bands were polymorphic, with an average of 60.16 percent polymorphism for a total of 10 RAPD primers used in their investigation.

### 2.1.3 Inter-simple sequence repeat (ISSR)

This marker system involves the amplification of DNA fragments present at an amplifiable location between two distinct microsatellite repeat regions oriented in opposite directions, which were first reported by (Zietkiewicz, Rafalski and Labuda, 1994). The repeating nucleotides of di, tri, and tetra even in the form of pentanucleotides are used as primers for amplification of tested DNA. These types of sequences are randomly distributed over the whole genome, and only a small quantity of DNA is required for amplification. The ISSRs are easily measurable by gel electrophoresis for a few to hundreds of samples, which requires low cost by researchers, especially with limited funding resources. Obtaining polymorphic bands is based on the difference in the number of repeats of the tested plant materials (Bornet and Branchard, 2001).

Regarding barley investigations in terms of diversity using this molecular marker, (Wang, Yu and Ding, 2009) conducted research on 90 barley accessions, which half of them originated from China's Tibetan region and the other half from the Middle East using ten ISSR primers. 79 and 66 polymorphic bands were detected among tested barley accessions for both barley regions, respectively. More recently, population differences among relatively higher barley genotypes (120) and higher ISSR primers (26) were carried out by (Ghomi *et al.*, 2021), and 85 polymorphic bands were determined among barley genotypes using this type of marker. The genetic diversity of 31 populations of Iranian wild barley, consisting of 27 populations from Fars province and four populations each from Alborz, Kermanshah, Khuzestan, and Lorestan provinces, was investigated using 15 ISSR markers. Cluster analysis showed that populations were divided into six main groups based on an 88.5% similarity level. Using 15 ISSR primers for determining the genetic diversity among 31 wild barley populations in Iran was investigated (Hosseini *et al.*, 2022). Genetic variation between populations in their study was 47%, while a higher percentage was recognized for the value of genetic dissimilarity within the populations at %53. Recently, genetic diversity in fifteen six-row barley cultivars grown in Egypt was evaluated using fifteen ISSR markers (Mohamed *et al.*, 2021). Using this technique, they revealed 62 polymorphic bands in their study and the highest value of PIC were observed by both primers, UBC 840 and UBC 814, with a value of 0.37, while the ability of the UBC 835 primer was less among tested primer in detecting the differences among tested cultivars, in which small value of PIC with a value of 0.26 was documented. Besides the use of this marker for detecting variances among cultivars, it is also used for identifying differences between isolates of pathogens. Recently, sixty-eight isolates from different regions of Pakistan that cause the kernel bunt infection in wheat were collected and subjected to twenty ISSR primers (Asad *et al.*, 2022). Surprisingly, a total of 215 bands among tested isolates were polymorphic using this type of primer and the average value of PIC was 0.902, indicating the significance of using the selected primers.

### 2.1.4 Conserved DNA-derived polymorphism (CDDP) marker

A single primer similar to ISSR and SCoT primers is required for amplification. Conserved DNA-derived Polymorphism markers are designed specifically to target conserved sequences of plant functional genes (Collard and Mackill, 2009). At multiple sites within plant genomes, these shorter conserved gene sequences should be present, providing multiple primer binding locations. In several plant species, to estimate genetic diversity, the use of this technique has already proved its effectiveness (Hajibarat *et al.*, 2015; Talebi *et al.*, 2018; Saidi, Jabalameli and Ghalamboran, 2018; Ajithan *et al.*, 2019). Using 13 CDDP primers, the population structure and genetic diversity of four populations of a rare tree species prevalent in China, Salix taishanensis, were determined by Liu *et al.* (2020), who found 128 distinct *loci* in these populations. Moreover, in the study of forty Iranian bread wheat varieties for identifying the diversity, a total of ten CDDP primers generated a total of 43 polymorphic bands (Hamidi, Talebi and Keshavarzi, 2014). To evaluate the genetic diversity, a total of forty-eight chickpea accessions from Iran were subjected to three different marker systems using fifteen SSR, nine SCoT, and ten CDDP primers (Hajibarat *et al.*, 2015). Almost equal portions of polymorphism information content were detected by three marker systems. In their investigation, CDDP showed superiority over SSR and SCoT markers, and five different clusters were stated among the tested accessions, while the cluster analysis revealed three clusters using SSR and SCoT markers. Talebi et al. (2018) estimated the genetic associations among forty-eight safflower accessions using 10 primers for each of the three gene-targeted markers: CDDP, SCoT, and CAAT box-derived polymorphism (CBDP). Although a high level of polymorphism was detected by all three marker systems, the markers of CDDP in their study produced a higher number of polymorphic bands in comparison with the other two marker systems. These results showed the effectiveness of this marker for genetic diversity analysis in diverse plant species and its potential for genome diversity and conservation of germplasm.

### 2.1.5 Start codon targeted (SCoT) marker

Based on the short conserved region in the plant genes surrounding the ATG translation start codon, this marker system was established. The primers target common plant genes that are responsible for plant development or related to biotic and abiotic stress (Collard and Mackill, 2009). In parallel with RAPD and ISSR markers, a single primer was used as reverse and forward in SCoT markers. This technique can be greatly used in bulk segregate analysis and quantitative trait loci (QTL) mapping because they are dominant markers in their nature, and much attention has been gained by researchers for these specific markers due to their good quality and better appearance over the other techniques (Collard and Mackill, 2009; Rahimi *et al.*, 2018). For identifying QTL associated with chlorophyll fluorescence, 27 SCoT primers were tested on 106 barley recombinant inbred lines by Makhtoum *et al.* (2021), then the linkage maps of barley chromosomes were constructed based on the SCoT marker. Nouri *et al.* (2021) used ISSRs and SCoT primers to determine the population structure and genetic diversity of 90 wild goatgrass (*Aegilops tauschii*). They found that SCoT markers showed the highest values of polymorphism information content (PIC) in comparison with ISSR markers. From five regions in Iran, forty Damask rose genotypes were collected to assess the genetic diversity of this collection using two marker systems: universal rice (URP) and SCoT primers (Mostafavi *et al.*, 2021), and 18.18 polymorphic bands were revealed using twelve SCoT primers with mean values of PIC of 0.37. A total of forty-three durum wheat germplasm samples were evaluated for indicating genetic variation using six SCoT and fifteen ISSR markers (Etminan *et al.*, 2016). A total of 54 bands produced from the SCoT primers were all polymorphic, which points to the usefulness of this marker for the detection of genetic variation.

### 2.1.6 Single nucleotide polymorphism (SNP)

This type of marker is considered the new generation in detecting polymorphism among or between individuals as a result of changing the position of a single nucleotide. In principle, the SNP approaches display differences between a probe of a known sequence and a target DNA holding the SNP location (Ramesh *et al.*, 2020). The target DNA sections are typically the products of PCR and mismatches with the probe discover SNPs within the amplified target DNA section. Then, to recognize SNP polymorphisms, the mismatched DNA sections can be sequenced (Amom and Nongdam, 2017; Ramesh *et al.*, 2020). Based on the availability of 1955 polymorphic SNP, Almerekova *et al.* (2021) assessed the genetic association of 68 two-rowed barley accessions from Kazakhstan, and then these accessions were compared with barley accessions that originated from six diverse regions around the world. A comprehensive study was conducted on 316 barley genotypes to estimate the genetic differences and population structure using the 50 K iSelect Illumina Barley SNP collection by (Verma *et al.*, 2021). The PIC and gene diversity values among studied barley genotypes were 0.289 and 0.362 respectively.

## 2.2 Barley Under Drought and Heavy Metal Stress Conditions

At the beginning of conducting any research related to identifying and selecting drought or heavy metal resistance accessions, the researchers must identify the traits that can be chosen to measure the effect of drought stress and other stressors on plants. The selected trait should distinguish the resistant and susceptible studied genotypes. In addition, the improvement in drought resistance should be in parallel with production as the growers require the ability to profitably produce their products under unfavorable drought conditions.

Regarding drought tolerance in barley, till the moment, many researches have been studied (Amini, 2013; Hagenblad *et al.*, 2019), but many obstacles are limiting the improvement of these crops for drought resistance, including: First, many genes are involved in controlling drought, hence it is a complex trait. However, to genetically improve drought resistance, these genes are important. Secondly, intense changes in the physiological parameters of the plant may be caused by drought stress, which needs to be understood and measured. Third, the interaction between genetics and the environment will affect selection of genotypes for drought resistance. Through morphological, physiological, biochemical, expression of genes or genetic studies, drought stress can be investigated. The use of many different methods and information from different studies is recommended to obtain a better understanding of drought tolerance.

Producing new barley varieties with a high degree of drought resistance can be achieved through a plant breeding program. Plant breeders must improve the outcome of yield products from any selected plants that also have the availability to resist drought conditions. Selection of potential germplasm is the first step that includes genetic differences for drought resistance (Ceccarelli *et al.*, 2000). Based on objective and climate conditions, the selection of germplasm will be tested at particular or multiple growth stages. The breeding program was initiated after the identification of a resistant group of varieties for their drought tolerance by crossing the selected varieties as donor parents. The selection of traits such as physiological, morphological, and yield by the researchers should be limited to drought resistance so they can distinguish between susceptible and resistant varieties, and the selected traits should have the features of high heritability in parallel with the significant final product of grain yield (Kosova *et al.*, 2014).

New approaches in genomic technology have accelerated the selection of usable traits. There is the possibility to sequence the whole genome of a particular genotype. Genotyping by sequencing, which produces large numbers of single nucleotide polymorphism (SNP) markers, is one of the most widely used techniques that covers the barley genomes (Thabet *et al.*, 2018; Tan *et al.*, 2020). Besides, the genome reference for barley is now available for use. Detecting the exact location and position on the chromosome for each SNP marker can be achieved by using this feature. To find genes or genomic regions that can control drought resistance, quantitative trait loci (QTL) mapping and genome-wide association studies (GWAS) are used to simplify the complexity of traits such as drought (Wójcik-Jagła *et al.*, 2018; Tarawneh *et al.*, 2020). It is very essential to recognize the number of genes controlling drought resistance in each used genotype. Many specific molecular markers for barley have been detected, such as *HvABI5* (Collin *et al.*, 2020), *HVMYB1* (Alexander, Wendelboe-Nelson and Morris, 2019), *HvAKT2*, and *HvHAK1* (Feng *et al.*, 2020). Thus, the absence or presence of these genes can be tested in any germplasm for drought resistance. In addition, drought tolerance is a qualitative resistance in plants, so engineering many genes in the same variety could be more profitable. This can be achieved through gene editing techniques such as transcription activator-like effector nucleases (TALENs), zinc finger nucleases (ZFNs), and more advanced gene editing techniques well-known as CRISPR/*Cas9* systems (Ilyas *et al.*, 2020). For those reasons, a detailed investigation of the morphological, biochemical reactions, and genetic studies on drought stress is essential to enable the great effects of different mechanisms to develop drought-tolerant varieties that may cope with drought conditions.

In general, all species of plants can persist under the stress of heavy metals at different rates depending on the variety of plants and the type of heavy metal. Metal-binding ligands play essential roles in plant metabolism as they are involved in the detoxification of heavy metals with naturally occurring ligands such as organic acids, amino acids, polypeptides, and peptides (Delangiz *et al.*, 2020). Phytochelatins are glutathione-derived peptides and they are responsible for reducing free metal concentration in plant tissues and protecting plant tissues from the damage that will take place in the presence of heavy metals (Ahmad *et al.*, 2019). This compound supports the detoxification of cells by creating stable complexes with metal ions and decreasing the contrary effects of heavy metal stress in the plant. Metallothioneins function similarly to Phytochelatins (Goldsbrough, 2020), having their own cysteine sulfhydryl groups in their structures, allowing storage of heavy metals in the cell wall and the vacuole by binding to heavy metals and establishing stable complexes (Kobayashi, Nozoye and Nishizawa, 2019).

Heavy metals cause many biochemical and physiological alterations, ranging from sub-cellular levels to the whole plant (Berni *et al.*, 2019). One of these modifications is the creation of reactive oxygen species (ROS), which harms plant metabolism. Plants naturally establish diverse pathways to resist this barrier (Sabagh *et al.*, 2019). For plant metabolism, a normal quantity of ROS is required, but if overdosed, the amount will harm the plants. Hence, plant cells have produced antioxidant defense systems for controlling oxidative damage and terminating reactive oxygen species. This mechanism consists of scavenging free radicals with regular antioxidants (Berwal *et al.*, 2021). The production of ROS is mainly organized by several enzymatic and non-enzymatic antioxidant defense systems, comprising ascorbate peroxidase (APX), catalase (CAT), superoxide dismutase (SOD), proline oxidase (POX), mono-dehydroascorbate reductase, and many others, whereas non-enzymatic catalysts, including glutathione, ascorbate, carotenoids, phenolic compounds, proline, glycine betaine, as well as sugars (Irato and Santovito, 2021).

One of the most important stages in a plant life is germination of seed, which is delicate to the physical and chemical conditions of the rhizosphere (Taiz and Zeiger, 2010). In response to heavy metals such as cadmium, most plant seeds and seedlings display a decline in germination and vigor, while to some extent, the coat of the seed can perform as a principle barrier limiting the destructive effects of heavy metals (Huybrechts *et al.*, 2019). From this point, the impacts of such a heavy metal on the seedling and germination growth stage are an essential research area that needs to be extensively conducted. Recent studies have well documented that through disturbance of cellular osmoregulation, inhibition of food storage mobilization, reduction in radical formation, and the degradation of the activities of proteolytic, cadmium can cause inhibition of the germination of barley seed and it is progressing (Lentini *et al.*, 2018; Demecsová *et al.*, 2020a; Vladimirovich, Grigorievich and Sergeevna, 2021).

## 2.3 Artificial Modification for Drought and Heavy Metals Resistance Genes

Drought is one of the most significant abiotic stress factors limiting the production of yields. To survive in unfavorable environments, plants need to provide rapid responses to drought stress factors. In general, there are some signaling sensors and receptors that could receive a change in environmental status, such as hydroxyprolinerich and cytoskeleton (Mahajan and Tuteja, 2005).

There are mainly two categories of genes that so far have been recognized as contributing to abiotic stress-controlling networks (Shinozaki, Yamaguchi-Shinozaki and Seki, 2003; Bashir *et al.*, 2019). The first group is functional genes, whose expression is changed or initiated by stress-related transcription factors (TFs), and there is direct involvement of these genes in physiological and biological changes required for stress adaptations, whereas the second group is regulatory genes, which include the number of genes encoding TFs that can bind to specific DNA sequences and control the expression of one or more genes by down- or up-regulation.To cope with unfavorable growth conditions in response to environmental stresses, the plant adopted effective strategies by the use of these regulations and significant changes in transcriptome level.

Generating stable expression and inheritance of drought-tolerant plants carrying single or multiple preferred traits in their next generation of offspring is one of the main objectives of genetic engineering. Transcription factors in transgenic plants, if overexpressed under the control of strong promoters, could enhance stress durability by regulating stress-responsive downstream genes. Nevertheless, severe growth delay and a decrease in the final product under normal growth conditions could be noticed as a result of overexpression of stress-related regulatory genes (Gürel *et al.*, 2016). Testing and finding suitable stress-inducible promoters for ideal expression levels of transgenesis is one of the critical techniques to reduce the harmful effects of overexpressed TFs on the growth of the plant and final yield of products. Having more *TaDREB2* and *TaDREB3* in barley transgenic lines made them more drought-resistant by protecting cells from damage and drying out (Khalid *et al.*, 2016; Niu *et al.*, 2020).

Under two abiotic stress conditions (cold and drought), to investigate stress expression in transgenic barley and wheat. Yang *et al.* (2020) isolated two representatives of the wheat transcription factors (*TdHDZipI-4* and *TdHDZipI-3*) and transgenic barley possessed *HDZI-3* promoters to optimize the expression of *TdHDZipI-3* in transgenic barley. They tested the transgenic lines in vegetative and reproductive stages by using two techniques of expression detection (northern blot hybridization as well as quantitative real-time PCR). They found that under strong water stress conditions to increase drought resistance at a vegetative stage, both mentioned promoters were effective, and overexpression of (*TdHDZipI-4*) and (*TdHDZipI-3*) could be used as an improvement for barley drought-sensitive genotypes for it is drought.

Transcriptome analysis on barley plants was carried out by Kintlová *et al.* (2021) to understand the molecular mechanisms of the barley (Morex) genotype in response to cadmium stress at a level of 80 µM. They recognized four diverse copies of the *HvPCR2* gene that are strongly upregulated by the presence of cadmium in the root and closely linked to the metabolism of reactive oxygen species. In eukaryotes, the methylation of RNA, especially the one known as *m6A*, is one of the most dominant mRNA modifications to regulate the expression of the gene in response to abiotic and biotic stress conditions (Zheng *et al.*, 2020). More recently, Su Su *et al.* (2022) in the roots of barley (Golden Promise) two weeks old were subjected to 5 µM cadmium for one week to examine the transcriptional changes using sequencing of *m6A* methylation. The presence of cadmium triggered the expression of more than three hundred hypermethylated genes, and the author stated the strong associations between the *m6A* methylation and ABC transporter. In addition, they found 5 *WRKY* and 4 *MYB* transcription factors that were linked to the expression of *m6A*.

## 2.4 Biochemical Traits Used for Detecting Tolerant Genotypes Under The Presence of Stress Conditions

There are many structures of reactive oxygen species (ROS) in plants; superoxide radicals ($O_2$), hydroxyl radicals (OH), singlet oxygen, and hydrogen peroxide ($H_2O_2$), which cause oxidative damage inside the plant cell. ROS are generated frequently by aerobic metabolism, which is restricted to the different plant cellular parts, like the mitochondria, chloroplast, and peroxisomes. In addition, as a site for ROS generation, other investigations shed light on the role of the apoplast (Qi *et al.*, 2017).

In plants, ROS is constantly being produced under favorable conditions and is unable to cause cellular damage for the reason that different antioxidant mechanisms are involved in scavenging the ROS complex (Torres, 2010). Different types of stress, such as drought and heavy metals, will disrupt the balance between scavenging and generation of ROS. Therefore, the plant's survival will highly depend on many factors, such as the duration and severity of stress, as well as the ability of the plant to adapt quickly to the surrounding changes. The normal amount of ROS is required for plant metabolism, but an excess quantity will harm the plants. Therefore, to limit oxidative damage and terminate active oxygen species, plant cells have produced antioxidant defense systems. This mechanism consists of scavenging free radicals with regular antioxidants. Several enzymatic and non-enzymatic antioxidant defense systems regulate ROS production, including ascorbate peroxidase (APX), catalase (CAT), proline oxidase (POX), superoxide dismutase (SOD), monodehydroascorbate reductase, and many others, while non-enzymatic catalysts contain ascorbate, glutathione, carotenoids, proline, phenolic compounds, sugars, and glycine betaine (Zhang *et al.*, 2015).

In the presence of abiotic stress such as drought and salinity, which usually companies together, a high accumulation of total flavonoid and phenol has been described in two Tibetan wild barleys (Ahmed *et al.*, 2015). Very recently, two varieties of spring barley were tested and exposed to drought stress in seedling stages (after 9 days) by a group of researchers (Kowalczewski *et al.*, 2020), and a significant increase in total phenolic profiling was detected. This was probably because of the role of antioxidants, especially phenolic compounds, which reduce capacities and availability of ROS under the circumstance of having oxidative stress such as drought (Kumar *et al.*, 2018). In addition, phenolic compounds include anthocyanin, which is usually considered to be resistant to drought stress by scavenging the activity of superoxide radicals as well as stabilizing water potential inside the plant cell (Wehner *et al.*, 2015). Catalase activity, an enzyme that is in charge of regulating levels of reactive oxygen species by scavenging hydrogen peroxide, was also measured for barley varieties in seedling stages (Harb, Awad and Samarah, 2015; Rohman *et al.*, 2020), and a high degree of catalase activity has been mentioned. As stated by Tamás *et al.* (2008), considerable changes in the accumulation of proline and the peroxidation of lipid (MDA) were detected after exposing barley plants for one day at the seedling stage to a cadmium concentration of 1 μM. In the presence of Cd stress, the MDA content in barley accessions was increased, representing the oxidative damage (Tamás *et al.*, 2009). Moreover, highland barley cultivars (Kunlun15) were subjected to 75 μM of cadmium and significant changes in the accumulations of proline and the activities of catalase were detected with respect to control conditions at the seedling stage (Wang *et al.*, 2017).

From the above point of view, when a plant is exposed to limited water supply and unfavorable conditions such as heavy metals, a remarkable increase in the activities of one or more antioxidant enzymes could be noted. To protect the cell from damage, these antioxidants could work collaboratively. In the breeding program, it is a primary goal to study many barley genotypes at the same time for a particular purpose, such as drought and heavy metals. Therefore, it is critical to identify genes responsible for such enzymatic activit. Few genotypes were investigated in previous research to identify the expression of genes related to antioxidant responses. Precise primers can be generated for these genes and they can be used in screening a large number of genotypes to improve drought and heavy-metal resistance in barley.

## 2.5 Phenotypic Traits Used for Detecting Tolerant Genotypes Under The Presence of Stress Conditions

The quick and simple approach to evaluate drought stress is morphological observations that can be used alongside genotypes for crop improvement. Several considerations should be taken into account before conducting a drought research program in any studied plant, and a useful protocol should be selected to achieve successful research outcomes as drought is a complicated trait and not easy to diagnose. The evaluations are most likely based on specific stages such as seedlings, the number of grains, or measurements taken during the period of grain filling, when drought is typically observed (Kumar *et al.*, 2008). The first step by the researchers or breeders is the selection of barley genotypes that possess drought resistance and then crossed with other favorable genotypes that have other useful traits, such as products to enhance drought resistance and attempting to cross several resistance genes for drought stress. Conventionally, phenotypic selection for the trait of interest was and still is the priority for plant breeders for the drought research program. The selection will be direct or indirect (related traits) that are easy to diagnose or have a greater chance of heritability. When crosses are made by the researchers, the elite offspring will be selected for drought resistance. However, drought resistance generally has low heritability, so the selection must be examined at least for two years or at different locations (Luo, Xia and Lu, 2019). Another obstacle is that multiple replications are required because the measurements of drought are usually affected by variation in the environment (Luo, Xia and Lu, 2019). In very recent years, a group of researchers conducted field observations for four barley varieties in Arizona to assess the ability of barley to resist drought, in which two of them had great performance for their productivities, and the rest were in the condition of low input productivity (Carter *et al.*, 2019). Remarkable efforts have been made in releasing the two resistant varieties as one of them originated from F6 selection, and their assumption was true when they found that the conventional varieties (low input) had better performance compared to the varieties with high yield outcomes under the condition of water limiting (Carter, Hawes and Ottman, 2019). In the presence of various environments for drought study, which is highly expected because of the differences between location to location and year to year, the performance of phenotypic selection for useful traits is affected by the interaction of genotypes and environments. Therefore, limited progress can be achieved in drought resistance if the level of interaction is high. Plant researchers have combined molecular marker technology into their investigations to overcome the limited heritability of drought resistance. With the advantage of quantitative trait *loci* (QTL) mapping, these researchers have identified genomic regions controlling drought tolerance and, by marker polymorphism, the genetic diversity among the elite genotypes has been discovered (Honsdorf *et al.*, 2014). High and slight genetic effects linked to drought resistance in cereals have been identified by QTL mapping techniques as they can detect many genomic locations related to drought. The great number of genotypes for the absence or presence of these genomic regions for improving drought resistance can be studied. Despite the high cost of this method, the outcome will save effort and time. QTL mapping highly relies on the type and number of the phenotypic assays and DNA markers used to identify the target genomic locations. The chance to spot many QTLs is likely for drought resistance in the case of having more markers DNA as it covers more genomic area. More than two hundred barley genotypes were tested for their responses to water deficit (Moualeu-Ngangue *et al.*, 2020). Most productivity-related traits were analyzed in different growth stages and the analysis of QTL was conducted to discover the suitable marker that was related to the studied traits with the help of genome-wide association studies when the genotypes were applied to water deficit, as 162 significant loci were detected at those stages and 110 *loci* were identified in the case of genotype x treatment interactions (Moualeu-Ngangue *et al.*, 2020).

## 2.6 Stress Responses at Different Growth Stages

The production of important cereal crops such as barley has been affected by climate change. There is a probability of observing the effect of drought at any growth stage in plants. For this reason, before planning a breeding program to improve drought tolerance, it is very critical to have earlier climatic data on the events of drought. The reason is that the drought stress severity is wholly rely on the environment in which the drought happens. Careful consideration should be taken into account for different growth stages when studying genotypes applied to water deficit regimes. For instance, if drought stress happens during the development stages and seedling growth, the development of seedling traits that are related to drought tolerance may be suitable. On the other hand, if the stress of drought only happens in later stages such as grain filling or flowering stage, studying seedling traits parameters will be unusable (Thabet *et al.*, 2018; Thabet *et al.*, 2020). In water-limited environments and field experiments, other limitations hindered breeding for drought resistance genotypes, such as unpredictable weather conditions, the stress of heat, the nutrition of soil, and so on. Therefore, it is highly recommended to establish control conditions such as greenhouses and/or growth chambers to study genetic variation for drought resistance. Weather conditions are highly affected by climate change, so there is a lack of ability to estimate the wheatear circumstances. Accordingly, drought stress may happen in the seedling stage instead of the flowering of grain filing stages. For those reasons, to select the superior genotypes for target traits such as drought resistance, it is quite recommendable to assess the same genotypes under controlled and field experiment conditions.

### 2.6.1 Germination test assays for drought and heavy metals resistance

Seed germination consists of complex events (biochemical and physiological changes) that are initiated with the absorption of water. After the radical appears from the seed coat, the process will end, resulting in the activation of the embryo (Hoyle *et al.*, 2015). Germination and emergence of seedlings in barley are very sensitive to water limitations. Many countries, for instance, Pakistan, Bangladesh, and India, as they are considered to be rain-fed dry areas, sow their plants depending on the chance of precipitation. As they have a phenomenon called the monsoon season, the germination will be reduced if rain does not fall immediately after sowing (Rehman *et al.*, 2015). As a result, Geravandi, Farshadfar and Kahrizi (2010) propose that one of the best approaches to studying the effects of drought stress on germination is to create drought stress conditions using different osmotic potentials..

In the lab, under controlled conditions, usually germination experiments are conducted. Researchers might want to apply a high osmotic potential condition to studied genotypes to find genetic differences and choose the best among them regarding resistance to drought (Aimar *et al.*, 2011). To induce drought stress at the germination stage, it is possible to apply a chemical compound of polyethylene glycol (PEG). This particular method greatly depends on measuring days after treating the seeds with the solution of PEG, depending on the plant species used for the study.

### 2.6.2 The impact of drought and heavy metals on flowering and grain filling stages

Besides the presence of drought or any stressor at the seedling growth stage, two main components of grain yield, including the number of seeds per spike and spike weight, could be negatively affected at the flowering and grain filling stage (Vaezi, Bavei and Shiran, 2010). For improving grain yield in dry environments, researchers more often practice indirect selection and use traits that are strongly associated with grain yield because of the complexity of indicating drought resistant genotypes straight away as many genes are involved in this process (Daryanto, Wang and Jacinthe, 2017). Many traits have been proven in wheat and barley that are well correlated with final yield outcomes, including; plant height, spike length, spike weight, seed number per plant, seed weight per plant, 1000 grain weight, biological yield, biomass production, and harvest index (Kosova *et al.*, 2014; Hasanuzzaman *et al.*, 2017).

Drought resistance performance or any plant stressor could be assessed based on the availability of well-collected data for the mentioned traits. Fernandez (1992) proposed two common ways of identifying drought resistance genotypes under drought and well-watered conditions: First, by calculating the reduction in studied traits due to the present drought for respective genotypes, the following formula:

$$\text{The decline in a trait caused by drought stress} = \frac{\text{Xn} - \text{Xd}}{\text{Xd}} * 100$$

Where (Xn) represents the main performance of genotypes for a particular trait under control conditions while (Xd) represents the main performance of genotypes for the same trait under drought conditions.

Second, the drought susceptibility index (DSI) for each studied genotype can be estimated using the following formula:

$$\mathrm{DSI} = \frac{1 - \frac{\mathrm{Xd}}{\mathrm{Xn}}}{\mathrm{DI}}$$

Were (DI) is the intensity of drought and can be calculated using the following formula:

$$\mathrm{DI} = 1 - \frac{\mathrm{Xd}}{\mathrm{Xn}}$$

Till the moment, these formulas are widely accepted by many researchers in demonstrating the performance of studied plant materials under drought or other stressor conditions (Zare *et al.*, 2021; Sarshad *et al.*, 2021; Mansour *et al.*, 2021). Researchers can conduct their research in a field where irrigation can be controlled under control and drought stress conditions. Carter, Hawes and Ottman (2019) studied the performance of four barley genotypes under controlled irrigation in the field to estimate the responses of the tested materials under well-watered and non-irrigated conditions. They used almost the same mentioned traits to identify resistant genotypes.

In a greenhouse, Romdhane *et al.* (2020) conducted research on two barley varieties at three different growth stages, including the heading stage under water deficit conditions. They found a considerable reduction of 1000 spike weight by half in this critical growth stage, while similar reductions were not documented at the other two stages, indicating the severity of tested varieties to water limitation at the heading stage. During grain filling of barley in the presence of drought, similar responses by barley were stated (Samarah, 2005; Samarah *et al.*, 2009).

For heavy metal contamination at different growth stages, Lobo (2013) conducted research on four barley varieties using a cadmium dose of (10,20, and 40) mg/L. They found a clear reduction in growth at the final dosage of treatment. Similarly, Ayachi *et al.* (2021) stated a clear reduction in the number of seeds per spike and spike weight in the presence of 10 ppm of cadmium in soil on tested barley cultivars at different growth stages. Cadmium solutions were foliar applied to winter wheat (Luohan 17) at concentrations of 10, 20, 30, and 40 mg/L every three days for a month from the flowering to the grain filling stages. This significantly decreased wheat grain yield and shoot biomass. The final cadmium dosage resulted in a 46.9 percent decrease in grain yield.

## 2.7 The Role of Plant Extract Under Natural and Stress Conditions

Barley in Iraq is mainly cultivated under rain-fed conditions where the precipitation is limited. Barley was grown because many farmers thought it was a good substitute for wheat cultivars, especially in the present dry year. Globally, commercial crops like wheat and rice are competing with each other, and the amount of land that can be used for farming is going down because of changes in the climate.

Therefore, an urgent need is required to discover sustainable techniques and different innovations by crop physiologists and researchers for improving the yield and growth parameters of barley. Although barley is considered to be stress resistant (Sabagh *et al.*, 2019; Kebede, Kang and Bekele, 2019), its productivity in severe environmental conditions is negatively affected by several factors such as water limitation, agronomic practices, heat stress, and so forth (Fahad *et al.*, 2017).

One of the most popular plant biostimulants that can be used as a substitute and natural source of mineral nutrition and fertilizer is Moringa Leaf Extract (MLE) gained from moringa (*Moringa oleifera*) as a result of having stimulant compounds in their content, for instance, cytokinins such as zeatin, antioxidants like ascorbic acid, flavonoid, amino acids, vitamins C and A, phenolics, and micro-as well as macronutrients (Nouman, Siddiqui and Basra, 2012). Besides, Yasmeen *et al.* (2013b) showed that the leaf extract of such a plant can also provide the balance among nutrients, phytohormones, and antioxidants. Zeatin is the main hormone detected in MLE and so far its concentration is thousands of folds higher compared to the most studied plants (Anwar *et al.*, 2007). To improve the productivity and growth of many plants grown in normal and stressful conditions, this kind of natural biostimulant has been applied as a foliar application (Jain *et al.*, 2020; Khan *et al.*, 2020).

The pollution of heavy toxic metals is increasing. As economies developed and industrialization advanced, biosorbents were generally improved with chemical treatments to increase sorption capacity (Jain, Malik and Yadav, 2016). In the process of saponification, the chemical reagent of sodium hydroxide has been widely used (Schulerud, 1963). Based on this process to absorb more heavy metal toxic elements, these chemical compounds treated with raw bio sorbent include; moringa leaf (Bello, Adegoke and Akinyunni, 2017), the bark of oak (Lee and Rowell, 2004), algae (Zeraatkar *et al.*, 2016) and other plant materials (Soon *et al.*, 2018) were used to generate carboxylate locations on the surface of bio sorbent that serve as the binding locations for heavy metal ions.

stress conditions.

# CHAPTER THREE
# MATERIALS AND METHODS

## 3.1 Molecular Study

### 3.1.2 Plant material and DNA extraction

A total of 59 barely accessions sourced from almost all research centers in Iraq (Table 3.1.1). These accessions are the most widely cultivated in Iraq. The total genomic DNA was extracted from young leaves (2-week-old seedlings) which were grown in a plastic house from each accession according to the CTAB protocol (Štorchová *et al.*, 2000). The quality of DNA was estimated and examined using a 1.5% agarose gel. Then, DNA samples were relatively diluted to 50 ng/µl using $ddH_2O$. The extracted genomic DNA was saved in freezer (-20 ºC) until used in a polymerase chain reaction.

**Table 3.1.1 Code and name of 59 barley accessions used in this research.**

| Accession Code | Source | Accession Name | Accession | Source | Accession Name |
|---|---|---|---|---|---|
| AC1 | South of Iraq | Shoaa | AC31 | Middle of Iraq | Scio/3 |
| AC2 | South of Iraq | Boraak | AC32 | Middle of Iraq | Victoria |
| AC3 | South of Iraq | Radical | AC33 | Middle of Iraq | Black-Bhoos-B |
| AC4 | South of Iraq | Arivat | AC34 | Middle of Iraq | Irani |
| AC5 | South of Iraq | 16 HB | AC35 | Middle of Iraq | A1 |
| AC6 | South of Iraq | Furat 9 | AC36 | Middle of Iraq | MORA |
| AC7 | South of Iraq | Al-warka | AC37 | Middle of Iraq | ABN |
| AC8 | South of Iraq | Numar | AC38 | Middle of Iraq | Arabi aswad |
| AC9 | South of Iraq | Al-amal | AC39 | Middle of Iraq | Clipper |
| AC10 | South of Iraq | Rafidain-1 | AC40 | Middle of Iraq | Bhoos-H1 |
| AC11 | South of Iraq | Al-khayr | AC41 | Middle of Iraq | BN2R |
| AC12 | South of Iraq | BN6 | AC42 | Middle of Iraq | BA4 |
| AC13 | South of Iraq | IBAA-99 | AC43 | North of Iraq | Qala-1 |
| AC14 | North of Iraq | Saydsadiq | AC44 | North of Iraq | Black-kalar |
| AC15 | Middle of Iraq | Bhoos-244 | AC45 | North of Iraq | White-kalar |
| AC16 | Middle of Iraq | IBAA-265 | AC46 | North of Iraq | Black-Akre |
| AC17 | North of Iraq | White-Akre | AC47 | North of Iraq | Black-Garmiyan |
| AC18 | North of Iraq | Black-Bhoos Akre | AC48 | North of Iraq | Black-Chiman |
| AC19 | North of Iraq | Black-Zaxo | AC49 | North of Iraq | Ukranian-Zarayan |
| AC20 | North of Iraq | White-Zaxo | AC50 | North of Iraq | White-Zarayan |
| AC21 | South of Iraq | Bhoos-912 | AC51 | North of Iraq | Abrash |
| AC22 | North of Iraq | White-Halabja | AC52 | North of Iraq | Bujayl 1-Shaqlawa |
| AC23 | South of Iraq | Samr | AC53 | North of Iraq | Bujayl 2-Shaqlawa |
| AC24 | South of Iraq | GOB | AC54 | North of Iraq | Bujayl 3-Shaqlawa |
| AC25 | South of Iraq | Abiad | AC55 | South of Iraq | Rehaan |
| AC26 | South of Iraq | CANELA | AC56 | South of Iraq | Sameer |
| AC27 | South of Iraq | MSEL | AC57 | South of Iraq | Warka-B12 |
| AC28 | South of Iraq | Acsad strain | AC58 | South of Iraq | Al-Hazzar |
| AC29 | South of Iraq | Acsad-14 | AC59 | South of Iraq | IBAA-995 |
| AC30 | South of Iraq | Gk-Omega | | | |

### 3.1.3 ISSR, CDDP, and SCoT assay

In the present investigation, forty-four ISSR primers were tested (Rasul *et al.,* 2022). The universal primers (UBC) designed at Biotechnology Laboratory, University of British Columbia, Canada and used by most researchers (Table 3.1.2). Nine specific primers for CDDP and twelve SCoT primers were also tested in our study (Rasul *et al.,* 2022) (Table 3.1.3). PCR reactions were performed at an overall quantity of 25 µL. The reaction mixture contained 4 µL of genomic DNA from each sample, 10 µL master mix (GoTaq® Green Master Mix, Promega, USA), 2 µL of forward and reverse primers, and the final volume of 25µl completed with de-ionized water ($dH_2O$). The PCR was performed following this protocol: denaturation (94ºC for 10 min), followed by 35 cycles of 1 min denaturing step at 94ºC, 1 min annealing temperatures, which ranged between (42ºC- 60ºC) and 2 min extension at 72ºC. Finally, post-extension was set up at 72ºC for 7 min. The amplification reaction products were detected and separated by 1.5% agarose gels (1xTBE buffer), stained with ethidium bromide and visualized under UV light.

### 3.1.4 Scoring and statistical data analysis

After amplification of the fragments, the scorable bands were manually coded by recording 0 and 1 for the absence and presence of bands, respectively. To calculate the similarity coefficient of Jaccard, the scored data matrices were subjected to statistical analysis using the XLSTAT 2016 computer software. To perform cluster analysis between accessions, the Jaccard coefficient was converted into a dissimilarity matrix using the unweighted pair-group technique with arithmetic averages (UPGMA). The binary data (0 and 1) was converted to A and T to create the dendrogram tree using CLC Sequence Viewer version 8. Polymorphism information content (PIC) allele frequency and gene diversity were calculated using Power Marker version 3.25 software. A model analysis was performed using the software STRUCTURE, version 2.3.4 (Pritchard, Stephens and Donnelly, 2000). The numbers of supposed populations (K) were set from one to ten, and the analysis was repeated three times. The burn-in and MCMC were fixed at 50,000 each for each category, and iterations were set at 5000. The run with the maximum likelihood was engaged to set accessions into populations.

**Table 3.1.2 Sequences and annealing temperature for the ISSR markers used in this investigation.**

| ISSR Markers | Sequences (5–3) | Anealing temp(Tm °C) | ISSR Markers | Sequences (5–3) | Anealing temp(Tm °C) |
|---|---|---|---|---|---|
| B-13 | CAACAACAACAACAA | 44.70 | UBC-810 | GAGAGAGAGAGAGAGAT | 50.00 |
| HB-10 | GAGAGAGAGAGACC | 50.00 | UBC-811 | GAGAGAGAGAGAGAGAC | 50.00 |
| HB-11 | GTGTGTGTGTGTCC | 50.00 | UBC-812 | GAGAGAGAGAGAGAGAA | 50.40 |
| HB-12 | CACCACCACGC | 48.50 | UBC-813 | CTCTCTCTCTCTCTCTT | 50.00 |
| HB-15 | GTGGTGGTGGC | 50.00 | UBC-814 | CTCTCTCTCTCTCTCTA | 50.00 |
| ISCS20 | DHBCGACGACGACGACGA | 60.00 | UBC-815 | CTCTCTCTCTCTCTCTG | 50.00 |
| ISCS21 | BDBACAACAACAACAACA | 55.00 | UBC-818 | CACACACACACACACAG | 52.80 |
| ISSR-1 | AGACACACACACACACAT | 50.00 | UBC-822 | TCTCTCTCTCTCTCTCA | 50.00 |
| ISSR-6 | GCCTCCTCCTCCTCCTCC | 50.00 | UBC-823 | TCTCTCTCTCTCTCTCC | 50.00 |
| ISSR-7 | AGATCCTCCTCCTCCTCC | 50.00 | UBC-825 | ACACAC ACACACACACT | 50.00 |
| ISSR-8 | ATCACACACACACACACA | 50.00 | UBC-826 | ACACACACACACACACC | 50.00 |
| ISSR-9 | CACACACACACACACATG | 50.00 | UBC-834 | AGAGAGAGAGAGAGAGGT | 50.00 |
| ISSR-10 | GAGAGAGAGAGAGAGAGG | 50.00 | UBC-841 | GAGAGAGAGAGAGAGACTC | 50.00 |
| ISSR-12 | AGAGAGAGAGAGAGAGCT | 50.00 | UBC-845 | CTCTCTCTCTCTCTCTTG | 50.00 |
| ISSR-14 | GAGAGAGAGAGAGAGAGC | 50.00 | UBC-846 | CACACACACACACACAAT | 50.00 |
| ISSR-16 | ACACACACACACACACGA | 50.00 | UBC-847 | CACACACACACACACAGC | 50.00 |
| ISSR-18 | TGTGTGTGTGTGTGTGG | 50.00 | UBC-849 | GTGTGTGTGTGTGTGTGA | 50.00 |
| ISSR-AGC6G | AGCAGCAGCAGCAGCAGCG | 55.00 | UBC-852 | GATAGATAGACAGACA | 48.00 |
| ISSR-CA | BDBCACACACACACA | 50.00 | UBC-856 | ACACACACACACACACYA | 42.00 |
| ISSR-CCA | DDBCCACCACCACCACCA | 55.00 | UBC-881 | GGGTGGGGTGGGGTG | 54.00 |
| PCP-3 | GTGCGTGCGTGCGTGC | 60.00 | UBC-888 | CGTCGTCGTCACACACACA CACA | 52.00 |
| UBC-808 | AGAGAG AGAGAGAGAGC | 50.00 | UBC-891 | ACTACTACTTGTGTGTGTGT GTG | 52.00 |

**Table 3.1.3 Sequences and annealing temperature for the SCoT and CDDP markers used in this investigation.**

| SCoT Markers | Sequences (5–3) | Annealing temp(Tm °C) | CDDP Markers | Sequence of primer (5–3) | Annealing temp (Tm °C) |
|---|---|---|---|---|---|
| SCoT2 | CAACAATGGCTACCACCC | 50.7 | ABP1-1 | ACSCCSATCCACCGC | 50 |
| SCoT3 | CAACAATGGCTACCACCG | 51.3 | ERF1 | CACTACCCCGGSCTSCG | 50 |
| SCoT6 | CAACAATGGCTACCACGC | 52.1 | ERF2 | GCSGAGATCCGSGACCC | 50 |
| SCoT7 | CAACAATGGCTACCACGG | 51.3 | Knox1 | AAGGGSAAGCTSCCSAAG | 50 |
| SCoT12 | ACGACATGGCGACCAACG | 55.9 | MADS-1 | ATGGGCCGSGGCAAGGTGC | 50 |
| SCoT13 | ACGACATGGCGACCATCG | 55.4 | Myb1 | GGCAAGGGCTGCCGC | 50 |
| SCoT16 | ACCATGGCTACCACCGAC | 54.1 | Myb2 | GGCAAGGGCTGCCGG | 50 |
| SCoT22 | AACCATGGCTACCACCAC | 51.9 | WRKYF1 | TGGCGSAAGTACGGCCAG | 50 |
| SCoT23 | CACCATGGCTACCACCAG | 52.4 | WRKY-R3 | GCASGTGTGCTCGCC | 50 |
| SCoT29 | CCATGGCTACCACCGGCC | 57.9 | | | |
| SCoT32 | CCATGGCTACCACCGCAC | 55.9 | | | |
| SCoT36 | GCAACAATGGCTACCACC | 51.5 | | | |

## 3.2 Drought Study or Experiment

### 3.2.1 Drought experiments under the seedling stage

#### 3.2.1.1 Test of germination and phenotypic traits

To determine the tolerance to water stress during germination, the grains were soaked in a 1% sodium hypochlorite solution for 11 min and then washed five times with purified water. The Petri dishes and the Whatman paper were all sterilized in autoclave. Twenty seeds of each accession with three replications were transferred to each disposable petri dish with a diameter of 9 cm, in which two filter papers were placed . Nine mL of distilled water was applied to each Petri dish as a control treatment. Subsequently, Nine ml of the solution was added to the Petri plates for each treatment with concentrations of (10.25 and 20.5% PEG). Then, the petri dishs were sealed with parafilm tape. This investigation was done in an incubator (Daihan LabTech Co., Ltd, Korea) with temperature maintained at about 20 ± 0.2°C. The grains with root lengths of two mm or more are measured as the germinated seeds. After 8 days of incubation, the seedlings were taken out to assess the morphological parameters, including germination percentage, root, and shoot length. This equation was used to calculate germination percentages:

$$\text{Germination (\%)} = \frac{\text{Number of germinated seeds}}{\text{Total number of seeds used}} \times 100$$

#### 3.2.1.2 Physiological tests

***Seed water content***

The initial weight of each individual seed was registered prior to start of the test to assess the quantity of water consumed by the seeds. At 40 and 64 h after sowing, the seeds (samples of 4 seeds with 3 replications) from treated samples, with or without PEG solution, were taken from each accession and weighted. This equation calculated the seed water uptake (WU):

$$\text{WU (\%)} = \frac{(\text{SWAS} - \text{SWBS})}{\text{SWBS}} \times 100$$

Where WU is the percentage of seed water uptake, SWAS is the seed weight after soaking with PEG or water (g), and SWBS is the seed mass before soaking (initial mass) (g). (El-Hendawy et al., 2019).

***Relative leaf water content and electrolyte leakage***

Seeds of 59 accessions were sown in the tray boxes (22 x 16 x 5 cm3, length x width x height) containing loamy soil in a greenhouse with the temperature sustained at 20 ± 0.5 °C. The field potential of the soil was 31%. After 18 days of sowing, the seedlings are exposed to water deficit until the field capacity reached 10%. At this point, three leaves from randomly chosen plants (12 plants) in each tray were taken. The top and bottom of all the leaves were cut off to obtain a 4.5 cm mild section. The samples were immediately placed into the pre-weighed tubes and sealed with the lid and stored in a cooled container (10-15 °C). All sample tubes were weighted (FW) and 2.5 mL of distilled water was applied to each sample tube. For the leaves to reach their full turgor, the sample tubes were stored in the dark at 4 °C in a refrigerator for 18 h. The leaf samples were taken out of the tube and gently dried with a paper towel. The leaf samples were weighted (TW) and placed in an oven at 69 °C for 87 h for drying. After drying, the leaf samples were weighted (DW) and the relative leaf water content (RWC) was estimated by the following equation (Pieczynski *et al.*, 2013).

$$\text{RWC (\%)} = \frac{(\text{FWS} - \text{D}\quad)}{(\text{TWS} - \text{DWS})} \times 100$$

Where FWS is the fresh weight of the sample (g), DWS is the dry weight of the sample (g), and TWS is the turgid weight of the sample (g). To measure the membrane permeability of the leaves, the electrolyte leakage was calculated using the process mentioned by (Dionisio-Sese and Tobita, 1998). Nine 4.5 cm mid-section parts of leaves with three replicates were put in test tubes comprising 9 mL of distilled deionized water. The sample tubes were held for 150 min at a temperature of 36 °C and the initial electrical conductivity of the medium (EC1) was tested with an electrical conductivity analyzer (IrDA Port 8303, China). The samples were boiled for 27 min at 98 °C to release the electrolytes, cooled to 23 °C and then the final electrical conductivity (EC2) was calculated. Leaf leakage was determined using the formula:

$$\text{EL (\%)} = \frac{\text{EC1}}{\text{EC2}} \times 100$$

Where EL is the electrolyte leakage, EC1 is the initial electrical conductivity, and EC2 is the final electrical conductivity.

### 3.2.1.3 Biochemical tests

***Extract preparation***

One gram of tissues was crushed in a mortar with a pestle using liquid nitrogen. The fresh powder (0.1 g) was extracted with 0.7 mL of 60% (v/v) acidic methanol (methanol + HCl with a ratio of 99:1) and incubated at 10 °C for 16 h. After extraction, the chlorophyll was removed with chloroform (0.7 mL) and centrifuged at 14000 rpm at 4 °C for 19 min. The upper layer solution was used to evaluate the TPC, TFC, and antioxidant capacity assays.

***Total phenolic content***

The content of total phenolic compounds in each extract was determined according to Djeridane *et al.* (2006) and Tahir *et al.* (2022) using the Folin-Ciocalteu method. An aliquot of 25 µL of each extract (Sample) or deionized water (Blank) was mixed with 2 mL of 1:9 Folin-Ciocalteu reagent: water (v/v) and allowed to react for 6 min. Then, 1.6 mL of 10% saturated $Na_2CO_3$ solution was added and allowed to stand for 53 min. in the dark at 38 °C. Then, the absorbance of the reaction mixture was read at 750 nm against the blank using a UV-visible spectrophotometer (UV-365, SHIMADZU, Japan). The standard gallic acid solution was prepared by dissolving 9 mg of it in 9 mL of methanol to attain a final concentration of 1 mg/mL. A sequence of serial dilutions of gallic acid (0, 50, 100, 150, 200, 250, 300 µg/mL) were used to obtain a standard curve and a linear association between the absorbance values at 750 nm and the gallic acid content was observed (Appendix 5.1). The total polyphenol content (TPC) of each extract was demonstrated as the equivalent of µg gallic acid (GAE) per gram of fresh matter by the following formula:

TPC (µg GAE/gm FM) = $\frac{V}{W}$ x C

Where V is the volume of extract (mL), W is the fresh weight of the sample (g), and C is the concentration of gallic acid collected from the standard curve. Each value reflects the mean of three measurements.

***Total flavonoid content***

Total flavonoid content in each extract was determined according to Rigane *et al.* (2017). A stock solution of the standard compound was prepared by dissolving 8 mg quercetin into 8 mL of methanol to achieve a final concentration of 1 mg/mL. A serieal dilution (0, 2.5, 5, 10, 20, 40, 80 µg/mL) of quercetin was prepared. A quercetin standard curve was created and a linear regression was found between the absorption values at 415 nm and the quercetin concentration (Appenix 5.1). Briefly, the extract solution or deionized water (40 µL) was mixed with 0.90 mL of methanol (80%), 0.30 mL of 2% aluminum chloride, 0.08 mL of 1 M potassium acetate, and 1.72 mL of deionized water. After incubation at room temperature under dark condition for 32 min, the absorbance of the solution was read at 415 nm with a UV-visible spectrophotometer. (UV-365, SHIMADZU, Japan). The total flavonoid content of each extract was expressed as the equivalent of µg quercetin (QE) per gram of fresh matter by the formula: TFC (µg QE/gm FM) = $\frac{V}{W}$ x C

Where V is the volume of extract (mL), W is the fresh weight of the sample (g), and C is the concentration of quercetin determined from the standard curve. Each value is the average of three measurements.

***Radical scavenging activity by DPPH***

The antioxidant potential of the extracts was determined according to the method of 1-diphenyl-2-picrylhydrazyl (DPPH) radical-scavenging as defined by Shimada *et al.* (1992) and Tahir *et al.* (2019). The standard compounds were: The 6-hydroxy-2,5,7,8-tetramethylchroman-2-carboxylic acid (Trolox) was used to construct the calibration curve by using a series of dilution solutions. Trolox of 11 mg was mixed with 250 ml of 75% ethanol solvent. The stock was subjected to serial dilutions to obtain 0.325, 0.650, 1.300, 1.950, 2.600, and 3.25 μg/mL. A linear regression between the absorbance values at 517 nm and the various concentrations of Trolox was identified (Appendix 5.1). Three groups of tubes were prepared. Group 1 included the mixing of 2 mL of DPPH solution (6 × 10-5 M) with a 10 μL sample. Group 2 is composed of a mixture of 10 μL methanol and 2 mL of DPPH solution. Group 3 consisted of the different concentrations of Trolox with 2 mL of DPPH solution. All tubes (sample extract, control, and Trolox) were incubated in the dark at 22 °C for 35 min and absorbed against a blank containing only methanol at a 517 nm UV-visible spectrophotometer (UV-365, SHIMADZU, Japan). All samples were measured with three replicates. The antioxidant capacity of different extracts was expressed as the equivalent of Trolox /g fresh matter (DM) by the formula below:

Antioxidant capacity by DPPH (μg Trolox/gm FM) = $\frac{V}{W} \times C$

Where V is the volume of extract (mL), W is the fresh weight of the sample (g), and C is the concentration of Trolox determined from the standard curve.

***Soluble sugar content***

Soluble sugar content was measured following the method described by Yemm and Willis (1954) and Zheng *et al.* (2008). A stock solution of standard compound (glucose, company name, country) was prepared by adding 10 mL of deionized water to 10 mg of glucose to get a final concentration of 1 mg/mL. A serial of dilutions of glucose (0, 4, 10, 20, 30, 50, 80, 160, 320, 640 μg) were prepared. Linear regression was observed between the absorbance values at 620 nm and the glucose concentrations (Appendix 5.1). Ground seedlings (0.1 g) were soaked in 800 μL deionized water. The solution mixture was boiled at 100 °C for 30 min. Then, it was cooled and centrifuged for 15 min at 4000 rpm. This residue was re-extracted two more times with deionized water. Three hundred μL of supernatant or standard compound (glucose) were mixed with 2.7 mL of anthrone reagent (0.15 g anthrone in 84 mL sulphuric acid and 16 mL deionized water).The mixture was boiled for 5 minutes. After cooling, the absorbance of the mixture was recorded at 620 nm. The content of soluble sugar was calculated from the standard curve of glucose at 620 nm using a UV-visible spectrophotometer (UV-365, SHIMADZU, Japan). The formula for calculating soluble sugar content was:

$$\text{SSC } (\mu g/gm \text{ FM}) = \frac{V}{W} \times C$$

Where V is the volume of extract (mL), W is the fresh weight of the sample (g), and C is the concentration of glucose determined from the standard curve.

***Proline content***

Proline content in seedling samples is defined following the method of Bates, Waldren and Teare (1973). Fresh material powder of 0.1g was homogenized in 3 mL 3% (w/v) sulphosalicylic acid and centrifuged at 4000 rpm for 28 min at 5 °C. 2 mL of supernatant was mixed with 2 mL of acid ninhydrin reagent (2.5 g ninhydrin in 60 mL of glacial acetic acid and 40 mL of 6 M phosphoric acid) and 2 mL of glacial acetic acid. The samples were subsequently incubated at 100 °C for 60 min. The sample materials were cooled and mixed with 4 mL of toluene. The toluene layer was read at 520 nm against the blank containing the toluene with a UV-visible spectrophotometer (UV-365, SHIMADZU, Japan). The standard curve of proline (1 mg/mL) was prepared, taking different concentrations of L-proline. The L-proline standard solutions (0.0, 50, 100, 150, 200, 250, 300, 350, 500, 700 µg) were applied to the stopper tubes and were all diluted with water up to 1 mL. A linear regression between the absorbance values at 520 nm and the L-proline content was detected (Appendix 5.1). The proline content of plant material has been determined from this typical curve. Values are the results of three replicates and are represented as µg/g of fresh tissue. The formula for the determination of the proline content was:

$$\text{PC } (\mu g/gm \text{ FM}) = \frac{V}{W} \times C$$

Where V is the volume of extract (mL), W is the fresh weight of the sample (g), and C is the concentration of proline determined from the standard curve.

***Guaiacol peroxidase activity***

Guaiacol peroxidase was tested using the formula mentioned in Zheng *et al.* (2008). The fine powdered tissue of seedlings (0.1 g) was combined with 1 mL of phosphate buffer (pH 7.0) and the mixture was separated at 4000 rpm at 4 °C for 35 min. The supernatant was taken for analysis. The reaction mixture contained 25 µL of 20 mM guaiacol, 2.4 mL of 10 mM phosphate buffer (pH 7.0), and 50 µL of extract. The reaction was started with the addition of 10 µL of 40 mM $H_2O_2$ and the absorbance was recorded five times at 470 with a 1 min interval using a UV-visible spectrophotometer (UV-365, SHIMADZU, Japan). The formula calculated the enzyme-specific activity:

$$\text{GPA (units/min/g FW)} = \left(\frac{35.86}{\Delta t}\right) \text{x} \left(\frac{1}{1000}\right) \text{x} \left(\frac{TV}{VU}\right) \text{x} \left(\frac{1}{FWT}\right)$$

Where extinction coefficient = 35.86 mM-1cm-1; Δt = time change in minute; TV = total volume of the extract (mL); VU = volume used (mL); FWT = weight of the fresh tissue (g)

***Catalase activity***

The catalase activity was determined using the formula described by Aebi (1984). Powder tissue of seedlings (0.1 g), which was homogenized in 2.5 mL of solution (50 mM Tris pH 8.0, 0.5% (v/v) Triton X-100, 2% (w/v) PVP, and 0.5 mM EDTA). The homogenate was centrifuged for 40 min at 4 °C at 4000 rpm, and the resultant supernatant was dialyzed prior to the enzyme assay. The reaction mixture containing 1.5 mL phosphate buffer (50 mM of phosphate buffer, pH 7) and 300 µL of enzyme extract was initiated by adding 1.2 mL 60 of mM of hydrogen peroxide. A spectrophotometer measured the absorbance at an observed rate of 240 nm for 3 min with a 1 min interval using a UV-visible spectrophotometer (UV-365, SHIMADZU, Japan). The formula calculated the enzyme-specific activity: CAT (units/min/g FW) =

$$\left(\frac{\text{Change in Absorbance/ min x Total volume of extraction (mL)}}{\text{Extinction coefficient x volume of sample taken (mL)}}\right) \text{x} \left(\frac{1}{FWT}\right)$$

Where extinction coefficient = 6.93 x 10-3 mM-1cm-1 and FWT = weight of the fresh tissue (g)

### 3.2.1.4 Statistical data analysis

One-way ANOVA-CRD and Duncan's new multiple range tests were used to assess significant variations ($p \leq 0.01$ and $p \leq 0.001$) among barley accessions using XLSTAT software version 2020 (Addinsoft, New York, USA). The box chart and principal component analysis plot were created using XLSTAT software. Correlation and key importance analysis were achieved by Q Research Software (Market Research Software, Australia). In addition, the ranking method was used to determine the best accessions as per the method suggested by Ketata, Yau and Nachit (1989), utilizing different calculated characters. The stress tolerance index (STI) and the average number of ranks (ASRs) were introduced to be used as the criterion for choosing the best accessions for all traits. In this assessment, each trait with greatest performance recorded the lowest rank; thus, the best accessions were known, with the maximum STI and lowest ASR values (Pour-Aboughadareh *et al.*, 2019; Rahim *et al.*, 2020).

### 3.2.2 Drought experiments under plastic house condition

#### 3.2.2.1 The layout of the experiment

The drought experiment was performed from November 2020 to July 2021. Based on a preliminary test conducted on 59 barley accessions at the seedling stage, four barley accessions (AC36, AC37, AC47, and AC53) were grown in a greenhouse at the College of Agricultural Engineering Sciences at the University of Sulaimani. To avoid the potential effect of nutrient availability on the treatments, no supplementary nutrients were given to the tested materials. Initially, five seeds were sown per pot, and after the first leaf appeared, plants were thinned to retain two seedlings per pot. The layout of the experiment was arranged according to a randomized complete block design, comprising eight treatments, including the control condition. Each barley accession had 10 pots, but only 2 representative pots with uniform growth were selected for analysis at different growth stages. The drought treatment was given to the plants by withholding water until the pots reached the desired moisture content (from 20% to 16%) at three critical growth stages: tillering (S1), flowering (S2), and anthesis (S3), as well as a combination of stress conditions (S1+S2), (S1+S3), (S2+S3), and (S1+S2+S3). The moisture of pots was regularly and carefully monitored by using specific moisture indicators (ECOWITT WH0291) to see whether they reached the point of drought condition.

#### 3.2.2.2 Morphological parameters

The parameters which have been confirmed to be significantly associated with drought resistance in barley were used, including; root length (cm), shoot length (cm), root and shoot fresh weights (g), root and shoot dry weight (g), relative water content %, Total Chlorophyll Content (spad unit), spike number per plant, spike length (cm), spike weight per plant (g), number of grains per spike, grain weight per spike (g), and total grain yield per plant (g).

#### 3.2.2.3 Biochemical parameters

The leaf organs from four barley accessions (AC36, AC37, AC47, and AC53) during diverse growth stages in a plastic house were collected for analysis. Similar protocols as displayed in section (3.2.1.3) performed to study the responses of barley accessions to biochemical traits. The only difference was the addition of a new biochemical marker (lipid peroxidation), which is discussed further below.

***Lipid peroxidation assays (POL)***

As a biomarker of membrane oxidative damage caused by the diverse concentrations of cadmium used in this research, the concentration of malondialdehyde (MDA), which is the final product of lipid peroxidation, was measured with little modifications, which was displayed by Juknys *et al.* (2012). This experiment was initiated by mixing a relatively larger amount of ground powder tissue (0.4) g compared to the previous experiment with a 2 mL Tris-HCl buffer solution comprising 1.5% of PVPP (pH 7.4). Then the mixture was shaken well for 10 minutes. Afterward, the solution mixture was centrifuged at 10,000 rpm for a half-hour. All the upper layers were then taken and transferred to the glass tube. Following that, 2 mL of 0.5% TBA in 20% trichloroacetic acid (TCA) (w/v) was mixed with the supernatant and boiled for 31 minutes at 95 °C in a water bath. After the heating, the samples were immediately placed in the cold water bath to stop the reactions, and a pinkish colour appeared among the samples. The reaction mixture, after centrifugation at 4000 rpm for 12 min, was measured at two different wavelengths (532 and 600) nm. The first measurement is the true measurement of the sample, while the second is for correcting unclear turbidity by subtracting the value of absorbance at 600 nm. The concentration of lipid peroxidation was stated in nmol $g^{-1}$ fresh weight using an extinction coefficient of 155 mM-1$cm^{-1}$ (Buege and Aust, 1978).

## 3.3 Heavy Metal Section

### 3.3.1 Cadmium experiments under the seedling stage

#### 3.3.1.1 Germination test and phenotypic characters

The seeds of the tested barley accessions were soaked in sodium hypochlorite solution for 11 min at a concentration of 1% and then washed five times with distilled water. Before conducting this experiment, all materials, including Whatman paper and Petri dishes, were sterilized in autoclave. Twenty seeds of each accession with three replications were transferred to each disposable Petri dish with a diameter of 9 cm in which two filter papers were embedded. As a control treatment, nine mL of distilled water was applied to each Petri dish. Then, three different cadmium chloride hemi-pentahydrate concentrations were prepared, namely (125, 250, and 500) µM. Subsequently, similar volumes of control from the prepared solution were added to Petri dishes for each separate treatment. This examination was carried out in an incubator with the temperature preserved at about 20 ± 0.2 °C for 8 days. Germinated seeds were measured and taken into account when the root length of seeds reached 2 mm or more. To evaluate the morphological parameters, including germination percentage, shoot, and root length, seeds were taken out after 8 days of placing the seeds on the Petri dishes ( Lateef *et al.,* 2021). This equation is considered for germination percentages:

$$\text{Germination (\%)} = \frac{\text{Number of germinated seeds}}{\text{Total number of seeds used}} \times 100$$

### 3.3.1.2 Biochemical Tests

When all 59 barley accessions were exposed to three regimes of cadmium exposure at seedling stages with concentrations of (125, 250, and 500) µM, similar protocols as shown in section 3.2.2.3 were performed.

### 3.3.2 Cadmium experiments under plastic house conditions

Two successive independent experiments of cadmium stress were performed from November 2020 to July 2021.

### 3.3.2.1 Barley accessions under the presence of cadmium

***Layout of experiment***

Regarding this part of the experiment, based on the initial seedling test on 59 barley accessions in response to cadmium with different concentrations, six barley accessions were exposed to heavy metal stress conditions AC29, AC37, AC38, AC47, AC48, and AC52. Similar to the drought investigation, the experiment was arranged according to a randomized complete block design, comprising four treatments, including the control condition. Initially, five seeds were sown per pot, and after the first leaf appeared, the plants were thinned out to maintain two plants per pot. Each barley accession had 9 pots, but only 3 representative pots with uniform growth were subjected to analysis at different growth stages. Cadmium concentrations of 500 µM were first applied at tillering stage (S1) three times, while when the plant reached further growth stage (S2) (flowering stage), again a triple cadmium dose was applied. In addition to all individual cadmium stress conditions, the tested materials with the same cadmium concentration were exposed to stress from the tillering stage till the end of flowering stage (S1+S2).

***Morphological parameters***

The parameters which have been confirmed to be significantly affected by the presence of cadmium heavy metals were investigated, including; plant height (cm), tiller number per plant, spike number per plant, spike length (cm), spike weight (g), number of grains per spike, grain weight per spike (g), total grain yield per plant (g), straw-weight per plant (g).

***Cadmium determination in seeds***

Harvested seeds from AC29, AC37, AC38, AC47, AC48, and AC52 grown in the presence of cadmium 500 µM at various growth stages (S1, S2, and S1+S2) were digested with a 5:1 concentrated HNO3: HClO4 solution, as stated by Chamon *et al.* (2005). To find out if the digested samples were contaminated with cadmium, the concentrations of Cd were measured four times with flame atomic absorption spectrometry (AAS).

**3.3.2.2 Seedling test in plastic house under the presence of cadmium and plant residues**

***The layout of the experiment and preparation of plant residues***

Seeds of four barley accessions (AC29, AC37, AC48, and AC52) were sown in pots in a plastic house depending on the results obtained from the test on 59 barley accessions in response to heavy metals with different concentrations. All treatments were arranged in a randomized complete block design. Each accession underwent four treatments: a negative control (only water was present), a positive control (cadmium + water), an oakleaf residues (*Quercus aegilops*) + cadmium treatment, and a gundelia plant parts residue (*Gundelia tournefortii*) + cadmium treatment. Each barley accession was organized into 12 plots (1 barley accession × 4 treatments × 3 replicates). In the beginning, 10 seeds were sown in each pot, and then, after the first leaf emerged, the plant was thinned-out to preserve 5 plants per pot. Four doses of cadmium with a concentration of 500 µM started to be applied from the first emergence of the leaf and continued till the appearance of five leaves. The plant residues were prepared according to the procedure provided by Bello, Adegoke and Akinyunni (2017) with a few modifications. Initially, the plant samples were ground, and then 300 g of each plant was dissolved with four liters of distilled water in the presence of 20 g of NaOH. The suspensions were shaken for 24 h and then filtered through fine mesh to remove the water and obtain the plant residues. After this process with deionized distilled water, the plant residues were repeatedly washed until the pH decreased to a near-neutral condition (7.0). Then, the residues dried at room temperature. The pre-treated 10 gm of dried plant extracts of oak and guandela were added separately to each representative pot at a depth of 6 cm in the soil profile. At the end of cadmium stress condition, all physiological and biomass measurements were documented, and the leaf samples were immediately frozen in liquid nitrogen and kept in the freezer for perceiving biochemical responses in the presence of cadmium stress and plant extract by tested barley accessions.

***Physiological parameters***

The plant physiological parameters which have been confirmed to be significantly affected by the presence of cadmium heavy metals were investigated, including; RL: Root length (cm), SL: Shoot length (cm), NL: Number of leaves, RFWP: Root fresh weight per plant (g), RDWP: Root dry weight per plant (g), SFWP: Shoot fresh weight of per plant (g), SDWP: Shoot dry weight per plant (g), TCC: Total chlorophyll content (spad units), and RWC relative water content %.

***Biochemical tests***

The same protocols as presented in section 3.2.2.3, the leaf organ of four barley accessions AC29, AC37, AC48, and AC52 under present and absence of plant residues were analyzed.

***Cadmium determination in roots and shoots***

Similar protocol of digestion to those presented in (3.3.2.1) for detecting cadmium in seeds were performed at seedling stages on the dry root and shoot of barley accessions AC29, AC37, AC48, and AC52.

After determining the cadmium concentration in roots and shoots, the percentage reduction of cadmium in roots and shoots due to the presence of cadmium and plant residues for respective accessions, the following formula was used.

$$\text{The percentage decline in cadimum accumilation } = \frac{\text{Xn} - \text{Xd}}{\text{Xn}} \times 100$$

Where (Xn) represents the main performance of barley accession under cadmium stress condition while (Xd) represents the main performance of same accession under cadmium and plant residues. The association between accumulated concentration of cadmium in the aerial part (shoots) and in the root of barley accessions was documented using a translocation factor (Mattina *et al.*, 2003), by dividing the obtained concentration of cadmium in the shoots by the concentration in the root of the treated barley accessions.

**3.4 Field Section**

**3.4.1 Experimental layout**

In this study, 59 barley accessions were grown in the field under rain-fed conditions at the Research Station of the College of Agricultural Engineering Sciences-University of Sulaimani (35° 34' 17.5"N 45° 22'01.0"E) in the 2019-2020 growing season. The layout of the experiment was arranged according to a randomized complete block design (RCBD) with two main blocks (3 replicates for each block). At the beginning of November, seeds of the tested barley accessions were sown into two-meter rows at a density of fifty seeds per meter, and the row interval of 30 cm. After the plant reached a reasonable growth stage at the beginning, each replicate was thinned to 20 plants. Standard agricultural practices were performed in the field, including weed control (by hand).

### 3.4.2 Preparation and application of moringa plant extract (MPE)

From young full-grown trees, Moringa plant parts (leaves, seeds, and roots) were harvested, which were planted in the nursery of the Faculty of Science, University of Sulaimani, and kindly provided by Dr. Jamal Saeed Rashid. For preparing MPE, the collected sample of moringa parts was dried in shade and then ground to a fine powder using the blender. After grinding, the extract was prepared by mixing 20 g of each part and macerated using 1 liter of distilled water. The mixture was shaken for 24 hours. Afterward, the mixture was centrifuged at 8000 x g for 30 min. To prevent any residue, the supernatant was then filtered using Whatman filter paper. To reach the required concentrations for foliar spray, the supernatant was diluted and mixed with distilled water (v:v) by adding 1 mL of supernatant to 30 mL of water. Henceforth, the solution is diluted, making MPE (1:30) treatment. To achieve optimal penetration into leaf tissues and prevent evaporation, foliar sprays were applied half an hour before sunset. At critical stages, foliar treatments were applied and repeated three times during the growth period, namely (before and after flowering time as well as the filling stage) at ten-day intervals. Control plants (compromising three replicates) were water sprayed at the same time as MPE treatments while the other treated blocks, which included the other three replicates treated with the MPE (Tahir *et al.,* 2022).

### 3.4.3 Plant measurements

From the last Moringa plant extract foliar application after three weeks when plant reached harvesting point, the biomass yield and growth parameters were measured, while the chlorophyll measurement using CCM-200 plus (spad unit) and (LA) leaf area $cm^2$ were recorded according to (Sestak, Catský and Jarvis, 1971) , after the second foliar application . Barley plants in each treated and untreated replicate were harvested at the end of the growing season, and the parameters of (PH) plant height (cm), (TNP) number of the tiller, (AL) awn length spike length (cm), (SL) spike length (cm), (SNP) spike number/plant, (1000 KW) grain weight (g), (SWS) grain weight/spike (g), (SNS) grain number/spike, (SW) spike weight (g), (TY) total yield (g), and (ST) straw weight (g) were recorded (Tahir *et al.,* 2022).

**Summary of the experiments conducted in this study**

**1-Molecular Experimnets**

Three marker system were tested on 59 barley accessions using (44 ISSR, 9 CDDP and 12 SCoT) primers.

**2-Seedling Experimnets**

**Regarding drought**

(Two concentration of PEG 6000: 10.25% and 20.50%) tested on all collected barley accessions.

**Regarding cadmium heavymetals**

(Three concentration 125, 250, and 500 µM $CdCl_2$) tested on 59 barley accessions.

**3-Plastic house Experiments**

**Regarding drought**

( 4 barley accessions with 8 treatments subjected to drought at three different growth stages (Tillering ,flowering, and anthesis).

**Regarding heavymetal**

( 6 barley accessions subjected to cadmium with concentration of 500 µM.

**Regarding detecting the roles of plant residues oakleaf residues (*Quercus aegilops*) and gundelia plant parts residue (*Gundelia tournefortii*) in accumulating cadmium**

( 4 barley accessions were subjected to cadmium with concentration of 500 µM.

**4- Application of moringa plant extract on barley accessions**

59 barley accessions were grown in the field under rain-fed conditions.
Foliar application of moringa were prepared with concentration of (1:30 v/v, MPE/Water).
Foliar treatments were used three times during the growth period, ten days apart, before and after the flowering and during the filling stage.

# CHAPTER FOUR
# RESULTS AND DISCUSSION

## 4.1 Molecular Section

### 4.1.1 Polymorphism parameters of ISSR markers

All ISSR primers produced scorable and well-defined amplification products and showed polymorphisms among the 59 analyzed barley accessions (Table 4.1.1). The forty-four ISSR primers used in this study generated 255 scorable polymorphic bands (Appendix 1.1). Number of amplified bands detected in our study ranged between 1 and 11 for the ISSR markers UBC-813 and ISSR-9, respectively. The major allele frequency ranged from 0.10 to 0.81, with 0.36 as the average allele per marker. Major alleles with the highest frequency (81%) were observed for the ISCS20 marker. The gene diversity values were observed in the variety of 0.32–0.96 with an average value of 0.77 per ISSR marker. The PIC showed that ISSR primers were completely different, which suggests that this DNA marker is a good way to find genetic differences among the barley accessions used in this study. Our results showed that the PIC values ranged from 0.96 (ISSR-8 and UBC-823) to 0.29 (ISCS20), with a mean of 0.74 (Table 4.1.1). This demonstrates the positive capability of ISSR-8 and UBC-823 primers to assess genotyping in barley germplasm, and thus provides a useful tool for analyzing population genetics in diverse plants and identifying populations. In the current study, 26 ISSR markers had PIC values larger than the average PIC value (0.74), which could be helpful for trait mapping and tagging studies in Iraqi barley accessions. The ability to determine genetic differences among different genotypes may be more directly related to the number of polymorphisms identified with each marker technique employed in diversity research. In a previous experiment conducted by a group of researchers, Yongcui, Zehong and Xiujin (2005) studied 60 barley accessions using two molecular marker techniques (RAMP and ISSR). For ISSR in their investigation, the PIC value ranged from (0.20 to 0.93) with an average of 0.676.

**Table 4.1.1 Sequences, annealing temperature, and polymorphism information parameters of the ISSR markers used in this investigation.**

| ISSR markers | Allele number | No. of polymorphic bands | Major allele frequency | Gene diversity | PIC | ISSR markers | Allele number | No. of polymorphic bands | Major allele frequency | Gene diversity | PIC |
|---|---|---|---|---|---|---|---|---|---|---|---|
| B-13 | 5.00 | 3.00 | 0.64 | 0.54 | 0.50 | UBC-810 | 21.00 | 6.00 | 0.24 | 0.89 | 0.88 |
| HB-10 | 21.00 | 7.00 | 0.25 | 0.88 | 0.87 | UBC-811 | 6.00 | 3.00 | 0.42 | 0.66 | 0.60 |
| HB-11 | 12.00 | 6.00 | 0.49 | 0.73 | 0.71 | UBC-812 | 33.00 | 8.00 | 0.12 | 0.95 | 0.95 |
| HB-12 | 5.00 | 3.00 | 0.68 | 0.50 | 0.47 | UBC-813 | 2.00 | 1.00 | 0.63 | 0.47 | 0.36 |
| HB-15 | 8.00 | 4.00 | 0.53 | 0.66 | 0.62 | UBC-814 | 10.00 | 5.00 | 0.36 | 0.79 | 0.76 |
| ISCS20 | 5.00 | 3.00 | 0.81 | 0.32 | 0.29 | UBC-815 | 21.00 | 8.00 | 0.17 | 0.92 | 0.91 |
| ISCS21 | 8.00 | 4.00 | 0.73 | 0.45 | 0.43 | UBC-818 | 22.00 | 9.00 | 0.36 | 0.84 | 0.83 |
| ISSR-1 | 7.00 | 3.00 | 0.32 | 0.76 | 0.73 | UBC-822 | 28.00 | 6.00 | 0.12 | 0.94 | 0.94 |
| ISSR-6 | 13.00 | 5.00 | 0.22 | 0.85 | 0.84 | UBC-823 | 40.00 | 9.00 | 0.10 | 0.96 | 0.96 |
| ISSR-7 | 12.00 | 6.00 | 0.54 | 0.68 | 0.67 | UBC-825 | 15.00 | 6.00 | 0.34 | 0.84 | 0.82 |
| ISSR-8 | 32.00 | 9.00 | 0.08 | 0.96 | 0.96 | UBC-826 | 22.00 | 9.00 | 0.15 | 0.92 | 0.91 |
| ISSR-9 | 27.00 | 11.00 | 0.14 | 0.93 | 0.93 | UBC-834 | 9.00 | 5.00 | 0.34 | 0.80 | 0.76 |
| ISSR-10 | 26.00 | 10.00 | 0.20 | 0.93 | 0.92 | UBC-841 | 12.00 | 4.00 | 0.66 | 0.55 | 0.53 |
| ISSR-12 | 8.00 | 6.00 | 0.56 | 0.63 | 0.58 | UBC-845 | 23.00 | 9.00 | 0.24 | 0.89 | 0.88 |
| ISSR-14 | 9.00 | 5.00 | 0.46 | 0.75 | 0.73 | UBC-846 | 34.00 | 10.00 | 0.20 | 0.93 | 0.93 |
| ISSR-16 | 7.00 | 3.00 | 0.46 | 0.69 | 0.65 | UBC-847 | 27.00 | 7.00 | 0.10 | 0.95 | 0.94 |
| ISSR-18 | 18.00 | 8.00 | 0.24 | 0.86 | 0.84 | UBC-849 | 20.00 | 8.00 | 0.34 | 0.84 | 0.83 |
| ISSR-AGC6G | 3.00 | 2.00 | 0.58 | 0.51 | 0.41 | UBC-852 | 4.00 | 2.00 | 0.53 | 0.59 | 0.51 |
| ISSR-CA | 9.00 | 4.00 | 0.31 | 0.80 | 0.77 | UBC-856 | 3.00 | 2.00 | 0.49 | 0.53 | 0.42 |
| ISSR-CCA | 6.00 | 4.00 | 0.71 | 0.47 | 0.45 | UBC-881 | 31.00 | 8.00 | 0.15 | 0.94 | 0.94 |
| PCP-3 | 16.00 | 5.00 | 0.36 | 0.82 | 0.81 | UBC-888 | 29.00 | 6.00 | 0.12 | 0.95 | 0.94 |
| UBC-808 | 11.00 | 5.00 | 0.24 | 0.86 | 0.84 | UBC-891 | 38.00 | 8.00 | 0.20 | 0.94 | 0.93 |
| Mean | | | | | | | 16.32 | 5.80 | 0.36 | 0.77 | 0.74 |
| Total | | | | | | | | 255.00 | | | |

### 4.1.2 Polymorphism information of CDDP markers

Genome-conserved sections in various plant species have helped to develop molecular markers, such as SCoT and CDDP. These markers use longer primers and higher annealing temperatures, making them more reliable, reproducible, and simple to create than other arbitrary markers like RAPD. Furthermore, they focus on gene domains, making them preferred to random markers in QTL mapping applications (Aouadi *et al.*, 2019). To study genetic diversity among 59 barley accessions, nine CDDP primers were tested. All primers produced scorable fragments (Appendix 1.2). Across all barley accessions, 35 polymorphic bands with an average of 3.89 bands per primer were generated (Table 4.1.2). The maximum and minimum number of polymorphic bands were obtained by MADS-1, WRKY-R3, and ERF2 (6 bands) and MYB1, ERF1, and KNOX1 (2 bands) respectively. The frequency of the main allele ranged from 0.34 to 0.75, with a mean value of 0.48. With a 0.75 frequency, ERF1 had the highest frequency of major alleles in the barley accessions. The gene diversity in the barley accessions collection had an average of 0.67 and ranged from 0.42 (ERF1) to 0.82 (WRKYF1). PIC values for nine primers ranged from 0.39 (ERF1) to 0.80 (WRKYF1), with an average value of 0.63 per primer. The PIC values detected in the CDDP primers were ranked in descending order as WRKY-F1 > ERF2 and WRKY-R3 > MYB2 > KNOX1 > ABP1-1 > MYB1 > MADS-1 > ERF1. WRKY-F1 is a transcription factor with developmental and physiological roles in plants (Xie *et al.*, 2005). The average polymorphic allele (3.89) found in this study was lower than the 4.60 alleles found in barley previously reported by Ahmed *et al.* (2021) using 10 CDDP primers across 82 barley genotypes. The average PIC value (0.63) in this study was high compared to another study finding documented by Ahmed *et al.* (2021), who mentioned a PIC mean value of 0.37.

**Table 4.1.2 Primer information and diversity parameters for the CDDP markers used in this investigation.**

| CDDP markers | No. of polymorphic bands | Major allele frequency | Gene diversity | PIC |
|---|---|---|---|---|
| ABP1-1 | 3.00 | 0.44 | 0.67 | 0.61 |
| WRKYF-1 | 4.00 | 0.34 | 0.82 | 0.80 |
| MYB1 | 2.00 | 0.54 | 0.60 | 0.54 |
| MYB2 | 4.00 | 0.44 | 0.74 | 0.71 |
| ERF1 | 2.00 | 0.75 | 0.42 | 0.39 |
| ERF2 | 6.00 | 0.39 | 0.77 | 0.74 |
| KNOX1 | 2.00 | 0.39 | 0.68 | 0.62 |
| MADS-1 | 6.00 | 0.63 | 0.55 | 0.51 |
| WRKY-R3 | 6.00 | 0.42 | 0.76 | 0.74 |
| Mean | 3.89 | 0.48 | 0.67 | 0.63 |
| Total | 35.00 | | | |

### 4.1.3 Polymorphism analysis of SCoT markers

For the genetic diversity analysis of 59 barley accessions, twelve SCoT markers were used. A total of 101 scoreable and sharp polymorphic bands were generated across all barley accessions (Appendix 1.3). The minimum and maximum number of polymorphic bands were obtained by SCoT12 (3 bands) and SCoT32 (15 bands), respectively. The average gene diversity was 0.81, with the SCoT3 primer having the lowest (0.48) and the SCoT16 and SCoT32 primers having the highest (0.97). PIC values for twelve SCoT primers ranged from 0.46 (SCoT3) to 0.97 (SCoT16 and SCoT32), with a mean value of 0.80 per primer (Table 4.1.3). The PIC values found in the SCoT primers were ranked in descending order as follows: SCoT16 and 32 > SCoT6 and 29 > SCoT23 > SCoT22 > SCoT7 > SCoT13 > SCoT2 > SCoT36 > SCoT12 > SCoT3. Recently, the relationship between 82 Iranian barley accessions was determined using 10 SCoT markers by Ahmed *et al.* (2021), who scored 54 polymorphic bands. The PIC value for the used marker ranged between 0.23 (SCoT11) and 0.43 (SCoT28), with an average value of 0.33 per primer. Lately, the relationship between 48 *Aegilops* *triuncialis* accessions was determined using 14 SCoT markers by Khodaee *et al.* (2021), who scored 147 polymorphic bands. The PIC value for the chosen marker ranged from 0.14 (SCoT3) to 0.42 (SCoT14), with a mean of 0.26 per primer, which was less than the PIC value in our study. Pour-Aboughadareh *et al.* (2018) similarly evaluated the molecular genetic diversity and relationships among some *Triticum* and *Aegilops* species by using 15 SCoT markers. In total, 164 polymorphic bands were detected in their investigation, and the PIC value ranged from (0.41) to (0.50) with an average of 0.48 per primer.

**Table 4.1.3 List of Scot primers used and the information obtained in the barley tested.**

| SCoT markers | No. of polymorphism | Major allele frequency | Gene diversity | PIC |
|---|---|---|---|---|
| SCoT2 | 4.00 | 0.53 | 0.68 | 0.66 |
| SCoT3 | 7.00 | 0.71 | 0.48 | 0.46 |
| SCoT6 | 8.00 | 0.24 | 0.89 | 0.89 |
| SCoT7 | 9.00 | 0.22 | 0.88 | 0.87 |
| SCoT12 | 3.00 | 0.46 | 0.66 | 0.59 |
| SCoT13 | 8.00 | 0.31 | 0.86 | 0.85 |
| SCoT16 | 13.00 | 0.10 | 0.97 | 0.97 |
| SCoT22 | 6.00 | 0.19 | 0.89 | 0.88 |
| SCoT23 | 8.00 | 0.17 | 0.90 | 0.89 |
| SCoT29 | 14.00 | 0.27 | 0.90 | 0.90 |
| SCoT32 | 15.00 | 0.05 | 0.97 | 0.97 |
| SCoT36 | 6.00 | 0.44 | 0.69 | 0.65 |
| Mean | 8.42 | 0.31 | 0.81 | 0.80 |
| Total | 101.00 | | | |

The allelic abundance of accessions in plants is a measure of their genetic variation resource, often exploited by informative molecular markers that identify populations for screening, breeding, and conservation (Igwe *et al.*, 2021; Vinceti *et al.*, 2013). The allelic count and frequency ranges produced in this research were significant, confirming the informative nature of this collection of primers in barley accessions. The major allele frequencies in three markers were in the following order: SCoT (0.31) < ISSR (0.36) < CDDP (0.48), demonstrating the usefulness of SCoT and ISSR markers for determining the allelic diversity of this important crop. The gene diversity reported was arranged in the following order: SCoT (0.81) > ISSR (0.77) > CDDP (0.67), proving the utility of SCoT and ISSR markers in estimating the allelic diversity of barley accessions. The PIC provided proficiency of marker systems, which helped determine the potential and value of the primers used in the process of fingerprinting (Serrote *et al.*, 2020). In the present investigation, the SCoT marker technique revealed a higher mean of PIC (0.80) than ISSR (0.74) and CDDP (0.63), which explains the precise application of the SCoT technique in the assessment of accessions diversity. Based on the above results, the discrimination power of the three approaches for assessing allelic diversity in barley accessions was as follows: SCoT > ISSR > CDDP, demonstrating that SCoT functional gene-based markers were informative and effective at assessing genetic diversity. In addition to SCoT and ISSR performance, CDDP markers demonstrated a moderate ability to differentiate barley accessions. Finally, it was suggested that genetic analyses based on SCoT and ISSR markers would be extremely useful for crop improvement programs, including QTL mapping, genetic diversity estimation, linking maps, and genotype identification, as SCoT markers were derived from the functional region of the genome. By having dominant and co-dominant bands, these characteristics increased the repeatability and resolution of gene-targeted markers (Abouseadaa *et al.*, 2020).

### 4.1.4 Clustering and population structure analysis of barley accessions

#### 4.1.4.1 Cluster analysis of different barley accessions

Multivariate statistical approaches are critical tools to study genetic diversity. Cluster analysis, one of the multivariate statistical techniques, separates individuals into graphs based on intervals. A marker profile data aims to maintain a genetic relationship between genotypes under investigation, using distance measures that express accessions' relationship. The dendrogram was constructed based on forty-four ISSR markers to better estimate the genetic distance among 59 barley accessions (Fig. 4.1.1). The UPGMA method and Jaccard coefficient dissimilarity were performed for analyzing the ISSR markers data set. Two major clades within 7 subgroups at the dissimilarity threshold were indicated (Fig. 4.1.1 A). The first clade included almost all barley accessions, which comprised 38 tested accessions. This group contains 5 subgroups, including the barley accessions: Bhoos-912, BA4, BN6, Bhoos-244, BN2R, CANELA, ABN, A1, MORA, MSEL) grouped in the first subgroup, and the second subgroup consisted of 10 barley accessions named Al-warka, 16 HB, GOB, Numar, Furat 9, Boraak, Radical, Arivat, Rafidain-1, and Al-khayr, which are mostly spread and cultivated in the south of Iraq. In addition, 13 barley accessions clustered in the third subgroup, including Black-Bhoos-B, Arabi aswad, Victoria, Scio/3, Acsad strain, Acsad strain, Samr, Irani, White-Halabja, Abiad, White-Zaxo, Black-Zaxo, Black-Bhoos Akre, while the fourth subgroup had four barley accessions: Al-amal, IBAA-99, IBAA-265, and White-Akre. However, Bhoos-H1 was isolated in the fifth subgroup, demonstrating its genetic divergence from the other accessions, whereas in the second clade, two main sub-clusters were exhibited. The first sub-cluster comprised five accessions (Rehaan, Sameer, Warka-B12, Al-Hazzar, IBAA-995), which are mainly distributed over the southern part of Iraq, and the rest of the barley accessions were grouped into the second sub-cluster. The clustering pattern for ISSR data showed that barley accessions can be separated into two major groups according to geographical origin. Therefore, to enhance the appearance of heterosis, the breeder may use genetic distance data to make informed decisions about crossing accessions from distinct groups or subgroups for population development or to promote the analysis of various parents to cross in hybrid combinations. The present outcome supports previous reports on the correlation between ISSR markers and the eco-geographical distribution of the accessions (Yongcui, Zehong and Xiujin, 2005; Wang, Yu and Ding, 2009; Etminan *et al.*, 2016). Our results confirm the effectiveness of ISSR markers in evaluating the genetic diversity reported previously by Wang, Yu and Ding (2009), who scored 145 polymorphic bands using 11 ISSR primers.

Regarding the clustering patterns for CDDP primers, two main groups were identified, but with different sub-clusters. In general, two main sub-clusters were determined in the first group. The first sub-cluster included only the Ukranian-Zarayan accession, while the second sub-cluster in the same group separated the barley accessions to the two sets of arrangements. The first sub-sub-cluster was composed of three accessions (Black-Akre, Black-Garmiyan, and Black-Chiman), which were cultivated by the farmer in the North of Iraq, and the remaining barley accessions were arranged in the second sub-sub-cluster based on genomic similarity. However, group 2 included only two barley accessions, namely Al-Hazzar and IBAA-995, indicating their genomic differences from the rest of the accessions (Fig. 4.1.1B). The number of groups detected by CDDP markers in this study was lower than in the report of Ahmed *et al.* (2021) who exhibited three clusters in 82 Iranian barley accessions. This could be due to the type of markers and the number of accessions used.

Clustering of barley accessions using the SCoT molecular dataset revealed two major classes (Fig. 4.1.1 C). The first class included most barley accessions categorized into two different sub-clusters. One of them comprised three barley accessions, namely (A1, Arabi Aswad, and Clipper), while the other sub-cluster distributes the most barley accessions into three sub-sub-clusters. The first sub-sub-cluster comprised four barley accessions (16 HB, Furat 9, Al-warka, and Black-kalar), which were mostly distributed and cultivated in the south of Iraq. The second sub-sub-cluster included five barley accessions (Black-Akre, Black-Garmiyan, Black-Chiman, White-Zarayan, and Bujayl 2-Shaqlawa). The farmer widely cultivated these barley accessions in the north of Iraq, while forty-two barley accessions based on genomic dissimilarity were arranged in the third sub-sub-cluster in this particular group. However, the second class involved only five distinct barley accessions, namely Rehaan, Sameer, Warka, B12, Al-Hazzar, and IBAA-995, which originated from the south of Iraq. The number of clusters formed by the SCoT approach in this study was smaller than that previously reported by Ahmed *et al.* (2021). Intriguingly, the general dendrogram constructed using the combined data of all molecular markers used in our investigation (ISSR, CDDP, and SCoT) (Fig. 4.1.1 D) divided barley accessions into two major clusters. Almost all barley accessions were well distributed in the first cluster, which included fifty barley accessions. In this particular cluster, it is clear that some barley accessions based on genetic distance were grouped, especially for the accessions of Arabi aswad and Clipper, while the rest of the forty-eight accessions were grouped into six sub-sub-clusters. Conversely, the second cluster consisted of nine barley accessions, namely Black-Akre, Black-Garmiyan, Black-Chiman, Bujayl2-Shaqlawa, Rehaan, Sameer, Warka-B12, Al-Hazzar, and IBAA-995. Accordingly, our finding supported the available suggestion that many molecular techniques could either be applied individually or in combination with other molecular marker techniques to find reliable information about genetic relationships and assess the genetic variation, which would support strategies for effective collection of barley germplasm and knowledge of its conservation. Insufficient genetic differentiation could imply a high level of gene flow. Stimulatingly, taking a close look at the dendrogram using two different marker systems, a similar pattern of alignments for most barley accessions can be found. Similarly, a group of researchers, Naceur *et al.* (2012), worked on 31 barley accessions to reveal genetic distance, and obtained 9 classes, demonstrating wide diversity among the studied accession. This is probably due to the collection of barley germplasm from three countries (Egypt, Algeria, and Tunisia). Whereas, compared to our previous work, seven clusters were obtained using the SSRs marker technique (Tahir *et al.*, 2021). This type of molecular technique is now widely used by other research groups to determine the genetic diversity of plant species (Liu *et al.*, 2020; Saidi, Jabalameli and Ghalamboran, 2018; Talebi *et al.*, 2018). For many plants, all three markers have been effectively used to determine genetic relationships and diversity. In addition, DNA analysis using three methods has proved to be an inexpensive and efficient way to provide molecular data for evaluating genetic differences (Hajibarat *et al.*, 2015; Tiwari *et al.*, 2016; Amom and Nongdam, 2017; Khodaee *et al.*, 2021). It has been noted that the larger the distance between accessions, the greater the likelihood of accumulating wider genetic diversity, which also defines their places in clusters (Skroch and Nienhuis, 1995).

#### 4.1.4.2 Population structure profile in barley accessions

STRUCTURE software and the Bayesian statistical index were used to perform effective population structure assessment, reliable population grouping, and the identification of mixed genotypes. To estimate the population structure of 59 barley accessions, separate and combined data from all molecular markers, ISSRs, CDDP, and SCoT were used.

The optimal value of k is 2 in all three used markers (Fig. 4.1.2, 4.1.3). The STRUCTURE outcomes suggested the population could be divided into two main populations in all case scenarios. The grouping was mildly following the geographic background of the barley accessions. Although two populations may be feasible for our panel due to the size of the population and the difference in the number of barley accessions representing the three main areas of Iraq from which they were collected. Cluster analysis was performed, as mentioned previously, to truly understand the genetic structure of this population. Based on the results of these analyses, it was observed that there were no differences between the two analyses (clustering and structure) and that they completely matched by analysis obtained from the molecular markers data set. This meets our expectations better, as it was observed that the accessions from the neighboring locations were mostly clustered together compared with the populations derived by STRUCTURE.

Remarkably, the result from STRUCTURE Harvester for all molecular markers demonstrated that the (k) value had the maximum peak at K =2, inferring that the probable number of genetic clusters in the population incorporates all individuals from 59 accessions with the highest likelihood (Fig. 4.1.2, 4.1.3). This was observed when the mean of the log of posterior probability was graphed, demonstrating that two populations can be observed, which were visualized in two distinct colors (green and red). Based on the membership fractions, the accessions with a probability of 80% or above were assigned to matching populations. characterized as an admixture, indicating the purity of tested materials. The combination of the two mentioned colors represents barley accessions in which they possess different genetic structures.

Concerning the separate and combined analysis of structure for all marker data set conducted in this investigation, the first population indicated in red color including individuals of pure genetic make-up (with the probability of ≥ 80% ), which comprised [16 (Bujayl 2-Shaqlawa, Sameer, Black-Akre, Black-Garmiyan, Black-Chiman, Warka-B12, IBAA-995, Bujayl 3-Shaqlawa, Rehaan, Bujayl 1-Shaqlawa, Ukranian-Zarayan, Al-Hazzar, Abrash, Shoaa, Gk-Omega, and Qala-1), 28 ( Boraak, Bhoos-244, Scio/3, Acsad strain, Al-warka, IBAA-99, Numar, Radical, IBAA-265, Victoria, Rafidain-1, Al-amal, Irani, White-Akre, 16 HB, Samr, Arivat, Black-Bhoos Akre, Black-Bhoos-B, Saydsadiq, Bhoos-H1, Furat 9, BN6, Gk-Omega, White-Halabja, Acsad-14, Black-Zaxo, and Bhoos-912) 12 (Sameer, A1, Clipper, Warka-B12, Black-Garmiyan, Black-Chiman, Arabi aswad, Rehaan, Al-Hazzar, IBAA-995, White-Zarayan, and Bujayl 2-Shaqlawa), and 15 (Sameer, Bujayl 2-Shaqlawa, Black-Akre, Black-Garmiyan, Warka-B12, IBAA-995, Rehaan, Al-Hazzar, Bujayl 3-Shaqlawa, Bujayl 1-Shaqlawa, Abrash, Black-Chiman, White-Zarayan, and Gk-Omega)] barley accessions for ISSR, CDDP, SCoT, and combined molecular dataset, respectively. Furthermore, the second population, depicted in green, included barley accessions with a pure genetic background (with the probability of ≥ 80%). This distinct population basically contained 22 [(CANELA, Scio/3, Acsad-14, Bhoos-244, IBAA-265, Victoria, ABN, Acsad strain, Bhoos-H1, Boraak, Al-amal, Bhoos-912, MSEL, Irani, BN2R, IBAA-99, White-Halabja, Rafidain-1, Black-Zaxo, Arabi aswad, Black-Bhoos-B, BA4, A1, Abiad, Black-Bhoos Akre, Al-khayr, BN6, MORA, White-Zaxo, and Numar), 26 (Ukranian-Zarayan, White-Zarayan, Al-Hazzar, Black-Garmiyan, Black-Chiman, Rehaan, Abrash, Bujayl 1-Shaqlawa, GOB, Warka-B12, Qala-1, Black-Akre, ABN, Bujayl 3-Shaqlawa, Black-kalar, IBAA-995, BA4, White-Zaxo, MORA, Bujayl 2-Shaqlawa, White-kalar, BN2R, Abiad, MSEL, Sameer, and CANELA), 38 (Bhoos-244, White-Akre, White-Halabja, Abiad, Qala-1, Black-Bhoos-B, IBAA-265, White-Zaxo, CANELA, MSEL, Scio/3, Victoria, Ukranian-Zarayan, Irani, BA4, Al-khayr, Saydsadiq, Black-Bhoos Akre, Black-Zaxo, Boraak, Samr, White-kalar, ABN, MORA, BN2R, Bujayl 1-Shaqlawa, Acsad strain, Rafidain-1, Bhoos-912, Numar, Al-amal, GOB, Shoaa, Radical, Bujayl 3-Shaqlawa, IBAA-99, Bhoos-H1, and Arivat), and 29 (Bhoos-244, IBAA-265, CANELA, Scio/3, Victoria, Al-amal, White-Halabja, Acsad strain, Boraak, MSEL, ABN, Rafidain-1, Acsad-14, IBAA-99, Bhoos-912, Black-Zaxo, Black Bhoos-B, Irani, Bhoos-H1, Al-khayr, BN2R, BA4, Abiad, Black-Bhoos Akre, BN6, MORA, Radical, White-Zaxo, and Numar)] barley accessions in ISSR, CDDP, SCoT as well as all markers data, respectively, while, the remaining barley accessions (with the probability of < 80%) are considered admixture of genome from other populations. Interestingly, in all cases, the combination of two populations in numbers of individuals was much higher than in the admixture form, which showed the uniformity of tested accessions. This finding confirms that the lemma and palea remain tightly closed during the period of pollen release in barley. This phenomenon is known as cleistogamy (Nair *et al.*, 2010). However, admixture probably occurred due to breeding lines developed through random mating by the breeders, for it is a specific trait improvement (Hernandez, Meints and Hayes, 2020).

### 4.1.5 Genetic differences within and among populations

The analysis of molecular variance (AMOVA) based on the results achieved from the molecular data from three different marker systems was suggested that 15,14, 9, and 14% of total variation were among the three populations (North, Middle, and South of Iraq) in which the sample was grown and collected (Table 3.1.1) based on the molecular data set from ISSR, CDDP, SCoT and the combined data of all three marker systems respectively, while the analysis was revealed the variation was much higher within the studied individuals, which were 85, 86, and 91% depending on the data obtained from ISSR, CDDP, and SCoT respectively (Table 4.1.4), whereas the collective data from all three marker sets propose 86% of total differences were inside the population. These outcomes suggest that barley accessions from Iraq shared a common ancestry and are highly admixed, with high variation within populations from all the marker systems in which this investigation was conducted. This result revealed a distinct genetic base for the 59 barley accessions. The AMOVA results confirmed clustering and structure analyses. The partitioning of molecular variance showed that the greatest divergence was detected among individuals in the same population as indicated in Table 4.1.4. Genetic diversity within and between populations improves the selection of populations that account for the vast majority of extant variations (Igwe *et al.*, 2021). If genetic diversity is largely found within a population, it means fewer populations are needed to conserve and sustain the differences in accessions or populations. Conversely, if genetic diversity across populations is preserved, a greater number of preserved across populations, more populations should be emphasized for maintenance and utilization (Igwe *et al.*, 2021).

**Table 4.1.4 Analysis of molecular variance (AMOVA) showing genetic diversity in the Iraqi barley accessions.**

| Method | Source | Df | SS | MS | Est. Var. | % | *P*-value |
|---|---|---|---|---|---|---|---|
| ISSR data | Among Pops | 2 | 320.99 | 160.49 | 6.59 | 15** | 0.001 |
| | Within Pops | 56 | 2012.1 | 35.93 | 35.93 | 85** | |
| | Total | 58 | 2333.09 | | 42.52 | 100 | |
| CDDP data | Among Pops | 2 | 36.76 | 18.38 | 0.74 | 14** | 0.001 |
| | Within Pops | 56 | 247.55 | 4.42 | 4.42 | 86** | |
| | Total | 58 | 284.31 | | 5.16 | 100 | |
| SCoT data | Among Pops | 2 | 73.72 | 36.86 | 1.29 | 9** | 0.001 |
| | Within Pops | 56 | 703.13 | 12.56 | 12.56 | 91** | |
| | Total | 58 | 776.85 | | 13.84 | 100 | |
| ISSR, CDDP, and SCoT data | Among Pops | 2 | 431.51 | 215.76 | 8.61 | 14** | 0.001 |
| | Within Pops | 56 | 2962.78 | 52.91 | 52.91 | 86** | |
| | Total | 58 | 3394.29 | | 61.52 | 100% | |

*Df: degree of freedom, SS: sum of aquare, MS: mean of square, Est. Var: estimated variance, *p*-value: probability value

### 4.1.6 Correlation analysis between genetic dissimilarity achieved by three sets of markers data

To compare the genetic distance matrices produced by three marker systems (ISSRs, CDDP, and SCoT), the Mantel test was performed. However, this analysis has its limitations, especially in the case of using two different markers (Vieira *et al.*, 2007). In the case of our investigation, 45 ISSR, 9 CDDP, and 12 SCoT primers were conducted. Remarkably, Mantel test correlation values revealed a positive significant correlation between and among all three different marker systems in which clusters with a general dendrogram were present (Table 4.1.5). The highest Mantel value was observed between ISSR and CDDP markers (0.49), while the minimum was displayed between CDDP and combined data markers (0.25). This demonstrates the novelty of current work and the possibility of composing reference collections of tested barley accessions using the information attained from genetic profiles tested by three different molecular methods. Natural selection could also clarify the appropriate relationship between the diverse patterns of these markers in the regions exacerbated by ISSR, CDDP, and SCoT markers. These findings are consistent with previous research (Ahmed *et al.*, 2021), which detected significant associations between CDDP and SCoT markers.

**Table 4.1.5 Mantel coefficient values for correlation between clusters obtained by three different molecular markers.**

| Markers | ISSR | CDDP | SCoT | Pooled markers data |
|---|---|---|---|---|
| ISSR | - | 0.49** | 0.37** | |
| CDDP | | - | 0.36** | 0.25** |
| Scot | | | - | |

### 4.2. Drought Section

### 4.2.1 Drought experiments at the seedling stage

#### 4.2.1.1 Responses of barley traits to drought stress

***Drought stress tolerance dependent on phenotypic characters***

Under control conditions as well as under induced drought (PEG 6000 treatment), a huge significant phenotypic diversity ($p \leq 0.001$) for all attributes was identified among the genotypes as indicated by the ANOVA analysis and box charts, and significant reductions were shown as the concentration of PEG increased (Table 4.2.1). The measured scores of the accessions under control condition were between 63.33-100.00% with an average of 84.34, 5.16-15.15 cm with a mean of 10.48 cm, 6.20-12.98 cm with an average of 9.87 cm, 11.41-27.00 cm with an average of 20.34 cm and 922.15-2371.18 with an average of 1729.21 for germination percentage (GP), root length (RL), shoot length (SL), seedling length (SEL), and seed vigor (SV), respectively (Table 4.2.1 and Fig.4.2.1). For detecting the influence of different PEG stress conditions and their interactions with the studied barley accessions, mean pairwise comparison showed significant impacts on the studied morphological traits (Appendix 2.1). For traits (RL, SL, SEL, WU-40, WU-64, GP and SV), the values ranged between (15.15 - T0.00 × AC56 and 1.28 - T20.50 × AC52), (12.98 - T0.00 × AC42 and 3.21 - T10.25 × AC5), (27.00 - T0.00 × AC56 and 1.16 - T20.50 * AC7), (210.83 - T0.00 * AC41 and 38.09 - T20.50 * AC23), (351.00 - T0.00 × AC40 and 51.47 - T20.50 × AC23), (100.00 - T0.00 × AC36 and 10.00 - T20.50 × AC48), and (2371.18 - T0.00 × AC30 and 16.59 - T20.50 × AC7), respectively. However, the same traits responded differently under the availability of all PEG stressor conditions by barley accessions as indicated in (Appendix 2.2). For traits (RL, SL, SEL, WU-40, WU-64, GP and SV), the values ranged between (12.29- AC56 and 4.30-AC47), (9.09-AC5 and 3.6-AC51), (20.89-AC56 and

11.36-AC53), (175.16-AC41 and 75.25-AC13), (246.83-AC41 and 87.00-AC56), (97.22-AC36,AC37 and 37.22-AC19) and (1619.57-AC36 and 508.64-AC18), respectively.

The vulnerability of barley to PEG was, also shown by all aspects of morphological data. At the T10.25 stress condition, the average values of GP, RL, SL, SEL, and SV were ranged between 30.00-100.00% with a mean of 71.98%, 4.52-13.87 cm with an average of 8.46 cm, 3.21-10.00 cm with a mean of 7.20 cm, 8.89-23.24 cm with a mean of 15.66 cm and 390.04-1853.03 with an average of 1150.80, respectively (Table 4.2.1 and Fig.4.2.2). The mean values of the accessions varied between 10.00-96.67% with an average of 50.20%, 0.86-9.24 cm with a mean of 4.87 cm, 0.30-5.60 cm with a mean of 2.68 cm, 1.16-14.21 cm with an average of 7.56 cm and 16.59-1165.13 with a mean of 418.76 for GP, RL, SL, SEL, and SV respectively, under induced drought stress (T20.50) (Table 4.2.3 and Fig.4.2.3). The box charts of all traits illustrated significant variations between T0.00 (control), T10.25, and T20.50 as shown for each characteristic by the lower and upper box plot limits (Fig.4.2.4). The barley accessions had significantly higher traits values in normal conditions compared to stressed plants. These reports indicated that, following exposure to T20.50, germination and seedling growth parameters depression were more serious for all accessions.

**Table 4.2.1 Descriptive statistics of different traits of different accession of barley in respect to drought stress at seedling stage.**

| | Traits | GP | RL | SL | SEL | SV | WU-40 | WU-64 | EL | RWC | TPC | TFC | DPPH | SSC | PC | GPA | CAT |
|---|---|---|---|---|---|---|---|---|---|---|---|---|---|---|---|---|---|
| T0.00 (Control) | Min. | 63.33 | 5.16 | 6.2 | 11.41 | 922.15 | 92.72 | 108.82 | 5.01 | 72.64 | 18.49 | 3.4 | 232.6 | 141.88 | 124.87 | 0.78 | 0.45 |
| | Max. | 100 | 15.15 | 12.98 | 27 | 2371.18 | 210.83 | 351 | 17.33 | 96.44 | 68.61 | 17.95 | 361.71 | 353.3 | 335.98 | 1.97 | 10.32 |
| | Mean | 84.34 | 10.48 | 9.87 | 20.34 | 1729.21 | 138.7 | 184.76 | 8.4 | 86.54 | 35.18 | 10.16 | 306.11 | 263.44 | 219.68 | 1.36 | 4.16 |
| | F | 7.19*** | 11.03*** | 7.54*** | 10.54*** | 9.89*** | 12.71*** | 29.63*** | 15.43*** | 2.24*** | 22.45*** | 6.56*** | 13.34*** | 6.59*** | 22.06*** | 2.12** | 24.09*** |
| | Pr > F | < 0.0001 | < 0.0001 | < 0.0001 | < 0.0001 | < 0.0001 | < 0.0001 | < 0.0001 | < 0.0001 | 0.0001 | < 0.0001 | < 0.0001 | < 0.0001 | < 0.0001 | < 0.0001 | 0 | < 0.0001 |
| T10.25 (10.25% PEG) | Min. | 20 | 3.56 | 1.95 | 7.56 | 217.5 | 52.74 | 36.76 | | | 37.27 | 1.41 | 322.14 | 223.01 | 190.51 | 0.38 | 0.9 |
| | Max. | 100 | 14.54 | 10.5 | 23.54 | 2158 | 198.14 | 383.11 | | | 177 | 15.24 | 565.69 | 604.74 | 650.77 | 2.91 | 23.57 |
| | Mean | 71.98 | 8.46 | 7.2 | 15.66 | 1150.8 | 112.26 | 147.03 | | | 90.82 | 6.26 | 467.36 | 434.73 | 396.65 | 1.6 | 7.43 |
| | F | 14.25*** | 17.84*** | 5.82*** | 10.31*** | 13.55*** | 13.52*** | 8.43*** | | | 73.19*** | 9.46*** | 31.65*** | 8.47*** | 55.87*** | 2.53*** | 143.28*** |
| | Pr > F | < 0.0001 | < 0.0001 | < 0.0001 | < 0.0001 | < 0.0001 | < 0.0001 | < 0.0001 | | | < 0.0001 | < 0.0001 | < 0.0001 | < 0.0001 | < 0.0001 | < 0.0001 | < 0.0001 |
| T20.50 | Min. | 10 | 0.86 | 0.3 | 1.16 | 16.59 | 38.09 | 51.47 | 6.11 | 43.67 | 31.65 | 2.48 | 368.04 | 269.6 | 347.61 | 1.25 | 0.94 |

| Pr > F | F | Mean | Max. |
|---|---|---|---|
| < 0.0001 | 27.80*** | 50.2 | 96.67 |
| < 0.0001 | 6.89*** | 4.87 | 9.24 |
| < 0.0001 | 5.25*** | 2.68 | 5.6 |
| < 0.0001 | 7.11*** | 7.56 | 14.21 |
| < 0.0001 | 23.28*** | 418.76 | 1165.13 |
| < 0.0001 | 6.51*** | 87.45 | 141.2 |
| < 0.0001 | 4.50*** | 102.94 | 216.98 |
| < 0.0001 | 52.73*** | 14.62 | 25 |
| < 0.0001 | 10.75*** | 64.37 | 84.33 |
| < 0.0001 | 57.87*** | 129.92 | 230.51 |
| < 0.0001 | 17.13*** | 8.17 | 19.17 |
| < 0.0001 | 75.63*** | 528.66 | 599.04 |
| < 0.0001 | 7.25*** | 447.18 | 741.7 |
| < 0.0001 | 290.60*** | 726.25 | 1347.07 |
| < 0.0001 | 5.02*** | 2.13 | 3.04 |
| < 0.0001 | 242.69*** | 20.1 | 65.67 |

***Drought stress tolerance reliant on physiological traits***

Drought stress has significantly impacted plant physiological characteristics. Significant differences in seed water uptake between the tested accessions were noticed (Table 4.2.1 and Figures 4.2.5, 4.2.6, and 4.2.7). The estimated scores of the controlled accessions were between 92.72-210.83% with an average of 138.70% and 108.82-351.00% with an average of 184.76% for WU-40 and WU-64, respectively. Under present of drought stress conditions, mean pairwise comparisons and their interactions for studied barley accessions revealed that both accessions AC59 and AC24 had the highest value of 25.00 for EL (%), while AC12 maintained the EL (%) as it recorded the lowest value of 6.11 under control conditions (Appendix 2.3). However, the highest RWC% was recorded under control conditions by barley accession AC21 with a value of 96.44, while AC55 had the lowest record of 43.67 under stress conditions. Mean pairwise comparisons among tested barley accessions indicated that AC57 had the maximum EL (%) with a value of 21.17, while the minimum EL (%) was recorded by AC1 with a value of 5.83, while for RWC% the values ranged between (85.09 and 64.70) for barley accessions AC28 and AC49, respectively (Appendix 2.4). The sensitivity of the barley to PEG was exposed by all characteristics of physiological data. At the T10.25 condition, the average values of WU-40 and WU-64 were extended between 67.49-197.22% with a mean of 112.26% and 67.97-288.49% with an average of 147.03%, respectively. The mean accession value varied between 38.09 and 141.20% with an average of 87.45% and between 51.47 and 216.98% with an average of 102.94% for WU-40 and WU-64 respectively under induced dry stress of T20.50. The trait charts showed significant differences between T0.00 (control), T10.25, and T20.50, as revealed by the lower and upper box limits for each physiological attribute (Fig.4.2.8). The barley accessions had significantly higher traits values in normal condition compared to stressed plants.

The difference between the leak values of leaf electrolytes (EL) and the relative water content of the leaves (RWC) is shown in Fig.4.2.8C and D for 21-day seedlings under normal conditions and dehydration (water deficiency). The determined mean under dehydration stress for RWC was drastically decreased compared to the control group. In the EL box plot, however, the range of the lower and upper deviation limits of the mean value is wider in the stressed condition than in the control condition.This is largely supported by the highly significant data analysis obtained by ANOVA and the Duncan test for the control vs. water deficiency treatment mentioned in Table 4.2.1. Drought stress had a significant effect on the physiological features of the leaves (EL and RWC). When treated with water stress, the EL in the stressed barley seedling was increased by 1.74% relative to the unstressed barley seedlings. The significant difference in RWC values for 21-day-old seedlings under normal conditions and stressed treatments is shown in Fig. 4.2.5. The calculated mean values for RWC were 64.37% and 86.54% respectively, for optimal and stress conditions. Under the water stress condition, RWC significantly declined by 26% compared to non-stressed plants.

***Drought stress tolerance through biochemical characteristics***

***Validation of drought tolerance by non-enzymatic stress markers***

Biochemical analysis of seedlings found that the biochemical profiles in the different accessions had the same patterns with regards to the treatments, but with substantial quantitative variations, and the results are illustrated in Table 4.2.1 and Figures 4.2.5, 4.2.6, 4.2.7, and 4.2.9. The stresses influenced biochemical parameters significantly, and the differences were mainly related to the accessions. In seedlings of all accessions under induced drought, TPC and DPPH were gradually affected by drought stress. In this study, TPC and antioxidant potential (DPPH) significantly improved with an increase in the severity of PEG stress in the sequence of T0.00 $<$ T10.25 $<$ T20.50. In T10.25 and T20.50, TPC and DPPH were augmented by 2.58% and 3.68%, respectively and 1.53% and 1.73%, respectively compared to T0.00 conditions. The present study found remarkable differences in TFC among the studied accessions. Broadly, the average TFC value in the control samples (T0.00) produced 10.16 mg QE/g FW, whereas, in the induced drought stress, the TFC reached 6.26 and 8.17 mg QE/g FW for T10.25 and T20.50, respectively.

***Validation of drought tolerance by osmolyte contents***

Under drought stress, plants generate and store suitable solutes, including carbohydrates, polyols, and amino acids, to promote osmotic equilibrium and the absorption and retention of water (Hussain *et al.*, 2018). The amount of SSC and PC content was measured to ascertain whether barley responds to the low water potential caused by PEG. Under control and stimulated stress conditions, SSC and PC had considerable variations in their content. As predicted, the concentration of SSC in the medium with 10.25% and 20.50% PEG increased significantly across all accessions by 1.65 and 1.70 times, respectively, compared to the PEG-free medium (Table 4.2.1 and Figures 4.2.5, 4.2.6, 4.2.7, and 4.2.10). The results disclosed that the PC in barley seedlings was significantly augmented by increasing the PEG concentrations, and the medium containing 20.50% of PEG displayed a maximum amount of proline compared to the PEG-free medium. Consequently, it improved by 3.31 times compared to the PEG-free medium. The amounts of SSC and PC varied from 263.44 to 447.18 µg/g FW and 219.68 to 726.25 µg/g FW. Nevertheless, the amounts of these compounds increased sharply and significantly from T0.00 to T10.25 and T20.50, as evidenced by the order: T0.00 < T10.25 < T20.50. In T10.25 and T20.50 stress conditions, SSC and PC were enhanced by 1.65 and 1.70% in T10.25 conditions and 1.81 and 3.31% in T20.50 treatments; respectively, when compared to normal condition.

***Accessions validation of the PEG tolerance by antioxidant enzymatic stress markers***

The antioxidant enzyme's results are presented in Table 4.2.1 and Figures 4.2.5, 4.2.6, 4.2.7, and 4.2.11. The enzyme's activities of barley seedlings were progressively affected by stimulated drought stress. However, GPA and CAT were sharply and statistically raised with the harshness of PEG stress, showing the order: T20.50 > T10.50 > T0.00. GPA and CAT ranged from 1.36 to 2.13 unit/min/g FW with the highest GPA and CAT was recorded in T20.50 stress and the lowest in control condition. Further, in T10.25 and T20.50, the increases in GPA and CAT activities were (1.18 and 1.57%) and (1.79 and 4.83%), respectively, over the control condition. These results are in accordance with the reports of (Hasanloo *et al.*, 2013), who reported that drought stress triggered a rise in GPA and CAT activities in barley genotypes. Mean pairwise comparisons and interactions in the case of using different concentrations of PEG on studied barley accessions showed significant impacts on the studied biochemical traits including (TPC, TFC, DPPH, SSC, PC, GPA, and CAT) (Appendix 2.5). Different values were documented and the values ranged between (230.51-T20.50*AC27 and 18.49-T0.00*AC1), (19.17-T20.50*AC45 and 1.56-T10.25*AC59), (599.04-T20.50*AC27-29 and 232.6-T0.00*AC54), (741.70-T20.50*AC45 and 141.88-T0.00*AC2), (1347.07-T20.50*AC38 and 124.87-T0.00* AC33), (3.04-T20.50*AC58 and 0.78-T0.00*AC4) and (65.67-T20.50*AC21 and 0.45-T0.00*AC48), respectively for stated traits. However, the same biochemical traits responded differently under the availability of all PEG stressor conditions by barley accessions as shown in (Appendix 2.6). For traits; (TPC, TFC, DPPH, SSC, PC, GPA, and CAT), the values ranged between (140.19-AC45 and 60.58-AC6), (13.57-AC45 and 3.05-AC36), (489.89-AC24 and 349.00-AC49), (478.68-AC40 and 286.06-AC2), (665.22-AC37 and 232.98-AC55) and (2.38-AC42 and 1.29-AC19), (28.99-AC16 and 2.35-AC47), respectively.

### 4.2.1.2 Ranking of genotypes for germination percentage and seedling elongation

The ranking method was performed on the basis of germination percentage and seedling length (root length + shoot length) to recognize the best accessions utilizing specific tested traits according to the procedure developed by (Ketata, Yau and Nachit, 1989). The average sum of ranks (ASR) has been used as a predictor for choosing the best accessions for germination and seedling growth (Pour-Aboughadareh *et al.*, 2019). The direction of accessions from PEG resistant > susceptible was MORA (AC36) > ABN (AC37) > MSEL (AC27) >… Abrash (AC51) > Black-Bhoos Akre (AC18) > Black-Zaxo (AC19) for T10.25 condition and ABN (AC37) > Bhoos-H1 (AC40) > MORA (AC36) > …White-Zarayan (AC50) > Black-Garmiyan (AC47) > Black-Zaxo (AC19) under T20.50 treatment (Tables 4.2.3 and 4.2.4). For both stress conditions (T10.25 and T20.50), the accessions resistance/susceptibility rating according to their germination parameters was as follows: PEG resistant > susceptible, ABN (AC37) > MORA (AC36) > MSEL (AC27) > … Bujayl 2-Shaqlawa (AC52) > Al-warka (AC7) > Black-Garmiyan (AC47). The mean value of all the investigated characters of both PEG stress states was measured and presented in Table 4.2.4 for comparison between the best performance variant (tolerant to PEG) and the lowest performance variant (sensitive to PEG). As shown in Table 4.2.5, all traits studied (phenotypic, physiological, and biochemical properties), except for TFC, were significantly higher in the highest performing ABN (AC37) accession than in the lowest-performing Black-Garmiyan (AC47) accession.

**Table 4.2.2 Rank of 59 barley accessions determined by stress tolerance index (STI) and the average number of ranks (ASR), depending on germination percentage and growth characteristics of seedlings under T10.25. Accessions with the highest STI and lowest ASR were considered the best accessions, and the lowest rank was assigned to the most consistent performance of each accession. The complete name of the accession is described in Table 3.1.1.**

| Accession Code | ASR | STI | Rank | Accession Code | ASR | STI | Rank | Accession Code | ASR | STI | Rank |
|---|---|---|---|---|---|---|---|---|---|---|---|
| AC41 | 22.73 | 0.94 | 23 | AC21 | 11.36 | 1.13 | 9 | AC1 | 37.45 | 0.68 | 38 |
| AC42 | 33.64 | 0.88 | 36 | AC22 | 40.18 | 0.67 | 42 | AC2 | 26.36 | 1.02 | 27 |
| AC43 | 30.00 | 0.91 | 32 | AC23 | 53.18 | 0.46 | 54 | AC3 | 18.09 | 1.00 | 17 |
| AC44 | 51.09 | 0.48 | 52 | AC24 | 23.18 | 0.94 | 24 | AC4 | 15.55 | 1.07 | 14 |
| AC45 | 8.18 | 1.12 | 6 | AC25 | 12.91 | 1.23 | 12 | AC5 | 27.55 | 0.89 | 28 |
| AC46 | 31.18 | 0.74 | 33 | AC26 | 7.45 | 1.22 | 5 | AC6 | 14.45 | 1.03 | 13 |
| AC47 | 46.73 | 0.49 | 49 | AC27 | 4.36 | 1.22 | 3 | AC7 | 48.36 | 0.63 | 50 |
| AC48 | 45.27 | 0.67 | 46 | AC28 | 31.45 | 0.85 | 34 | AC8 | 41.18 | 0.52 | 43 |
| AC49 | 44.91 | 0.66 | 45 | AC29 | 28.36 | 0.94 | 30 | AC9 | 29.45 | 0.98 | 31 |
| AC50 | 12.45 | 1.03 | 10 | AC30 | 6.73 | 1.24 | 4 | AC10 | 20.73 | 1.02 | 19 |
| AC51 | 56.36 | 0.41 | 57 | AC31 | 21.00 | 1.00 | 20 | AC11 | 25.91 | 0.85 | 25 |
| AC52 | 54.91 | 0.42 | 56 | AC32 | 26.09 | 0.99 | 26 | AC12 | 20.27 | 0.92 | 18 |
| AC53 | 53.00 | 0.43 | 53 | AC33 | 43.73 | 0.66 | 44 | AC13 | 36.82 | 0.87 | 37 |
| AC54 | 38.55 | 0.79 | 39 | AC34 | 33.09 | 0.89 | 35 | AC14 | 46.27 | 0.57 | 47 |
| AC55 | 54.18 | 0.43 | 55 | AC35 | 8.73 | 1.13 | 7 | AC15 | 40.18 | 0.76 | 41 |
| AC56 | 27.91 | 0.97 | 29 | AC36 | 2.55 | 1.33 | 1 | AC16 | 39.18 | 0.83 | 40 |
| AC57 | 22.45 | 0.82 | 22 | AC37 | 3.55 | 1.27 | 2 | AC17 | 16.82 | 1.12 | 16 |
| AC58 | 46.36 | 0.62 | 48 | AC38 | 16.09 | 1.07 | 15 | AC18 | 57.27 | 0.38 | 58 |
| AC59 | 51.00 | 0.49 | 51 | AC39 | 10.27 | 1.24 | 8 | AC19 | 58.64 | 0.36 | 59 |
| | | | | AC40 | 12.82 | 1.10 | 11 | AC20 | 21.45 | 1.07 | 21 |

**Table 4.2.3 Rank of barley accessions calculated by stress tolerance index (STI) and the average number of ranks (ASR) based on germination percentage and growth characteristics of seedlings under stress condition T20.50. The accessions with the highest STI and the lowest rank were assigned to the highest performance of each accession. The full name of the accession is given in Table 3.1.1.**

| Accession Code | ASR | STI | Rank | Accession Code | ASR | STI | Rank | Accession Code | ASR | STI | Rank |
|---|---|---|---|---|---|---|---|---|---|---|---|
| AC1 | 42.00 | 0.27 | 42 | AC21 | 17.09 | 0.86 | 18 | AC41 | 6.73 | 0.94 | 6 |
| AC2 | 33.00 | 0.49 | 32 | AC22 | 33.45 | 0.42 | 33 | AC42 | 7.91 | 0.94 | 7 |
| AC3 | 15.36 | 0.84 | 15 | AC23 | 50.27 | 0.21 | 51 | AC43 | 28.36 | 0.64 | 29 |
| AC4 | 21.82 | 0.79 | 22 | AC24 | 12.27 | 0.87 | 11 | AC44 | 47.09 | 0.23 | 48 |
| AC5 | 15.91 | 0.83 | 16 | AC25 | 22.45 | 0.83 | 23 | AC45 | 14.64 | 0.87 | 14 |
| AC6 | 27.36 | 0.67 | 28 | AC26 | 6.00 | 1.08 | 4 | AC46 | 45.45 | 0.24 | 46 |
| AC7 | 52.64 | 0.18 | 55 | AC27 | 6.64 | 1.04 | 5 | AC47 | 56.64 | 0.12 | 58 |
| AC8 | 45.09 | 0.22 | 45 | AC28 | 30.27 | 0.57 | 30 | AC48 | 54.18 | 0.17 | 56 |
| AC9 | 40.00 | 0.34 | 39 | AC29 | 17.18 | 0.83 | 19 | AC49 | 35.00 | 0.40 | 34 |
| AC10 | 38.73 | 0.37 | 38 | AC30 | 14.18 | 0.91 | 12 | AC50 | 19.36 | 0.81 | 21 |
| AC11 | 36.27 | 0.37 | 36 | AC31 | 25.36 | 0.71 | 26 | AC51 | 40.09 | 0.28 | 40 |
| AC12 | 10.27 | 0.87 | 8 | AC32 | 14.45 | 0.88 | 13 | AC52 | 50.73 | 0.19 | 52 |
| AC13 | 30.82 | 0.53 | 31 | AC33 | 48.09 | 0.23 | 50 | AC53 | 54.55 | 0.15 | 57 |
| AC14 | 37.73 | 0.31 | 37 | AC34 | 51.18 | 0.19 | 53 | AC54 | 12.00 | 0.86 | 10 |
| AC15 | 25.73 | 0.66 | 27 | AC35 | 10.64 | 0.94 | 9 | AC55 | 44.82 | 0.24 | 44 |
| AC16 | 35.36 | 0.42 | 35 | AC36 | 5.45 | 1.12 | 3 | AC56 | 23.73 | 0.76 | 25 |
| AC17 | 23.55 | 0.78 | 24 | AC37 | 3.09 | 1.16 | 1 | AC57 | 47.36 | 0.23 | 49 |
| AC18 | 52.18 | 0.19 | 54 | AC38 | 18.91 | 0.83 | 20 | AC58 | 44.45 | 0.25 | 43 |
| AC19 | 56.82 | 0.13 | 59 | AC39 | 16.36 | 0.90 | 17 | AC59 | 41.82 | 0.27 | 41 |
| AC20 | 46.45 | 0.24 | 47 | AC40 | 4.64 | 1.04 | 2 | | | | |

**Table 4.2.4 Rank of barley accessions measured by stress tolerance index (STI) and the average number of ranks (ASR) relying on germination percentage and seedling growth characteristics under both stress conditions (T10.25 and T20.50). Accessions with the highest STI and lowest ASR values were deemed to be the best accessions and the lowest rank was granted to the best performance of each accession. The complete name of the accession is mentioned in Table 3.1.1.**

| Accession Code | ASR | STI | Rank | Accession Code | ASR | STI | Rank | Accession Code | ASR | STI | Rank |
|---|---|---|---|---|---|---|---|---|---|---|---|
| AC1 | 42.00 | 0.47 | 42 | AC21 | 13.91 | 0.99 | 12 | AC41 | 10.36 | 0.94 | 8 |
| AC2 | 30.91 | 0.75 | 31 | AC22 | 35.82 | 0.54 | 35 | AC42 | 16.18 | 0.91 | 16 |
| AC3 | 16.55 | 0.92 | 17 | AC23 | 53.27 | 0.33 | 55 | AC43 | 28.55 | 0.78 | 28 |
| AC4 | 18.27 | 0.93 | 20 | AC24 | 15.00 | 0.90 | 13 | AC44 | 49.45 | 0.35 | 50 |
| AC5 | 18.45 | 0.86 | 21 | AC25 | 17.27 | 1.03 | 18 | AC45 | 11.82 | 1.00 | 9 |
| AC6 | 25.00 | 0.85 | 27 | AC26 | 4.73 | 1.15 | 4 | AC46 | 40.64 | 0.49 | 41 |
| AC7 | 56.64 | 0.29 | 58 | AC27 | 4.00 | 1.13 | 3 | AC47 | 58.55 | 0.24 | 59 |
| AC8 | 44.18 | 0.37 | 45 | AC28 | 29.45 | 0.71 | 29 | AC48 | 48.73 | 0.42 | 48 |
| AC9 | 36.45 | 0.66 | 36 | AC29 | 20.73 | 0.89 | 23 | AC49 | 40.27 | 0.53 | 40 |
| AC10 | 33.09 | 0.69 | 33 | AC30 | 10.27 | 1.07 | 7 | AC50 | 15.91 | 0.92 | 15 |
| AC11 | 34.82 | 0.61 | 34 | AC31 | 24.09 | 0.85 | 25 | AC51 | 51.00 | 0.34 | 52 |
| AC12 | 13.45 | 0.90 | 11 | AC32 | 18.00 | 0.94 | 19 | AC52 | 54.82 | 0.31 | 56 |
| AC13 | 32.27 | 0.70 | 32 | AC33 | 52.18 | 0.30 | 53 | AC53 | 55.55 | 0.29 | 57 |
| AC14 | 43.18 | 0.44 | 44 | AC34 | 42.18 | 0.54 | 43 | AC54 | 22.73 | 0.82 | 24 |
| AC15 | 30.36 | 0.71 | 30 | AC35 | 9.36 | 1.03 | 6 | AC55 | 52.18 | 0.33 | 54 |
| AC16 | 37.91 | 0.62 | 39 | AC36 | 3.91 | 1.22 | 2 | AC56 | 24.64 | 0.86 | 26 |
| AC17 | 19.36 | 0.95 | 22 | AC37 | 2.73 | 1.21 | 1 | AC57 | 37.64 | 0.52 | 37 |
| AC18 | 50.27 | 0.41 | 51 | AC38 | 15.18 | 0.95 | 14 | AC58 | 45.91 | 0.44 | 47 |
| AC19 | 44.45 | 0.44 | 46 | AC39 | 12.73 | 1.07 | 10 | AC59 | 48.91 | 0.38 | 49 |
| AC20 | 37.73 | 0.66 | 38 | AC40 | 6.00 | 1.07 | 5 | | | | |

**Table 4.2.5 Average values of studied traits under control, T10.25 and T20.50 conditions for best performance accession and least performance accession.**

| Trait | Best performance accession | Least performance accession |
|---|---|---|
| **Phenotypic trait** | ABN | Black-Garmiyan |
| GP | 97.22a | 44.44b |
| RL | 8.52a | 4.30b |
| SL | 7.99a | 5.68b |
| SEL | 16.51a | 9.98b |
| SV | 1607.74a | 549.71b |
| **Physiological trait** | | |
| WU-40 | 123.30a | 91.36b |
| WU-64 | 224.54a | 101.10b |
| **Biochemical trait** | | |
| TPC | 91.91a | 45.79b |
| TFC | 5.89b | 9.72a |
| DPPH | 444.09a | 407.84b |
| SSC | 473.77a | 325.59b |
| PC | 665.22a | 364.73b |
| GPA | 1.89a | 1.77b |
| CAT | 26.53a | 2.35b |

### 4.2.1.3 Relationship analysis of traits under control and induced drought conditions

#### *Association among different characters*

The Pearson correlations (r) of the studied traits under non-stressed and stressed conditions are determined from their mean values and described in Fig.4.2.12. Under the control conditions, 31 significant positive and negative r values (between 0.28 and 0.89) have been registered, 2 for WU-40, 5 for WU-64, 4 for GP, 5 for RL, 4 for SL, 3 for SEL, 2 for SV, 3 for TPC, 1 for SSC, and 2 for PC of the 8-day-old seedlings. Positive correlations were noted for most characteristics tested. EL displayed significant and negative associations with the RL, SL, SEL, SV, PC, and CAT traits (r = -0.29*, $p$ = 0.027; r = -0.28*, $p$ = 0.033; r = -0.33**, $p$ = 0.009; r = -0.28*, $p$ = 0.03; r = 0.32*, $p$ = 0.012; r = -0.63***, $p$ = 0.000 and r = -0.35**, p = 0.007, respectively). Similarly, SEL stated strong positive correlations with SV (r = 0.89***, $p$ = 0.0001) and RL (r = 0.89***, $p$ = 0.0001). There are positive and negative associations between the characters studied under the induced drought of T10.25 treatment. Some phenotypic and biochemical characteristics are significantly correlated. Furthermore, 39 important r-values (significant and highly significant) with a diversity of 0.26-0.88 were reported. Out of 39 significant r-values, only 10 revealed highly significant r values were ≥ 0.5 for GP, SEL, SV, DPPH, and CAT traits. GP and WU-64 also correlated significantly with most of the traits. GP and SV had the highest positive significant r-value (r = 0.88***, $p$ = 0.000), followed by SL and SEL (r = 0.78***, *p* = 0.000), SV and SEL (r = 0.76***, *p* = 0.000). The negative correlation between WU-64 and TFC (r= -0.32*, *p* = 0.013) was significant.

Correlation coefficient calculations under greater osmotic stress (T20.50) showed there were significant positive and negative correlations among observed traits. A total of 81 significant r values were detected among the observed characters. The r values varied from 0.26 to 0.96. GP correlated significantly with 12 characters, WU-40, WU-64, GP, RL, SL, SEL, SV, TPC, DPPH, SSC, PC, GPA, CAT, EL, and RWC, followed by RL, which was aligned with 10 traits, WU-40, WU-64, GP, RL, SL, SEL, SV, TPC, SSC, PC, GPA, CAT, EL, and RWC. A strong positive significant association between SEL and RL traits (r = 0.96***, p = 0.000) followed by SEL and SL (r = 0.91***, p = 0.000), SV and GP (r = 0.91***, p = 0.000) was observed.

***Relationship between different traits and accessions***

Statistical methods like drought tolerance indices, the correlation between biomass and physiological characteristics, and biplot analysis have been used to classify the genotypes that react to drought stress conditions (Grzesiak *et al.*, 2018).

To further clarify the quantitative relationship between morphological, physiological, and biochemical parameters, PCA was conducted using the stressed and unstressed plant values of morphological, biochemical, and physiological characteristics to assess their involvement in drought tolerance. In each principal component, the differential influence of the variables is calculated by the association between each variable and the main component. As marked in Fig. 4.2.13A, under the optimal condition, the first primary component (PC1) described roughly 28.40% of the variation and was positively influenced by phenotypic traits (GP, RL, SL, SEL, and SV), physiological traits (WU-64), and biochemical parameters (CAT), which were positively correlated with the first component. The second main component (PC2) explained 12.25% of the variance, which was positively correlated with TPC and negatively associated with PC trait.

At T10.25 stress, PC1 and PC2 were able to hold together 48.55% (32.43% by PC1 and 16.12% by PC2) of the initial variation (Fig. 4.2.13B). Among the measured traits in PC1, GP, SL, SEL, SV, WU-64, and CAT were responsible for the maximum variation, while RL, TPC, DPPH, and PC were liable for the highest variances in PC2. GP, SL, SEL, SV, WU-64, and CAT were responsible for the differentiation of MORA (AC36), ABN (AC37), Bhoos-H1 (AC40), BN2R (AC41), BN6 (AC12), 16 HB (AC5), Bhoos-912 (AC21), CANELA (AC26), and MSEL (AC27), situated on the right side of the PC1. Those accessions were characterized by the highest values of GP, SL, SEL, SV, WU-64, and CAT and were considered the most tolerant plants. Alternatively, the accessions located on the left side of the PCA plot, especially Shoaa (AC1), Al-warka (AC7), Black-Bhoos Akre (AC18),

Black-Zaxo (AC19), Black -Kalar (AC44), Abrash (AC51), Black-Garmiyan (AC47), Rehaan (AC55), Black-Chiman (AC48), and Bujayl 2-Shaqlawa (AC53), were categorized by the lowest values of GP, SL, SEL, SV, WU-64, TPC, DPPH, PC, SSC, GPA, and CAT, considered the lowest tolerant accessions under T10.25 conditions. Under the T20.50 condition, the first two components jointly explained 53.03% of the observed variation; therefore, they are represented in a two-dimensional space (Fig. 4.2.13C). PC1 plotted on the horizontal axis accounted for the largest proportion of variance (41.10%), while PC2 plotted on the vertical axis represented an additional 11.93% of the variance. The maximum variation along PC1 was explained by GP, RL, SL, SEL, SV, TPC, SSC, PC, CAT, and RWC, which appeared to be positively associated with drought stress, as they were placed on the right side of the horizontal axis reflecting PC1. GP, RL, SL, SEL, SV, TPC, SSC, PC, CAT, and RWC were then considered responsible for the discrimination of the tolerant-plant group, located on the right side of PC1. Therefore, this group was distinguished by accessions with a higher potential to germinate under adverse conditions and with the greatest values of GP, RL, SL, SEL, SV, TPC, SSC, PC, CAT, and RWC. Furthermore, the PCA graph discriminated the accessions that tended to the left side of the biplot, namely Al-warka (AC7), Black-Garmiyan (AC47), Samr (AC23), Black-Bhoos Baghdad (AC33), Black-Bhoos Akre (AC18), Irani (AC34), Bujayl 1-Shaqlawa (AC52), and IBAA-995 (AC59), whose accessions were considered to be the group of sensitive accessions and were described as having the lowest ability to germinate under low water availability. Moreover, the loading plot of various variables indicated positive correlations between growth, biochemical, and RWC traits, which correlated negatively with the EL character.

#### 4.2.1.4 Key driver analysis of different studied characters

The key driver analysis, or analysis of relative importance, determines the importance of each physiological and biochemical trait (independent variable) in predicting a phenotypic trait (dependent variable). Each predictor is commonly referred to as a driver and explores the correlations between the variables.

***Importance of the physiological and biochemical traits under T10.25 stress***

WU-64, PC, GPA, and CAT rated favorable relative values for GP in order from highest to lowest: WU-64 > PC > GPA > CAT, while WU-40 and TPC had slightly negative scores (Fig. 4.2.14A). WU-64 and TFC had the highest positive (47.73%) and negative (-4.68%) values for seed germination, respectively, indicating that WU-64 had the maximum influence in determining GP and increasing TFC levels, thereby reducing GP germination rates (Winkel-Shirley, 1998, Winkel-Shirley, 2001). Regarding the RL characteristic, WU-64, TPC, SSC, PC, GPA, and CAT evidenced positive relative significance, while WU-40, TFC, and DPPH expressed negative values (Fig. 4.2.14B). CAT reported the largest positive effects (33.87%), followed by SSC (16.08%), while WU-40 (-13.15%) documented the highest negative impact. Four positive and five negative relative importance scores were reported as effective physiological and biochemical characteristics of the SL trait (Fig. 4.2.14C). The greatest positive effects on SL were observed in WU-64, CAT, and GPA, where the negative impacts of WU-40 and PC on the SL character were detected.

***Effectiveness of the physiological and biochemical characters under T20.50 stress***

All physiological and biochemical characteristics had a positive effect on GP.The ranking of characteristics from the most important to the least important for the GP trait was as follows: PC > CAT > TPC > WU-64 > DPPH > SSC > WU-40 > GPA > TFC (Fig. 4.2.14A). Six positive and three negative values were observed for RL characteristics. The order of the positive values from the highest to the minimum effect was as follows: CAT > GPA > TPC > PC > SSC > WU-40 (Fig. 4.2.14B). Conversely, the rank of independent variables, in terms of their effects on RL, ranged from the greatest negative influence to the lowest negative influence as follows: WU-64 > TFC > DPPH. Relating SL, seven positive importance values for WU-40, WU-64, TPC, SSC, GPA, CAT, and PC were detected (Fig. 4.2.14C). TFC and DPPH stated the negative impacts on SL. The ranking of characters from the positive greatest importance to the positive lowest importance for SL was as follows: WU-40 > WU-64 > TPC SSC > GPA > CAT > PC. These results suggest that PC, CAT, TPC, and WU traits were more important for GP, RL, and SL characters under the T20.50 condition. Effective screening to differentiate between drought-tolerant and drought-sensitive genotypes based on easily measurable traits can help to investigate the stress capacity of barley accessions. The reaction of plants to drought is the product of dynamic and diverse adaptations. Consequently, it should be studied at a level ranging from morphological and physiological changes in organs to complex reactions to gene expression and regulation rates and biochemical reactions at the cell and organ level. The present study showed natural variations between the studied accessions, which exhibited different growth characteristics in the control and stress conditions of the fifty-nine accessions, and significant decreases were noted as the concentration of PEG increased. These results are in line with previous studies and provide further evidence of the suitability of PEG as a molecule for simulating droughts under in vitro conditions. These findings are consistent with previous research and show the response to droughts depends on both genotypes and levels of stress (Cai *et al.*, 2020). It is generally known that the first step of drought-imposed moisture deficiency is disrupted germination, resulting in poor plant standing at the early seedling level and impeding early crop development. It was noticed that a decrease in germination percentage and seedling growth with a rise in PEG concentrations (reduction in water potential) may be attributed to low environmental hydraulic conductivity and the high viscosity of the medium in which the solubility and diffusion of oxygen have been decreased, whereby PEG 6000 causes water to be unavailable to seeds and disrupts seed imbibition processes that are important for germination. The reduction of seedling growth is the result of cell division and enlargement restrictions, which are due to reduced water absorption (Fraser, Silk and Rost, 1990; Wu *et al.*, 2019; Basal, Szabó and Veres, 2020; Koskosidis *et al.*, 2020). The outcomes of the present research were in line with the reporting of the (Hellal *et al.*, 2018; Thabet *et al.*, 2018; Xue *et al.*, 2019). For both stress states, the resistance/susceptibility of the accessions was assessed according to their germination parameters as follows: PEG-resistant > susceptible, ABN (AC37) > MORA (AC36) > MSEL (AC27) > …. > Bujayl 2-Shaqlawa (AC53) > Al-warkaa (V7) > Black-Garmiyan (AC47). Under both stress environments, ABN grew better than the other accessions, exhibiting greater tolerance to osmotic stress, while Black-Garmiyan displayed the lowest germination and growth performance accession under drought stress treatments. Many methods are available to investigate the reactions of plants to water shortages. One of these methods is the water absorption process, which plays a role in breaking down stored starch and protein into soluble sugars to provide energy and nutrients for germinating seeds. So, we hypothesized the water uptake trend during seed germination could provide valuable information on early growth stage mechanisms of PEG tolerance in barley. It has been observed that a decline in seed water uptake with an improvement in PEG concentrations can be due to the high viscosity of the medium in which the diffusion of water has decreased, whereby PEG 6000 renders water inaccessible to seeds and interrupts seed imbibition processes.

Dehydration of plant tissues has a detrimental effect on many processes in biosynthetic pathways, like chlorophyll synthesis, functional and structural adjustments in chloroplast, disruption in the aggregation and distribution of assimilation products, and even on some biological activities, in specific changes in the capacity of leaf water and photosynthesis. The reduction in RWC in this study initially leads to a stomatal closure, resulting in a decrease in the supply of $CO_2$ to the mesophyll cells and, subsequently, a decrease in photosynthesis rates (Flexas and Medrano, 2002; Lawlor and Cornic, 2002; Zhang, 2011). Measuring electrolyte leakage in leaves is an effective way to evaluate the sensitivity to drought stress because drought induces the loosening of lamellar membranes in the chloroplast and raises coarse grain matrix rates (Palta, 1990; Grzesiak *et al.*, 2018). Oxidative stress is one of the significant causes of reduced development under stress conditions. Lipid peroxidation, also used to measure oxidative stress damage in plants, is one of the major damages caused by oxidative stress. Membrane lipid peroxidation typically weakens the membranes, which allows minerals and other essential cellular metabolites to leak out of the cells, leading to cell death (Cruz

de Carvalho, 2008). Thus, a significant increase in EL, especially in plants under severe water stress, where the rate was about 3-fold relative to control plants, could be the main cause of growth impairment in barley under water deficit conditions.

TPC in this study was 2.58 and 3.69 times higher in T10.25 and T20.50 compared to normal conditions. Several researchers have documented elevated levels of antioxidant activity, polyphenols and flavonoids, with extreme drought stress. The bioactivity of phenolic molecules is considered a signal trigger that leads to protective mechanisms against ROS attack by acting as an antioxidant agent to reduce the generation of cell-damaging oxidants like free radicals (Akula and Ravishankar, 2011; Kooyers, 2015; Quan *et al.*, 2016; Tani *et al.*, 2019; Piasecka *et al.*, 2020).

In our results, SSC was boosted by 1.65-fold and 1.70-fold under T10.25 and T20.50 treatments, compared to T0.00. In osmotic adjustment, proline and soluble sugars were inferred as an adaptive reaction to drought stress (Bandurska *et al.*, 2017). SSC has several roles, including osmotic modification, carbon preservation, detoxification of reactive oxygen species, defense of membrane integrity, safety of DNA structures, and protein stabilization. In serious dehydrated conditions, sugars become an important water substitute, perhaps more than proline, for the hydration of proteins (Bowne *et al.*, 2012). In this study, PC was extended 1.81 and 3.31 times in both states of stress, compared to the optimal state. Proline is osmotic, plays a significant role in the stabilization of the membrane. It also works by scavenging free radicals and syncing the redox ability of the cells, which allows the plants to fight abiotic stress (Marcińska *et al.*, 2013).

As the results show, CAT activity improved 1.79 times in T10.25 and 4.83 times in T20.50 compared to non-stress conditions. One of the first reactions of the plant to abiotic stress is the accumulation of ROS, which acts as a signal molecule inside the dynamic network of the process of stress response in the plant. Prolonged stress, however, may induce ROS aggregation at the plasma membrane, and resultant damage to cells. Therefore, to minimize ROS generation, the plant needs the up-regulation of antioxidant/detoxifying systems including APX, SOD, CAT, and POD (Barna *et al.*, 2003; Zhang *et al.*, 2015).

As suggested in this study, all traits (phenotypic, physiological, and biochemical characteristics) except TFC in the best performing accession ABN (AC37) were significantly greater than those obtained from the lowest performing accession Black-Garmiyan (AC47). High associations for WU-64, GP, RL, and SL, with the most traits, have been reported in this analysis under normal and stressful conditions, suggesting that these traits are, to an enormous degree, genetically regulated. Thus, concentrating on these variants will provide knowledge to determine the genetic diversity of seedling traits in barley accessions and to accurately test many genotypes in a limited period. Another primary goal of this research is to classify barley accessions according to their response to dry stress.

The 59 accessions were represented in the PCA scattered plot with the values of all the characters studied. The classification of the accessions of the PCA plot showed that the ABN and MORA accessions with the highest values of the best-studied traits and the Black-Garmiyan and Alwarka accessions with the lowest values of the parameters studied were grouped as two distinct classes. According to the results of the key importance analysis, WU, TPC, SSC, and CAT are rated as the most favorable relative importance traits for GP, RL, and SL under stress conditions. The positive and negative effects of proline in plants under stress conditions depend on the severity of the drought stress and the amount of proline. Proline at low concentrations had beneficial effects on seedling production, enhancing metablite process, providing proline exogeniously with higher doses acts as a detrimental factor and slows plant growth. (Roy *et al.*, 1993; Hayat *et al.*, 2012).

### 4.2.2 Confirmation of drought tolerance under plastic house condition

#### 4.2.2.1 Morphological traits responses

For all studies concerning drought resistance, identifying and assessing the genotypes for it is drought resistance is very essential (Kebede, Kang and Bekele, 2019). For this purpose, accurate phenotyping approaches, different genetic resources, and appropriate growth stages are crucial when drought stress may happen. To improve drought resistance in the collection of barley varieties under the present drought stress condition, the variations' responses morphologically and physiologically are the reflection of plant genetic diversity for drought resistance, and those genotypes that have adaptability to this type of stress can be chosen as candidate genetic resources that cope with the drought stress condition (Sallam *et al.*, 2019). For screening drought resistance, a series of morphological, physiological, and biochemical traits have been widely conducted, including root and shoot length (Mahalingam and Bregitzer, 2019), root and shoot fresh and dry weight (Boudiar *et al.*, 2020), relative water content (RWC) (Abdelaal *et al.*, 2020), chlorophyll fluorescence (Kalaji *et al.*, 2018), and the formation of yield (Feiziasl *et al.*, 2022).

All these parameters have been confirmed to be significantly associated with drought resistance in barley, and they have been conducted for the purpose of identifying drought-resistant barley genotypes in a breeding program (Sabagh *et al.*, 2019). From this point, these traits in this study were selected.

Screening drought resistance genotypes in most studies confirmed their outcomes by either generating different osmotic potential at the early plant growth stage using polyethylene glycol (PEG) (Hellal *et al.*, 2018; Rahim *et al.*, 2020) or by stopping the irrigation of water to plants grown in pots with soil in control conditions (plastic house) (Mahalingam and Bregitzer, 2019; Stevens, Jones and Lawson, 2021). In this investigation, both strategies were conducted to select the most appropriate barley accessions that could eliminate the adverse effect of drought stress conditions. Four barley accessions based on the outcomes from the screening of drought resistance barley accessions under PEG treatments (Lateef, Mustafa and Tahir, 2021) were selected for further verification of their response to drought stress (resistance and susceptibility) under plastic house conditions in which they showed dependable performance under the mentioned circumstances. In this regard, barley accessions AC53 and AC47 have been considered drought susceptible accessions, while both barley accessions AC36 and AC37 performed as drought resistant accessions.

Tillering stage (S1), flowering stage (S2), anthesis stage (S3), and all combinations of these stages were carefully chosen for testing the performance of selected barley accessions. ANOVA analysis was used to confirm the significance of the means, and Duncan test analysis was used to compare the significance of differences..

Regarding the analysis of differences under all stress treatments with respect to control conditions, the negative values were considered as percentage increments, while positive values reflected the decrease by the percentage of the studied trait. As indicated by ANOVA analysis in our investigation (Appendixes 2.7, 2.8, and 2.9), significant differences were observed for most studied traits under all individual stress treatments. In the presence of drought stress in the early growth stage (S1), significant differences were detected for most studied parameters, with the exception of relative water content (RWC), tiller no./plants, and spike no./plants. When plant materials were subjected to stress at the (flowering-S2), a similar response by two studied traits (RWC) and tiller no./plants was detected. When the four barley accessions reached the further growth stage (S3), again no significant differences were observed for (RWC), while no clear differences were shown for the values of Total Chlorophyll Content compared to the previous two growth stages.

Comparing the percentage values of selected barley accessions under drought and control conditions at the (S1) stage, considerable responses can be seen by barley accessions for studied traits (Table 4.2.6). As expected, an essential increase of the percentage value by the two barley accessions AC36 and AC37 under this particular stress condition was observed for the (length, fresh and dry weight) of root and shoot compared to their control. On the contrary, the two barley accessions AC47 and AC53 possessed no obvious changes in response to drought stress for the mentioned trait. Both barley accession AC36 and AC37, with a value of 7.46% and 6.92% , respectively, reduced the length of the spike, while higher reductions by the other two accessions, AC53 39.75% and AC47 24.31%, were documented. As can be clearly seen from (Table 4.2.6), the most critical investigated parameters that are closely related to the final yield productivity include spike weight, number of grains/spike, grain weight/spike, and finally the total yield productivity reduced but not as much as with control condition for both barley accessions AC36 and AC37, which are considered drought resistant barley accessions in comparison with the other two drought susceptible accessions, indicating strong establishment by AC36 and AC37 barley accessions after re-watering at this particular growth stage when exposed to stress condition.

When plant materials were exposed to stress conditions during the flowering stage (S2), a similar pattern of responses by barley accessions AC36 and AC37 were recorded for studied parameters including (root length, fresh and dry weight of shoot and root) as shown in (Table 4.2.7), in which all these parameters increased dramatically in comparison to untreated conditions. On the other hand, the barley accessions AC47 and AC53 negatively responded. However, reductions in the length of the shoot were detected in all studied accessions. For instance, a slight reduction in length was noticed for AC36 1.11% followed by AC37 3.70% while greater reductions by AC47 and AC53 were observed with values 8.43% and 7.52% respectively. The content of chlorophyll in this specific stage was reduced by all barley accessions, but AC36 9.80% and AC37 14.24% maintained higher chlorophyll content in comparison with AC47 72.19% and AC53 57.08% in control condition. In parallel with the previous stress stage, favorable responses were identified by barley accessions AC36 and AC37 for studied traits that were strongly related to final yield outcomes when accessions were studied in the case of withholding irrigation. Both barley accessions AC36 and AC37 dropped by %25 for spike number/plant in comparison with the well-watered condition, while a dramatic reduction for AC53 50% was detected, followed by AC47 36.11%. Severe reductions in spike length were observed by AC53 46.04% and AC47 19.81% whereas AC36 and AC37 somehow reserved more length with value 19.7% and 12.81%, respectively. Despite the fact that under the presence of water limiting conditions in this particular stage, an enormous decline in spike weight/plant was documented by AC47  41.40% and A53  37.93%, followed by AC37  15.37% and AC36  10.15%. The data presented in Table 4.2.7 revealed a significant decline in grain numbers for AC47 and AC53 with values 35.40% and 34.96%, respectively, and followed by AC36 21.92%, while a slight reduction in this trait was documented by AC37 8.47%. The weight of grain/spike was similarly reduced by AC47 46.62% and A53 37.29%. On the contrary, a small portion of reduction was perceived by AC36 9.44% and AC37 5.42%. At this particular growth stage, the stress condition caused the reduction of final yield by 31% for AC37 and 33.34% for AC36, while AC47 and AC53 were highly sensitive to drought stress conditions in which almost double redaction in final yield production was recorded compared to both AC36 and AC37 (Table 4.2.7).

After the flowering stage (Anthesis stage-S3), the selected barley accessions were exposed to drought stress conditions. The data presented in Table 4.2.8 revealed significant differences between studied accessions in response to water limiting conditions. The root length was boosted by 64.41% and 49.91% by both barley accessions AC36 and AC37, correspondingly, while a considerable decrease was noted by AC47 43.39%, and AC53, with a value of 14.95%. Among tested barley accessions, AC37 shortened it is own shoot length with a value of 19.66% in respect to the untreated condition under the presence of drought, followed by AC36 12.61%, AC47 3.65%, and AC53 0.91%. In comparison with the two previously mentioned conditions, the fresh weight of the root decreased in all studied accessions. A slight decrease in root length was spotted for both barley accessions AC36 6.92% and AC37 11.32%, while AC53 and AC47 possessed higher reductions for the studied parameter with a value of (71.52-62.27)% respectively. The dry weight in root of barley accessions AC36 and AC37 increased by 124.25%, followed by AC38 29.34%, and decreased by (65.98-39.49)% for AC53 and AC47, respectively. In this particular stress condition, the fresh weight of the shoot was reduced by all barley accessions. Barley accessions were reduced less than the control condition at AC36 10.88% and AC37 15.53%, whereas AC53 39.16% and AC47 25.49% were reduced more. A similar pattern of response was observed for shoot dry weight by studied materials in which AC37 had the lowest reduction of 3.09%, followed by AC36 4.11%, and AC47 11.24%, while the highest reduction was discovered by AC53 62.21%. Regarding the analysis of differences under the presence of water limiting conditions for tillering number by barley accessions, more than 50% reductions were observed by AC53, followed by AC37 41.33%, AC47 32.78% and AC36 25%. All accessions reduced the spike number/plant, as in (S2). AC36 and AC37 responded similarly in reduction for this trait, with values of (26.67–25.00)%, respectively, while AC47 33.33% and AC53 46.67% demonstrated a slightly higher reduction Table 4.2.11. A slight reduction in spike length was spotted by AC36 3.05% and AC37 4.74%, compared to their control. Meanwhile, a higher decrease for the length of the spike was observed by AC53 35.63% followed by AC47 15.76%. In the case of the existing drought, AC53 displayed the highest value of reduction for the weight of spike of 39.56%, followed by AC47 with a value of 34.09%. On the other hand, AC36, with a value of 4.69%, had the lowest reduction, and AC37, 19.83% came in second. In the presence of drought stress, AC37 6.19% and AC36 7.65% conserved more grain in their spikes than AC52 33.58% and AC47 27.58%. Drought stress negatively impacted grain weight/spike for both barley accessions AC47 42.11% and AC53 40.16%, while the negative effect was reduced for both barley accessions AC36 3.76% and AC37 7.75%. The total yield was well-preserved by both barley accessions AC37 17.70% and AC36 24.68% in respect to their control, while the greater reduction in total yield was detected in this particular growth stage for AC47 45.08% and AC53 36.70% (Table 4.2.8).

As indicated in our analysis, both barley accessions AC36 and AC37 maintained their performance and showed closer values to control levels under all individual stress conditions (S1, S2, and S3). In addition, they exhibited a minimum percentage of changes in related yield parameters compared to AC47 and AC53.

While studied materials experienced double stress conditions (S1+S2), (S1+S3), significant differences as shown by ANOVA (Appendices 2.10, and 2.11) were detected for most studied parameters. Under the presence of drought stress in (S1 and S2), the only traits which showed no significant differences among the tested barley accessions with respect to their control was tiller no./ plants. However, in response to this combination of stress conditions, the rest of the investigated traits showed clear differences. In the case of the present different stressor conditions (S1+S2), a similar pattern of responses by barley accessions as (S1) was identified for both traits (Tiller No./plants and Spike No./plants) in which no observable variations were noticed. On the other hand, positive variations for other studied traits were documented by barley accessions. In the case of withholding water at two particular growth stages (S1+S3), the relative water content again in comparison with (S1, S2, and S3) showed no significant differences. The Spike No./Plants was another trait similar to (S1) that did not show significant variations for this specific stress condition.

When four barley accessions were exposed to stress conditions into two growth stages, S1 and S2, a dramatic increase in some studied traits compared to (S1) and (S2) was responded to by two barley accessions, AC36 and AC37, including; (root length, fresh and dry weight of root, and fresh and dry weight of shoot) as stated in (Table 4.2.9), in which all these parameters were boosted dramatically in comparison with untreated conditions. On the other hand, the barley accessions AC47 and AC53 negatively responded, with the only exception of root length, in which slight increases were detected. However, shoot length was reduced in all studied accessions. For example, the higher reduction of shoot length stated by AC37 %23.62 followed by AC53 14.47%, while AC47 and AC36 reduced it by 2.03% and 3.08%, respectively. In contrast to the individual stress conditions, obvious changes in response to the combination of stress S1 and S2 by barley accessions were observed for the relative water content parameter in comparison to the control condition. Both accessions AC36 3.16% and AC37 5.07% maintained higher relative water content in comparison to the sensitive group, while AC53 8.38% and AC47 8.27% showed higher reduction for this specific trait. Similar to relative water content, despite the presence of reductions in the content of Chlorophyll by all barley accessions, AC36 and AC37 with values of (16.42–21.35)% respectively preserved more content of Chlorophyll, while AC47 75.61% and AC53 56.59% experienced greater reductions in this specific stress condition. For studied traits that are strongly linked to final yield outcomes, positive responses were recognized by barley accessions AC36 and AC37 in maintaining their performance under the present combination of drought stress S1 and S2 with respect to their control. Barley accessions AC36 and AC37 experienced a drop of (23.19-30.41)% for spike number/plant in comparison with the untreated condition, while a greater reduction in spike length was noticed by AC53 48.78% and AC47 36.11%. Barley accessions AC53 and AC47 were more sensitive to the combination of drought conditions and weight of spike than AC36 and AC37. The highest reduction of studied traits was recorded by AC53, in which it lost more than 50% of its spike weight, and AC47 came second with a value of 33.50%. On the contrary, both barley accessions AC36 and AC37 maintained a higher reduction of the studied trait by a value of (11.27-19.8) %, respectively. As stated in Table 4.2.9, clear effects of combinations of drought can be identified for both AC53 and AC47 for grain no./spike. They lost (39.54–27.94)% of their grains, indicating sensitivity of both accessions to drought stress, whereas AC36 and AC37 were kept at the level of (15.15-18.64)% of grain reduction. The percentage change of decline for spike weight by both accessions AC53 and AC47 in respect to the well-watered condition was high compared to AC36 and AC37. For instance, AC53 lost 48.61% of its own spike weight and AC47 by 33.25%, while minor reductions were documented by AC36 at 13.07% and AC37 at 19.75%. Regarding the analysis of total yield performance under the combination of stress conditions S1 and S2, interestingly small effects of reduction can be seen for both accessions AC36 29.98% and AC37 24.81%, while the effects were higher for AC47 47.24% and AC53 45.90% (Table 4.2.9).

Significant changes between studied accessions in response to water holding conditions were revealed when the plant materials were targeted twice for stress conditions at S1 and S3 (Table 4.2.10). The length of the root increased by 63.78% for AC36 and 42.75% for AC37, while a huge decrease was observed for AC47 35.49% and AC53 25.13%. The shoot length reduction by AC37 among tested barley accessions was 29.87%, and AC36 came in second with a value of 16.71%, while the reduction of shoot length by barley accession AC47 was 7.36%. Under the availability of stress conditions S1 and S3, AC36 increased the root fresh weight by 7.66% while the rest of the accessions responded in a way that reduced the weight of the shoot by 18.71%, 67.30%, and 83.32% for the barley accessions AC37, AC47, and AC53, respectively. The dry weight of the root is similar to all individual stress conditions increased by AC36 and AC37. AC36 was boosted by 32.64% and AC37 by 24.24%, while AC47 and AC53 had a lack of increase in the weight of accumulation of dry matter in the root. Instead, they reduced its accumulation by 40.08% and 83.13%, respectively. The fresh weight of the shoot was reduced by all barley accessions, similar to the S3 stress condition. Barley accessions had less reduction in comparison to the control condition: AC36 5.66% and AC37 23.00%, whereas AC53, with a value of 75.60%, and AC47, with a value of 37.57%, had more reduction. The dry weight of the shoot, as shown in Table 4.2.10, decreased by most barley accessions AC53 77.71%, AC47 15.44%, and AC37 8.04% with the only exception of AC36, in which it increased by 13.69%. For the content of total chlorophyll, AC37 had a greater reduction of 35.07%, followed by AC53 26.22%,

AC47 22.84%, and AC36 21.66%. A negative impact can be seen on the Tiller No./plant by all accessions under the presence of combination drought stress S1 and S3. AC36 was slightly affected by drought for the studied trait as the redaction was %23.33, followed by AC53 33.33%, AC37 41.67%, and AC47 64.81%. Likewise, all accessions possessed a reduction in spike length in which AC36 shortened its length by 4.91% and AC47 by 11.57% followed by AC37 18.68% and AC53 53.71%. Unlike previous stress conditions, AC36 increased the weight of the spike by 1.94% when compared to its relative control, whereas the reduction for this trait began to decrease by the other three accessions, with values of 35.51% for AC47, 36.91% for AC47, and 52.21% for AC53. AC36 maintained its grain loss per spike by the value 9.14% under combined drought conditions S1 and S3 compared to the other three accessions in which more than twenty-two percent of the studied parameter was reduced by AC37 and AC47 while the reduction increased by AC53 to reach 43.45%. AC53 and AC47 dropped the grain significantly under this stress condition, with values of (50.78-34.77)%, respectively. However, no huge differences in reduction were noted for AC36 and AC37, in which they possessed the redaction for this trait by 2.89% and 21.19% respectively. AC36 and AC37 significantly maintained total yield, which is the primary goal of any drought resistance study, with the former showing a 22.34% total yield reduction and the latter by 31.53%, while both accessions, AC53 and AC47, were considered very sensitive barley accessions for the combination of drought S1 and S3.As shown in Table 4.2.10, barley accession AC47 lost it is total yield by 64.28% and AC53 by nearly half compared to their control condition.

In the case of present double (S2+S3) and triple (S1+S2+S3) stress conditions, significant differences as shown by ANOVA (Appendixes 2.12, and 2.13) were discovered for most studied traits. In both case scenarios similar to (S1), three traits (Tiller No./plants, Spike No./plants, and relative water content) showed no significant differences by tested barley accessions in respect to their own untreated condition. However, in response to these combinations of stress conditions, the rest of the investigated traits displayed clear variances.

When plant materials were directed to double stress conditions S2 and S3 (Table 4.2.11), significant changes between studied barley accessions in response to control conditions were discovered. Similar to all previously mentioned stressor conditions, AC36 and AC37 showed an enormous increase in root length by the values of 142.11% and 58.80%, respectively, while contrary responses by AC53 and AC47 were detected, in which they both decreased root length by 40.06% and 35.45%, respectively. AC37, with an almost similar value as (S1+S3), shortens its own shoot length by 30.72% and AC36 comes after with a value of 19.44%, while no considerable reduction by AC53 and AC47 was detected, in which they both reduced the length of the shoot by 8.00%. Similar to S1, S2, and (S1+S2) stress conditions, the fresh weight of the root was improved by both accessions AC36 and AC37 by (31.86-29.50)% respectively. On the other hand, AC53 was more sensitive to this type of drought in which it experienced a great deduction in this trait by 94.47% and AC47 69.75% came after (Table 4.2.11). The resistant barley accessions established much better accumulations of dry matter after being exposed to this drought regime in parallel to all previous drought systems. AC36 almost nearly one-fold increased dry matter accumulations in the root compared to its control and AC37 exhibited a similar response as AC36 with an increase of 87.67%, while the sensitive group was more delicate to the combination of water limiting in which a notable reduction in root weight was observed by AC53 82.31% and AC47 46.44%. The fresh shoot weight by AC37 and AC36 started to decline from (S3) and (S1+S3), indicating the severity of these stages for this trait. Similar to these stages, both accessions did not preserve the shoot's fresh weight. AC37 reduced the shoot weight by more than half a fold compared to control, while AC36 was less affected for the studied trait in which it reduced by 18.12%. AC53 reduced shoot weight by 81.47% and AC47 by 26.64%. The dry weight of the shoot in AC53 was reduced in association with it is control by 81.27% and followed by AC37 49.10%, AC36 3.10%, and AC47 1.16%. Nearly half a fold of reductions were noted by AC37 and AC47 for Total Chlorophyll Content and AC53, with a value of 34.07% and AC36, with a value of 30.50%, came after. AC53 presented the greatest reduction in length of a spike by a value of 58.19%, followed by AC37 30.03%, AC47 14.99%, and AC36 9.16% (Table 4.2.11). AC53 is again considered as the most sensitive to this particular stress condition for losing the weight of spike since it reduced spike weight by 61.00% while the minimum reduction was documented by AC36 14.20 and an almost similar drop by AC47 32.35% and AC37 37.09% were detected. AC36 is less affected by the combination of S2 and S3 drought stress conditions.It dropped the grain number per spike by 9.25% next, AC37 with a value of 22.80%, AC47 33.30%, and AC53 46.87% came after. In comparisons with AC53 and AC47, differences in reduction for grain weight were not great in AC36 and AC37 as they possessed the redaction of 11.45% and 23.67%, respectively, while AC53 had the reduction value of 62.24%, followed by AC47 32.04%. The total yield was well-preserved by both barley accessions AC36 34.55% and AC37 36.51% in respect to their own control, while the greater reduction in total yield was noticed in this particular growth stage by AC53 66.47% and AC47 48.33% (Table 4.2.11).

When the plant materials were exposed to extremely serious, critical, and harsh drought stress conditions at (S1, S2, and S3) (Appendix 2.13), it was found that there were significant differences between the studied traits of barley accessions in how they responded to water holding conditions.

Among tested barley accessions, the length of root increased by 136.29% for AC36 and 56.65% for AC37, while a huge decrease was observed by AC47 52.39% and AC53 35.38%. The biggest reduction of shoot length was possessed by AC37, in which it declined its own high by 32.42% and AC36 was considered the second studied accession that was affected by these multiple stress conditions in reducing this trait, in which 19.2% of reduction was noted in comparison with its own control while the third barley accession was AC53, with a value of 7.65% and the last accession was AC47, which performed by 10.19% of shoot reduction. Unlike all previous drought stress conditions, negative influence by both accessions AC36 and AC37 started to appear, in which both reduced the fresh weight of the root by 14.25% and 47.97%, respectively, while in comparison with AC53 92.28% and AC47 75.01%, both considered drought resistance accessions maintained their root fresh weight in considerable ways (Table 4.2.12). The most performed barley accession in terms of accumulation of dry matter was AC36, which increased the dry weight of root by 46.25% whereas the rest of the studied accessions were greatly reduced in the presence of multiple stress conditions. The reduction by AC37 of 25.74% was somehow much lighter than AC53 75.96% and AC47 58.59% for this trait. The dry weight of shoots similar to S2 and S3 was reduced by all studied plants, with AC53 experiencing the greatest reduction at 89.31% and AC37 experiencing a half-reduction in shoot dry weight in comparison to their own control (Table 4.2.12). As stated in the analysis for the total content of chlorophyll in this particular stress condition, all barley accessions were negatively affected. The maximum reduction in this trait was documented by AC37 %59.42, followed by AC47 61.57%, AC53 45.42%, and lastly, AC36 41.01%. The length of spike similar to S1, S2, S3, and (S1+S2) negatively affected by all accessions, AC53 with value 62.27% reached the peak of reduction while AC37, AC36 and AC47 come after with value (42.04,26.10 and 25.78)% respectively. All barley accessions under availability of multiple stress conditions dramatically reduced the weight of the spike. AC53 achieved the greatest reduction of 67.39%, followed by AC37, AC47, and AC36 with values of (51.78, 38.24, and 25.69)%. All accessions similar to all individual and combined stress conditions have reduced the grain number per plant. Accessions AC36 and AC37, in comparison with AC47 and AC53, are relatively less affected by the triple stress conditions. AC36 had a 27.85%, AC37 had a 32.21%, and AC53 and AC47 had a (56.06 and 38.62)% decrease, respectively. The grain weight per spike at multiple stress conditions enormously dropped by AC53 and AC47, with values of (67.53 and 40.36)%, respectively. However, no huge differences in reduction were noted for AC36 and AC37, in which they possessed the redaction for this trait by 16.81% and 25.65%, respectively.

As shown in Table 4.2.12, barley accession AC53 lost it is total yield by 74.00% and AC47 by 68.99% compared to their own control condition. However, similar to all stress conditions barley accessions AC36 and AC37 maintained a higher proportion of yield lost by the value (44.06 and 45.10) %, respectively. All these results revealed that AC36 and AC37 again similar to all individual stress conditions under the combination of drought stress even under the availability of multiple stress conditions tolerated more drought stress in comparison with AC53 and AC47. Significant differences as shown by ANOVA were discovered for most studied physiological traits based on all individuals and combinations of drought stress conditions, while RWC, Total Chlorophyll Content, and Tiller No./p showed no significant differences based on the data obtained for four barley accessions. The slight changes were observed when the data was subjected to analysis to detect the interactions between all stages and studied barley accessions, as most of the studied traits showed variations in response and were significant, except for four traits: RWC, Tiller No./p, Spike No./p., and No. of grain/spike (Appendix 2.14). Based on all available data obtained from all drought stressors conditions for studied physiological traits in respect to control conditions (Table 4.2.13), AC36 with value of 67.16% showed the maximum increase for root length, while AC47 was most sensitive accession as 18.90% of this trait was reduced under all stressor conditions. Regarding the interactions between studied barley accessions and drought stressor conditions, the maximum reduction of root length was documented under stage 3*AC47 with a value of 43.39%, while the interactions between stage (2+3) and V36 showed the highest increase of the length of root by 142.11%. Shoot length under all stressor conditions was reduced by all studied barley accessions and the values ranged between (14.86 and 5.46)% for AC37 and AC36, respectively. The combination of stage (2+3) caused the highest reduction by studied barley accessions for this trait with a value of 16.56%, while under the availability of water stress in stage 1, only 3.43% of reduction was noted. The interactions between Stage 1 and AC53 showed the maximum decrease of shoot by 43.28%. On the other hand, the shoot length increased by 18.42% in the case of interactions between stage 1 and AC37. Root fresh weight increased by 81.64% for accession AC37, while among all four barley accessions, AC53 had the maximum reduction of its root fresh weight as 52.27% of reduction was documented. The value was boosted to reach a value of 170.72% of increase in the case of stage (1+2) with AC37, while stage (2+3) with AC53 had the highest influence in reducing the root fresh weight to reach a value of 94.47% of reduction. The rest of physiological parameters (Root dry weight, shoot fresh weight, shoot dry weight, RWC %, Total chlorophyll content, Tiller number/plant, spike number/plant, spike length, grain weight/spike, spike weight, number of grain/spike and total yield/plant) for the percentage responses by the four barley accessions under availability of all drought stress conditions in respect to control conditions, the values for accessions, drought stages and the interactions among them ranged between (-89.15-AC37, -57.26 - S. 1, -193.81-S. 1+2 × AC37 and 63.03-AC53, 16.58 - S. 1+3, 83.13 - S. 1+3 × AC53), (-17.29-AC37, -33.74 - S. 1, -142.20 - S. 1 × AC37 and 61.10-AC53, 45.7 - S. 2, 81.47 - S. 2+3 × AC53), (-74.02-AC37, -40.24 - S. 1, S. -213.64 – S. 1+2 × AC37 and 62.81-AC53, 33.66 - S. 2+3, 79.58 - S. 1+2 × AC53), (4.44-AC36, 3.15 - S. 1, 2.04 - S. 2 × AC53 and 5.03-AC53, 6.22-S. 1+2, 8.38 - S. 1+2 × AC53), (13.81-AC36, -52.78 - S. 1, -121.07 - S. 1 × AC47 and 21.21-AC53, 42.49 - S. 1+2, 75.61 - S. 1+2 × AC47), (39.76-AC37,

44.72-S. 1+2, 64.81 - S. 1+3 × AC47 and 31.30-AC36, 16.90 - S. 1, 8.33 - S. 1 × AC53), (39.26-AC53, 39.17 - S. 2+3, 50.00 - S. 2 × AC53 and 26.30-AC36, 20.56 - S. 1, 16.67 - S. 1 × AC36), (47.02-AC53, 34.62 - S. 1+2, 58.19 - S. 2+3 × AC53 and 11.16-AC36, 14.80 - S. 3, 4.91 - S. 1+3 × AC36), (45.09-AC53, 36.16 - S. 2+3, 61.00 - S. 2+3 × AC53 and 7.15-AC36, 15.06 - S. 1, -1.94 - S. 1+3 × AC36), (37.38-AC53, 28.05 - S. 2+3, 43.45 - S. 1+3 × AC53 and 12.25-AC36, 16.91 - S. 1, 6.19 - S. 3 × AC37), (44.99-AC53, 32.35 - S. 2+3, 62.24 - S. 2+3 × AC53 and 7.11-AC36, 16.29 - S. 1, 2.03 - S. 1 × AC36), and (52.85-AC47, 48.93 - S. 2, 67.15 - S. 2 × AC47 and 27.08-AC36, 31.04 - S. 3, 17.58 - S. 1 × AC36), respectively.

**Table 4.2.6 The percentage responses of some physiological traits in drought stress condition (S1) with respect to control condition in four barley accessions based on Multiple Rang Duncan's test at *p* value < 0.05. Any values of mean holding a common letter are not significantly different.**

| Physiological traits | AC53 | AC47 | AC36 | AC37 | Significant |
|---|---|---|---|---|---|
| Root length-(cm) | 14.93 a | 4.34 a | -35.19 b | -49.18 b | Yes |
| Shoot length-(cm) | 43.28 a | 9.07 b | -20.20 c | -18.42 c | Yes |
| Root fresh weight-(g) | -75.49 a | -81.01 a | -121.36 a | -267.58 b | Yes |
| Root dry weight-(g) | 4.06 a | -26.45 ab | -92.14 bc | -114.51 c | Yes |
| Shoot fresh weight-(g) | 46.04 a | 2.92 b | -41.74 c | -142.20 d | Yes |
| Shoot dry weight-(g) | 39.38 a | 12.68 b | -48.11 c | -164.91 d | Yes |
| RWC % | 4.42 a | 3.67 a | 2.23 a | 2.26 a | No |
| Total Chlorophyll Content | -67.97 b | -121.07 c | -8.67 a | -13.42 a | yes |
| Number of tillers/plant | 8.33 a | 9.26 a | 25.00 a | 25.00 a | no |
| Number of spike/plant | 19.44 a | 27.78 a | 16.67 a | 18.33 a | no |
| Spike length-(cm) | 39.75 a | 24.31 b | 7.46 c | 6.92 c | Yes |
| Spike weight-(g) | 29.47 a | 17.10 b | 4.55 c | 9.13 c | Yes |
| Number of grain/spike | 25.86 a | 22.18 a | 10.39 b | 9.22 b | Yes |
| Grain weight/spike-(g) | 30.85 a | 22.87 ab | 2.03 c | 9.41 bc | Yes |
| Total yield/plant-(g) | 48.22 a | 45.00 a | 17.58 b | 22.68 b | Yes |

**Table 4.2.7 The parentage responses of some physiological traits in drought stress condition (S2) in four barley accessions with respect to control condition based on Multiple Rang Duncan's test at *p* value < 0.05. Any values holding a common letter are not significantly different.**

| Physiological traits | AC47 | AC53 | AC36 | AC37 | Pr > F(Model) | Significant |
|---|---|---|---|---|---|---|
| Root length-(cm) | 5.81 a | 3.62 a | -45.83 b | -44.95 b | < 0.0001 | Yes |
| Shoot length-(cm) | 8.43 a | 7.52 a | 1.11 b | 3.70 ab | 0.04 | Yes |
| Root fresh weight-(g) | 55.55 a | 58.54 a | -22.84 b | -52.06 c | < 0.0001 | Yes |
| Root dry weight-(g) | 60.92 a | 60.26 a | -63.59 b | -85.31 c | < 0.0001 | Yes |
| Shoot fresh weight-(g) | 31.61 b | 41.83 a | -12.19 c | -13.21 c | < 0.0001 | Yes |
| Shoot dry weight-(g) | 51.88 a | 36.73 b | -79.10 c | -125.80 d | < 0.0001 | Yes |
| RWC % | 3.11 a | 2.04 a | 3.01 a | 5.38 a | 0.35 | No |
| Total Chlorophyll Content | 72.19 a | 57.08 b | 9.80 c | 14.24 c | < 0.0001 | Yes |
| Number of tillers/plant | 56.48 a | 41.67 a | 41.67 a | 25.00 a | 0.41 | No |
| Number of spike/plant | 36.11 b | 50.00 a | 25.00 b | 25.00 b | 0 | Yes |
| Spike length-(cm) | 19.81 b | 46.07 a | 19.17 b | 12.81 b | 0.01 | Yes |
| Spike weight-(g) | 41.40 a | 37.93 a | 10.15 b | 15.37 b | 0.01 | Yes |
| Number of grain/spike | 35.40 a | 34.96 a | 21.92 ab | 8.47 b | 0.02 | Yes |
| Grain weight/spike-(g) | 46.62 a | 37.29 b | 9.44 c | 5.42 c | < 0.0001 | Yes |
| Total yield/plant-(g) | 67.15 a | 64.22 a | 33.34 b | 31.00 b | 0 | Yes |

**Table 4.2.8 The percentage responses of some physiological traits in drought stress condition (S3) with respect to control condition in four barley accessions based on Multiple Rang Duncan's test at *p* value < 0.05. Any values of mean holding a common letter are not significantly different.**

| Physiological traits | AC53 | AC47 | AC37 | AC36 | Pr > F (Model) | Significant |
|---|---|---|---|---|---|---|
| Root length-(cm) | 14.95 a | 43.39 a | -49.91 b | -64.41 b | 0 | Yes |
| Shoot length-(cm) | 0.91 c | 3.65 c | 19.66 a | 12.61 b | < 0.0001 | Yes |
| Root fresh weight-(g) | 71.52 a | 62.27 a | 11.32 b | 6.92 b | < 0.0001 | Yes |
| Root dry weight-(g) | 65.98 a | 39.49 b | -29.34 c | -124.25 d | < 0.0001 | Yes |
| Shoot fresh weight-(g) | 39.16 a | 25.49 b | 15.53 c | 10.88 c | 0 | Yes |
| Shoot dry weight-(g) | 62.21 a | 11.24 b | 3.09 c | 4.11 c | < 0.0001 | Yes |
| RWC % | 4.20 a | 2.77 a | 3.40 a | 6.99 a | 0.57 | No |
| Total Chlorophyll Content | 21.28 a | 15.60 a | 19.79 a | 13.17 a | 0.19 | No |
| Number of tillers/plant | 52.38 a | 32.78 b | 41.33 ab | 25.00 b | 0.02 | Yes |
| Number of spike/plant | 46.67 a | 33.33 ab | 25.00 b | 26.67 b | 0.04 | Yes |
| Spike length-(cm) | 35.63 a | 15.76 b | 4.74 c | 3.05 c | 0 | Yes |
| Spike weight-(g) | 39.56 a | 34.09 a | 19.83 b | 4.69 c | 0 | Yes |
| Number of grain/spike | 33.58 a | 27.58 a | 6.19 b | 7.65 b | 0 | Yes |
| Grain weight/spike-(g) | 40.16 a | 42.11 a | 7.75 b | 3.76 b | < 0.0001 | Yes |
| Total yield/plant-(g) | 36.70 a | 45.08 a | 17.70 b | 24.68 b | 0 | Yes |

**Table 4.2.9 The percentage responses of some physiological traits in combination of drought stress conditions (S1+S2) with respect to control condition in four barley accessions based on Multiple Rang Duncan's test at *p* value < 0.05. Any values of mean holding a common letter are not significant.**

| Physiological traits | AC53 | AC47 | AC37 | AC36 |
|---|---|---|---|---|
| Root length-(cm) | -2.06 a | -11.10 a | -85.84 c | -51.67 b |
| Shoot length-(cm) | 14.47 b | 2.03 c | 23.62 a | 3.08 c |
| Root fresh weight-(g) | 81.28 a | 76.04 a | -170.72 c | -51.26 b |
| Root dry weight-(g) | 82.42 a | 80.32 a | -193.81 c | -81.92 b |
| Shoot fresh weight-(g) | 82.51 a | 52.79 b | -43.43 d | -38.68 c |
| Shoot dry weight-(g) | 79.58 a | 69.30 b | -213.64 d | -154.72 c |
| RWC % | 8.38 a | 8.27 a | 5.07 ab | 3.16 b |
| Total Chlorophyll Content | 56.59 b | 75.61 a | 21.35 c | 16.42 c |
| Number of tillers/plant | 41.67 a | 40.00 a | 55.56 a | 41.67 a |
| Number of spike/plant | 41.67 a | 44.44 a | 25.00 a | 33.33 a |
| Spike length-(cm) | 48.78 a | 36.11 ab | 30.41 b | 23.19 b |
| Spike weight-(g) | 50.37 a | 33.50 b | 19.18 c | 11.27 c |
| Number of grain/spike | 39.54 a | 27.94 ab | 18.64 bc | 15.15 c |
| Grain weight/spike-(g) | 48.61 a | 33.25 ab | 19.75 bc | 13.07 c |
| Total yield/plant-(g) | 45.90 a | 47.24 a | 24.81 b | 29.98 b |

**Table 4.2.10 The percentage responses of some physiological traits in combination of drought stress conditions (S1+S3) with respect to control condition in four barley accessions based on Multiple Rang Duncan's test at *p* value < 0.05. Any values of mean holding a common letter are not significant.**

| Physiological traits | AC53 | AC47 | AC37 | AC36 |
|---|---|---|---|---|
| Root length-(cm) | 25.13 b | 35.49 a | -42.75 c | -63.78 d |
| Shoot length-(cm) | 9.10 c | 7.36 c | 29.87 a | 16.71 b |
| Root fresh weight-(g) | 83.32 a | 67.30 b | 18.71 c | -7.66 d |
| Root dry weight-(g) | 83.13 a | 40.08 b | -24.24 c | -32.64 c |
| Shoot fresh weight-(g) | 75.60 a | 37.57 b | 23.00 c | 5.66 d |
| Shoot dry weight-(g) | 77.71 a | 15.44 b | 8.04 c | -13.69 d |
| RWC % | 6.26 a | 4.67 a | 7.28 a | 6.73 a |
| Total Chlorophyll Content | 26.22 ab | 22.84 b | 35.07 a | 21.66 b |
| Number of tillers/plant | 33.33 b | 64.81 a | 41.67 ab | 23.33 b |
| Number of spike/plant | 41.67 a | 47.22 a | 33.33 a | 21.67 a |
| Spike length-(cm) | 53.71 a | 11.57 bc | 18.68 b | 4.91 c |
| Spike weight-(g) | 52.21 a | 35.51 b | 36.91 b | -1.94 c |
| Number of grain/spike | 43.45 a | 22.46 b | 22.74 b | 9.14 b |
| Grain weight/spike-(g) | 50.78 a | 34.77 b | 21.19 c | 2.89 d |
| Total yield/plant-(g) | 49.83 ab | 64.28 a | 31.53 bc | 22.34 c |

**Table 4.2.11 The percentage responses of some physiological traits in combination of drought stress conditions (S2+S3) with respect to control condition in four barley accessions based on Multiple Rang Duncan's test at *p* value < 0.05. Any values of mean holding a common letter are not significant.**

| Physiological traits | AC53 | AC47 | AC37 | AC36 |
|---|---|---|---|---|
| Root length-(cm) | 40.06 a | 35.45 a | -58.80 b | -142.11 c |
| Shoot length-(cm) | 8.09 c | 8.00 c | 30.72 a | 19.44 b |
| Root fresh weight-(g) | 94.47 a | 69.75 b | -29.50 c | -31.86 c |
| Root dry weight-(g) | 82.31 a | 46.44 b | -87.67 c | -97.25 c |
| Shoot fresh weight-(g) | 81.47 a | 26.64 c | 56.59 b | 18.12 d |
| Shoot dry weight-(g) | 81.27 a | 1.16 d | 49.10 b | 3.10 c |
| RWC % | 4.86 a | 3.78 a | 5.84 a | 4.55 a |
| Total Chlorophyll Content | 34.07 b | 50.77 a | 47.19 a | 30.50 b |
| Number of tillers/plant | 24.44 a | 29.63 a | 50.00 a | 31.11 a |
| Number of spike/plant | 36.11 a | 44.44 a | 41.67 a | 34.44 a |
| Spike length-(cm) | 58.19 a | 14.99 bc | 30.03 b | 9.16 c |
| Spike weight-(g) | 61.00 a | 32.35 b | 37.09 b | 14.20 c |
| Number of grain/spike | 46.87 a | 33.30 b | 22.80 b | 9.25 c |
| Grain weight/spike-(g) | 62.24 a | 32.04 b | 23.67 bc | 11.45 c |
| Total yield/plant-(g) | 66.47 a | 48.33 b | 36.51 c | 34.55 c |

**Table 4.2.12 The percentage responses of some physiological traits in combination of drought stress conditions (S1+S2+S3) with respect to control condition in four barley accessions based on Multiple Rang Duncan's test at *p* value < 0.05. Any values of mean holding a common letter are not significant**

| Physiological traits | AC53 | AC47 | AC37 | AC36 |
|---|---|---|---|---|
| Root length-(cm) | 35.38 b | 52.39 a | -56.65 c | -136.29 d |
| Shoot length-(cm) | 7.65 c | 10.19 c | 32.42 a | 19.21 b |
| Root fresh weight-(g) | 92.28 a | 75.01 b | 47.97 c | 14.25 d |
| Root dry weight-(g) | 75.96 a | 58.59 b | 25.74 c | -46.25 d |
| Shoot fresh weight-(g) | 88.36 a | 60.52 c | 74.07 b | 57.72 c |
| Shoot dry weight-(g) | 89.31 a | 50.29 c | 71.27 b | 51.04 c |
| RWC % | 10.54 a | 7.62 a | 7.88 a | 9.96 a |
| Total Chlorophyll Content | 45.42 bc | 61.57 a | 59.42 ab | 41.01 c |
| Number of tillers/plant | 55.56 a | 46.30 a | 50.00 a | 31.67 a |
| Number of spike/plant | 41.67 a | 61.11 a | 41.67 a | 50.00 a |
| Spike length-(cm) | 62.27 a | 25.78 b | 42.04 b | 26.10 b |
| Spike weight-(g) | 67.39 a | 38.24 c | 51.78 b | 25.69 d |
| Number of grain/spike | 56.06 a | 38.62 b | 32.21 b | 27.85 b |
| Grain weight/spike-(g) | 67.53 a | 40.36 b | 25.65 c | 16.81 c |
| Total yield/plant-(g) | 74.00 a | 68.99 a | 45.10 b | 44.06 b |

**Table 4.2.13 The percentage responses of some physiological traits under availability of all drought stress conditions in four barley accessions and inactions between accessions and those drought stages based on Multiple Rang Duncan's test at *p* value < 0.05. Any values of mean holding a common letter are not significant.**

| Root length-(cm) | | | | | | Shoot length-(cm) | | | Root fresh weight-(g) | | | Root dry weight-(g) | | |
|---|---|---|---|---|---|---|---|---|---|---|---|---|---|---|
| Accessions (AC) | | Stages (S) | | S. * AC | | Ac | S | S*AC | Ac | S | S*AC | Ac | S | S*AC |
| AC53 | 16.11 a | S. 1 | -16.27 a | S. 1*AC36 | -35.19 f | 13.89 a | 3.43 e | -20.20 g | 52.27 a | -136.36e | -121.36 i | 63.03 a | -57.26 d | -92.14 e-g |
| | | | | S. 1*AC37 | -49.18 fg | | | -18.42 g | | | -267.58 k | | | -114.51 gh |
| | | | | S. 1*AC47 | 4.34 c-e | | | 9.07 d-f | | | -81.01 h | | | -26.45 d |
| | | | | S. 1*AC53 | 14.93 b-d | | | 43.28 a | | | -75.49 h | | | 4.06 c |
| | | S. 1+2 | -37.67 b | S.1+2*AC36 | -51.67 fg | | 10.80bc | 3.08 ef | | -16.16 d | -51.26 g | | -28.25 c | -81.92 ef |
| | | | | S. 1+2*AC37 | -85.84 h | | | 23.62 bc | | | -170.72 j | | | -193.81 i |
| AC47 | 18.90 a | | | S. 1+2*AC47 | -11.10 e | 6.42 b | | 2.03 ef | 41.65 b | | 76.04 a-c | 40.13 b | | 80.32 a |
| | | | | S. 1+2*AC53 | -2.06 de | | | 14.47 c-e | | | 81.28 a-c | | | 82.42 a |
| | | S. 1+3 | -11.48 a | S. 1+3*AC36 | -63.78 g | | 15.76ab | 16.71 cd | | 40.42 a | -7.66 ef | | 16.58 a | -32.64 d |
| | | | | S. 1+3*AC37 | -42.75 fg | | | 29.87 b | | | 18.71 d | | | -24.24 d |
| | | | | S. 1+3*AC47 | 35.49 ab | | | 7.36 d-f | | | 67.30 bc | | | 40.08 b |
| | | | | S. 1+3*AC53 | 25.13 a-c | | | 9.10 d-f | | | 83.32 ab | | | 83.13 a |
| AC37 | -55.24 b | S. 2 | -20.34 a | S. 2*AC36 | -45.83 fg | 14.86 a | 5.19 de | 1.11 f | -81.64 d | 9.80 c | -22.84 f | -89.15 c | -6.93 b | -63.59 e |
| | | | | S. 2*AC37 | -44.95 fg | | | 3.70 ef | | | -52.06 g | | | -85.31 e-g |
| | | | | S. 2*AC47 | 5.81 c-e | | | 8.43 def | | | 55.55 c | | | 60.92 ab |
| | | | | S. 2*AC53 | 3.62 c-e | | | 7.52 d-f | | | 58.54 bc | | | 60.26 ab |
| | | S. 2+3 | -31.35 b | S. 2+3*AC36 | -142.11 i | | 16.56 a | 19.44 b-d | | 25.72 b | -31.86 fg | | -14.04 b | -97.25 fgh |
| | | | | S. 2+3*AC37 | -58.80 g | | | 30.72 b | | | -29.50 fg | | | -87.67 efg |
| AC36 | -67.16 c | | | S. 2+3*AC47 | 35.45 ab | 5.46 b | | 8.00 d-f | -38.01 c | | 69.75 abc | -81.96 c | | 46.44 b |
| | | | | S. 2+3*AC53 | 40.06 a | | | 8.09 d-f | | | 94.47 a | | | 82.31 a |
| | | S. 3 | -13.99 a | S. 3*AC36 | -64.41 g | | 9.21 cd | 12.61 c-f | | 38.01 a | 6.92 de | | -12.03 b | -124.25 h |
| | | | | S. 3*AC37 | -49.91 fg | | | 19.66 b-d | | | 11.32 de | | | -29.34 d |
| | | | | S. 3*AC47 | 43.39 a | | | 3.65 ef | | | 62.27 bc | | | 39.49 b |
| | | | | S. 3*AC53 | 14.95 b-d | | | 0.91 f | | | 71.52 abc | | | 65.98 ab |

**Continue**

| Shoot fresh weight-(g) | Shoot dry weight (g) | RWC % | Total Chlorophyll Content (Spad unit) |
|---|---|---|---|

| Accessions (AC) | | Stages (S) | | S. * AC | | AC | S | S*AC | AC | S | S*AC | AC | S | S*AC |
|---|---|---|---|---|---|---|---|---|---|---|---|---|---|---|
| AC53 | 61.10 a | S. 1 | -33.74 e | S. 1*AC36 | -41.74 m | 62.81 a | -40.24 d | -48.11 k | 5.03 a | 3.15 b | 2.23 b | 21.21 a | -52.78 d | -8.67 i |
| | | | | S. 1*AC37 | -142.20 n | | | -164.91 o | | | 2.26 b | | | -13.42 i |
| | | | | S. 1*AC47 | 2.92 k | | | 12.68 fg | | | 3.67 ab | | | -121.07 k |
| | | | | S. 1*AC53 | 46.04 b-d | | | 39.38 e | | | 4.42 ab | | | -67.97 j |
| | | S. 1+2 | 13.30 d | S.1+2*AC36 | -38.68 m | | -54.87 e | -154.72 n | | 6.22 a | 3.16 ab | | 42.49 a | 16.42 fgh |
| | | | | S. 1+2*AC37 | -43.43 m | | | -213.64 p | | | 5.07 ab | | | 21.35 fgh |
| AC47 | 29.50 b | | | S. 1+2*AC47 | 52.79 bc | 26.95 b | | 69.30 b | 4.38 a | | 8.27 a | 19.32 ab | | 75.61 a |
| | | | | S. 1+2*AC53 | 82.51 a | | | 79.58 a | | | 8.38 a | | | 56.59 bc |
| | | S. 1+3 | 35.45 b | S. 1+3*AC36 | 5.66 jk | | 21.88 b | -13.69 j | | 6.23 a | 6.73 ab | | 26.45 b | 21.66 fgh |
| | | | | S. 1+3*AC37 | 23.00 gh | | | 8.04 gh | | | 7.28 ab | | | 35.07 def |
| | | | | S. 1+3*AC47 | 37.57 d-f | | | 15.44 f | | | 4.67 ab | | | 22.84 fgh |
| | | | | S. 1+3*AC53 | 75.60 a | | | 77.71 a | | | 6.26 ab | | | 26.22 fgh |
| AC37 | -17.29 d | S. 2 | 45.70 a | S. 2*AC36 | -12.19 l | -74.02 d | -29.07 c | -79.10 l | 4.87 a | 3.39 b | 3.01 ab | 20.70 ab | 38.33 a | 9.80 h |
| | | | | S. 2*AC37 | -13.21 l | | | -125.80 m | | | 5.38 ab | | | 14.24 gh |
| | | | | S. 2*AC47 | 31.61 e-g | | | 51.88 d | | | 3.11 ab | | | 72.19 ab |
| | | | | S. 2*AC53 | 41.83 c-e | | | 36.73 e | | | 2.04 b | | | 57.08 bc |
| | | S. 2+3 | 12.01 d | S. 2+3*AC36 | 18.12 hi | | 33.66 a | 3.10 hi | | 4.76 ab | 4.55 ab | | 40.63 a | 30.50 efg |
| | | | | S. 2+3*AC37 | 56.59 b | | | 49.10 d | | | 5.84 ab | | | 47.19 cde |
| AC36 | -9.66 c | | | S. 2+3*AC47 | 26.64 f-h | -48.07c | | 1.16 i | 4.44 a | | 3.78 ab | 13.81 b | | 50.77 cd |
| | | | | S. 2+3*AC53 | 81.47 a | | | 81.27 a | | | 4.86 ab | | | 34.07 def |
| | | S. 3 | 22.77 c | S. 3*AC36 | 10.88 i-k | | 20.16 b | 4.11 hi | | 4.34 ab | 6.99 ab | | 17.46c | 13.17 gh |
| | | | | S. 3*AC37 | 15.53 h-j | | | 3.09 hi | | | 3.40 ab | | | 19.79 fgh |
| | | | | S. 3*AC47 | 25.49 gh | | | 11.24 fg | | | 2.77 b | | | 15.60 fgh |
| | | | | S. 3*AC53 | 39.16 de | | | 62.21 c | | | 4.20 ab | | | 21.28 fgh |
| **Number of tillers/plant** | | | | | | **Number of spike/plant** | | | **Spike length-(cm)** | | | **Spike weight (g)** | | |
| **Accessions (AC)** | | **Stages (S)** | | **S. * AC** | | **AC** | **S** | **S*AC** | **AC** | **S** | **S*AC** | **AC** | **S** | **S*AC** |
| AC53 | 33.64 a | S. 1 | 16.90 b | S. 1*AC36 | 25.00 ab | 39.26 a | 20.56 b | 16.67 e | 47.02 a | 19.61 cd | 7.46 hi | 45.09 a | 15.06 d | 4.55 gh |
| | | | | S. 1*AC37 | 25.00 ab | | | 18.33 de | | | 6.92 hi | | | 9.13 fgh |
| | | | | S. 1*AC47 | 9.26 b | | | 27.78 a-e | | | 24.31 efg | | | 17.10 fg |
| | | | | S. 1*AC53 | 8.33 b | | | 19.44 de | | | 39.75 bcd | | | 29.47 de |
| | | S. 1+2 | 44.72 a | S.1+2*AC36 | 41.67 ab | | 36.11 a | 33.33 a-e | | 34.62 a | 23.19 efg | | 28.58 bc | 11.27 fg |
| | | | | S. 1+2*AC37 | 55.56 a | | | 25.00 b-e | | | 30.41 def | | | 19.18 ef |
| AC47 | 38.83 a | | | S. 1+2*AC47 | 40.00 ab | 38.89 a | | 44.44 a-c | 20.42 b | | 36.11 cde | 32.32 b | | 33.50 d |
| | | | | S. 1+2*AC53 | 41.67 ab | | | 41.67 a-d | | | 48.78 ab | | | 50.37 abc |
| | | S. 1+3 | 40.79 a | S. 1+3*AC36 | 23.33 ab | | 35.97 a | 21.67 c-e | | 22.22 bc | 4.91 i | | 30.67 ab | -1.94 h |
| | | | | S. 1+3*AC37 | 41.67 ab | | | 33.33 a-e | | | 18.68 fgh | | | 36.91 d |
| | | | | S. 1+3*AC47 | 64.81 a | | | 47.22 ab | | | 11.57 ghi | | | 35.51 d |
| | | | | S. 1+3*AC53 | 33.33 ab | | | 41.67a-d | | | 53.71 a | | | 52.21 ab |
| AC37 | 39.76 a | S. 2 | 41.20 a | S. 2*AC36 | 41.67 ab | 28.06 b | 34.03 a | 25.00 b-e | 17.26 b | 24.46 bc | 19.17 fgh | 22.92 c | 26.21 bc | 10.15 fgh |
| | | | | S. 2*AC37 | 25.00 ab | | | 25.00 b-e | | | 12.81 ghi | | | 15.37 fg |
| | | | | S. 2*AC47 | 56.48 a | | | 36.11 a-e | | | 19.81 fgh | | | 41.40 bcd |
| | | | | S. 2*AC53 | 41.67 ab | | | 50.00 a | | | 46.07 abc | | | 37.93 d |
| | | S. 2+3 | 33.80 ab | S. 2+3*AC36 | 31.11 ab | | 39.17 a | 34.44 a-e | | 28.09 b | 9.16 hi | | 36.16 a | 14.20 fg |
| | | | | S. 2+3*AC37 | 50.00 ab | | | 41.67 a-d | | | 30.03 def | | | 37.09 d |
| AC36 | 31.30 a | | | S. 2+3*AC47 | 29.63 ab | 26.30 b | | 44.44 a-c | 11.16 c | | 14.99 ghi | 7.15 d | | 32.35 d |
| | | | | S. 2+3*AC53 | 24.44 ab | | | 36.11 a-e | | | 58.19 a | | | 61.00 a |
| | | S. 3 | 37.87 a | S. 3*AC36 | 25.00 ab | | 32.92 a | 26.67 a-e | | 14.80 d | 3.05 i | | 24.54 c | 4.69 gh |
| | | | | S. 3*AC37 | 41.33 ab | | | 25.00 b-e | | | 4.74 i | | | 19.83 ef |
| | | | | S. 3*AC47 | 32.78 ab | | | 33.33 a-e | | | 15.76 ghi | | | 34.09 d |
| | | | | S. 3*AC53 | 52.38 a | | | 46.67 ab | | | 35.63 cde | | | 39.56 cd |

**Continue**

| Number of grains/spike | Grain weight/spike-(g) | Total grain yield/plant-(g) |
|---|---|---|

| Accessions (AC) | | Stages (S) | | S. * AC | | AC | S | S*AC | AC | S | S*AC |
|---|---|---|---|---|---|---|---|---|---|---|---|
| AC53 | 37.38 a | S. 1 | 16.91 c | S. 1*AC36 | 10.39 e-g | 44.99 a | 16.29 c | 2.03 m | 51.89 a | 33.37 d | 17.58 f |
| | | | | S. 1*AC37 | 9.22 e-g | | | 9.41 k-m | | | 22.68 ef |
| | | | | S. 1*AC47 | 22.18 c-f | | | 22.87 f-j | | | 45.00 b-d |
| | | | | S. 1*AC53 | 25.86 b-d | | | 30.85 e-h | | | 48.22 bc |
| | | S. 1+2 | 25.32 ab | S.1+2*AC36 | 15.15 d-g | | 28.67 ab | 13.07 i-m | | 36.98 cd | 29.98 ef |
| | | | | S. 1+2*AC37 | 18.64 d-g | | | 19.75 h-l | | | 24.81 ef |
| AC47 | 28.14 b | | | S. 1+2*AC47 | 27.94 b-d | 35.28 b | | 33.25 e-g | 52.85 a | | 47.24 bc |
| | | | | S. 1+2*AC53 | 39.54 ab | | | 48.61 bc | | | 45.90 b-d |
| | | S. 1+3 | 24.45 ab | S. 1+3*AC36 | 9.14 e-g | | 27.41 ab | 2.89 m | | 41.99 bc | 22.34 ef |
| | | | | S. 1+3*AC37 | 22.74 c-e | | | 21.19 g-k | | | 31.53 d-f |
| | | | | S. 1+3*AC47 | 22.46 c-e | | | 34.77 d-f | | | 64.28 a |
| | | | | S. 1+3*AC53 | 43.45 a | | | 50.78 b | | | 49.83 b |
| AC37 | 14.67 c | S. 2 | 25.19 ab | S. 2*AC36 | 21.92 c-f | 14.53 c | 24.69 b | 9.44 k-m | 27.37 b | 48.93 a | 33.34 c-e |
| | | | | S. 2*AC37 | 8.47 e-g | | | 5.42 m | | | 31.00 d-f |
| | | | | S. 2*AC47 | 35.40 a-c | | | 46.62 b-d | | | 67.15 a |
| | | | | S. 2*AC53 | 34.96 a-c | | | 37.29 c-e | | | 64.22 a |
| | | S. 2+3 | 28.05 a | S. 2+3*AC36 | 9.25 e-g | | 32.35 a | 11.45 j-m | | 46.46 ab | 34.55 c-e |
| | | | | S. 2+3*AC37 | 22.80 c-e | | | 23.67 f-i | | | 36.51b-e |
| AC36 | 12.25 c | | | S. 2+3*AC47 | 33.30 a-c | 7.11 d | | 32.04 e-g | 27.08 b | | 48.33 bc |
| | | | | S. 2+3*AC53 | 46.87 a | | | 62.24 a | | | 66.47 a |
| | | S. 3 | 18.75 bc | S. 3*AC36 | 7.65 fg | | 23.45 b | 3.76 m | | 31.04 d | 24.68 ef |
| | | | | S. 3*AC37 | 6.19 g | | | 7.75 lm | | | 17.70 f |
| | | | | S. 3*AC47 | 27.58 b-d | | | 42.11 b-e | | | 45.08 b-d |
| | | | | S. 3*AC53 | 33.58 a-c | | | 40.16 b-e | | | 36.70 b-e |

### 4.2.2.2 Biochemical traits responses

After conducting a plastic house experiment, the difference in drought tolerance between sensitive and resistant barley accessions was further investigated on leaf organs, which were collected during measuring the morphological traits, by using seven biochemical markers including; Proline Content (PC), Total Phenolic Content (TPC), Anti-oxidant (AC), Soluble Sugar Content (SSC), Catalase (CAT), Guaiacol peroxidase activity (GPA) and Lipid Peroxidation (LP). Under all individual stress treatments (S1, S2, and S3), the responses by all used traits were highly different among the barley accessions, and the interactions were significantly different, as indicated by ANOVA analysis in Appendixes (2.15, 2.16, and 2.17).

The results presented in Tables (4.2.15, 4.2.17, and 4.2.19) showed that water limiting in all individual growth stages (S1, S2, and S3) caused a considerable increase in the activities of studied traits in comparison with control conditions. In the growth (S1) condition, the total activities of studied traits; (PC, SSC, TPC, AC, GPA, CAT, and LP) under present drought stress were increased with mean values (1355.73, 307.47, 205.07, 1050.07, 0.37, 66.02 and 19.13) respectively while not considerable changes for same traits were observed at control condition as the mean values were less than stressed conditions (1043.60, 269.27, 189.52, 979.46, 0.19, 38.96 and 14.73) respectively by studied traits (Table 4.2.15). When plant materials faced drought conditions at a (S2) growth stage similar to (S1), the total activities of studied traits (PC, SSC, TPC, AC, GPA, CAT, and LP) under present drought stress were improved with mean values (1866.63, 194.35, 209.28, 1001.76, 0.28, 89.83, and 12.55), respectively, whereas the changes by studied traits were not considerable at control conditions as the mean values were less than stressed conditions (1631.69, 153.35, 168.92, 939.59, 0.15, 47.62, and 10.33) respectively by studied traits (Table 4.2.17). At a growth stage (S3) similar to two previous growth stages, the total activities of studied traits; (PC, SSC, TPC, AC, GPA, CAT, and LP) under present drought stress were enhanced with mean values (3147.88, 163.77, 216.20, 941.11, 0.24, 74.13, and 15.21) respectively, while no substantial alterations for the same traits were observed at control conditions as the mean values were fewer than stressed conditions (2553.10, 138.04, 195.81, 880.74, 0.14, 54.11, and 12.23) respectively by studied traits (Table 4.2.19).

Comparing the mean values of selected barley accessions under drought and control conditions at the (S1) stage, considerable responses can be seen by barley accessions for studied traits (Table 4.2.14). As expected, an essential increase of the mean value by the two barley accessions AC36 and AC37 under this particular stress condition was observed for all studied traits, with the only exception of soluble sugar content compared to their control. The two barley accessions AC47 and AC53 possessed no obvious increase in response to drought stress for any mentioned trait, with the only exception of lipid peroxidation, in which high activity by both accessions was detected.

AC37 and AC36, with mean values of 2176.82 and 1081.05, respectively, increased the accumulation of total proline content, while the other two accessions, AC47 (785.50) and AC53 (755.28) showed only minor increases. As can be clearly seen from (Table 4.2.14), the soluble sugar content was slightly increased by both barley accessions AC53 and AC47 with mean values of (315.34 and 309.01) respectively compared with AC36 and AC37 with mean values of (271.82 and 257.31) respectively in comparison with their control. When barley accessions were exposed to drought at this particular growth stage, the total phenolic content was increased by all accessions but the maximum increase was documented by AC36 (217.19), AC37 (214.38), and AC53 with a mean value of (180.39), followed by AC47, which recorded the lowest change with a mean value of 177.21 in respect to control conditions. The capacity of antioxidants determined by DPPH was much higher in AC36 (1109.53) and AC37 (1075.41), while the smaller activity of this trait was detected in AC53 (1001.08) and AC47 (873.04). The activity of Guaiacol peroxidase was measured by all barley accessions under the presence of drought stress at the early growth stage (S1) and a considerable increase by AC37 with a value of 0.40 was detected, followed by AC36 with a value of 0.39, while the activity was reduced by AC47 and AC53 with values of (0.12, 0.22) respectively. Accession AC36 (82.25) had the highest activities of catalyze in response to drought, followed by AC37 (51.95), while the activity of this critical enzyme dropped by both barley accessions AC47 and AC53 to reach the minimum activities of (30.30 and 45.45) respectively. Unlike all biochemical traits, a considerable increase in lipid peroxidation was detected by AC47 (20.20) and AC53 (19.91) in respect to their well-watered condition, while the accession AC37 was less affected by stress as the value of 12.54 was detected in comparison with its control and followed by AC36 with a value of 15.06 (Table 4.2.14).

In the case of subjecting four barley accessions to stress conditions at the flowering stage (S2), a similar pattern of responses by barley accessions AC36 and AC37 were recorded for studied parameters including (PC, TPC, AC, CAT, and GPA) as shown in (Table 4.2.16) in which all these parameters improved dramatically in comparison to untreated conditions. On the other hand, the barley accessions AC47 and AC53 displayed a slight increase in respect to their control for the mentioned traits. Regarding the accumulation of PC in leaf organs under the presence of drought at the flowering stage, AC37 had the maximum response for accumulating PC content with a mean value of (2038.62), and AC36 placed in the next position with a mean value of (1747.33), followed by AC53 and AC47 with a mean value of (1568.62-1642.08) respectively. The circumstances of responses by barley accessions for studying the accumulation and response of SSC content in comparisons with (S1) were totally inversed. The highest responses were documented by AC36 (234.78) and AC37 (180.23), while few reactions were observed by AC47 (158.09) and AC53 (122.28) (Table 4.2.16). Similarly, both accessions AC37 and AC36 were ranked as the most accessions in accumulating the TPC content with a mean value of (249.21–214.01), while AC47 (144.16) and AC53 (149.03) came after. Both accessions (AC37 and AC36) dramatically increased their AC content in response to drought stress at this stage, with relatively similar boosting patterns, while AC53 and AC47 were the third and last responses in comparison to their control, with mean values of (892.30 and 804.80) respectively. As presented in Table 4.2.16, a similar pattern of responses to the enzymatic activities of GPA and CAT by four barley accessions was documented. For GPA, AC37 had the most superior activity with a value of 0.40, and AC36 (0.22) was placed next, while the activities of this particular enzyme declined by AC47 (0.14) and AC53 (0.11). Regarding the CAT activates, similar to the previous enzyme, AC37 was placed first, owning the value (93.07) in respect to its control, and AC36, with a value of (82.25), was placed in the second position, while no considerable increase in the activity of this enzyme was documented by AC47 (56.28) and AC53 (43.29). As stated in (Table 4.2.16) for lipid peroxidation response by barley accessions under drought stress conditions at the flowering stage, AC47 was taken into account as the most sensitive accesson as it showed the highest value of lipid peroxidation, and somehow AC36 placed next with a value of (11.25), while AC53 slightly maintained the lipid peroxidation in comparison to other barley accessions responses in the (S1) growth stage, owning the value of (10.49) and AC37 (8.91), considered as an accession that is less affected by water limiting for this trait.

After the flowering stage (Anthesis stage-S3), the selected barley accessions were exposed to drought stress conditions. The data presented in (Table 4.2.18) discovered significant differences between studied accessions in response to water holding conditions. A critical increase of the mean value by the two barley accessions AC36 and AC37 under this particular stress condition was observed for all studied biochemical parameters, excluding lipid peroxidation, compared to their control at this growth stage. On the other hand, the two barley accessions AC47 and AC53 possessed no obvious increase in response to drought stress for the mentioned traits except for the peroxidation of lipid. The drought stress increased the (PC) content in all studied barley accessions, but extensive increases were documented by AC36 (3802.97) and AC37 (2906.88), followed by AC47 (2475.79) and AC53 (2216.31). The soluble sugar content in the same way as (S2) was stimulated by both barley accessions AC36 (193.58) and AC37 (160.25). On the other hand, the stimulation of this trait was not as much in comparison with it is control by both accessions AC53 (152.20) and AC47 (97.59). Also, the TPC content of accessions AC37 and AC36 increased a lot in response to a lack of water, with a mean value of (232.45–226.09), while the SSC content of accessions AC53 (191.37) and AC47 (174.12) increased less at this stage of growth. Regarding the analysis in (Table 4.2.18) for activities of anti-oxidant in relation to own control condition, AC36 and AC37 were found to have stimulated the activity of this biochemical trait the most out of all the tested barley accessions, with a mean value (1086.89–1070.34) that was significantly higher than AC47 (755.34) and AC53 (731.15).The increase in GPA activity was highly observed by AC37 (0.38) and AC36 (0.20), while AC53 was experienced (0.10) for GPA activities and almost non-activities were detected by AC47 (0.08) in respect to its own control. Substantial activation of the CAT enzyme was also stated by AC37 (96.32) and AC36 (88.74), while no high motivations were identified by AC53 (41.13) and AC47 (30.30). Similar to the (S1) growth stage under existing water stress conditions, the same performance in response by the barley accessions to lipid peroxidation was documented. For instance, the highest peroxidation of lipid was explored by AC53 (20.48) and AC47 (13.95), while no substantial changes were indicated by AC37 (10.79) and AC36 (9.68). After re-watering at the individual plant developmental stages (S1), (S2), and (S3), when four barley accessions were exposed to stress conditions, AC36 and AC37 performed better than AC53 and AC47 in accumulating and increasing the activities of most studied traits (Table 4.2.18).

Under all combination stress treatments (S1+S2), (S1+S3), (S2+S3) and (S1+S2+S3) present in our investigation, the responses by all used traits were highly different among the barley accessions and the interactions were highly significantly different as indicated by ANOVA analysis in Appendices (2.18,2.19,2.20, and 2.21).

The results presented in Tables (4.2.21, 4.2.23, 4.2.25, and 4.2.27) displayed that in all combinations of growth stages (S1+S2, S1+S3, S2+S3, and S1+S2+S3), the drought resulted in a substantial increase in the activities of studied traits in comparison with control conditions. In growth (S1+S2) conditions, the total performance of studied traits (PC, SSC, TPC, AC, GPA, CAT, and LP) under current drought stress was increased with mean values (2425.99, 242.89, 275.22, 1021.01, 0.41, 107.14, and 14.21), whereas no extensive changes for the same traits were perceived at control conditions as the mean values were less than stressed conditions (1584.19, 136.68, 146.42). When plant materials were exposed to drought conditions at (S1+S3) growth stages, similar to (S1+S2), the total activities of studied traits (PC, SSC, TPC, AC, GPA, CAT, and LP) under present drought stress were enhanced with mean values (3805.67, 167.22, 244.20, 1006.32, 0.36, 106.06, and 16.64) respectively, whereas the modifications by studied traits were not considerable at control conditions as the mean values were less than stressed conditions (2528.10, 130.29, 205.81, 950.74, 0.14, 54.11, and 12.23) respectively by studied traits (Table 4.2.23). At the growth stage (S2+S3), similar to two previous combination growth stages, the total activities of studied traits (PC, SSC, TPC, AC, GPA, CAT, and LP) under present drought stress were enhanced with mean values (3612.14, 157.64, 250.81, 1010.03, 0.47, 115.80, and 18.71) respectively, while no extensive alterations for the same traits were detected at control conditions as the mean values were fewer than stressed conditions (2578.10, 125.29, 205.81, 950.74, 0.14, 54.11, and 12.23) respectively by studied biochemical traits (Table 4.2.25). When four barley accessions faced triple water holding conditions at (S1+S2+S3) similar to all previous double stress conditions, the total activities of studied traits (PC, SSC, TPC, AC, GPA, CAT, and LP) under present drought stress were enhanced with mean values (3901.24, 138.33, 221.34, 969.32, 0.33, 134.20, and 21.44) respectively, while the adjustments by studied traits were not substantial at control conditions as the mean values were less than stressed conditions (2653.10, 125.29, 202.48, 950.74, 0.14, 54.11, and 12.23) respectively by studied traits (Table 4.2.27).

Significant changes between studied accessions in response to water holding conditions were revealed when the plant materials were targeted twice at (S1 + S2) (Table 4.2.20). As expected, both barley accessions AC47 and AC53 showed similar responses for studied traits in the presence of more severe drought conditions when compared to all individual stressors, whereas the performance of both accessions AC36 and AC37, surprisingly, showed more activity for studied traits except lipid peroxidation in the presence of current combination stress conditions.The minimum mean value for (PC) content which accumulated in the leaf organ was presented by AC53 (1697.59) and AC47 (1801.69), while AC36 (1974.51) and AC37 (2546.56) preserved more (PC) content. All barley accessions under the presence of double stress conditions (S1+S2) experienced similar trends of responses for two studied traits (SSC) and (TPC). In both traits, AC36 experienced the peak value of (SSC) and (TPC) with values of (269.35-230.26) respectively. In addition to AC36, AC37 was placed next with a mean value of (212.69–205.96) respectively. On the contrary, for (SSC) and (TPC), small portions of increase by AC47 with a mean value of (140.62–198.65) and AC53 with a mean value of (136.48–208.43) were stated (Table 4.2.20). The combination of drought stressors (S1+S2) highly stimulated the activities of antioxidants with almost similar responses by AC36 and AC37, while lighter stimulation for this trait was found by AC53 (898.04) and AC47 (809.19). GPA activities were lowest in both barley accessions AC53 and AC47, with values of (0.12-0.13), respectively, while GPA enzyme activities peaked in barley accession AC37 (0.46) and AC36 (0.42) in comparison to their control condition. Similar to the mentioned enzyme, both accessions AC53 and AC47 exhibited the lowest actions for CAT with a mean value of (49.78–54.11) correspondingly. However, the highest activities of CAT were documented by AC36 (103.90) and AC37 (101.73). The highest oxidative stress of lipid peroxidation was observed by AC47 (16.68) and AC53 (12.18), while both barley accessions AC37 and AC36 were maintained and detected little change with respect to their own control with a mean value of (9.07-10.65) respectively (Table 4.2.20).

When plant materials were directed to double stress conditions (S1+S3) (Table 4.2.22), significant changes between studied barley accessions in response to drought with respect to control conditions were discovered by all biochemical traits. Similar to all previously mentioned stressor conditions, AC36 and AC37 showed an enormous increase in most traits in the presence of this combination of drought stress conditions, while contrary responses by AC53 and AC47 were detected in which they both showed a slight increase in response to drought stress. The extreme increase in proline content was perceived by AC36 (4053.36) and AC37 (3557.97) and then followed by AC47 (2746.56) and AC53 (2309.64). In addition to (PC), all barley accessions increased soluble sugar content, but the highest accumulations were documented by AC36 and AC37 with mean values of (215.49-160.74), respectively, whereas the other two barley accessions showed little response (119.51), AC47 (99.29) in comparison to the untreated condition. Drought stress (S1+S3) caused both barley accessions AC37 and AC36 to produce more total phenolic content with a value of (267.47-234.42), respectively, whereas this phenomenon was not observed in AC53 and AC47, which both experienced a light increase for the mentioned trait with a mean value of (217.85-180.30). AC53 with a mean value (802.43) was considered as the least accession in response to a combination of drought (S1+S3) for the activities of antioxidants, followed by AC47 with a mean value (900.41) while substantial increases were practiced by AC36 and AC37 with mean value (1108.18-1103.11) respectively in respect to well-watered condition (Table 4.2.22). Both barley accessions AC47 and AC53 had the lowest GPA enzyme activities with values of (0.10-0.12), while barley accession AC37 had the highest (0.42) and AC36 had the lowest (0.36) in comparison to the untreated condition.Similar to the GPA enzyme, both accessions AC47 and AC53 displayed the lowest actions for CAT with a mean value of (41.13–49.78) correspondingly. However, the highest activities of CAT were documented by AC37 (129.87) and AC36 (99.57). Under existing water shortage conditions in (S1+S3) growth stages, the same performances as previous stress conditions were documented in response by barley accessions to lipid peroxidation. For instance, the highest peroxidation of lipid was explored by AC53 (22.32) and AC47 (14.58), while no extensive changes were indicated by AC36 (9.94) and AC37 (10.90) in respect to untreated conditions (Table 4.2.22). In response to water limiting conditions when plant materials were directed to double stress conditions at both (S2+S3) growth rates (Table 4.2.24), significant variations between studied accessions were discovered. Similar responses by both barley accessions AC47 and AC53 were documented for studied traits in the presence of drought conditions (S2+S3) in comparison with two previous combined stress conditions, while the performance of both accessions AC36 and AC37 under the present combination stress conditions indicated more activity of studied traits except for the lipid peroxidation in respect to the control condition. The minimum mean value for (PC) content was displayed by AC53 (2301.56) and AC47 (2578.87), while AC36 (3903.49) and AC37 (3596.56) preserved more (PC) content. The soluble sugar content in AC36 (186.02) was displayed as the highest value among tested accessions in the presence of this particular stress condition, followed by AC37 with a mean value of (149.60), while no considerable response by AC47 and AC53 was documented as they produced soluble sugar by (102.99, and 127.25) respectively in respect to the control condition. The minimum responses were recorded by AC47 (178.43) and AC53 (218.97) for total proline content, while the responses by AC37 and AC36 were high as they were stored more (TPC) in response to drought by the mean value (265.88-249.96) respectively. Similar responses by AC36 and AC37 were detected in the synthesis of anti-oxidants. The top activities of anti-oxidants were stated by AC36 (1110.88) and AC37 (1099.05), while fewer activities of anti-oxidants were observed by AC47 and AC53, as AC47 was placed third with a mean value of 904.12, followed by AC53 (807.50). Similar trends of enzymatic responses for two studied traits (GPA) and (CAT) by four barley accessions under the presence of double stress conditions (S2+S3) were experienced. In both traits, AC37 experienced the highest values of (GPA) and (CAT) with mean values of 0.54–138.53, respectively. In addition to AC37, AC36 was placed next with a mean value (0.45–110.39) respectively. On the contrary, little enzymatic activity by AC47, with a mean value of (0.11-43.29) and AC53, with a mean value of (0.12-47.62) was detected for (GPA) and (CAT), respectively (Table 4.2.24). In the presence of water deficit conditions in (S2+S3) growth stages, the same performances as previous stress conditions were recognized in response by four barley accessions for peroxidation of lipid. In respect to control conditions, the oxidative stress of lipid peroxidation was highly maintained by AC36 (10.32) and AC37 (11.41), while higher lipid peroxidations were stated by both accessions AC53 and AC47 with a mean value of (24.06-16.08) respectively (Table 4.2.24).

When the plant materials faced triple stress conditions (S1+S2+S3), significant differences between studied traits by barley accessions in response to water holding conditions were revealed (Table 4.2.26). Similar to all previously mentioned stressor conditions, AC36 and AC37 demonstrated a massive increase in most traits in the presence of these triple drought stress conditions, whereas AC53 and AC47 demonstrated opposite responses, both achieving a slight increase in response to drought stress. Regarding the analysis of proline content, AC36 and AC37, with a mean value of (4334.26-3831.69), accumulated more (PC) while slight accumulations were detected by AC47 (2832.33) and AC53 (2110.41). A dramatic increase in soluble sugar content was observed by AC36 (177.38) and AC37 (161.33), whereas the content of this trait was perceived to be little increased by AC53 (98.52) and AC47 (90.03). AC47 was the least performed accession in the accumulation of TPC in the presence of triple drought stress, with a mean value of (170.75), followed by AC53 with a mean value of (204.06), while both accessions AC37 and AC36 performed better under this critical stress condition, with mean values of (245.84–226.99) respectively. The activities of antioxidants were highly stimulated in the presence of multiple drought conditions by both accessions AC37 and AC36, with a mean value of (1099.73–1086.55), respectively. However, the stimulation of this trait was not as much with respect to untreated conditions by AC47 and AC53 as they were promoted by (870.68-783.18) respectively. Similar patterns as (S2+S3) drought conditions for enzymatic responses for two studied traits (GPA) and (CAT) by four barley accessions under the availability of triple stress conditions (S1+S2+S3) were documented (Table 4.2.26). In both traits, AC47 experienced the lowest value of (GPA) and (CAT) enzymatic activates with a mean value of 0.10-41.13, followed by AC53 with a mean value of 0.11-45.45, respectively. On the contrary, high enzymatic activities by AC37 with a mean value of (0.42-160.17) and AC36 with a mean value of (0.31-129.87) were spotted for (GPA) and (CAT) respectively (Table 4.2.26). The same performance as all combinations of stress conditions was documented in response by four barley accessions for peroxidation of lipid under existing drought stress in (S1+S2+S3) growth stages. The oxidative stress of lipid peroxidation was highly preserved by AC36 (10.60) and AC37 (11.46) while both accessions AC53 and AC47 were more sensitive to the triple stress conditions as they were stated to have the highest activities of lipid peroxidation with a mean value of (26.07-19.22) respectively in respect to the control condition (Table 4.2.26).

**Table 4.2.14 The mean comparisons of seven biochemical traits in drought stress condition (S1) with respect to control condition in four barley accessions based on Multiple Rang Duncan's test at *p* value < 0.01. Any values of mean holding a common letter are not significant.**

| | PC-Con. µg/g | SSC-Con. µg/g | TPC µg gallic acid per gram of fresh seedling | Antioxidant capacity by DPPH (µg Trolox/g FM) | GPA (units/min/g FW) | CAT (units/min/g FW) | Lipid peroxidation (nmol/g FW) |
|---|---|---|---|---|---|---|---|
| **AC36** | 1081.05 b | 271.82 b | 217.19 a | 1109.53 a | 0.39 a | 82.25 a | 15.06 c |
| **AC37** | 2176.82 a | 257.31 b | 214.38 b | 1075.41 b | 0.40 a | 51.95 b | 12.54 d |
| **AC53** | 755.28 d | 315.34 a | 180.39 c | 1001.08 c | 0.22 b | 30.30 c | 19.91 b |
| **AC47** | 785.50 c | 309.01 a | 177.21 d | 873.04 d | 0.12 c | 45.45 b | 20.20 a |

**Table 4.2.15 Overall mean comparisons responses by four barley accessions in respect to their control under present drought stress at (S1) by seven biochemical markers.**

| | PC-Con. µg/g | SSC-Con. µg/g | TPC µg gallic acid per gram of fresh seedling | Antioxidant capacity by DPPH (µg Trolox/g FM) | GPA (units/min/g FW) | CAT (units/min/g FW) | Lipid peroxidation (nmol/g FW) |
|---|---|---|---|---|---|---|---|
| **S1** | 1355.73 a | 307.47 a | 205.07 a | 1050.07 a | 0.37 a | 66.02 a | 19.13 a |
| **C1** | 1043.60 b | 269.27 b | 189.52 b | 979.46 b | 0.19 b | 38.96 b | 14.73 b |

**Table 4.2.16 The mean comparisons of seven biochemical traits in drought stress condition (S2) with respect to control condition in four barley accessions based on Multiple Rang Duncan's test at *p* value < 0.01. Any values of mean holding a common letter are not significant.**

| | PC-Con. µg/g | SSC-Con. µg/g | TPC µg gallic acid per gram of fresh seedling | Antioxidant capacity by DPPH (µg Trolox/g FM) | GPA (units/min/g FW) | CAT (units/min/g FW) | Lipid peroxidation (nmol/g FW) |
|---|---|---|---|---|---|---|---|
| **AC36** | 1747.33 b | 234.78 a | 214.01 b | 1093.65 a | 0.22 b | 82.25 b | 11.25 b |
| **AC37** | 2038.62 a | 180.23 b | 249.21 a | 1091.96 a | 0.40 a | 93.07 a | 8.91 d |
| **AC47** | 1642.08 c | 158.09 c | 144.16 d | 804.80 c | 0.14 c | 56.28 c | 14.65 a |
| **AC53** | 1568.62 d | 122.28 d | 149.03 c | 892.30 b | 0.11 d | 43.29 d | 10.94 c |

**Table 4.2.17 Overall mean comparisons responses by four barley accessions in respect to their control under present drought stress at (S2) by seven biochemical markers.**

| | PC-Con. µg/g | SSC-Con. µg/g | TPC µg gallic acid per gram of fresh seedling | Antioxidant capacity by DPPH (µg Trolox/g FM) | GPA (units/min/g FW) | CAT (units/min/g FW) | Lipid peroxidation (nmol/g FW) |
|---|---|---|---|---|---|---|---|
| **S2** | 1866.63 a | 194.35 a | 209.28 a | 1001.76 a | 0.28 a | 89.83 a | 12.55 a |
| **Control (C2)** | 1631.69 b | 153.35 b | 168.92 b | 939.59 b | 0.15 b | 47.62 b | 10.33 b |

**Table 4.2.18 The mean comparisons of seven biochemical traits in drought stress condition (S3) with respect to control condition in four barley accessions based on Multiple Rang Duncan's test at *p* value < 0.01. Any values of mean holding a common letter are not significant.**

| | PC-Con. µg/g | SSC-Con. µg/g | TPC µg gallic acid per gram of fresh seedling | Antioxidant capacity by DPPH (µg Trolox/g FM) | GPA (units/min/g FW) | CAT (units/min/g FW) | Lipid peroxidation (nmol/g FW) |
|---|---|---|---|---|---|---|---|
| **AC37** | 2906.88 b | 160.25 b | 232.45 a | 1070.34 b | 0.38 a | 96.32 a | 10.79 c |
| **AC36** | 3802.97 a | 193.58 a | 226.09 b | 1086.89 a | 0.20 b | 88.74 a | 9.68 d |
| **AC53** | 2216.31 d | 152.20 c | 191.37 c | 731.15 d | 0.10 c | 41.13 b | 20.48 a |
| **AC47** | 2475.79 c | 97.59 d | 174.12 d | 755.34 c | 0.08 c | 30.30 c | 13.95 b |

**Table 4.2.19 Overall mean comparisons responses by four barley accessions in respect to their control under present drought stress at (S3) by seven biochemical markers.**

| | PC-Con. µg/g | SSC-Con. µg/g | TPC µg gallic acid per gram of fresh seedling | Antioxidant capacity by DPPH (µg Trolox/g FM) | GPA (units/min/g FW) | CAT (units/min/g FW) | Lipid peroxidation (nmol/g FW) |
|---|---|---|---|---|---|---|---|
| **S3** | 3147.88 a | 163.77 a | 216.20 a | 941.11 a | 0.24 a | 74.13 a | 15.21 a |
| **Control (C3)** | 2553.10 b | 138.04 b | 195.81 b | 880.74 b | 0.14 b | 54.11 b | 12.23 b |

**Table 4.2.20 The mean comparisons of seven biochemical traits in drought stress condition (S1+S2) with respect to control condition in four barley accessions based on Multiple Rang Duncan's test at *p* value < 0.01. Any values of mean holding a common letter are not significant.**

| | PC-Con. µg/g | SSC-Con. µg/g | TPC µg gallic acid per gram of fresh seedling | Antioxidant capacity by DPPH (µg Trolox/g FM) | GPA (units/min/g FW) | CAT (units/min/g FW) | Lipid peroxidation (nmol/g FW) |
|---|---|---|---|---|---|---|---|
| **AC36** | 1974.51 b | 269.35 a | 230.26 a | 1106.82 a | 0.42 b | 103.90 a | 10.65 c |
| **AC37** | 2546.56 a | 212.69 b | 205.96 b | 1107.16 a | 0.46 a | 101.73 a | 9.07 d |
| **AC47** | 1801.69 c | 140.62 c | 198.65 c | 809.19 c | 0.13 c | 54.11 b | 16.68 a |
| **AC53** | 1697.59 d | 136.48 d | 208.43 b | 898.04 b | 0.12 c | 49.78 b | 12.18 b |

**Table 4.2.21 Overall mean comparisons responses by four barley accessions in respect to their control under present drought stress at (S1+S2) by seven biochemical markers.**

| | PC-Con. µg/g | SSC-Con. µg/g | TPC µg gallic acid per gram of fresh seedling | Antioxidant capacity by DPPH (µg Trolox/g FM) | GPA (units/min/g FW) | CAT (units/min/g FW) | Lipid peroxidation (nmol/g FW) |
|---|---|---|---|---|---|---|---|
| **S1,S2** | 2425.99 a | 242.89 a | 275.22 a | 1021.01 a | 0.41 a | 107.14 a | 14.21 a |
| **Control (C2)** | 1584.19 b | 136.68 b | 146.42 b | 939.59 b | 0.15 b | 47.62 b | 10.08 b |

**Table 4.2.22 The mean comparisons of seven biochemical traits in drought stress condition (S1+S3) with respect to control condition in four barley accessions based on Multiple Rang Duncan's test at *p* value < 0.01. Any values of mean holding a common letter are not significant.**

| | PC-Con. µg/g | SSC-Con. µg/g | TPC µg gallic acid per gram of fresh seedling | Antioxidant capacity by DPPH (µg Trolox/g FM) | GPA (units/min/g FW) | CAT (units/min/g FW) | Lipid peroxidation (nmol/g FW) |
|---|---|---|---|---|---|---|---|
| **AC37** | 3557.97 b | 160.74 b | 267.47 a | 1103.11 b | 0.42 a | 129.87 a | 10.90 c |
| **AC36** | 4053.36 a | 215.49 a | 234.42 b | 1108.18 a | 0.36 b | 99.57 b | 9.94 d |
| **AC53** | 2309.64 d | 119.51 c | 217.85 c | 802.43 d | 0.12 c | 49.78 c | 22.32 a |
| **AC47** | 2746.56 c | 99.29 d | 180.30 d | 900.41 c | 0.10 c | 41.13 d | 14.58 b |

**Table 4.2.23 Overall mean comparisons responses by four barley accessions in respect to their control under present drought stress at (S1+S3) by seven biochemical markers.**

| | PC-Con. µg/g | SSC-Con. µg/g | TPC µg gallic acid per gram of | Antioxidant capacity by DPPH (µg | GPA (units/min/g FW) | CAT (units/min/g FW) | Lipid peroxidation (nmol/g FW) |
|---|---|---|---|---|---|---|---|

| | | | fresh seedling | Trolox/g FM) | | | |
|---|---|---|---|---|---|---|---|
| **S1,S3** | 3805.67 a | 167.22 a | 244.20 a | 1006.32 a | 0.36 a | 106.06 a | 16.64 a |
| **Control (C3)** | 2528.10 b | 130.29 b | 205.81 b | 950.74 b | 0.14 b | 54.11 b | 12.23 b |

**Table 4.2.24 The mean comparisons of seven biochemical traits in drought stress condition (S2+S3) with respect to control condition in four barley accessions based on Multiple Rang Duncan's test at *p* value < 0.01. Any values of mean holding a common letter are not significant.**

| | **PC-Con. µg/g** | **SSC-Con. µg/g** | **TPC µg gallic acid per gram of fresh seedling** | **Antioxidant capacity by DPPH (µg Trolox/g FM)** | **GPA (units/min/g FW)** | **CAT (units/min/g FW)** | **Lipid peroxidation (nmol/g FW)** |
|---|---|---|---|---|---|---|---|
| **AC37** | 3596.56 b | 149.60 b | 265.88 a | 1099.05 b | 0.54 a | 138.53 a | 11.41 c |
| **AC36** | 3903.49 a | 186.02 a | 249.96 b | 1110.88 a | 0.45 b | 110.39 b | 10.32 d |
| **AC53** | 2301.56 d | 127.25 c | 218.97 c | 807.50 d | 0.12 c | 47.62 c | 24.06 a |
| **AC47** | 2578.87 c | 102.99 d | 178.43 d | 904.12 c | 0.11 c | 43.29 c | 16.08 b |

**Table 4.2.25 Overall mean comparisons responses by four barley accessions in respect to their control under present drought stress at (S2+S3) by seven biochemical markers.**

| | **PC-Con. µg/g** | **SSC-Con. µg/g** | **TPC µg gallic acid per gram of fresh seedling** | **Antioxidant capacity by DPPH (µg Trolox/g FM)** | **GPA (units/min/g FW)** | **CAT (units/min/g FW)** | **Lipid peroxidation (nmol/g FW)** |
|---|---|---|---|---|---|---|---|
| **S2,S3** | 3612.14 a | 157.64 a | 250.81 a | 1010.03 a | 0.47 a | 115.80 a | 18.71 a |
| **Control (C3)** | 2578.10 b | 125.29 b | 205.81 b | 950.74 b | 0.14 b | 54.11 b | 12.23 b |

**Table 4.2.26 The mean comparisons of seven biochemical traits in drought stress condition (S1+S2+S3) with respect to control condition in four barley accessions based on Multiple Rang Duncan's test at *p* value < 0.01. Any values of mean holding a common letter are not significant.**

| | **PC-Con. µg/g** | **SSC-Con. µg/g** | **TPC µg gallic acid per gram of fresh seedling** | **Antioxidant capacity by DPPH (µg Trolox/g FM)** | **GPA (units/min/g FW)** | **CAT (units/min/g FW)** | **Lipid peroxidation (nmol/g FW)** |
|---|---|---|---|---|---|---|---|
| **AC37** | 3831.69 b | 161.33 b | 245.84 a | 1099.73 a | 0.42 a | 160.17 a | 11.46 c |
| **AC36** | 4334.26 a | 177.38 a | 226.99 b | 1086.55 b | 0.31 b | 129.87 b | 10.60 d |
| **AC53** | 2110.41 d | 98.52 c | 204.06 c | 783.18 d | 0.11 c | 45.45 c | 26.07 a |
| **AC47** | 2832.33 c | 90.03 d | 170.75 d | 870.68 c | 0.10 c | 41.13 c | 19.22 b |

**Table 4.2.27 Overall mean comparisons responses by four barley accessions in respect to their control under present drought stress at (S1+S2+S3) by seven biochemical markers.**

| | **PC-Con. µg/g** | **SSC-Con. µg/g** | **TPC µg gallic acid per gram of fresh seedling** | **Antioxidant capacity by DPPH (µg Trolox/g FM)** | **GPA (units/min/g FW)** | **CAT (units/min/g FW)** | **Lipid peroxidation (nmol/g FW)** |
|---|---|---|---|---|---|---|---|
| **S1,S2,S3** | 3901.24 a | 138.33 a | 221.34 a | 969.32 a | 0.33 a | 134.20 a | 21.44 a |
| **Control (C3)** | 2653.10 b | 125.29 b | 202.48 b | 950.74 b | 0.14 b | 54.11 b | 12.23 b |

The difference in drought resistance between the sensitive and resistant accessions was comparatively investigated with their control. The growth of biomass and yield-related traits in drought-sensitive barley accessions AC47 and AC53 was greatly repressed in all individual and combined stress conditions, while no such phenomenon was observed for AC36 and AC37, in which the effects of drought were lighter.

The most critical growth stage of crop development that is adversely affected by drought stress in terms of yield is anthesis (Luo, Xia and Lu, 2019). When drought is present during this particular growth stage, the fertility of seeds is reduced, which, as a consequence, the yield of grain is reduced. While in the presence of drought at vegetative stages, the yield of grain is reduced as a consequence of the considerable reduction in the weight of grain (Kebede, Kang and Bekele, 2019).

In the current study, a considerable reduction in most traits under drought stress conditions compared to control conditions was detected. These results showed that drought stress reduced total yield by decreasing seed number per spike, fertile tiller, and other related yield traits. Our outcomes are supported by the results of (Al-Ajlouni *et al.*, 2016), who studied the response of eleven barley genotypes to stress at vegetative and reproductive growth stages and stated great yield reduction due to the presence of water deficit at these stages by reducing fertile spikes and seed number per plant.

Our results are in agreement with Dodig *et al.* (2020), who found that drought stress declined the yield of grain by reducing spikes and grains per plant in barley as a consequence of the early ripeness of barley spikes and shortened the period of the filling of grain at the present of water holding at vegetative and flowering stages.

The parameters include; shoot fresh and dry weight, fluorescence, and chlorophyll content. Relative water content is measured as a quick trait for detecting drought-resistant genotypes. Unexpectedly, the sensitive barley accessions AC47 and AC53 showed quite a similar ratio of response for relative water content and chlorophyll content to the resistance barley accessions AC36 and AC37, indicating the change between the susceptible group and the resistant one was not significant. Similarly, a group of researchers (Hasanuzzaman *et al.*, 2017) conducted their research on six barley genotypes in the plastic house to assess their performance under drought conditions, and no obvious significant changes were detected among sensitive and resistant genotypes for chlorophyll content and relative water content. Munns *et al.* (2010) stated that the records are generally made on a single leaf and could not reflect the performance of the entire plant. From their viewpoint, the measurements of chlorophyll are not a suitable indicator for detecting genotypic alterations in the growth response to water stress conditions. However, in our research, the measurements were taken carefully by setting up the SPAD meter on an average of 5 readings to obtain one record, and the triple measurements were then taken over the replications. Furthermore, all records were taken at the middle of the leaf to overcome the mentioned limitation. Regarding the analysis of barley genotypes under stress conditions, Cai *et al.* (2020) also reported the same responses to the chlorophyll content by tested materials. The performance of plant height under drought stress compared to under control conditions is also another indicator for identifying drought resistant genotypes by way of allocation of carbohydrate reserves to maintain the filling of grain (Kumar *et al.*, 2008; Kebede, Kang and Bekele, 2019). The percentage of shoot length was higher for both AC53 and AC47 as stated in the analysis under all drought conditions in comparison with AC36 and AC37 in respect to their untreated condition.

A multisite study by Carter, Hawes and Ottman (2019) revealed positive and significant associations between the performance of plant height with the outcomes of barley grain yields under drought stress conditions. In parallel with our findings, they observed that the susceptible varieties maintained the plant height under moderate and high irrigation systems while the resistant varieties were significantly affected by these systems in reducing their height.

As stated by Kebede, Kang and Bekele (2019), the performance of root traits was adversely related to the plant height under the limited availability of water. This statement is clearly documented in our results. When the length and dry accumulation of roots increased by AC36 and AC37, the length of the shoot decreased in response to drought conditions, while the opposite was true for AC47 and AC53.

Two types of tillers are recognized in cereals (fertile and non-fertile). The first, which is also known as the productive tiller, leads to the formation of spikes and therefore is most essential for the seed yield (Sreenivasulu and Schnurbusch, 2012). The total number of tillers can be increased after drought (Daryanto, Wang and Jacinthe, 2017). However, as a consequence of drought, especially at the flowering stage, a considerable reduction in the number of productive tillers (fertile) was detected, and it is strongly associated with a decrease in grain number and spike number per treated plant (Duggan *et al.*, 2005; Barati *et al.*, 2018). Strong reductions in biomass and tiller number similar to our results regarding the analysis of all stress conditions were discovered by (Dhanagond *et al.*, 2019) under the presence of drought stress with respect to control conditions. Across the study conducted by (Dhanagond *et al.*, 2019), six QTL regions were localized and among them, the *HvPPD-H1* region was the most important location related to the recovery phase for barley materials after re-watering, while this region was not significant in control conditions. Accordingly, they assumed that this region is correlated in maintaining the performance of tiller and biomass traits.

Under two irrigation conditions, field studies at Arizona University were conducted by Carter, Hawes and Ottman (2019) on four barley cultivars: Solar and Solum (considered drought resistant) as well as Kopiousand and Cochise (considered drought susceptible) to assess their performance for yield and root architecture. They found that both resistance cultivars were well-adapted to limiting water availability as the reduction in yield was much more similar to our outcomes than susceptible cultivars.

For exploiting more water uptake and nutrients from the soil texture and improving the total performance of studied cultivars under the presence of drought stress, it is generally believed that more prolific root architectures are the essential traits. The same research group in a different study (Carter, Hawes and Ottman, 2019) researched the same barley cultivars to assess the performance of roots in the presence of drought. In agreement with our results, they revealed a significant increase in root length and dry accumulation in roots by resistant cultivars.

In agreement with our outcomes, similar results were also found in the drought-tolerant genotypes by (Samarah *et al.*, 2009). They investigated the performance of growth and yield under three watering conditions of four barley cultivars under field and plastic house conditions. They found clear reductions in field-related traits for both susceptible cultivars while the resistant cultivars maintained higher proportions of yield under severe drought stresscondition.

The most abundant plant proteins formed in response to abiotic stresses are dehydrins. This type of protein has a high tendency to be attracted to water, and it is considered to have a high content of ion charge and polar residues. These characteristics permit dehydrins to co-operate well with water and to bind a large number of solute ions (Riyazuddin *et al.*, 2021). Many researchers have reported a significant association between the accumulation of Dhns gene families and resistance to abiotic stresses in different species. In wheat (Yu *et al.*, 2019b), barley (Vítámvás *et al.*, 2019), and Arabidopsis (Cui *et al.*, 2020).

Drine *et al.* (2018) conducted research on eight barley genotypes at the two-week growth seedling stage that originated from different geographic origins (Tunisia, Algeria, Jourdan, and UK). The plant materials were subjected to stress conditions by using 20% (w/v) of polyethylene glycol with a similar molecular weight as in our investigations. The expression of the *Dhn6* gene was determined after 6, 12, 24, and 48 hours of stress treatment. The resistance genotypes showed a rapid stimulation of *Dhn6* expression. On the contrary, for susceptible genotypes, delays in response and little expression of this particular gene were observed. Iqbal (2018) also stated that the expression level of this gene is strongly related to drought resistance in barley. The responses by both barley accessions AC36 and AC37 to studied traits were changed during drought stress conditions but rapidly re-established once accessions were re-watered. This was most likely due to increased *Dhn6* gene expression under drought stressor conditions.

The accumulation of proline at the various growth stages of four barley accessions when exposed to water stress conditions was estimated. Plants alter their physiology and modify root growth to adapt to the available moisture content in the soil profile (Luo, Xia and Lu, 2019). In our investigation, the resistant accessions exhibited more proline concentration, dry root accumulation, and root elongation compared to susceptible accessions. This agrees well with the findings of, Bandurska *et al.* (2017), who proposed that the accumulation of proline in the roots improves the performance of barley genotypes under water stress conditions. With limited water availability, osmotic adjustment of water potential as a result of the accumulation of proline has been identified to play an essential role in maintaining the elongation of roots (Ghoulam, Foursy and Fares, 2002; Abdelaal *et al.*, 2020). The increase in proline content and root-related traits may be due to the fact that the expression of *P5cs1* is most highly induced in root and leaf tissue organs.

It has been demonstrated that soluble sugar content increases osmoprotectant synthesis as a metabolic adaptation to help the plant cope with drought or heavy metal stress conditions (Dien, Mochizuki and Yamakawa, 2019; Janeczko *et al.*, 2016; Souahi *et al.*, 2021). A considerable increase in the levels of total soluble sugars was stated in our analysis in response to all types of stress by both accessions AC36 and AC37, On the contrary, the levels of lipid peroxidation, which is measured as a suitable indicator of oxidative stress in barley (Torun, 2019), were much lower in the AC36 and AC37 barley accessions than in AC47 and AC53. Similar to our finding under plastic house conditions (Dbira *et al.*, 2018), nine barley accessions were tested for drought resistance using SSC and TPC and lipid peroxidation traits on dry barley leaves. They stated a significant increase in these traits except lipid peroxidation under the presence of drought by two barley tolerant accessions.

Similarly, to manage the accumulation of reactive oxygen species, the activity of antioxidant enzymes such as catalase (CAT) and guaiacol peroxidase activity (GPA) is considerably increased under present drought conditions (Zhanassova *et al.*, 2021). The maximum levels of both enzyme activities were perceived in both barley accessions AC36 and AC37 compared with control well-watered conditions. In parallel with our finding, Kaur *et al.* (2021) tested the catalytic activities on flag leaf and seeds of four barley genotypes under the presence of drought at the stage of grain filling. They found a considerable increase in this trait in two barley genotypes.

Phenolic compounds play an enormous role in eliminating oxidative stress since they are involved in the process of ROS detoxification (Akula and Ravishankar, 2011). AC36 and AC37 exhibited the highest values of phenolic compounds and therefore seem to have a greater ability to eliminate ROS. Similar to our outcomes, Ahmed *et al.* (2015) exposed three barley cultivars (two resistant and one susceptible) to drought, salinity, and combinations of stress under plastic house conditions. They found a considerable increase in this trait as well as a secondary metabolite of an antioxidant under the presence of drought and other types of stressors by resistant cultivars. Likewise, Sehar *et al.* (2021) also stated a remarkable increase in the presence of drought by tolerant barley accessions for increasing the concentration of total phenolic compounds as well as anti-oxidant.

Recently, three barley accessions, including Tibetan wild barley, which is considered drought resistant and subjected to drought and salinity stress conditions (Ahmed *et al.*, 2020). They found a considerable increase in the enzymatic activities of CAT and GPA by these accessions under the presence of stress conditions. Besides, the activities of these enzymes were reduced in the case of the silencing of *HvSAMS3*. From this point, it is possible to conclude that the expression of this particular protein was increased by both barley accessions AC36 and AC37 in response to drought stress in our investigation. For the synthesis of salicylic acid (SA) in plants, the synthase of isochorismate (*ICS*) is considered an essential enzyme (Wildermuth *et al.*, 2001). A group of researchers generated transgenic lines from barley, Golden Promise spring barley, owning the *ICS* gene (Hao *et al.*, 2018). They confirmed that these lines can resist Fusarium disease in barley. Based on this discovery, Wang *et al.* (2021) tested these lines in the presence of drought with respective wild types of barley. Interestingly, transgenic lines and wild type at the early growth stage when subjected to PEG (6000)-20% for one week showed a higher expression level of the *ICS* gene. In addition, the enzymatic activities of CAT and GPA were high under the present drought stress conditions in transgenic lines, while no considerable changes in these lines were found for lipid peroxidation. In our study, the expression level of this particular gene was probably upregulated under the presence of drought by both accessions AC37 and AC36. In addition to the *ICS* gene, Yuan *et al.* (2019) found that regulation of *miR393* could confer resistance to drought in barley . In their study, they faced drought conditions on two transgenic lines (overexpressed and knocked down with *miR393*) and wild relative barley. They revealed that knocked down lines enhanced drought resistance and fewer activities by these lines were detected for lipid peroxidation in comparison with wild type and overexpressed one.

Improving a plant for it is drought resistant is no easy task as many genes are involved in controlling drought and were considered a complex trait. Here, many genes that till the moment showed strong associations with drought morphologically and biochemically in barley were addressed. These pathways or/and expressions of mentioned genes are most likely involved in drought tolerance by both barley accessions AC36 and AC37.

Taking all morphological and biochemical outcomes into consideration, it revealed the novelty of this investigation conducted in plastic house conditions for selecting AC36 and AC37 as resistance barley accessions and AC47 and AC53 as drought susceptible accessions.

## 4.3 Heavy Metal Section

### 4.3.1 Cadmium experiments at seedling stage

#### 4.3.1.1 Performance of some morphological traits under cadmium stress conditions

Cadmium accumulation in plant tissues inhibits growth and several toxicity symptoms are induced. The first symptom of Cd accumulation is a decrease in plant growth and development (Shiyu *et al.*, 2020). At different levels, this effect is observed, such as the elongation of roots and shoots, germination, and the accumulation of cadmium in plant tissue organs by different plant species (El Rasafi *et al.*, 2021).

Research conducted by Kintlová *et al.* (2017) showed the inhibitory effects of cadmium on barley growth, particularly on root length, using eleven different concentrations of cadmium ranging between 0.01 µM and 1000 µM. It is worth mentioning that even with using a high concentration of cadmium, the barley roots survived in their research. In this investigation, barley accessions were cultivated in a control medium and a medium supplemented with three different levels of cadmium concentrations (125, 250 and 500) µM to determine the cadmium effects of the tested barley accession on growth traits, mainly (RL) root length (cm), (SL) shoot length (cm), (GP) germination percentage %, (FWS) fresh weight of seedling (g) and (DWS) dry weight of seedling (g). Eight days after treatments, typical symptoms of Cd toxicities among barley accessions were developed, where clear necrotic lesions, and dark-brown dots were observed on both roots and shoots.

As indicated by the ANOVA analysis and box charts, the plant growth traits under both untreated and treated with different concentrations of Cd a tremendous significant phenotypic diversity at level ($p \leq 0.001$) were observed among tested barley accessions (Table 4.3.1), and significant reductions in root and shoot length, germination percentage, and fresh weight of seedling were inhibited when the barley accessions were exposed to the high dosage of cadmium. The reduction started and continued from the lowest dose of 125 µM to the highest dose of 500 µM while the dry weight of barley seedling exhibited more accumulations with the increase of cadmium dosage (Fig. 4.3.1). Under availability of all cadmium stress conditions, mean pairwise comparisons showed that barley accessions responded differently for morphological traits (GP, RL, SL, FWS, DWS and WU) (Appendix 3.1). For these traits respectively the values ranged between (99.17-AC2 and 39.58-AC48)%, (7.18-AC1 and 2.99-AC57)cm, (9.85-AC1 and 4.84-AC57)cm, (0.34-AC58 and 0.16-AC7)g, (0.04-AC29 and 0.02-AC3)g and (281.34-AC3 and 90.23-AC52).

The box charts of all morphological traits showed significant variations between Cd-0 (control), Cd-125, Cd-250, and Cd-500, as shown for each studied trait by the lower and upper box plot limits (Figure 4.3.1). Compared to stressed conditions, higher trait values in normal conditions in overall tested barley accessions were detected in almost all studied parameters, with the only exception of DWS. These outcomes indicated that following exposure to Cd-500, germination and seedling growth parameter depression were more severe for all accessions (Appendix 3.2).

Regarding the analysis for GP under control conditions, the performance of this trait ranged between %(50 to 100) with a mean of 88.19. The thirteen barley accessions among all tested accessions were fully germinated while poor performance by the three accessions, namely; AC 51, 53, and 57, was observed (Table 4.3.1). While the performance of barley accession AC48 for this trait was less in all cadmium-treated conditions (125, 250, and 500) µM with the values of (25.00, 36.67, and 26.67)% and the mean value were (87.18, 86.61, and 84.80)% respectively.

The measured scores of the accessions under control conditions regarding the analysis of root length ranged between (1.59 and 15.85)cm, with the mean value of 9.16 cm for barley accessions AC3 and AC1 respectively. Barley accession AC15 exhibited the lowest length of 3.50 cm compared to AC39, in which the highest root length was documented by this accession with the value of 9.08 cm under treatment of Cd-125, and the mean value in this particular state was 5. 77cm. Following the increased concentration of Cd-250, the length of the studied trait (RL) ranged between (2.29 and 5.29)cm with the mean value of 3.68 cm for barley accessions AC34 and AC8 respectively.

As observed for root length under control conditions, the barley accession AC1 had better performance over other studied accessions for its shoot length (SL), having a value of 13.19 cm, while the accession AC3 was characterized as the shortest barley accessions for a studied trait with a value of 1.27 cm, and the mean value for this particular condition was 9.83 cm. AC24 had more potential to elongate in the case of the present lowest dose of cadmium applied in our investigation, with a value of 11.80 cm. By contrast, the barley accession AC57 exhibited the shortest shoot length of 5.17 cm, compared to other accessions, and the mean value was 7.99. The lowest value of shoot length was recorded by AC7 (4.41) cm while the accession AC23 performed well as the length of 9.84 cm was documented by this accession under the second higher exposure of cadmium with the mean value of 6.77 cm as shown in (Table 4.3.1). A notable reduction was observed for the studied trait in the case of the highest dose, in which the mean value was 5.49 cm. Over again, the barley accession AC52 followed the same trend as detected for root length, retaining the lowest length for it is shoot at 2.66 cm, while the barley accession AC8 performed better over the rest, having a 7.64 cm shoot length.

In the present investigation, fresh weight seedling (FWS) was performed for the purpose of assessing barley accessions under different concentrations of cadmium. As it is clear from the mean analysis presented in (Table 4.3.1) and box chart form (Fig. 4.3.1), the cadmium exposures were negatively affected (FWS). The overall performance of barley accessions as indicated in (Figure 4.3.1) revealed a gradual decrease in the studied trait starting from the lowest Cd-0 to the highest Cd-500 dose with mean values of (0.29, 0.27, 0.22, and 0.18)g, respectively. Under control conditions, the lowest fresh weight was detected by barley accession AC47 with a value of 0.15 g. By contrast, the two barley accessions AC24 and AC58, with a value of 0.40 g, recorded the highest FWS (Table 4.3.1). The barley accession AC58, in comparison with the rest of the accessions, had a superior performance for the studied trait, having a value of 0.41 g under the dose of 125 µM of cadmium exposure, while both accessions AC7 and AC9 recorded the lowest value of 0.16 g. A remarkable reduction in fresh weight was detected as presented in (Table 4.3.1) for barley accession AC3 with a value of (0.14 and 0.09)g when accessions were subjected to (250-500) µM of cadmium exposure, respectively. By contrast, under the same condition, barley accessions AC41, with a value of 0.31 g, and AC58, with a value of 0.27 g, recorded the highest value.

Barley accessions were completely dried in the oven to assess the dry weight of the seedling after measuring the fresh weight. Under control and Cd-125, there were not many differences in the overall performance of barley accessions for (DWS) as both shared the same value of 0.030 g (Table 4.3.1), while a slight increase in mean value was detected by all tested accessions in the case of present Cd-250 (0.031) g and Cd-500 (0.033) g. Under untreated conditions, dry accumulation for barley accession AC34 was highest with a value of 0.042 g compared to AC5 with a value of 0.018 g. Similar to FWS, both accessions AC7 and AC9, with a value of 0.019 g, possessed the lowest rate of dry accumulation under Cd-125, whereas barley accession AC29 possessed the highest value of dry weight of 0.042 g (Table 4.3.1). In the presence of the higher dose of cadmium, AC34 collected more dry matter compared to AC5 with a value of (0.043-0.018) respectively, while barley accession AC29, similar to Cd-125 condition-owned the highest accumulation of 0.05 g compared to AC3 with the value of 0.02 g under the final cadmium dose exposure.

Both studies by Kalai et al. (Kalai *et al.*, 2014; Kalai *et al.*, 2016) show that roots and shoots of barley plants grew slower when exposed to cadmium at three different concentrations: 25, 50, and 100 M, compared to untreated plants. This was true even with the lowest dose of cadmium used in their research, which is similar from what we found. In addition, they found clear inhibition of seed germination after cadmium exposure because of failure in reserve mobilization from the endosperm, not a consequence of limiting water uptake by barley seeds. Through diverse mechanisms, cadmium is known to inhibit the germination of seeds (Huybrechts *et al.*, 2019). Very recently, Kintlová *et al.* (2021) from the Czech Academy of Sciences, apart from detailed research on transcriptomic responses of barley to cadmium, showed a significant reduction in barley growth at the seedling stage for both shoot and root underexposure of barley plant at 80 µM of cadmium.

Wang *et al.* (2019) developed 108 doubled haploid (DH) barley lines that were generated from a cross between the Cd-tolerant parent (Weisuobuzhi) and the Cd-sensitive parent (Suyinmai 2) to identify the genes that are associated with Cd accumulation and resistance. As shown in their investigation, a novel *HvPAA1* and p-type ATPase were identified and cloned to examine it is function and they found a critical role for this gene in the detoxification of cadmium in barley under a cadmium concentration of 10 µM. The transmembrane proteins that are critical for the homeostasis of ions and detoxification of heavy metals are P-type ATPases that play an essential role in the transport of a wide variety of cations across membranes (Lei *et al.*, 2020). From this point, the probable reason behind the diverse variation and better performance by some barley accessions under the severe concentration of cadmium exposure as indicated in (Table 4.3.1) was due to the expression of these particular genes in excessive quantity.

**Table 4.3.1 Effect of the different Cd treatments on plant growth traits of barley seedlings.**

| | | | | | | | | | | |
|---|---|---|---|---|---|---|---|---|---|---|
| **Cd-0** | | | | | | | | | | |
| | **Germination percentage** | **Accessions** | **Root length (cm)** | **Accessions** | **Shoot lenght (cm)** | **Accessions** | **Fresh weight of seedling (g)** | **Accessions** | **Dry weight of seedling (g)** | **Accessions** |
| Min. | 50.00 | AC:51,53 and 57 | 1.59 | AC3 | 1.27 | AC3 | 0.15 | AC47 | 0.018 | AC5 |
| Max. | 100.00 | AC: 2, 4, 9, 21, 24, 27, 30, 31, 36, 38, 40, 41 and 42 | 15.85 | AC1 | 13.19 | AC1 | 0.40 | AC: 24 and 58 | 0.042 | AC34 |
| Mean | 88.19 | | 9.16 | | 9.83 | | 0.29 | | 0.030 | |
| Std. | 12.27 | | 2.65 | | 1.88 | | 0.06 | | 0.006 | |
| F | 127.95*** | | 19793*** | | 11648.15* | | 5.42*** | | 4.48*** | |
| Pr | < 0.0001 | | < 0.0001 | | < 0.0001 | | < 0.0001 | | < 0.0001 | |
| **Cd-125** | | | | | | | | | | |
| | **GP** | | **RL** | | **SL** | | **FWS** | | **DWS** | |
| Min. | 25.00 | AC48 | 3.50 | AC15 | 5.17 | AC57 | 0.16 | AC: 7 and 9 | 0.019 | AC: 7 and 9 |
| Max. | 100.00 | AC25 | 9.08 | AC39 | 11.80 | AC24 | 0.41 | AC58 | 0.042 | AC29 |
| Mean | 87.18 | | 5.77 | | 7.99 | | 0.27 | | 0.030 | |
| Std. | 13.61 | | 1.21 | | 1.24 | | 0.06 | | 0.005 | |
| F | 16.40*** | | 15.52*** | | 13.77*** | | 6.44*** | | 4.25*** | |
| Pr | < 0.0001 | | < 0.0001 | | < 0.0001 | | < 0.0001 | | < 0.0001 | |
| **Cd-250** | | | | | | | | | | |
| | **GP** | | **RL** | | **SL** | | **FWS** | | **DWS** | |
| Min. | 36.67 | AC48 | 2.29 | AC34 | 4.41 | AC7 | 0.14 | AC3 | 0.018 | AC5 |
| Max. | 100.00 | AC: 2, 9, 11, 28, 29, 36, and 38 | 5.29 | AC8 | 9.84 | AC23 | 0.31 | AC41 | 0.043 | AC34 |
| Mean | 86.61 | | 3.68 | | 6.77 | | 0.22 | | 0.031 | |
| Std. | 13.96 | | 0.74 | | 1.31 | | 0.05 | | 0.006 | |
| F | 15.06*** | | 5.68*** | | 6.84*** | | 4.92*** | | 6.751*** | |
| Pr | < 0.0001 | | < 0.0001 | | < 0.0001 | | < 0.0001 | | < 0.0001 | |
| **Cd-500** | | | | | | | | | | |
| | **GP** | | **RL** | | **SL** | | **FWS** | | **DWS** | |
| Min. | 26.67 | AC48 | 0.70 | AC52 | 2.66 | AC52 | 0.09 | AC3 | 0.02 | AC3 |
| Max. | 100.00 | AC: 2, 3, 25 and 31 | 3.65 | AC39 | 7.64 | AC8 | 0.27 | AC58 | 0.05 | AC29 |
| Mean | 84.80 | | 2.03 | | 5.49 | | 0.18 | | 0.033 | |
| Std. | 17.69 | | 0.72 | | 1.15 | | 0.04 | | 0.006 | |
| F | 22.39*** | | 7.41*** | | 7.82*** | | 6.20*** | | 5.834*** | |
| Pr | < 0.0001 | | < 0.0001 | | < 0.0001 | | < 0.0001 | | < 0.0001 | |

### 4.3.1.2 Biochemical markers assays

Through secondary metabolism in plants, a large number of compounds are provided that mostly function to enhance plants' tolerance to different stress conditions (Berni *et al.*, 2019; Ishtiyaq *et al.*, 2021). Secondary metabolites and phytohormones play a precise role in reducing the adverse effects of heavy metals by chelating metal ions of cadmium and other heavy metal forms, reducing the level of ROS, limiting the synthesis of free radicals, and providing an osmotic homeostasis balance of nutrients (Ashfaque *et al.*, 2020). One critical marker that affects the response of plants to secondary metabolism production is the concentration of heavy metals. As mentioned by Jain, Khatana and Vijayvergia (2019), a low level of heavy metals increases the production of secondary metabolites, while the synthesis and its production will be reduced in the plant in the case of a higher dose of heavy metals. In our investigation, seven biochemical markers, including; Proline Content (PC), Total Phenolic Content (TPC), Anti-oxidant (AC), Soluble Sugar Content (SSC), Catalase (CAT), Guaiacol peroxidase activity (GPA), and Lipid Peroxidation (LP), were conducted to observe the response of tested barley accessions to different concentrations of cadmium.

Among the tested barley accessions (Table 4.3.2), highly significant biochemical responses at a level ($p \leq 0.001$) were observed under both control and treatment with different concentrations of Cd. Mean pairwise comparisons and interactions in the case of using different concentrations of cadmium stress on studied barley accessions showed significant impacts on the studied biochemical traits including (PC, SSC, TPC, AC, LP, APX and CAT) (Appendix 3.3). Different values were documented and the values ranged between (9390.41 - AC40 × Cd-500 and 294.00 - AC5×Cd-0), (1185.86 - AC33 × Cd-500), (323.93 - AC25 × Cd-250 and 68.88 - AC27 × Cd-0), (1188.92 - AC3 × Cd-250 and 552.43 - AC3 × Cd-0), (29.40 - AC42 × Cd-500 and 2.08 - AC45 × Cd-0), (0.96 - AC50 × Cd-0 and 0.02 - AC44 × Cd-250) and (174.03 - AC12 × Cd-500 and 12.99 - AC18 × Cd-125), for mentioned traits

respectively. However, the same biochemical traits responded differently by barley accessions under the availability of all cadmium stressor conditions as shown in (Appendix 3.4). For traits (PC, SSC, TPC, AC, LP, APX and CAT), the values ranged between (4940.92-AC38 and 841.44-AC12), (520.33-AC39 and 102.76-AC54), (221.59-AC25 and 93.14-AC53), (1033.51-AC51 and 711.72-AC45), (26.11-AC42 and 6.51-AC1), (0.58-AC38 and 0.12-AC51), and (112.34-AC25 and 35.06-AC22) respectively.

As shown by the ANOVA analysis and box charts (Fig. 4.3.2), a sharp increase was observed in the case of the highest dose of cadmium treatment, especially for PC and SSC, and followed by TPC, AC, CAT, GPA, and LP, while the barley accessions in the second exposure of cadmium similarly responded greatly to this particular condition. For instance, the highest accumulations were detected for TPC, AC, LP, and CAT, followed by PC and SSC. In the case of the lowest dose of cadmium, the incensement was less in comparison with other highest doses of cadmium exposure, with the only exception of AC, in which, similar to the second dose, barley accessions collect more AC in their cells to prevent the toxicity effects of cadmium. These outcomes point out that following exposure to cadmium, the accumulation of secondary metabolites and phytohormones increased in tested barley accessions.

The uptake of water is the first step for seed germination and ends with the start of elongation by the radicle. The dynamics of seed germination are altered under several physical and heavy metal stress conditions. There has been much research on the effect of heavy metals on germination mechanisms and water uptake (Nouri, El Rasafi and Haddioui, 2019; Asif, Ali and Malik, 2020; Wang *et al.*, 2017).

Regarding the analysis of seed water uptakes, as indicated by the boxplot in the first and second doses of cadmium treatment, the barley accessions tended to accumulate more water or cadmium in this particular growth stage, while no clear differences were observed for the third dose in comparison with the respective control conditions (Fig. 4.3.2).

In three treatment conditions, the minimum value was recorded by accession AC52 Cd-0 (71.16), Cd-125 (86.92), and Cd-500 (87.53) with the mean value of (173.12, 246.21, and 164.45), respectively, while the highest values were detected by AC37 (368.17), AC4 (395.08), and AC6 (236.91) for the three mentioned conditions, respectively. In addition to these situations, the minimum record was exhibited by barley accession AC15 (87.68) in contrast to barley accession AC3 (349.26) under the condition of Cd-250 (Table 4.3.2).

A question arose from these outcomes, that is, whether the notable increase in seed weight after the application of cadmium was due to the absorption of water or cadmium accumulation. Research conducted by (Kalai *et al.*, 2014) on barley to know the response of plant materials in the presence of two types of heavy metals, namely, copper and cadmium, under three different doses of cadmium (25, 50, and 100) µM . Similar to our investigation, barley seeds were exposed to heavy metals, but no obvious water uptake was detected by stressed seeds. One among various mechanisms that are proposed to inhibit cadmium uptake and protect the plant from toxicity is the biosynthesis of phytosiderophore chelation (Bali, Sidhu and Kumar, 2020). This type of chelator, as stated and tested on barley by Kudo, Kudo and Kawai (2007) could mediate cadmium mobilization around the root and cause the improvement of cadmium aggregation in the radical. From those points, it can be assumed that increasing the weight of the seedling, especially for the first and second dose of cadmium application by most accessions, is related to this mechanism.

In the presence of heavy metals, proline can directly chelate these substrates and, accordingly, reduce the toxic effects of metals on plant organs (Ahmad *et al.*, 2019). The accumulation of proline is also related to carbohydrate metabolism. Alhasnawi (2019) stated that the accumulation of proline requires carbohydrates. Interestingly, in our investigation, a similar pattern of responses by barley accessions was stated for both proline and soluble sugar content accumulations.

When the PC data were analyzed under untreated conditions, the mean value was 1303.15, with a range of AC52 (294.00) to AC33 (2907.33). In the subsequent condition, AC35 (3598.36) collected more PC in comparison with AC49 (367.59) with a mean value of 1371.76 whereas different patterns of response for this particular trait by the barley accessions were observed under Cd-250 in which mean values were almost doubled compared with previous cadmium supply and barley accession AC55 aggregate more PC compared to AC6 with the value of 5836.56 and 769.64 while slight increase by all accessions was documented for PC under the final concentration of cadmium exposure compared to intermediate cadmium condition with the mean value of 2913.45 and the performance of barley accession were ranged between (9390.41-358.62) for barley accessions AC40 and AC48 respectively (Table 4.3.2). Bandurska *et al.* (2017) tested the synthesis of proline under drought conditions for two barley genotypes. They found a strong association between the expression of the *P5CS* gene and proline accumulation under water deficit conditions. From this point, it is possible to conclude that other types of stress conditions such as heavy metals will regulate and enhance the expression of this particular gene to store more proline.

In response to cadmium by tested accessions, TPC traits were performed and the data were analyzed as shown in (Table 4.3.2). At the second dose of cadmium exposure, the overall barley accessions stored more phenolic compounds with a mean value of 172.21, followed by the last exposure of cadmium and the first dose, with a mean value of 161.05 and 149.05, respectively, in association with the control condition 107.42. Under intermediate cadmium conditions, the maximum storage for the studied trait was documented by AC25 (323.93). On the other hand, barely accession AC53 stored less TPC (102.21). In the presence of the extreme cadmium exposure in our experiment, AC25 again showed superior performance for TPC storage with a value of 307.27 compared to AC18 with a value of 86.10. However, a different pattern of TPC accumulation was documented under first cadmium exposure, and the highest and lowest TPC were recognized by barley accessions AC57 (234.42) and AC56 (91.80). In the absence of cadmium, the TPC accumulation ranged between 177.12 and 68.88 for barley accessions AC57 and AC27, respectively (Table 4.3.2).

The results displayed that the effect of the provided treatments of cadmium was significant on antioxidants (AC), especially when the barley accessions were treated with the first and second dose of cadmium and followed by the third dose, with a mean value of 930.74, 930.85, and 910.95, respectively, while the mean value of the control condition was less with a value of 799.91. The barley accession AC3 in control and the first dose of cadmium treatment is considered as an accession that retains the lowest AC with a value of 552.43 and 686.89, respectively, while in the case of the present second dose, the accumulation of this trait by the same accession is boosted to become the superior accession among others with a value of 1188.92. Under control conditions, as shown in (Table 4.3.2), barley accession AC58, with a value of 994.32, owned the greatest value for AC accumulation, while in the first dose of treatment of cadmium, the barley accession AC39, with a value of 1118.65, was similarly considered as the well-performed accession for AC storage compared to the rest. The subsequent increase in treatment revealed that barley accession AC45, with a value of 678.11, had a limited ability to collect AC. Regarding the accumulation of AC under the severe concentration of heavy metals in our study, The values ranged between 557.16 and 1179.46 for barley accessions AC48 and AC37, respectively.

The functions of soluble sugars are signaling and sensing molecules in plants and, in that way, regulate and activate several genes that are involved in metabolic and protection activities against diverse stress conditions, including heavy metals (Yu *et al.*, 2019a). The enzyme that plays a fundamental role in controlling the carbohydrate pathway and sugar signaling is cell wall invertase (CWI). In many plants under stressed environments, this particular enzyme is involved in a variety of metabolic and signaling processes (Jing *et al.*, 2020; Arnao *et al.*, 2021). Related to our investigation, probably introducing the toxic level of cadmium enhanced those pathways, and in return, most barley accessions accumulated more soluble sugar.

A similar pattern of responses by the barley accessions in comparison with PC was detected for SSC (Table 4.3.2). Under extreme cadmium application for this trait, the maximum scores were achieved for mean value by overall accessions 379.39 followed by the intermediate and first dose of cadmium with values of 340.21 and 272.74 respectively, whereas the accumulations of soluble sugar were not as much by the accessions compared with untreated condition 191.01. Under all cadmium treatments, the barley accession AC54 showed an almost similar pattern of SSC accumulation and was considered the worst accession in comparison with the rest of the accessions for collecting soluble sugar in their cell organ. The value of these accessions from the first to final cadmium exposure was 94.20, 82.16, and 90.80 respectively. By contrast, AC20 (406.85), AC17 (809.01), and AC33 (1185.86) under three cadmium applications (Cd-125, Cd-250, and Cd-500) respectively reached the pick regarding the gathering of SSC, while different distribution by all accessions was noted under control conditions in which the value ranged between 105.10 and 374.34 for barley accessions AC53 and AC19, respectively. Lim et al. (2020) who worked on barley wild type stated that the overexpression of *HvCslF6* precisely in embryos improved the levels of soluble carbohydrates. Probably under cadmium stress conditions, this particular gene was expressed, thus more soluble sugar was accumulated.

Under intermediate cadmium contamination, the catalyze (CAT) activity significantly increased with a mean value of 71.36 compared with the other treatments, while similar responses by all barley accessions were documented for both treatments (Control and Cd-500) with mean values of 65.67 and 64.67, respectively (Table 4.3.2), followed by the first treatment of cadmium with a mean value of 61.29. Under non-cadmium treatment, the activity of catalase ranged between 12.99 and 127.27 for both barley accessions AC6 and AC20, respectively, while different responses by barley accession to the activity of this enzyme were detected under different concentrations of cadmium. Under continuous dose of cadmium, AC55 (106.88), AC27 (148.05), and AC25 (174.03) were recorded as the highest activities for this particular enzyme activity, while AC18, AC22, and AC16, with values of 12.99, 20.78, and 15.58 respectively, were considered as accessions for fewer CAT activities.

Diverse responses by barley accessions were obtained for (GPA) enzyme activities in the presence of toxic cadmium conditions. Under control conditions, the mean value was 0.34 and the responses for GPA activities ranged between 0.09 and 0.96 for AC23, AC30, and AC50, while under the first cadmium application, the minimum response was observed compared with the second and third dose, with a mean value of 0.23, 0.30, and 0.32. The maximum GPA activities were detected by AC53 (0.43) compared to AC58 (0.04) under Cd-125. Following the next two doses of cadmium, the barley accession AC38 achieved the most possible GPA activities with values of 0.82 and 0.75, respectively, while the least activities in the same order of cadmium application were spotted by barley accession AC44 (0.02) and AC52 (0.05).

Strong associations were identified in the case of the analysis of our data in response to cadmium heavy metal by tested accessions between TPC and MDA. As presented by the boxplot in (Fig. 4.3.2) similar distributions and responses were observed for all treatment conditions. There was a significant increase in lipid peroxidation at the second level of cadmium, followed by the third level and the first level, which had mean values of 13.253, 12.201, and 10.995, respectively. At the other levels of cadmium, all of the barley accessions had lower lipid peroxidation activities, with a mean value of 9.103.Among all tested barley accessions, AC42 in all experiment conditions starting from control to the maximum level of cadmium reached the pick activities of lipid peroxidation with values of 22.290, 25.274, 27.484, and 29.403 respectively. In the same chronological order, the lowest activities of MDA were observed by barley accessions AC45 (2.081), AC8 (5.177), AC55 (7.548), and AC59 (3.435) (Table 4.3.2).

In regulating the growth and development of plants, hormones play an essential role. In recent years, many types of research have been conducted, especially on one particular hormone named Strigolactones (SLs) that prevent the adverse role of stress conditions, and the interest in this hormone is quickly growing (Emamverdian, Ding and Xie, 2020; Bhoi *et al.*, 2021; Mostofa *et al.*, 2021). The application of strigol analogue (GR24) by inhibiting the uptake of cadmium enhanced the toxic effects of this type of heavy metal on barley genotypes, as stated by (Qiu *et al.*, 2021). Although there were no conclusive explanations by this research group indicating the molecular mechanism involved in the enhancement, Marzec *et al.* (2020) on rice found that plants naturally produce a wide variety of strigolactones and they also found several transcription factors in promoter regions of *MAX1* homologs that are expressed in response to cadmium and other types of stress. Similar to rice, these transcription factors are probably expressed in the presence of cadmium on barley accessions showing great responses by our materials for studied traits.

**Table 4.3.2 Effect of the different Cd treatments on water uptake and plant biochemical traits of barley seedlings.**

| | | Min. | Max. | Mean | Std. | F | *Pr* | | | Min. | Max. | Mean | Std. | F | *Pr* |
|---|---|---|---|---|---|---|---|---|---|---|---|---|---|---|---|
| **Cd-0** | **WU** | 71.16 | 368.17 | 173.12 | 61.54 | 4.35*** | < 0.0001 | **Cd-125** | **WU** | 86.92 | 395.08 | 246.21 | 67.9 | 5.62*** | < 0.0001 |
| | **Accessions** | AC52 | AC37 | | | | | | | AC52 | AC4 | | | | |
| | **PC** | 294 | 2907.33 | 1303.15 | 661.23 | 69.81*** | < 0.0001 | | **PC** | 367.59 | 3598.36 | 1371.76 | 657.16 | 166.74*** | < 0.0001 |
| | **Accessions** | AC5 | AC33 | | | | | | | AC49 | AC35 | | | | |
| | **TPC** | 68.88 | 177.12 | 107.42 | 23.24 | 392.67*** | < 0.0001 | | **TPC** | 91.8 | 234.42 | 149.05 | 32.55 | 958.04*** | < 0.0001 |
| | **Accessions** | AC27 | AC57 | | | | | | | AC56 | AC57 | | | | |
| | **AC** | 552.43 | 994.32 | 799.91 | 86.39 | 636.90*** | < 0.0001 | | **AC** | 686.89 | 1118.65 | 930.74 | 100.54 | 238.67*** | < 0.0001 |
| | **Accessions** | AC3 | AC58 | | | | | | | AC3 | AC39 | | | | |
| | **SSC** | 105.1 | 374.34 | 191.01 | 53.03 | 469.63*** | < 0.0001 | | **SSC** | 94.2 | 406.85 | 272.74 | 72.6 | 216.37*** | < 0.0001 |
| | **Accessions** | AC53 | AC19 | | | | | | | AC54 | AC20 | | | | |
| | **CAT** | 12.99 | 127.27 | 65.67 | 24.39 | 66.58*** | < 0.0001 | | **CAT** | 12.99 | 106.88 | 61.29 | 22.64 | 61.54*** | < 0.0001 |
| | **Accessions** | AC6 | AC20 | | | | | | | AC18 | AC55 | | | | |
| | **GPA** | 0.09 | 0.96 | 0.34 | 0.14 | 64.70*** | < 0.0001 | | **GPA** | 0.04 | 0.43 | 0.23 | 0.1 | 40.57*** | < 0.0001 |
| | **Accessions** | AC23,AC30 | AC50 | | | | | | | AC58 | AC53 | | | | |
| | **LP** | 2.081 | 22.29 | 9.103 | 3.577 | 879.64*** | < 0.0001 | | **LP** | 5.177 | 25.274 | 10.995 | 3.411 | 2043.50*** | < 0.0001 |
| | **Accessions** | AC45 | AC42 | | | | | | | AC8 | AC42 | | | | |
| **Cd-250** | | **Min.** | **Max.** | **Mean** | **Std.** | **F** | ***Pr*** | **Cd- 500** | | **Min.** | **Max.** | **Mean** | **Std.** | **F** | ***Pr*** |
| | **WU** | 87.68 | 349.26 | 209.03 | 54.4 | 5.51*** | < 0.0001 | | **WU** | 87.53 | 236.91 | 164.45 | 37.52 | 4.37*** | < 0.0001 |
| | | AC15 | AC3 | | | | | | | AC52 | AC6 | | | | |
| | **PC** | 769.64 | 5836.56 | 2442.8 | 1249.94 | 188.69*** | < 0.0001 | | **PC** | 358.62 | 9390.41 | 2913.45 | 2163.47 | 1748.41*** | < 0.0001 |
| | | AC6 | AC55 | | | | | | | AC48 | AC40 | | | | |
| | **TPC** | 102.21 | 323.93 | 172.21 | 40.48 | 921.50*** | < 0.0001 | | **TPC** | 86.1 | 307.27 | 161.05 | 51.92 | 1637.91*** | < 0.0001 |
| | | AC53 | AC25 | | | | | | | AC18 | AC25 | | | | |
| | **AC** | 678.11 | 1188.92 | 930.85 | 127.32 | 1755.22*** | < 0.0001 | | **AC** | 557.16 | 1179.46 | 910.95 | 171.25 | 552.33*** | < 0.0001 |
| | | AC45 | AC3 | | | | | | | AC48 | AC37 | | | | |
| | **SSC** | 82.16 | 809.01 | 340.21 | 128.27 | 871.32*** | < 0.0001 | | **SSC** | 90.8 | 1185.86 | 379.39 | 213.54 | 2210.86*** | < 0.0001 |
| | | AC54 | AC17 | | | | | | | AC54 | AC33 | | | | |
| | **CAT** | 20.78 | 148.05 | 71.36 | 32.96 | 74.41*** | < 0.0001 | | **CAT** | 15.58 | 174.03 | 64.67 | 33.16 | 103.42*** | < 0.0001 |
| | | AC22 | AC27 | | | | | | | AC16 | AC25 | | | | |
| | **GPA** | 0.02 | 0.82 | 0.3 | 0.18 | 74.99*** | < 0.0001 | | **GPA** | 0.05 | 0.75 | 0.32 | 0.21 | 137.91*** | < 0.0001 |
| | | AC44 | AC38 | | | | | | | AC52 | AC38 | | | | |
| | **LP** | 7.548 | 27.484 | 13.253 | 3.227 | 775.40*** | < 0.0001 | | **LP** | 3.435 | 29.403 | 12.201 | 4.724 | 975.66*** | < 0.0001 |
| | | AC55 | AC42 | | | | | | | AC59 | AC42 | | | | |

### 4.3.1.3 Ranking of barley accessions for morphological traits under cadmium stress conditions

According to the procedure established by Ketata, Yau and Nachit (1989), the ranking technique was conducted based on all morphological parameters, including germination percentage and (root and shoot) length, as well as the fresh and dry weight of the seedling that was measured for all barley accessions under all treatment conditions. As an indicator to choose the best barley accessions based on this method, the average rank (AR) and the stress tolerance index (STI) were used. In view of this, the best performance accession for studied traits obtained the lowest (AR) and heights (STI) and was considered as a tolerant accession against different concentrations of cadmium heavy metal stress, while the contrary was true for susceptible barley accessions.

Based on the available data in our study, three barley accessions (AC29, AC24, and AC25) as indicated in (Table 4.3.3) under the first cadmium exposure (Cd-125) showed a low AR value. Therefore, these accessions can be recommended as the cadmium resistant accessions under this particular dose of cadmium. In contrast, AC48, AC58, and AC52 revealed high (AR) values, which point to their susceptibility to cadmium under the same condition.

Under intermediate cadmium exposure (Cd-250) as shown in (Table 4.3.4), the responses by barley accessions were changed. The best-performed barley accessions for studied traits based on the lowest values of (AR) can be selected for their tolerance, which includes AC36, AC38, and AC11. On the contrary, both barley accessions AC48 and AC52, similar to the first dose of cadmium, were indicated as the least performed accessions, followed by AC47.

In the case of the present final dose of cadmium (Cd-500) for studied traits by all accessions, AC38 greatly responded to this cadmium exposure similar to Cd-250. It was among the best three performed barley accessions, owning a low (AR) value. While for indicating susceptible accessions, a similar pattern of response by two barley accessions, AC48 and AC52, was noted in comparison with the first and intermediate dose of cadmium conducted in our investigation, and AC53 came after (Table 4.3.5).

The mean value of the investigated traits for all three cadmium stress conditions was measured as displayed in (Table 4.3.6) for comparison between the best-performed barley accessions resistant to cadmium and the least-performed barley accessions which had a sensitive response to cadmium stress. In this regard, AC29 is considered the best barley accession in response to all used cadmium treatments, followed by AC38 and AC37, whereas the least performed barley accession in response to cadmium exposure was AC48, followed by the other two barley accessions, AC52 and AC44.

**Table 4.3.3 Rank of barley accessions calculated by an average number of ranks (ASR) and stress tolerance index (STI) based on morphological characteristics of seedlings under Cd-125 µM conditions.**

| Accession code | ARS | STI | rank | Accession code | ARS | STI | rank | Accession code | ARS | STI | rank |
|---|---|---|---|---|---|---|---|---|---|---|---|
| AC29 | 10.09 | 1.16 | 1 | AC26 | 22.91 | 1.06 | 21 | AC14 | 37.00 | 0.96 | 41 |
| AC24 | 10.64 | 1.17 | 2 | AC7 | 23.82 | 1.07 | 22 | AC53 | 37.73 | 0.43 | 42 |
| AC25 | 10.73 | 1.14 | 3 | AC21 | 24.36 | 1.13 | 23 | AC11 | 38.18 | 0.92 | 43 |
| AC37 | 11.82 | 1.11 | 4 | AC50 | 24.45 | 0.93 | 24 | AC56 | 39.00 | 1.02 | 44 |
| AC23 | 11.91 | 1.13 | 5 | AC10 | 24.91 | 1.08 | 25 | AC51 | 39.09 | 0.47 | 45 |
| AC13 | 13.00 | 1.16 | 6 | AC43 | 25.27 | 0.91 | 26 | AC42 | 39.27 | 0.97 | 46 |
| AC8 | 14.64 | 1.13 | 7 | AC31 | 26.00 | 1.14 | 27 | AC49 | 42.73 | 0.88 | 47 |
| AC16 | 14.64 | 1.15 | 8 | AC34 | 27.27 | 0.95 | 28 | AC33 | 42.82 | 0.91 | 48 |
| AC2 | 15.45 | 1.22 | 9 | AC3 | 28.18 | 0.86 | 29 | AC18 | 43.36 | 0.79 | 49 |
| AC28 | 16.00 | 1.16 | 10 | AC12 | 28.45 | 0.95 | 30 | AC59 | 43.73 | 0.88 | 50 |
| AC36 | 16.36 | 1.15 | 11 | AC17 | 29.00 | 1.05 | 31 | AC55 | 48.55 | 0.82 | 51 |
| AC41 | 16.36 | 1.09 | 12 | AC32 | 29.09 | 1.11 | 32 | AC44 | 50.18 | 0.78 | 52 |
| AC40 | 17.27 | 1.15 | 13 | AC45 | 29.45 | 0.93 | 33 | AC19 | 50.73 | 0.75 | 53 |
| AC38 | 17.64 | 1.15 | 14 | AC15 | 30.00 | 0.97 | 34 | AC46 | 51.27 | 0.77 | 54 |
| AC30 | 17.91 | 1.17 | 15 | AC1 | 31.09 | 1.06 | 35 | AC47 | 51.64 | 0.57 | 55 |
| AC27 | 18.09 | 1.15 | 16 | AC22 | 33.27 | 0.82 | 36 | AC54 | 53.55 | 0.65 | 56 |
| AC39 | 19.73 | 1.09 | 17 | AC20 | 33.91 | 0.90 | 37 | AC52 | 53.82 | 0.58 | 57 |
| AC35 | 20.82 | 1.00 | 18 | AC57 | 34.09 | 0.50 | 38 | AC58 | 54.09 | 0.66 | 58 |
| AC6 | 21.45 | 1.02 | 19 | AC5 | 35.73 | 1.00 | 39 | AC48 | 58.55 | 0.30 | 59 |
| AC9 | 22.18 | 1.12 | 20 | AC4 | 36.73 | 1.03 | 40 | | | | |

**Table 4.3.4 Rank of barley accessions calculated by an average number of ranks (ASR) and stress tolerance index (STI) depending on morphological characteristics of seedlings under Cd-250 µM conditions.**

| Accession code | ARS | STI | rank | Accession code | ARS | STI | rank | Accession code | ARS | STI | rank |
|---|---|---|---|---|---|---|---|---|---|---|---|
| AC36 | 7.18 | 1.15 | 1 | AC26 | 22.45 | 1.02 | 21 | AC53 | 36.55 | 0.45 | 41 |
| AC38 | 9.45 | 1.15 | 2 | AC16 | 23.00 | 1.09 | 22 | AC20 | 37.00 | 0.85 | 42 |
| AC11 | 12.00 | 1.05 | 3 | AC14 | 23.55 | 0.99 | 23 | AC4 | 37.00 | 0.99 | 43 |
| AC30 | 12.27 | 1.17 | 4 | AC6 | 23.82 | 0.97 | 24 | AC57 | 38.36 | 0.34 | 44 |
| AC29 | 12.45 | 1.11 | 5 | AC3 | 24.45 | 0.86 | 25 | AC18 | 38.73 | 0.79 | 45 |
| AC21 | 12.91 | 1.14 | 6 | AC10 | 25.18 | 1.04 | 26 | AC12 | 39.91 | 0.86 | 46 |
| AC37 | 12.91 | 1.05 | 7 | AC13 | 26.36 | 1.07 | 27 | AC54 | 40.91 | 0.91 | 47 |
| AC28 | 13.45 | 1.14 | 8 | AC8 | 27.45 | 1.04 | 28 | AC1 | 41.91 | 0.94 | 48 |
| AC31 | 13.91 | 1.16 | 9 | AC41 | 28.00 | 0.99 | 29 | AC49 | 43.36 | 0.85 | 49 |
| AC24 | 14.55 | 1.12 | 10 | AC33 | 29.82 | 0.94 | 30 | AC55 | 44.00 | 0.82 | 50 |
| AC39 | 14.55 | 1.07 | 11 | AC50 | 30.36 | 0.84 | 31 | AC19 | 46.45 | 0.75 | 51 |
| AC9 | 15.09 | 1.10 | 12 | AC45 | 30.64 | 0.88 | 32 | AC15 | 49.64 | 0.79 | 52 |
| AC42 | 19.27 | 1.03 | 13 | AC25 | 31.27 | 0.99 | 33 | AC58 | 50.73 | 0.68 | 53 |
| AC32 | 19.82 | 1.11 | 14 | AC7 | 31.82 | 1.00 | 34 | AC46 | 51.45 | 0.71 | 54 |
| AC35 | 19.91 | 0.98 | 15 | AC23 | 32.18 | 1.00 | 35 | AC22 | 52.09 | 0.66 | 55 |
| AC2 | 20.18 | 1.17 | 16 | AC51 | 34.09 | 0.53 | 36 | AC44 | 54.82 | 0.58 | 56 |
| AC27 | 20.18 | 1.10 | 17 | AC59 | 34.18 | 0.90 | 37 | AC47 | 54.82 | 0.50 | 57 |
| AC5 | 20.55 | 1.04 | 18 | AC17 | 34.27 | 1.00 | 38 | AC52 | 56.64 | 0.47 | 58 |
| AC34 | 22.00 | 0.97 | 19 | AC43 | 34.82 | 0.77 | 39 | AC48 | 58.27 | 0.37 | 59 |
| AC40 | 22.18 | 1.09 | 20 | AC56 | 34.82 | 1.01 | 40 | | | | |

**Table 4.3.5 Rank of barley accessions calculated by an average number of ranks (ASR) and stress tolerance index (STI) depending on morphological characteristics of seedlings under Cd-500 µM conditions.**

| Accession code | ARS | STI | rank | Accession code | ARS | STI | rank | Accession code | ARS | STI | rank |
|---|---|---|---|---|---|---|---|---|---|---|---|
| AC31 | 10.91 | 1.14 | 1 | AC13 | 21.36 | 1.07 | 21 | AC17 | 34.64 | 0.97 | 41 |
| AC38 | 11.00 | 1.11 | 2 | AC7 | 21.55 | 1.02 | 22 | AC22 | 35.00 | 0.74 | 42 |
| AC29 | 12.36 | 1.09 | 3 | AC34 | 22.73 | 0.93 | 23 | AC51 | 36.00 | 0.44 | 43 |
| AC37 | 12.73 | 1.02 | 4 | AC24 | 23.55 | 1.07 | 24 | AC57 | 36.27 | 0.33 | 44 |
| AC39 | 13.18 | 1.06 | 5 | AC41 | 25.18 | 0.99 | 25 | AC54 | 37.73 | 0.91 | 45 |
| AC30 | 13.36 | 1.14 | 6 | AC35 | 25.55 | 0.91 | 26 | AC56 | 40.91 | 0.89 | 46 |
| AC25 | 13.73 | 1.05 | 7 | AC33 | 25.73 | 0.95 | 27 | AC15 | 45.91 | 0.75 | 47 |
| AC2 | 14.45 | 1.17 | 8 | AC23 | 26.55 | 1.00 | 28 | AC18 | 46.73 | 0.68 | 48 |
| AC8 | 14.64 | 1.07 | 9 | AC21 | 26.64 | 1.05 | 29 | AC19 | 48.00 | 0.67 | 49 |
| AC32 | 14.91 | 1.10 | 10 | AC10 | 26.91 | 1.02 | 30 | AC55 | 48.55 | 0.68 | 50 |
| AC5 | 15.73 | 1.04 | 11 | AC14 | 28.91 | 0.96 | 31 | AC12 | 49.09 | 0.68 | 51 |
| AC16 | 16.00 | 1.09 | 12 | AC36 | 29.36 | 1.03 | 32 | AC58 | 49.64 | 0.63 | 52 |
| AC9 | 16.64 | 1.08 | 13 | AC43 | 30.45 | 0.78 | 33 | AC59 | 50.00 | 0.68 | 53 |
| AC11 | 17.36 | 0.99 | 14 | AC50 | 30.91 | 0.81 | 34 | AC47 | 53.18 | 0.46 | 54 |
| AC26 | 17.55 | 1.03 | 15 | AC4 | 31.36 | 1.00 | 35 | AC46 | 53.45 | 0.52 | 55 |
| AC27 | 19.73 | 1.08 | 16 | AC45 | 31.82 | 0.86 | 36 | AC44 | 54.45 | 0.48 | 56 |
| AC28 | 19.91 | 1.09 | 17 | AC42 | 32.91 | 0.96 | 37 | AC53 | 56.64 | 0.24 | 57 |
| AC3 | 20.00 | 0.88 | 18 | AC49 | 34.09 | 0.87 | 38 | AC52 | 57.00 | 0.33 | 58 |
| AC40 | 20.00 | 1.07 | 19 | AC20 | 34.18 | 0.84 | 39 | AC48 | 58.18 | 0.28 | 59 |
| AC6 | 20.27 | 0.98 | 20 | AC1 | 34.45 | 0.98 | 40 | | | | |

**Table 4.3.6 Rank of barley accessions calculated by an average number of ranks (ASR) and stress tolerance index (STI) depending on morphological characteristics of seedlings under all cadmium stress conditions.**

| Accession code | ARS | STI | rank | Accession code | ARS | STI | rank | Accession code | ARS | STI | rank |
|---|---|---|---|---|---|---|---|---|---|---|---|
| AC29 | 10.09 | 1.12 | 1 | AC6 | 21.91 | 0.99 | 21 | AC1 | 35.82 | 0.99 | 41 |
| AC38 | 10.09 | 1.14 | 2 | AC41 | 22.09 | 1.02 | 22 | AC57 | 36.27 | 0.39 | 42 |
| AC37 | 11.09 | 1.06 | 3 | AC11 | 22.45 | 0.99 | 23 | AC56 | 39.27 | 0.97 | 43 |
| AC24 | 13.27 | 1.12 | 4 | AC21 | 22.55 | 1.11 | 24 | AC49 | 39.82 | 0.87 | 44 |
| AC39 | 14.64 | 1.07 | 5 | AC32 | 22.55 | 1.10 | 25 | AC53 | 40.00 | 0.37 | 45 |
| AC30 | 15.36 | 1.16 | 6 | AC3 | 23.36 | 0.87 | 26 | AC22 | 42.36 | 0.74 | 46 |
| AC25 | 15.55 | 1.06 | 7 | AC5 | 24.00 | 1.03 | 27 | AC12 | 43.82 | 0.83 | 47 |
| AC28 | 16.18 | 1.13 | 8 | AC34 | 24.73 | 0.95 | 28 | AC15 | 45.45 | 0.83 | 48 |
| AC31 | 16.36 | 1.15 | 9 | AC7 | 25.45 | 1.03 | 29 | AC59 | 46.00 | 0.82 | 49 |
| AC36 | 16.45 | 1.11 | 10 | AC10 | 25.55 | 1.05 | 30 | AC18 | 46.09 | 0.75 | 50 |
| AC2 | 16.64 | 1.19 | 11 | AC50 | 28.09 | 0.86 | 31 | AC54 | 47.73 | 0.82 | 51 |
| AC16 | 16.73 | 1.11 | 12 | AC45 | 29.27 | 0.89 | 32 | AC55 | 47.91 | 0.77 | 52 |
| AC8 | 17.55 | 1.08 | 13 | AC14 | 29.82 | 0.97 | 33 | AC19 | 48.91 | 0.73 | 53 |
| AC9 | 17.73 | 1.10 | 14 | AC43 | 30.27 | 0.82 | 34 | AC58 | 52.09 | 0.66 | 54 |
| AC27 | 19.18 | 1.11 | 15 | AC33 | 31.91 | 0.93 | 35 | AC46 | 52.73 | 0.67 | 55 |
| AC40 | 19.64 | 1.10 | 16 | AC42 | 32.27 | 0.98 | 36 | AC47 | 54.27 | 0.51 | 56 |
| AC26 | 20.27 | 1.04 | 17 | AC20 | 32.82 | 0.86 | 37 | AC44 | 54.45 | 0.61 | 57 |
| AC23 | 20.64 | 1.04 | 18 | AC17 | 33.09 | 1.01 | 38 | AC52 | 56.64 | 0.46 | 58 |
| AC13 | 20.91 | 1.10 | 19 | AC4 | 34.00 | 1.01 | 39 | AC48 | 58.55 | 0.32 | 59 |
| AC35 | 21.55 | 0.96 | 20 | AC51 | 35.73 | 0.48 | 40 | | | | |

Statistical differential expression (Omics approach) was used in the fields of genomics (Tong *et al.*, 2017) and biochemistry (Bianchi *et al.*, 2021) to help identify features that are affected by descriptive variables. In our case scenario, the degree of tolerance and susceptibility were indicated by using the mean value of physio-chemical traits presented in (Table 4.3.7) that were obtained in response to different exposures of cadmium. In this particular analysis, three levels of responses were set. For indicating high-tolerant responses by all barley accessions, the range of stress tolerance index was set between (0.95-1.20), while for moderate-tolerant ranged between (0.60-0.94). For the third level, indicating the low tolerant, the prediction was set up between (0.10-0.59).

In the first application of cadmium (Cd-125), the only traits that showed significant responses to altering the effect of cadmium were seed water uptakes by seeds, and the mean value observed for these particular traits ranged between (273.39 and 129.67) for the highly tolerant and low-tolerant responses, respectively, while no significant change was noted for other biochemical traits studied. Following the increment of cadmium (Cd-250), three traits started to respond to barley accessions in which their *p-value* ≤ 0.01 the mean values for the mentioned traits ranged between (229.18 and 151.43), (83.56 and 47.12), and (184.80 and 148.25) respectively. In our study, the catalyze activity decreased when the barley accessions were exposed to excessive cadmium (Cd-500), as stated in (Table 4.3.7), no obvious differences were observed between the mean of this trait and the *p-value* was greater than 0.01 while the other conducted traits responded clearly by all barley accessions, indicating the overall value of preferring toleration and susceptibility for cadmium exposure.

**Table 4.3.7 Statistical Omics analysis for integrating the responses of tested materials by different biochemical traits in the presence of three different treatments of cadmium.**

| Cd-125 | | | | | |
|---|---|---|---|---|---|
| **Features** | ***p*-value** | **Significant** | **High tolerance** | **Moderate tolerance** | **Low tolerance** |
| WU | 0.00 | Yes | 273.39 (a) | 229.81 (b) | 129.67 (c) |
| AC | 0.61 | No | 946.52 (a) | 912.26 (a) | 888.43 (a) |
| CAT | 0.61 | No | 64.94 (a) | 55.74 (a) | 55.11 (a) |
| SSC | 0.78 | No | 275.63 (a) | 277.90 (a) | 240.78 (a) |
| TPC | 0.78 | No | 151.90 (a) | 146.29 (a) | 139.81 (a) |
| PC | 0.85 | No | 1410.00 (a) | 1325.00 (a) | 1274.00 (a) |
| **Cd-250** | | | | | |
| Features | *p*-value | Significant | High tolerance | Moderate tolerance | Low tolerance |
| WU | 0.00 | Yes | 229.18 (a) | 191.28 (b) | 151.43 (b) |
| CAT | 0.00 | Yes | 83.56 (a) | 56.23 (b) | 47.12 (b) |
| TPC | 0.02 | Yes | 184.80 (a) | 156.15 (b) | 148.25 (b) |
| SSC | 0.20 | No | 367.77 (a) | 305.40 (a) | 286.97 (a) |
| AC | 0.24 | No | 950.15 (a) | 921.59 (a) | 856.87 (a) |
| PC | 0.63 | No | 2530.00 (a) | 2437.00 (a) | 2022.00 (a) |
| **Cd-500** | | | | | |
| Features | *p*-value | Significant | High tolerance | Moderate tolerance | Low tolerance |
| WU | 0.00 | Yes | 181.08 (a) | 155.79 (b) | 115.35 (c) |
| TPC | 0.00 | Yes | 183.47 (a) | 138.30 (b) | 119.77 (b) |
| SSC | 0.00 | Yes | 473.20 (a) | 270.21 (b) | 238.06 (b) |
| AC | 0.00 | Yes | 975.98 (a) | 843.01 (b) | 795.59 (b) |
| PC | 0.00 | Yes | 3711.00 (a) | 2130.00 (b) | 1388.00 (b) |
| CAT | 0.08 | No | 73.12 (a) | 56.13 (a) | 49.03 (a) |

#### 4.3.1.4 Principal component analysis and correlations among measured traits under three different concentrations of cadmium

Many techniques were conducted by researchers to classify the genotypes under particular circumstances, such as stress conditions. Among those techniques, multivariate analysis was the most suitable. The main advantage of this analysis is to develop an easier and new model with a smaller number of artificial analyses that account for most of the variance in the data set. In this regard, principal component analysis (PCA), has been used to obtain the quantitative association between morphological, physiological, and biochemical traits responded by barley accessions. The data gained from the normal and three cadmium stress treatments was used to produce the PCA plot to assess the involvement of studied traits in cadmium resistance. Regarding the distribution of studied traits on the PCA plot, traits placed far away from the scattering point of the plot in the positive patterns of distinct traits showed the best performance under the particular circumstance, whereas the traits near the scattering point of the plot revealed poor results in the negative orientation of studied traits.

Based on differences and similarities in the correlation and variance for fourteen studied traits, four well-defined clusters were detected. The Bi-plot divided four treatments into four separate groups, demonstrating that the outcomes of these treatments for different barley accessions responses varied from each other for studied traits. The analysis of PCA showed that the first two components (PC1 and PC2) captured 91.78% of the total variance. The first major component (PC1) describes 68.68% of the total variance, while the second component (PC2) describes 23.10% (Fig.4.3.3).

PCA grouped the morphological traits (FWS, GP, SL, and RL) and (GPA) under optimum conditions, as shown in the first and third clades, indicating their significant contribution and strong correlation between these traits responded by all barley accessions. Further, the water uptake (WU) was similarly increased by barley accessions in response to 125 µM cadmium exposure. In the second clade, three biochemical traits (AC, TPC, and LP) plus (STI) were grouped together, showing their similar boosting response under intermediate cadmium exposure (Cd-250). While under heavy cadmium exposure (Cd-500), a strong correlation was observed among (SSC, PC, and DWS) in which placed near each other, showing that their variation may follow a similar pattern that could be interpreted as having a similar meaning in the context. The CAT trait, under the same circumstance, followed the same direction corresponding to variables but with less impact compared to the three mentioned traits.

A simple Pearson correlation (r) analysis between (STI) and other biochemical studied traits under three stressed cadmium conditions was conducted. Under the first exposure (Cd-125), an equal portion of six positive and negative correlations were found (Fig. 4.3.4, A). The positive correlations among studied traits for r value ranged between 0.61 and 0.29, for it is the relationship between (STI*WU) and (STI*GPA) and (AC*LP) respectively. The (STI) in this particular cadmium condition showed only two positive correlations with WU (r = 0.61***, $p < 0.0001$) and GPA (r = 0.29*, $p < 0.05$). PC, GPA, SSC, and AC documented only one positive association in comparison with the rest of the studied traits, having an r-value of 0.39, 0.37, 0.34, and 0.29), while no positive correlations were observed for TPC, LP, and CAT. In contrast, two negative correlations were found by both SSC and AC in associations with GPA and CAT, followed by TPC * LP, which had only a negative linkage with CAT. Under the cadmium condition of Cd-250, sixteen positive and five negative correlations were revealed (Fig. 4.3.4, B). The r-value for positive correlation ranged between 0.65 and 0.26 for its associations between (GPA*CAT) and (STI*TPC). (STI) discovered four positive relationships with (GPA, r = 0.56**), (WU, r = 0.46**), (CAT, r = 0.43**), and (TPC, r = 0.26*), followed by TPC and AC, each of which had three positive correlations with (CAT, AC, and GPA) and (WU, CAT, and GPA). After that, SSC traits show two positive correlations with GPA and TPC, with r values of 0.39* and 0.27*, respectively, while only one positive correlation was found between GPA and CAT (r = 0.65**), PC and TPC (r = 0.37*), LP and CAT (r = 0.35**), and CAT and WU (r = 0.28*). Conversely, the five negative correlations were distributed over the three studied traits: PC, TPC, and AC. PC under Cd-250 discovered three negative relations with SSC, LP, and WU, while the TPC and AC showed negative correlations with LP. Under the final cadmium treatment, Cd-500, positive correlations among almost all studied traits were detected, with the only exception of CAT, which showed no positive linkage, and no negative correlations were found in this condition (Fig. 4.3.4, C). A total of 31 significant r values were detected among the observed traits. 7 for (STI), 6 for (PC), 5 for each (SSC * TPC), 4 for (AC), 2 for each (LP *GPA). The r values varied from (0.67 to 0.29) for associations between PC with SSC and AC with WU, respectively. Impressive positive significant linkage between PC and SSC was detected (r = 0.67**, $p < 0.0001$) and then the relationship between TPC and GPA came after with a value of (r = 0.62**, $p < 0.0001$) followed by SSC and GPA (r = 0.62**, $p < 0.0001$). These findings display that a positive and significant association between STI and other studied parameters was observed and improved by the tested barley accessions under the increment of cadmium conditions.

A significant genetic distance among 59 barley accessions in the presence of different cadmium exposures (Cd125, Cd-250, and Cd500) µM exists. The responses by barley accessions to studied traits varied based on the amount of cadmium exposure. As indicated in our results, the intermediate exposure of cadmium (Cd-250) µM stimulated the activities of most biochemical traits tested by most accessions. Favorable behavior under all cadmium exposure was found by barley accessions AC29, AC38, and AC37, respectively.

### 4.3.2 Confirmation of cadmium resistance under plastic house condition

#### 4.3.2.1 Evaluation of morphological markers

Six barley accessions based on the results from the screening of cadmium tolerant barley accessions under three different cadmium exposures (125,250, and 500) µM were selected for further confirmation of their response to cadmium stress (resistance and susceptibility) under plastic house conditions in which they showed reliable performance under the mentioned conditions. In this context, barley accessions AC47, AC52, and AC48 are considered cadmium susceptible accessions, while barley accessions AC29, AC37, and AC38 performed as cadmium-resistant. For testing the performance of selected barley accessions, two essential growth stages were chosen: tillering (S1) and flowering (S2), as well as a combination of these two stages (S1+S2).

To confirm the significance of the means, the ANOVA analysis was used, and the significance of differences was compared by Duncan's test analysis. Regarding the analysis of differences under all stress treatments with respect to untreated conditions, the negative values were considered as percentage increments while positive values represented the decrease by the percentage of studied parameters.

ANOVA analysis in our investigation showed significant differences for most studied traits under both individual stresses (S1 and S2), even under the presence of severe exposure to cadmium stress at (S1+S2) Appendices (3.5,3.6, and 3.7). A similar trend of responses by studied barley accessions was observed when cadmium was applied at the flowering stage (S2) and two growth stages (S1+S2) for both traits (Tiller No./plants and Spike No./plants) as they were not significant. In addition to the performance of these two traits in (S2), no significant changes in straw weight per plant were detected (Appendix 3.6). However, under the presence of cadmium at the early growth stage (S1), the only parameter that showed no obvious differences was spike number/plants (Appendix 3.5).

When plant materials were exposed to cadmium stress conditions at tillering stage (S1), a similar pattern of responses by all barley accessions was confirmed for studied parameters as shown in (Table 4.3.8), in which all these parameters decreased in comparison to untreated conditions. Three barley accessions, AC29, AC37, and AC38, showed significant reductions in shoot length with values of (20.56, 11.57, and 8.04)%, respectively, while AC48 2.86%, AC52 4.09%, and AC47 6.21% showed only a slight reduction. The accessions AC47, AC52, and AC48 were severely affected by the presence of cadmium at this particular growth stage for reducing the number of tillers per plant as the value of reductions were (57.14, 44.44, and 41.67) %, respectively, while the rest of the tested accessions maintained the reduction of this trait. In respect to the performance of studied accessions for the number of spikes, a strong reduction by all accessions was detected. AC38 conserved the reduction of spike length with respect to its own control with a value of 6.21%, followed by AC37 10.53% and AC29 16.39%. On the contrary, AC52 possessed the maximum reduction in this trait by 39.82%, followed by AC47 37.94% and AC48 20.87%. The highest drop was stated by barley accession AC47 for the weight of spike with a value of 60.47%, followed by AC52 47.94% and AC48 28.02%, while AC38 was less affected by cadmium for this trait in this growth stage, in which 10.07% of reduction was observed in comparison with a control condition, and AC29 and AC37 came after with values of (18.45 and 20.08)%, respectively. In respect to control condition, grain number per spike were maintained by AC38, AC29, and AC37 for it is a reduction with value (6.39,16.32 and 18.70)%, respectively. However, AC47 was more sensitive to cadmium, which preserved the maximum decrease for this trait with a value of 37.13%. AC52 and AC48, similar to AC47, possessed high reductions of grain number per spike with a value of (33.59 and 30.43) %, respectively. AC38 was the first to maintain the weight of grain per spike reduction with a value of 19.43, AC37 was second with a lighter reduction, and AC29 was third with a value of 26.03 among tested barley accessions, while both AC47 and AC52 showed nearly a half-fold reduction. In addition, AC48 experienced a 34.08% reduction in weight of grain with respect to untreated conditions. According to the analysis presented in Table 4.3.8, a similar pattern of responses was observed by all barley accessions in terms of study total yield per plant and straw weight per plant in respect to the control condition. A slight reduction in the presence of cadmium at this particular growth stage was documented by AC38, AC37, and AC29 for both traits, while the reductions were higher by AC47, AC52, and AC48. AC47 achieved the highest yield reduction among rest barley accessions with a value of 65.62%, followed by AC52 58.94% and AC48 46.88%, while AC38 practiced the lowest yield reduction with a value of 17.53%, followed by AC37 and AC29 with values of (20.48 and 25.31)%, respectively. AC48 experienced an extreme reduction of straw weight per plant with a value of 34.80%. Further, AC47, with a value of 32.39, and AC52, with a value of 28.77% came after while no considerable changes were documented for this trait by AC37, AC29, and AC38 in which they owned (7.61, 9.42, and 15.38)% of decreases, respectively (Table 4.3.8).

The data presented in Table 4.3.9 revealed significant differences between studied accessions in response to cadmium stress when the selected barley accessions were exposed to particular stress conditions at the flowering stage (S2). Similar reductions as (S1) for all studied traits were observed by all barley accessions. The presence of cadmium at this stage resulted in a great decline in the plant heights for barley accessions AC29 18.01%, AC37 16.97%, and AC38 16.83%, while fewer drops were shown by AC48, AC52, and AC47 with the values of (4.15, 6.79, and 12.19)%, respectively. Both studied traits (Tiller and spike number per plant) declined in all barley accessions, with no considerable differences among barley accessions. The maximum reductions in terms of studying the length of spike were documented by AC47, AC52, and AC48 with the value of (39.58, 37.35, and 22.59)%, respectively. On the other hand, AC37 was less affected by the cadmium in reducing it is own spike length in comparison with it is control in which only 1.28% of length was reduced. Similarly, both accessions AC29 4.69% and AC38 6.58% were slightly maintained it is a reduction for the mentioned parameter. Severe declines in terms of weight of spike were noted by AC47 %63.87.AC52 and AC48 come after with values of (45.82 and 35.20)%, respectively. Conversely, AC37, with a value of 4.43%, is considered the most accession that the weight of the spike was not altered significantly in respect to the control condition. AC29 9.85% and AC38 10.52% come after AC37 in maintaining the reduction of spike weight under the presence of cadmium. The number of grains per spike was another trait that was negatively affected by the presence of cadmium, especially the barley accessions AC52 33.52%, AC47 30.68%, and AC48 21.62%, while small portions of decline by AC38, AC29, and AC37 were stated with values of (4.31,5.56 and 5.92)%, respectively. More than half a fold of grain weight per spike was reduced by barley accession AC47, followed by AC52 and AC48 with values of (46.70 and 40.28)% respectively, while the minimum reduction among tested plant materials was AC37 with a value of %9.18, followed by AC28 21.34% and AC29 25.47%. The results for yield per plant in the presence of cadmium at the (S2) growth stage were strongly influenced on barley accessions AC53, AC47, and AC48, with more than half of the total yield reduced in comparison to their control.As an alternative, AC37 and AC29 were less affected by the presence of cadmium in which they preserved (19.68 and 22.26)% of reduction by this trait. AC38, in terms of the yield performance, was more affected by cadmium in this particular growth stage in comparison to the S1 stage, in which a 39.75% reduction was observed. In addition to all the previously mentioned traits, the minimum decline in the weight of straw was detected by AC37 with a value of 13.48%, while AC48 had the poorest performance in preserving the weight of this trait in which a 34.39% reduction was noted in comparison with the control condition.

Six barley accessions when faced with double exposure of cadmium in two different growth stages (S1+S2), significant differences between studied traits by barley accessions in response to this type of stressor were revealed as stated in Table 4.3.10. Severe reductions in almost all traits were detected. Among the studied barley accessions, AC38, AC37, and AC29 were reduced in own height by (29.89, 27.90, and 21.61)%, respectively, while no observable changes were documented by the other three accessions, in which only 3.52% of shoot reduction was stated by AC52, followed by AC48 6.76%, and AC47 14.83% in association with their control. In terms of tiller and spike number, similar to (S2), no significant changes were observed in which all accessions dramatically reduced the number of these traits with respect to control conditions under the presence of double cadmium stress conditions. Similar to the previous stress condition, the presence of cadmium resulted in an enormous reduction in the spike length of barley accessions AC47, AC52, and AC48. AC47 is considered as the accession that is highly affected by this type of stress condition in which it shortens its own spike length by more than half in comparison with the untreated condition. The reduction of spike length (32.91 and 16.04)% was stated by both accessions AC52 and AC48, respectively. However, an almost similar trend of spike length drop was stated for AC29, AC37, and AC38 with values of (8.24, 9.47, and 9.95)%, correspondingly. Unlike all previous cadmium stress conditions, the negative influence of cadmium by accessions AC29, AC37, and AC38 started to appear (Table 4.3.10). The lowest decrease in spike weight was stated by AC37 4.46% followed by AC29 16.19% and AC38 20.30% while AC47 exhibited the highest reduction for this trait by the value of %69.01 followed by AC52 59.0% and AC48 27.43%. The responses by barley accessions AC38, AC29, and AC37 to double stress conditions of cadmium were light in decreasing the number of grains per spike in which the values of the reduction in respect to the control condition were (3.19, 6.14, and 8.63)%, respectively, whereas AC47, AC52, and AC48, especially the first two accessions, experienced the maximum decrease with a value of (45.60 and 41.38)%, respectively. The maximum and minimum declines in grain weight per spike were (65.19 and 7.30)% for accessions AC47 and AC37, respectively, while the responses of other barley accessions were (61.70, 33.42, 22.92, and 21.59)% for AC52, AC48, AC29, and AC38, respectively. Compared to two previous cadmium stress conditions, the decline of total yield was increased by AC29, AC37, and AC38 but still maintained the reduction compared with the other barley accessions in respect to their control. AC47 stated a 68.38% drop in total yield, followed by AC52 59.95% and AC48 55.62%, while AC29 performed better than the rest of the accessions, losing only 31.01% of total yield, while AC38 and AC37 increased this value for this trait, indicating (36.77 and 39.97)% of reduction, respectively. Interestingly, different responses by both accessions AC37 and AC52 were stated for the weight of straw, in which both increased the weight compared to the control condition, while the rest of the barley accessions, similar to both individual cadmium stress conditions, reduced their straw weight in association with untreated conditions (Table 4.3.10).

Analysis of variance reveals that the six barley accessions (AC29, AC37, AC38, AC47, AC48, and AC52) under cadmium stress in different growth stages under plastic house conditions, were not significant for most studied physiological traits, with the exception of plant height, straw Wt./plants, and number of grains/spike, while strong variation in response by all studied barley accessions was observed for most studied traits and was highly significant except for tiller and spike number. Similar to the impact of stages on physiological traits responded by barley accessions, the interactions between stages and accessions were not significant for most studied traits (Appendix 3.8).

Based on all available data obtained from all cadmium stressors conditions for studied physiological traits in respect to control conditions, the plant height was reduced sharply in the case of the double dose of cadmium in both stages (S1 and S2), with a value of 17.42%, while all studied barley accessions were less affected by cadmium in reducing their own height in respect to control conditions at S1 stage, where only 8.89% was reduced. On the other hand, AC29 with a value of 20.06%, reduced its height. Instead, only 4.59% of plant height was reduced by AC48. Regarding the interactions between studied barley accessions and cadmium stressor conditions, the maximum reduction of plant height was documented under the availability of a double dose of cadmium (S1+S2) stages *AC38 with a value of 29.89%, while the interactions between stage S1 and AC48 showed the least reduction in height by %2.86. AC47 is highly affected by the presence of cadmium for the spike length, which %43.07 of this trait was reduced, while AC37 reserved its spike length, which only 7.09% was documented. Regarding the interactions between the barley accessions and different stages of cadmium applications for this trait, the maximum and minimum reductions ranged between (51.69- S1+S2 × AC47 and 9.95- S1+S2 × AC38)% (Table 4.3.11). The weight of the spike was reduced by all barley accessions in all stages when cadmium was present. However, AC37 maintained its spike weight and only a 9.66% reduction was noticed. AC47, on the other hand, reduced its spike weight by 64.45%. A severe reduction of this trait was observed in the case of interaction (S1+S2) with AC47 with a value of 69.01%, while a minor decrease in spike weight was documented between S2 and AC37 at 4.43%. All studied barley accessions were more sensitive to cadmium in S1, where 23.76% of grain number per spike was reduced, while a slighter reduction in S2 was stated with a value of 16.93%. AC47 was again considered the most sensitive to cadmium for this trait, with a value of 37.80%, while AC29 persevered in the reduction with a value of 9.34%. The maximum and minimum values for interactions ranged between (45.60 and 3.19)% for (S1+S2 × AC47 and S1+S2 × AC38), respectively. Regarding the grain weight per spike in response to cadmium stress by all studied barley accessions in different growth stages, a considerable reduction was stated. More than half a fold of reduction was displayed by AC47 for the mentioned trait, while AC37, with a value of 12.04%, was considered the least affected by cadmium in reducing its grain weight. The highest and lowest values for interactions ranged between (65.19 and 7.30)% for (S1+S2 × AC47 and S1+S2 × AC37), respectively. The availability of cadmium stress in respect to studying the total yield for different growth stages was dramatically reduced. AC47, with a value of 62.33% was measured as the most sensitive accession in reducing the total yield, while AC29 showed only 26.19% of reduction. The total yield reduction reached 68.38 in the case of interaction of (S1+S2) with AC47. S1 with AC38, however, recorded the lowest yield loss with a value of 17.53%. The responses of all studied barley accessions to cadmium stress in various growth stages ranged between (24.34 and -0.85)% for (S2 and S1+S2). Depending on the available data for all six barley accessions for this trait, the values ranged between (28.52 and -13.76)% for AC48 and AC37, respectively. In the case of interaction, the values ranged between (34.90 and -62.38)% for (S2 × AC38 and S1+S2 × AC37), respectively.

**Table 4.3.8 The percentage responses of some physiological traits in cadmium stress condition (S1) with respect to control condition in six barley accessions based on Multiple Rang Duncan's test at *p*-value < 0.05. Any values of mean holding a common letter are not significant.**

| Accessions | Plant height (cm) | Number of tillers/plant | Number of spikes/plant | Spike length (cm) | Spike weight (g) | Number of grain/spike | Grain weight/spike (g) | Total yield/plant (g) | Straw weightplants (g) |
|---|---|---|---|---|---|---|---|---|---|
| **AC47** | 6.21 b | 57.14 a | 38.89 a | 37.94 a | 60.48 a | 37.13 a | 54.00 a | 65.62 a | 32.39 a |
| **AC52** | 4.09 b | 44.44 ab | 0.00 a | 39.82 a | 47.94 a | 33.59 a | 49.15 ab | 58.94 a | 28.77 a |
| **AC48** | 2.86 b | 41.67 ab | 33.33 a | 20.87 b | 28.02 b | 30.43 a | 34.08 bc | 46.88 a | 34.80 a |
| **AC29** | 20.56 a | 23.33 bc | 8.33 a | 16.39 bc | 18.45 bc | 16.32 b | 26.03 c | 25.31 b | 9.42 b |
| **AC37** | 11.57 b | 0.00 c | 8.33 a | 10.53 bc | 20.08 bc | 18.70 b | 19.63 c | 20.48 b | 7.61 b |
| **AC38** | 8.04 b | 16.67 bc | 16.67 a | 6.21 c | 10.07 c | 6.39 c | 19.43 c | 17.53 b | 15.38 b |

**Table 4.3.9 The percentage responses of some physiological traits in cadmium stress condition (S2) with respect to control condition in six barley accessions based on Multiple Rang Duncan's test at *p*-value < 0.05. Any values of mean holding a common letter are not significant.**

| Accessions | Plant height (cm) | Number of tillers/plant | Number of spikes/plant | Spike length (cm) | Spike weight (g) | Number of grain/spike | Grain weight/spike (g) | Total yield/plant-(g) | Straw weight/plants (g) |
|---|---|---|---|---|---|---|---|---|---|
| **AC47** | 12.19 ab | 22.62 a | 8.33 a | 39.58 a | 63.87 a | 30.68 ab | 59.36 a | 52.99 a | 20.44 ab |
| **AC52** | 6.79 ab | 33.33 a | 11.11 a | 37.35 a | 45.82 b | 33.52 a | 46.70 ab | 53.10 a | 19.96 ab |
| **AC48** | 4.15 b | 27.78 a | 30.56 a | 22.59 b | 35.20 b | 21.62 b | 40.28 bc | 52.84 a | 34.39 a |
| **AC29** | 18.01 a | 16.67 a | 16.67 a | 4.69 c | 9.85 c | 5.56 c | 25.47 cd | 22.26 b | 22.86 ab |
| **AC37** | 16.97 a | 27.78 a | 22.22 a | 1.28 c | 4.43 c | 5.92 c | 9.18 d | 19.68 b | 13.48 b |
| **AC38** | 16.83 a | 27.78 a | 27.78 a | 6.58 c | 10.52 c | 4.31 c | 21.34 d | 39.75 ab | 34.90 a |

**Table 4.3.10 The percentage responses of some physiological traits in cadmium stress condition (S1+S2) with respect to control condition in six barley accessions based on Multiple Rang Duncan's tests at *p*-value < 0.05. Any values of mean holding a common letter are not significant.**

| Accessions | Plant height (cm) | Number of tillers/plant | Number of spikes/plant | Spike length (cm) | Spike weight (g) | Number of grain/spike | Grain weight/spike (g) | Total yield/plant-(g) | Straw weight/plants (g) |
|---|---|---|---|---|---|---|---|---|---|
| **AC47** | 14.83 c | 19.05 a | 8.33 a | 51.69 a | 69.01 a | 45.60 a | 65.19 a | 68.38 a | 18.51 a |
| **AC52** | 3.52 d | 16.67 a | 19.44 a | 32.91 b | 59.01 a | 41.38 a | 61.70 a | 59.95 ab | -26.67 b |
| **AC48** | 6.76 d | 22.22 a | 33.33 a | 16.04 c | 27.43 b | 15.96 b | 33.42 b | 55.62 abc | 16.36 a |
| **AC29** | 21.61 b | 16.67 a | 16.67 a | 8.24 c | 16.19 bc | 6.14 b | 22.92 bc | 31.01 c | 31.75 a |
| **AC37** | 27.90 ab | 8.33 a | 24.44 a | 9.47 c | 4.46 c | 8.63 b | 7.30 c | 39.97 bc | -62.38 c |
| **AC38** | 29.89 a | 16.67 a | 25.00 a | 9.95 c | 20.30 bc | 3.19 b | 21.59 bc | 36.77 bc | 17.31 a |

**Table 4.3.11 The percentage responses of some physiological traits under availability of cadmium stress in different growth stages in four barley accessions under plastic house and inactions between accessions and those cadmium stages based on Multiple Rang Duncan's test at *p* value < 0.05. Any values of means holding common letter are not significant.**

| **Plant height (cm)** | | | | | | **Number of tiller** | | | **Number of spike** | | |
|---|---|---|---|---|---|---|---|---|---|---|---|
| **Stage ( S.)** | | **Accessions (AC)** | | **S.*AC.** | | **S.** | **AC.** | **S.*AC.** | **S.** | **AC.** | **S.*AC.** |
| **S1** | 8.89 c | **AC47** | 11.08 b | S1*AC29 | 20.56 b-d | 30.54 a | 32.94 a | 23.33 a-c | 21.20 a | 18.52 a | 8.33 a |
| | | | | S1*AC37 | 11.57 d-g | | | 0.00 c | | | 8.33 a |
| | | | | S1*AC38 | 8.04 e-g | | | 16.67 bc | | | 16.67 a |
| | | **AC48** | 4.59 c | S1*AC47 | 6.21 fg | | 30.56 a | 57.14 a | | 17.59 a | 38.89 a |
| | | | | S1*AC48 | 2.86 g | | | 41.67 b | | | 33.33 a |
| | | | | S1*AC52 | 4.09 g | | | 44.44 b | | | 0.00 a |
| **S2** | 12.49 b | **AC52** | 4.80 c | S2*AC29 | 18.01 cd | 25.99 a | 31.48 a | 16.67 bc | 19.44 a | 21.20 a | 16.67 a |
| | | | | S2*AC37 | 16.97 c-e | | | 27.78 a-c | | | 22.22 a |
| | | | | S2*AC38 | 16.83 c-e | | | 27.78 a-c | | | 27.78 a |
| | | **AC38** | 18.25 a | S2*AC47 | 12.19 d-g | | 23.15 a | 22.62 a-c | | 20.37 a | 8.33 a |
| | | | | S2*AC48 | 4.15 g | | | 27.78 a-c | | | 30.56 a |
| | | | | S2*AC52 | 6.79 g | | | 33.33 a-c | | | 11.11 a |
| **S1+S2** | 17.42 a | **AC29** | 20.06 a | S1+S2*AC29 | 21.61 a-c | 17.59 a | 18.89 a | 16.67 bc | 16.60 a | 13.89 a | 16.67 a |
| | | | | S1+S2*AC37 | 27.90 ab | | | 8.33 bc | | | 24.44 a |
| | | | | S1+S2*AC38 | 29.89 a | | | 16.67 bc | | | 25.00 a |
| | | **AC37** | 18.81 a | S1+S2*AC47 | 14.83 c-f | | 18.33 a | 19.05 a-c | | 12.04 a | 8.33 a |
| | | | | S1+S2*AC48 | 6.76 fg | | | 22.22 a-c | | | 33.33 a |
| | | | | S1+S2*AC52 | 3.52 g | | | 16.67 bc | | | 11.11 a |
| **Spike length (cm)** | | | | | | **Spike weight (g)** | | | **Number of grain/spike** | | |
| **Stage ( S.)** | | **Accessions (AC)** | | **S.*AC.** | | **S.** | **AC.** | **S.*AC.** | **S.** | **AC.** | **S.*AC.** |
| **S1** | 21.96 a | **AC47** | 43.07 a | S1*AC29 | 16.39 c-e | 30.84 a | 64.45 a | 18.45 fg | 23.76 a | 37.80 a | 16.32 d-f |
| | | | | S1*AC37 | 10.53 d-g | | | 20.08 e-g | | | 18.70 de |
| | | | | S1*AC38 | 6.21 e-g | | | 10.07 g | | | 6.39 fg |
| | | **AC48** | 19.83 c | S1*AC47 | 37.94 b | | 30.22 c | 60.48 a-g | | 22.67 b | 37.13 ab |
| | | | | S1*AC48 | 20.87 cd | | | 28.02 e-f | | | 30.43 bc |
| | | | | S1*AC52 | 39.82 b | | | 47.94 b-d | | | 33.59 a-c |
| **S2** | 18.68 a | **AC52** | 36.69 b | S2*AC29 | 4.69 fg | 28.28 a | 50.92 b | 10.07 g | 16.93b | 36.16 a | 5.56 fg |
| | | | | S2*AC37 | 1.28 g | | | 60.48 a-g | | | 5.92 fg |
| | | | | S2*AC38 | 6.58 e-g | | | 28.02 e-f | | | 4.31 fg |
| | | **AC38** | 7.58 d | S2*AC47 | 39.58 b | | 13.63 d | 47.94 b-d | | 4.63 c | 30.68 bc |
| | | | | S2*AC48 | 22.59 c | | | 16.19 fg | | | 21.62 cd |
| | | | | S2*AC52 | 37.35 b | | | 4.46 g | | | 33.52 a-c |
| **S1+S2** | 21.38 a | **AC29** | 9.77 d | S1+S2*AC29 | 8.24 e-g | 32.73 a | 14.83 d | 16.19 fg | 20.15 ab | 9.34 c | 6.14 fg |
| | | | | S1+S2*AC37 | 9.47 e-g | | | 4.46 g | | | 8.63 e-g |
| | | | | S1+S2*AC38 | 9.95 d-g | | | 20.30 e-g | | | 3.19 g |
| | | **AC37** | 7.09 d | S1+S2*AC47 | 51.69 a | | 9.66 d | 69.01 a | | 11.08 c | 45.60 a |
| | | | | S1+S2*AC48 | 16.04 c-f | | | 18.45 fg | | | 15.96 d-f |
| | | | | S1+S2*AC52 | 32.91 b | | | 20.08 e-g | | | 41.38 ab |
| **Grain weight /spike ( g)** | | | | | | **Total yield/Plant (g)** | | | **Straw weight/plant (g)** | | |
| **Stage ( S.)** | | **Accessions (AC)** | | **S.*AC.** | | **S.** | **AC.** | **S.*AC.** | **S.** | **AC.** | **S.*AC.** |
| **S1** | 33.72 a | **AC47** | 59.52 a | S1*AC29 | 26.03 e-g | 39.12 b | 62.33 a | 25.31 de | 21.0 a | 23.78 a | 9.42 b |
| | | | | S1*AC37 | 19.63 fg | | | 20.48 e | | | 7.61 b |
| | | | | S1*AC38 | 19.43 fg | | | 17.53 e | | | 15.38 ab |
| | | AC48 | 35.93 b | S1*AC47 | 54.00 a-c | | 51.78 a | 65.62 a | | 28.52 a | 32.39 a |
| | | | | S1*AC48 | 34.08 d-f | | | 46.88 a-d | | | 34.80 a |
| | | | | S1*AC52 | 49.15 a-d | | | 58.94 ab | | | 28.77 ab |
| **S2** | 33.72 a | **AC52** | 52.52 a | S2*AC29 | 25.47 e-g | 40.10 ab | 57.33 a | 22.26 de | 24.34 a | 7.35 b | 22.86 ab |
| | | | | S2*AC37 | 9.18 g | | | 19.68 e | | | 13.48 ab |
| | | | | S2*AC38 | 21.34 fg | | | 39.75 b-e | | | 34.90 a |
| | | **AC38** | 20.79 cd | S2*AC47 | 59.36 ab | | 31.35 b | 52.99 a-c | | 22.53 a | 20.44 ab |
| | | | | S2*AC48 | 40.28 c-e | | | 52.84 a-c | | | 34.39 a |
| | | | | S2*AC52 | 46.70 b-d | | | 53.10 a-c | | | 19.96 ab |
| **S1+S2** | 35.35 a | **AC29** | 24.80 c | S1+S2*AC29 | 22.92 e-g | 48.61 a | 26.19 b | 31.01 c-e | -0.85 b | 21.34 a | 31.75 a |
| | | | | S1+S2*AC37 | 7.30 g | | | 39.97 b-e | | | -62.38 d |
| | | | | S1+S2*AC38 | 21.59 fg | | | 36.77 b-e | | | 17.31 ab |
| | | **AC37** | 12.04 d | S1+S2*AC47 | 65.19 a | | 26.71 b | 68.38 a | | -13.76 c | 18.51 ab |
| | | | | S1+S2*AC48 | 33.42 d-f | | | 55.62 a-c | | | 16.36 ab |
| | | | | S1+S2*AC52 | 61.70 ab | | | 59.95 ab | | | -26.67 c |

### 4.3.2.2 Cadmium concentration evaluation in grains

The cadmium concentrations of 500 µM were first exposed at the tillering stage (S1) three times on six barley accessions: AC29, AC37, AC38, AC47, AC48, and AC52, while when the plant reached the further growth stage (S2) (flowering stage), again a triple cadmium dose was applied. The tested materials with the same cadmium concentration were exposed to stress from the tillering stage till the end of the flowering stage (S1+S2). The cadmium concentration was detected in grains using flame atomic absorption spectrometry.

In respect to considering barley accessions for its heavy metal resistance, the concentrations of cadmium ranged between 0.08 and 0.59 ppm. Regarding the S1 stage, the cadmium concentrations values of 0.10, 0.11, and 0.53 ppm were respectively observed for barley accessions AC29, AC37, and AC38 and almost similar accumulation patterns in stage 2 were detected for the same accessions with values of 0.11, 0.13, and 0.13 respectively. Except for the barley accessions AC38, which had a value of 0.59 ppm, cadmium at S1 and S2 caused little accumulations in the grains of barley accessions AC29 and AC37, which had values of 0.12 and 0.13 ppm, respectively.

In respect to considering barley accessions for its susceptibility to cadmium, the accumulations of cadmium ranged between 0.09 and 2.00 ppm. Regarding the S1 stage, the cadmium concentration values of 0.12, 0.39, and 0.48 ppm were respectively perceived for barley accessions AC47, AC48, and AC52 and 0.11, 0.10, and 0.53 ppm were the cadmium storage in grains at S2 for the mentioned barley accessions, respectively. When the barley accessions AC47, AC48, and AC53 experienced double exposures of cadmium at S1 and S2, the trends of responses were changed by those accessions as they were more sensitive to the presence of cadmium and accumulated relatively more cadmium in the grain organ with values of 0.47, 2.00, and 1.20, respectively (Table 4.3.12).

The European Union has recently presented a law defining the maximum permissible levels for cadmium concentrations in a range of plants, including barley (*Hordeum vulgare* L.) grain. The maximum acceptable cadmium concentration is between 0.1 and 0.2 ppm for human consumption (Hayes, Carrijo and Meints, 2020). This low value of cadmium reflects the end uses of barley products. For instance, as much as possible reduction is required to minimize the cadmium content, especially in the case of using these products for brewing and malting, for the reason that a large portion of the cadmium in the barley grains is bound to cell walls and remains attached to the grain after the process of smashing, while if the case was using it for animal feeding, which takes account about 75% of barley production, this limit border increased to a range between 0.5 to 10 ppm (Hayes, Carrijo and Meints, 2020).

These results indicated that the total amount of cadmium present in the grains of barley accessions was negatively affected by the cadmium element in different growth stages. Compared to control conditions, the amount of cadmium accumulations increased, particularly in AC47, AC48, and AC53, while the barely accessible AC29, AC37, and AC38 maintained cadmium accumulations in grains to some extent. It is possible to conclude that exceeding the European limit of cadmium, which ranged between 0.1 and 0.2 ppm, can be used for animal feeding rather than human consumption. Hopefully, in our investigations, the values of cadmium accumulations of grains did not exceed even the maximum limit of animal use of 10 ppm.

It has been observed that barley can naturally accumulate less cadmium in its grain organ compared to other cereal plants. Baghaie and Aghili (2019) found that the amount of Cd in wheat grains was higher than in barley grains. They did this by randomly picking samples from 60 sites, including 30 fields of barley and wheat, around the Shahin mine in Iran. Gray *et al.* (1999) in ten distinct plant species studied cadmium concentrations in ten primary soil types in New Zealand and discovered that Cd concentrations dropped in the following order: lettuce > carrot canopy > root of carrot > lucerne > cabbage > wheat > maize > ryegrass > clover and finally barley.

**Table 4.3.12 Cadmium determinations in grains of six barley accessions in different growth stages (S1, S2, and S1+S2) using atomic absorption spectrometry.**

| Accessions | Status | Concentrations (PPM) | Accessions | Status | Concentrations (PPM) |
|---|---|---|---|---|---|
| AC29 | C | 0.12 | AC47 | C | 0.09 |
| AC29 | S1 | 0.10 | AC47 | S1 | 0.12 |
| AC29 | S2 | 0.11 | AC47 | S2 | 0.11 |
| AC29 | S1+S2 | 0.12 | AC47 | S1+S2 | 0.47 |
| AC37 | C | 0.09 | AC48 | C | 0.11 |
| AC37 | S1 | 0.11 | AC48 | S1 | 0.39 |
| AC37 | S2 | 0.13 | AC48 | S2 | 0.10 |
| AC37 | S1+S2 | 0.13 | AC48 | S1+S2 | 2.00 |
| AC38 | C | 0.08 | AC52 | C | 0.18 |
| AC38 | S1 | 0.53 | AC52 | S1 | 0.48 |
| AC38 | S2 | 0.13 | AC52 | S2 | 0.53 |
| AC38 | S1+S2 | 0.59 | AC52 | S1+S2 | 1.20 |

### 4.3.3 Seedling test in the plastic house under the presence of cadmium and plant residues

#### 4.3.3.1 Morphological markers assessment

In parallel with indicating the performance of six barley accessions in the plastic house under the presence of cadmium stress, four barley accessions (AC29, AC37, AC48, and AC52) were similarly subjected to cadmium stress but this time under the presence of two types of plant residue: oak leaves and *Gundelia tournefortii*, which is locally named as Kangar with little modification. They were mixed as plant residues with soil to observe the alterations that could result in the presence of these residues morphologically and biochemically at early growth stages. The responses by all used morphological traits were highly different among the barley accessions, and the interactions were significantly different as indicated by ANOVA analysis in (Appendix 3.9).

The overall percentage comparisons responded by all barley accessions were improved under the presence of Oak leaf and Kangar residues in respect to control condition (Table 4.3.13). The length of root and root fresh and dry weight increased dramatically under the presence of oak as a residue in the soil. Similarly, a considerable increase in root fresh and dry weight was observed by Kangar. In addition, both plant residues maintained the reduction of relative water content and total chlorophyll content with respect to control conditions, while no observable changes were stated by other traits. To be more specific, oak residue (SOC) increased root length by 26.62% whereas kangar (SKC) increased root length by 0.57% in comparison to the control condition. Cadmium alone (SC), as stated by analysis, increased the root length by 7% for all barley accessions. The length of the shoot was boosted by %13.55 under the presence of cadmium, while the percentage of increase was reduced under the presence of oak, in which 2.79% of increase was responded by barley accessions, while the presence of kangar caused the reduction of the shoot by 3.39%. The highest average leaf number was stated by (SK) treatment followed by (SOC) in which a 7.17% number of leaves increased in respect to control condition by all barely accessions, while (SKC) resulted in the reduction of this trait by 0.84%. The fresh weight of the root under the presence of (SOC) and (SKC) was increased by the values of (25.62 and 14.45)% respectively, while the absence of these residues caused the reduction of this trait by 3.44%. The accumulation of dry matter was similarly improved by (SKC) and (SOC) with a value of (33.57 and 31.58)% respectively, while in the case of the presence of cadmium alone (SC), only a small portion of accumulation was documented at 2.53%. Both treatments (SKC and SOC) reduced the weight of the shoot by (18.31 and 4.76)%, respectively, while 14.75% of fresh shoot weight was increased in the case of the (SC) treatment. Similarly, the dry accumulation in the shoot dropped by both treatments (SKC and SOC) by (14.71 and 0.46)%, respectively, whereas the presence of cadmium alone caused an increase of 12.76% in accumulating the dry matter in the shoot. In the absence of these two plant residues, all barley accessions reported nearly half-fold reductions in total chlorophyll content compared to their control, with (SOC) maintaining the reduction by 11.37%, followed by (SKC) with a value of 24.15%. The relative water content was similarly maintained under the presence of (SOC and SKC) in respect to control conditions with values of (0.36 and 2.55)% respectively, while 9.12% of reduction for this trait was responded by barley accessions in the absence of two plant residues (Table 4.3.13). The percentage interaction comparisons between four barley accessions and the presence of cadmium and plant residues of nine physiological markers (average root length, average shoot length, average number of leafs, root fresh weight, root dry weight, shoot fresh weight, shoot dry weight, total Chlorophyll Content, and relative water content) (Appendix 3.11) were ranged between (24.27 - S+G+Cd × AC37 and -45.14 - S+O+Cd × AC48), (13.84 - S+Cd × AC48 and -52.87 - S+Cd × AC37), (30.15 - S+G+Cd × AC48 and -72.84 - S+Cd × AC37), (49.56 - S+O+Cd × AC48 and -68.46 - S+O+Cd × AC29), (49.04 - S+Cd × AC48 and -140.98 - S+G+Cd × AC29), (43.36 - S+Cd × AC48 and -74.96 - S+Cd × AC37), (33.66 - S+Cd × AC48 and -70.52 - S+Cd × AC37), (55.53 - S+Cd × AC52 and -12.62 - S+O+Cd × AC29), (18.20 - S+Cd × AC52 and -3.40 - S+G+Cd × AC37), respectively.

Principal component analysis (PCA) has been tested to achieve the quantitative association between physiological and biochemical traits used in this part of the investigation, which is responded by selected barley accessions. The data obtained from the normal (C) and three cadmium stress conditions with and without plant residues were used to produce the PCA plot to assess the involvement of studied traits in cadmium resistance. Regarding the distribution of studied traits on the PCA plot, traits placed far away from the scattering point of the plot in the positive patterns of distinct traits showed the best performance under particular conditions, while the traits near the scattering point of the plot revealed poor results in the negative orientation of studied traits.

Based on similarity and differences in the correlation and variance of nine physiological traits, four well-defined groups were identified, demonstrating that the outcomes of conducted treatments for different barley accessions responses varied from each other. In respect to physiological data, the analysis of PCA showed that the first two components (PC1 and PC2) were perfectly fitted in capturing the total differences, in which 100% of the total differences were seized. The first major component (PC1) describes 78.08% of the total variance, while the second component (PC2) describes 21.92% (Fig. 4.3.5).

In this context, PCA grouped five physiological traits (RL, TCC, RFWP, RWC, and RDWP) into clade 2, indicating their significant contribution and a strong correlation between these traits with the presence of cadmium alone, as responded by all barley accessions. Further, in the first clade, the rest of the morphological traits were grouped, including SDWP, NL, SFWP, and SL, which were placed close to each other and showed strong associations with the presence of kanger, displaying that their variation may follow a similar pattern that could be interpreted as owning a similar meaning in the context.

In respect to determining the associations between studied barley accessions and physiological data, the analysis of PCA displayed that the first two components (PC1 and PC2) arrested 85.37% of the total variances. The first main component (PC1) describes 68.04% of the total variance, while the second component (PC2) describes 17.33 percent (Fig. 4.3.6).

Interestingly, AC37 with AC52 and AC29 with AC48 were placed in different orientations, indicating different performances of these accessions in terms of responses to physiological traits. Most of the studied traits were located and grouped on the positive side of both factors. Clade 2 comprised five traits (RWC, SL, NL, SFWP, and SDWP) and was strongly associated with AC52, while three other physiological traits were placed in clade 4, including (TCC, RFWP, and RDWP) and exhibited great linkage with AC48. RL alone was placed in Clade 1 and showed great accession with AC29 (Fig. 4.3.6).

**Table 4.3.13 Overall percentage comparisons responses by four barley accessions in respect to their control under the presence of cadmium and plant residue by nine physiological markers.**

| | Root length (cm) | Shoot length (Cm) | Number of leaves | Root fresh weight/plant (g) | Root dry weight/plant (g) | Shoot fresh weight/plant (g) | Shoot dry weight/plant (g) | Total chlorophyll content | RWC % |
|---|---|---|---|---|---|---|---|---|---|
| **SKC** | -0.57 a | 3.39 a | 0.84 a | -14.45 ab | -33.57 b | 18.31 a | 14.71 a | 24.15 b | 2.55 b |
| **SC** | -7.00 a | -13.55 b | -19.88 b | 3.44 a | -2.53 a | -14.75 b | -12.76 b | 43.46 a | 9.12 a |
| **SOC** | -26.62 b | -2.79 a | -7.17 ab | -25.62 b | -31.58 b | 4.76 ab | 0.46 ab | 11.37 c | 0.36 b |

* SC, SKC, and SOC represents the treatments of soil + cadmium, soil+ kanger + cadmium, and soil + oak + cadmium. Regarding the analysis of differences under all stress treatments, the negative values were considered as percentage increments while positive values represented the decrease by the percentage of studied parameters.

### 4.3.3.2 Biochemical markers assessments

The differences in cadmium tolerance between sensitive and resistant barley accessions were further investigated on leaf organs, which were collected after measuring all morphological traits at the early growth stage, by using seven biochemical markers including; (PC) Proline Content (µg/g), (TPC) Total Phenolic Content (µg GAE/gm), (AC) Anti-oxidant (µg Trolox/gm), (SSC) Soluble Sugar

Content (μg/g), (CAT) Catalase (units/min/g), (GPA) Guaiacol peroxidase activity (units/min/g), and (LP) Lipid Peroxidation nmol /g.

Under all treatments, the responses to all biochemical traits were highly different among the barley accessions, and the interactions were significantly different, as indicated by ANOVA analysis in (Appendix 3.10). Comparing the mean values of selected barley accessions under the presence of cadmium and two plant residues, significant responses can be perceived by barley accessions for studied traits (Table 4.3.14). An essential increase of the mean value by the two barley accessions AC29 and AC37 under these particular stress conditions was observed for all studied traits compared to their control. On the contrary, the two barley accessions AC52 and AC48 possessed no obvious increase in response to cadmium stress even with the availability of plant residues detected for the mentioned trait, with only the exception of lipid peroxidation, in which high activity by both accessions was documented (Table 4.3.14).

Both barley accessions AC29 and AC37, with an almost similar mean value of 2436.18 and 2401.63, respectively, increased the accumulation of total proline content, while slight increases by the other two accessions, AC52 (988.62) and AC48 (2054.32), were stated. As can be seen from (Table 4.3.14), the soluble sugar content was slightly increased by both barley accessions AC52 and AC48, with mean values of (239.34 and 185.32), respectively, compared with AC37 and AC29, with mean values of (179.07 and 146.20) in comparison with their control. When barley accessions were exposed to cadmium with and without the availability of plant residue at this particular growth stage, the total phenolic content was increased by all accessions but the maximum increase was documented by AC37 (179.78) and AC29 (174.82) and AC52 with a mean value of (172.53) came after, while AC48 was recorded the lowest change with a mean value of 168.27 in respect to control conditions. The capacity of antioxidants, which was determined by DPPH, was much higher in AC37 (1107.50) and AC29 (1073.38), while the smaller activity of this trait was spotted in AC52 (948.89) and AC48 (902.26). In the presence of cadmium and plant residues, the activity of Guaiacol peroxidase was measured by all barley accessions. Considerable increases by AC37 with a value of 0.47 were detected, followed by AC29 and AC52 with a similar value of increase (0.38), while the activity of this enzyme decreased by AC48 (0.24). Accession AC37, with a value of (95.24), had the highest activities of catalyze in response to available treatment conditions, followed by AC29 (75.76), while the activity of this critical enzyme dropped by both barley accessions AC48 and AC52 to reach the minimum activities of (51.95 and 61.69), respectively. An extensive increase in lipid peroxidation was detected by AC48 (12.05) and AC52 (11.45) in respect to their control conditions, while the accession AC29 was less affected by stress as the value of (8.58) was discovered in comparison with its control, and followed by AC37 with a value of 9.72 (Table 4.3.14).

The presence of both plant residues altered the adverse effects of cadmium in which both improved the performance of all biochemical traits as indicated in (Table 4.3.15). Based on multiple Rang Duncan's tests at $p$-value < 0.05, the mean comparisons of seven biochemical traits under the presence of cadmium and plant residue with respect to control conditions in four barley accessions were analyzed. The maximum activities of (PC, SSC, TPC, AC, GPA, and CAT) were stated under the presence of soil +Oak +cadimum (SOC) with mean values of (2538.49, 268.50, 206.67, 1056.32, 0.64, and 116.88) respectively. Kangar similar to Oak, has improved the activities of these traits and placed next after Oak in which mean values of (2067.33, 183.63, 191.54, 1014.76, 0.43, and 71.43) were documented respectively for (PC, SSC, TPC, AC, GPA, and CAT). However, no obvious changes were experienced by all barley accessions for mentioned traits in the absence of the two plant residues, in which mean values of (1672.91, 160.63, 155.44, 995.51, 0.22, and 50.87) were observed for mentioned traits in chronological order in respect to the control condition, under the availability of cadmium alone, in which the mean values of (1602.01, 137.18, 141.75, 965.44, 0.18, and 45.45) in case of the control condition were indicated respectively for mentioned traits. Under the presence of cadmium alone, the highest activities of lipid peroxidation with a mean value of 14.92 were stated, while when oak and kanger were available in the soil, the oxidative stress of lipid peroxidation was diminished by oak and kangar, in which the mean values of 8.83 and 10.05 respectively were stated in respect to control condition 8.01 (Table 4.3.15).The mean interaction comparisons between four barley accessions and the presence of cadmium and plant residues of seven biochemical markers (PC, SSC, TPC, AC, GPA, CAT, and LP) (Appendix 3.12) were ranged between (3226.82 - S+O+Cd × AC29 and 543.49 - S+Cd × AC52), (413.33-S+O+Cd × AC52 and 101.60 – Co × AC29), (221.91 - S+O+Cd × AC52 and 136.48 – Co × AC29), (1150.41 - S+O+Cd × AC37 and 867.30 – Co × AC48), (0.74 - S+O+Cd × AC37 and 0.10 – Co × AC48, S+Cd × AC48 and S+Cd × AC52), (160.17 - S+O+Cd × AC37 and 34.63 - S+Cd × AC52), and (19.23 - S+Cd × AC52 and 6.36 – Co × AC29), respectively.

Based on differences and similarities in the correlation and variance of seven biochemical traits (Fig. 4.3.7), four well-defined groups were recognized, demonstrating that the outcomes of conducted treatments for different barley accessions responses varied from each other. In respect to biochemical data, the analysis of PCA presented that the first two components (PC1 and PC2) were collected for 98.27% of the total differences. The first major component (PC1) describes %84.48 of the total variance, while the second component (PC2) represents 13.79% (Fig. 4.3.7). In this regard, PCA grouped six biochemical traits (PC, SSC, TPC, AC, GPA, and CAT) over Clades 2 and 4 and significantly correlated with the presence of Oak and Kangar. However, there is a strong association between lipid peroxidation and the presence of cadmium alone, which is found in different orientations (Clade1) with other biochemical traits, with no correlation with rest parameters.Besides, the control condition is located far away from the scattering point on the negative side of both factors, indicating no correlations with studied biochemical traits (Fig. 4.3.7).

The analysis of PCA confirmed that the first two components (PC1 and PC2) arrested 98.27% of the total variances in the case of indicating the association between studied barley accessions with biochemical data under various cadmium stress conditions. The first main component (PC1) describes 84.48% of the total variance, while the second component (PC2) describes %13.79 (Fig. 4.3.8). AC37 with AC48 and AC29 with AC52 were placed in different alignments, indicating the different performance of these accessions in terms of responses to biochemical traits. Most of the biochemical traits were placed and grouped on the positive side of both factors. Clade 2 included five traits (GPA, TPC, CAT, and AC) and was strongly associated with AC37, while PC in Clade 4 showed strong linkage with AC29 and was opposite to SSC and LP in Clade 1. SSC and LP were strongly associated with AC52 and negative correlations with respect to the activities of GPA, TPC, CAT, and AC with AC48 can be observed (Fig. 4.3.8).

**Table 4.3.14 The mean comparisons of seven biochemical traits under the presence of cadmium and plant residue with respect to control conditions in four barley accessions based on Multiple Rang Duncan's tests at *p*-value < 0.05. Any values of mean holding a common letter are not significant.**

| Accessions | PC | SSC | TPC | AC | GPA | CAT | LP |
|---|---|---|---|---|---|---|---|
| **AC37** | 2401.63 b | 179.07 c | 179.78 a | 1107.50 a | 0.47 a | 95.24 a | 9.72 c |
| **AC29** | 2436.18 a | 146.20 d | 174.82 b | 1073.38 b | 0.38 b | 75.76 b | 8.58 d |
| **AC52** | 988.62 d | 239.34 a | 172.53 c | 948.89 c | 0.38 b | 61.69 c | 11.45 b |
| **AC48** | 2054.32 c | 185.32 b | 168.27 d | 902.26 d | 0.24 c | 51.95 d | 12.05 a |

**Table 4.3.15 Overall mean comparisons responses by four barley accessions in respect to their control under the presence of cadmium and plant residues by seven biochemical markers.**

| Statues | PC | SSC | TPC | AC | GPA | CAT | LP |
|---|---|---|---|---|---|---|---|
| **SOC** | 2538.49 a | 268.50 a | 206.67 a | 1056.32 a | 0.64 a | 116.88 a | 8.83 c |
| **SKC** | 2067.33 b | 183.63 b | 191.54 b | 1014.76 b | 0.43 b | 71.43 b | 10.05 b |
| **SC** | 1672.91 c | 160.63 c | 155.44 c | 995.51 c | 0.22 c | 50.87 c | 14.92 a |
| **C** | 1602.01 d | 137.18 d | 141.75 d | 965.44 d | 0.18 d | 45.45 c | 8.01 d |

* C, SC, SKC, and SOC represents the treatments of (soil, soil + cadmium, soil+ kanger + cadmium, and soil + oak + cadmium).

#### 4.3.3.3 Cadmium concentration evaluation in roots and shoots

Grains of four barley accessions (AC29, AC37, AC48, and AC52) were sown in pots in a plastic house depending on the results obtained from the test on 59 barley accessions in response to heavy metals with different concentrations. Four doses of cadmium with a concentration of 500 µM started to be applied from the first emergence of the leaf and continued till the appearance of five leaves. After reaching this stage, the collected roots and shoots are subjected to atomic absorption to detect the accumulated cadmium in these organs under the present and absence of plant residues of oak leaf and guandela (kanger).

As a result of the experiment; cadmium accumulation values in ppm were 12.27, 14.93, 28.93, and 35.60 in the root, 5.20, 10.40, 7.73, and 7.74 in the shoot, respectively for AC29, AC37, AC48, and AC52 in the presence of oak leaf residue in the soil profile, while in the presence of Guandela, cadmium accumulation values in ppm were 16.27, 33.47, 34.47, and 34.27 in the root, 5.87, 11.87, 9.20, and 8.53 in the shoot, respectively for the mentioned barley accessions (Table 4.3.16). Accordingly, the cadmium accumulation was in the order of root > shoot. A similar pattern of distribution between cadmium accumulations in roots and shoots of barley was stated by Özyigit *et al.* (2021).

According to the results of the analysis, there was a significant decrease in cadmium accumulations in the presence of both plant residues, particularly oak leaves, followed by Guandela. Oak leaves caused (34.24, 56.92, 22.50, and 14.42)% reductions in cadmium accumulations in roots for barley accessions AC29, AC37, AC48, and AC52, respectively compared to their positive control conditions. Similarly, in shoots, oak leaves minimized the accumulations of cadmium compared to positive control by (40.91- AC29, 33.33- AC37, 17.15- AC48, and 9.26- AC52)%. The availability of Guandela similarly, but with fewer effects than oak leaves, reduced the cadmium accumulations in roots and shoots. The roots of barley accessions AC29, AC37, AC48, and AC52 accumulated less cadmium compared to their positive controls, respectively, by (12.81, 3.43, 7.66, and 17.62)% while in shoots the reductions were (33.30, 23.91, 1.39, and 39.04)% in chronological order for AC29, AC37, AC48, and AC52.

The results showed that the highest uptake of cadmium occurred in the roots in comparison with the accumulations in the shoots. Barley accessions in soil with plant residues accumulated more concentrations of cadmium. This was demonstrated further with the Translocation Factor (TF), by way of it being considered as an indicator for measuring the performance of heavy metal accumulation in plants, and it shows the association between the concentration of an ironic element in the aerial part, including shoots, and the concentration in the roots (Zakaria *et al.*, 2021). The findings of translocation factor (TF) in barley accessions planted in soil with available plant residues showed significant differences between the positive control and all treatments with cadmium addition. Because the TF values were less than one, it means that the plant stopped more cadmium from moving from the roots to the shoots. The TF values with respect to positive control were (0.47, 0.45, 0.25, and 0.21), respectively, for AC29, AC37, AC48, and AC52. While under the availability of Guandela and oak leaf, a reduction in TF values was observed, which was (0.36, 0.35, 0.27, and 0.15) for Guandela, (0.42, 0.70, 0.27, and 0.22) for oak leaf for barley accessions AC29, AC37, AC48, and AC52, respectively.

**Table 4.3.16 Cadmium levels in the roots and shoots of four barley accessions at the seedling growth stage were measured using atomic absorption spectrometry with and without plant residues. The percentage reductions of cadmium were calculated and compared to the amount of cadmium absorbed in their own positive control.**

| Accessions | Status | Presence of cadmium in root (PPM) | Cadmium reduction % | Accessions | Status | Presence of cadmium in shoot (PPM) | Cadmium reduction % | Translocation factor (TF) |
|---|---|---|---|---|---|---|---|---|
| **AC29** | **C** | 2.00 | - | **AC29** | **C** | 0.93 | - | - |
| **AC29** | **SC** | 18.66 | - | **AC29** | **SC** | 8.80 | - | 0.47 |
| **AC29** | **SKC** | 16.27 | 12.81 | **AC29** | **SKC** | 5.87 | 33.30 | 0.36 |
| **AC29** | **SOC** | 12.27 | 34.24 | **AC29** | **SOC** | 5.20 | 40.91 | 0.42 |
| **AC37** | **C** | 2.40 | - | **AC37** | **C** | 0.13 | - | - |
| **AC37** | **SC** | 34.66 | - | **AC37** | **SC** | 15.60 | - | 0.45 |
| **AC37** | **SKC** | 33.47 | 3.43 | **AC37** | **SKC** | 11.87 | 23.91 | 0.35 |
| **AC37** | **SOC** | 14.93 | 56.92 | **AC37** | **SOC** | 10.40 | 33.33 | 0.70 |
| **AC48** | **C** | 1.27 | - | **AC48** | **C** | 1.20 | - | - |
| **AC48** | **SC** | 37.33 | - | **AC48** | **SC** | 9.33 | - | 0.25 |
| **AC48** | **SKC** | 34.47 | 7.66 | **AC48** | **SKC** | 9.20 | 1.39 | 0.27 |
| **AC48** | **SOC** | 28.93 | 22.50 | **AC48** | **SOC** | 7.73 | 17.15 | 0.27 |
| **AC52** | **C** | 2.80 | - | **AC52** | **C** | 0.12 | - | - |
| **AC52** | **SC** | 41.60 | - | **AC52** | **SC** | 8.53 | - | 0.21 |
| **AC52** | **SKC** | 34.27 | 17.62 | **AC52** | **SKC** | 5.20 | 39.04 | 0.15 |
| **AC52** | **SOC** | 35.60 | 14.42 | **AC52** | **SOC** | 7.74 | 9.26 | 0.22 |

* C, SC, SKC, and SOC represents the treatments of (soil, soil + cadmium, soil+ kanger + cadmium, and soil + oak + cadmium).

The presence of heavy metals inside the plant harms a variety of morphological, biochemical, and physiological traits, resulting in decreased crop productivity. However, the harmful impact of heavy metals strictly depends on the duration of exposure, plant growth stage, the stress intensity and concentration, and the type of variety exposed to heavy metal stress, even the plant parts within one plant. Together with nutrients from the soil solution, toxic metal ions enter plant cell tissues (Nagajyoti, Lee and Sreekanth, 2010). Growth reduction is the most physiological consequence when a plant is exposed to heavy metal stress conditions.

The process of cell metabolism is negatively affected and an observable reduction in energy can be noted as a result of the impact of heavy metals on changing the structure of the leaf; the physiological alteration of two critical processes (respiration and photosynthesis) (Souri *et al.*, 2019). The capacity of roots under metal stress for water uptake and nutrients is likewise reduced (Khan *et al.*, 2015). From these points, as a result of changes in the functioning of the leaf and root, many growing processes, including seedling, flowering, and anthesis, are unfavorably affected. Guala, Vega and Covelo (2010) previously stated, the toxic impact of heavy metals on plants is by producing phytotoxicity inside the cells, whereby chlorosis is initiated and the growth of the plant is reduced, nutrient uptake by roots is limited and, as a consequence, the production of yield will decline. Cadmium is measured as a very toxic element for plants in comparison with other heavy metals (El Rasafi *et al.*, 2021) by reducing photosynthesis and inhibiting the growth of shoots and roots. Besides, the activities of many enzymes, which are essential for maintaining homeostasis balance within the cell under present stressor conditions (Gupta *et al.*, 2017).

Keshavarz *et al.* (2022) demonstrated that the exposure of cadmium at a level of 120 mg kg-1 significantly reduced yield-related traits including the shoot length, number of seeds per spike, and the weight of 1000 seeds as well as dry accumulation in barley, which is mainly cultivated in Iran.

Cadmium is absorbed by the roots and relocated to the vegetative upper parts of the organ of the plant, reducing the quality of yield crops and is considered a nonessential component for plant growth (Huybrechts *et al.*, 2019). In the presence of cadmium, nutrient availability, including macro and micronutrients, will be inhibited in the plant species (Nazar *et al.*, 2012). Paunov *et al.* (2018) revealed a clear reduction in photosynthesis under the presence of cadmium in wheat and therefore slows down the metabolism of carbohydrates.

For removing heavy metals in wastewater, the industrial remaining materials were characterized as effective bio-sorbent and at the same time two critical issues can be solved (treatment of heavy metals, waste disposal) (Salam, Reiad and ElShafei, 2011). Many reports stated the possibility of eliminating heavy metals by using agricultural leftovers from both soil and water for the reason of owning the potential priorities of lignin, cellulose, and holding polar functional groups such as carbonyl, phenol, and many other groups, which have a high tendency to chelate with metal ions (Malik, Jain and Yadav, 2017; Thakare *et al.*, 2021).

A free electron pair was supplied by the mentioned groups. Thus, a complex compound with metal ions present in the soil or solution will be formed (Singh *et al.*, 2020). The use of agricultural waste seems to be a feasible option for heavy metal removal as a result of possessing the unique chemical composition present in the plant as well as its availability. Considerable portions of lignin and hemicelluloses were present in the leaves of Oaktree, which is about %15 as stated by (Şen *et al.*, 2015). Argun *et al.* (2007) studied the performance of modified oak (*Quercus coccifera*) sawdust for removing the concentrations of three diverse heavy metals from aquatic solutions. They found a significant reduction in these metals using modified oak sawdust with HCL, besides the properties of lignin increased by the modification. This statement has also been proven earlier by (Gaballah and Kilbertus, 1998), who showed a similar pattern of response on modified oak barks. In addition, the sawdust of oak was polarized with lanthanum using the application of biochar as a modification to increase the chelating features against heavy metals (Wang *et al.*, 2015).

Although Egypt is not a natural habitat for oak trees, Abd El-Latif and Ibrahim (2009) conducted research to see if modified oak sawdust with 0.1 N sodium hydroxide and sulphuric acid can absorb the toxic methylene blue dye from aqueous solutions.Progressive changes were observed by scanning electron micrographs on the surface of particles of modified oak sawdust and the properties of removing this toxic improved.

Two toxic elements, mercury (Hg) and cesium (Cs) with concentrations of 1000 ppm were subjected to 2.5% of the powder of oak galls and their fruits (Latifi *et al.*, 2012). Their results showed that eliminating the mercury by gall nuts and oak powder were 96% and 94.8%, respectively, while no significant reduction for cesium was noted. They assumed that the high absorption capacity was solely due to the presence of the tannin compound, which is abundant in an oak tree.

*Gundelia tournefortii* L. is considered a medicinal plant. It is noted that all above-ground portions of this species are used as an important food source, and it is widespread in Asian-temperate zones, including our region (Coruh *et al.*, 2007). Four major classes of phenolic compounds were detected in this species, including (cryptochlorogenic, neochlorogenic, chlorogenic, and caffeic) acid (Haghi, Hatami and Arshi, 2011). In addition, the total phenolic content is similar to our investigation conducted using foline reagent, and a significant portion of this trait was detected in the leaf at the flowering and anthesis stages with values of (128.4 and 103.8) µg/mg respectively (Haghi, Hatami and Arshi, 2011). Moreover, eighteen plant species, including *Gundelia tournefortii*, were tested for their ability to accumulate heavy metals, which are abundant near the zinc and lead mines in Angouran, which is located in Iran (Chehregani, Noori and Yazdi, 2009). In the shoots and leaves of Gundelia, the concentrations of Cd, Cu, Fe, Ni, Pb, and Zn were (2.30, 24.00, 1952, 8.40, 652.00, and 820.00) mg/kg, respectively. Thus, in such a contaminated area, this species may significantly contribute to restoring the natural environment and eliminating the toxic impact of heavy metals. Similarly, bio-absorbent Gundelia, which was collected from Kermanshah, Iran, was exposed to Pb (II) solution with a concentration of 1000 mg/L (Rahimpour, Shojaeimehr and Sadeghi, 2017). They discovered excessive absorption of this metal in the fibril structure and porous surface of Guanila. In this context, both Oakleaf and Gundelia plant parts received our attention in removing the toxic impact of cadmium, and a similar pattern of response by our analysis was detected, besides these two species naturally available in our area (Khwarahm, 2020; Coruh *et al.*, 2007). They are sources that can be used again and again, so they don't need to be regenerated after being used to collect heavy metals. Guo *et al.* (2018) studied the impact of contaminated soil with heavy metals on the yield of wheat and rice and the ways of improvement using two different inorganic ingredients (hydroxyapatite and hydrated lime) and organic fertilizer. The soil profile of two different locations (Dayu and Yixing) in Jiangxi province in China that were contaminated with heavy metals was used in their investigation. They found considerable amounts of Cd, Cu, and Pb when the soil was subjected to analysis in both locations. Clear reductions in both crops in terms of yield were documented when exposed to contaminated soils. On the contrary, even under the present low dosage of these organic and inorganic compounds, especially hydrated lime in the soils, the yield of rice and wheat improved in both soil conditions.

Biochar, similar to plant residues, is currently receiving considerable interest from researchers and it is being effectively evaluated for its ability to eliminate heavy metals through thermochemical treatment on organic material under a restricted oxygen environment (Puga *et al.*, 2015; Liu *et al.*, 2022).

To assess the impact of biochar on the uptake of cadmium in wheat cultivar, Abbas *et al.* (2018) subjected the wheat cultivar to two different stress conditions. First, the soil used in their experiment was contaminated with cadmium. Second, besides the availability of cadmium, they exposed the wheat cultivar to drought stress with three different regimes (well-watered, mild, and extreme) stress conditions. As a treatment of biochar, three levels were performed: % (0, 3, and 5) w/w after incubation of 15 days. Drought and cadmium reduced most critical yield parameters, including plant height, the dry accumulation of biomass, the length of the spike, as well as the final yield, while under the presence of biochar, all studied parameters improved, including physiological and morphological traits, and the enzymatic activities of antioxidants improved in a great manner. In the same way, in our study, the presence of plant residues improved these traits in response to cadmium.

Under natural conditions, silicon (Si) is generally present as silica ($SiO_2$) and is considered the most prevalent element in the soil profile after oxygen. The concentration of silicon has ranged between %25 and %35 depending on the soil types (Tubaña and Heckman, 2015). Despite the fact that silicon is not widely recognized as a necessary nutrient for vascular plants, it has been implicated in increasing crop yield (Vasanthi, Saleena and Raj, 2014), improving root architecture (Tripathi *et al.*, 2021), and regulating the activity of antioxidants exposed to biotic and abiotic stresses (Hossain *et al.*, 2018). For (Si) absorption and transportation, roots played a major role in these actions. Many passive transporters facilitate the uptake of Si from the root to other parts of a plant organ. For instance, in rice, the presence of the cadmium *Lsi1* transporter was highly expressed and the presence of this transporter enhanced the activities of antioxidants (Lin *et al.*, 2017). Similarly, *HvLsi1* and *HvLsi2* transporters in barley were identified that take a similar role as *Lsi1* (Talakayala, Ankanagari and Garladinne, 2020). Based on the outcomes they obtained, they showed that the expression level of these transporters mainly happened in the root rather than other plant parts, indicating the essential role of the root for uptaking the (Si). Recently, two barley cultivars were subjected to aluminum stress for two days, and the application of silicon with a concentration of two mM was used. The sensitive cultivar under the availability of silicone was enhanced in terms of increasing radical scavenging activity and phenolic compounds (Vega *et al.*, 2019). The influence of wood waste using three different trees, including oak (*Quercus robur*), on the biological and mechanical properties of silicone-based composites was investigated by a group of researchers from Poland (Mrówka, Szymiczek and Skonieczna, 2021). They found that oak waste, under the availability of silicon, had a high tendency to accumulate silicon in their cellular profile, and as a consequence, the rigidity or hardness of this particular waste increased. Based on this finding, there is the possibility that oak trees during their life cycle accumulated and converted $SiO_2$ from soil to an available form of silicon (Si) to make it stored in their organs. The results of our study showed that under the presence of oak residues, the performance of enzymatic activities as well as root architectures were improved similar to conducted researches.

In plants, chelation has been characterized as a great physiological response to heavy metals by transporting these metals to the safe side, which is vascular inside the cell to make sure that the concentrations stay in a narrow range. To deal with heavy metals, several natural biomolecules are owned by the plant to bind with metals such as metallothioneins, and histidine phytochelatins (Yadav, 2010).

Heavy metal transporters belonging to the ATPase subfamily, including *AtHMA4* and *AtHMA2*, play essential roles in *Arabidopsis thaliana* in exporting both cadmium and zinc to vascular cells (Verret *et al.*, 2004). Similarly, the same phenomenon on rice was observed (Takahashi *et al.*, 2012). Based on this, Mikkelsen *et al.* (2012) have cloned *HvHMA1*, which is an orthologue to *AtHMA1* in barley by *Agrobacterium transformation*. During grain filling, when the expression level was estimated, they found strong upregulation of *HvHMA1* in response to cadmium. Zhang *et al.* (2021a) discovered a higher expression level of *HvHMA1* as well as an increase in expression in two additional genes, *HvHMA3* and *HvHMA4*, under cadmium stress conditions on barley ZJU3 variety. Probably AC29, AC37, and AC38 possessed the same expression pattern under the presence of cadmium and, in return, the impact of cadmium was less on these accessions by taking action in exporting cadmium to the safe side of the cell.

Gene expression level of phytochelatin synthase on seedling wild type wheat and mutant allele carried *Rht3* dwarf lines after one week treated with cadmium concentration of 50-μM were estimated (Szalai *et al.*, 2020) and a considerable increase in the phytochelatin synthase detected in dwarf lines that considered as tolerant to cadmium. In our investigation, a great reduction in shoot length was documented by AC37, AC29, and AC38. Presumably, the presence of cadmium increased the expression level of *Dwf2*, which is said to have a similar response to *Rht3* in barley for shortening shoot length (Khlestkina *et al.*, 2020), and as a result, phytochelatin synthase was activated.

For the growth and development of the plant, respiration metabolism plays a crucial role. In addition to the cyanide-sensitive cytochrome pathway (CP), plant mitochondria have an alternative pathway (AP), which is placed in the mitochondria's internal membrane and is involved in improving plant tolerance to environmental stresses (Scheibe, 2019). He *et al.* (2021) recently studied the relationship and role of $H_2O_2$ and alternative pathways in cadmium resistance and susceptible varieties. After two days of exposing the plant material at the seedling stage with a concentration of 150 µM $CdCl_2$, they found a high-level expression of alternative oxidase (*AOX*) genes, especially *AOX1a*, in cadmium-resistant varieties. In our study, this pathway was probably similarly activated in response to cadmium by considering a cadmium-resistant variety.

Labudda *et al.* (2020) measured the activity of glutathione reductase (GR) in barley cultivar Airway that had been exposed to 5 M $CdCl_2$ for two weeks.The activity of this enzyme is highly stimulated under cadmium stress compared to control conditions. This enzyme catalyzed the transformation of oxidized glutathione to reduced glutathione. There is the possibility of chelating cadmium ions with reduced glutathione (GSH) and forming the complex compound of $Cd(GS)_2$ that is likely seized into the vacuoles for detoxification, indicating that the performance of studied varieties for this enzyme activity in the presence of cadmium is very important (Rao and Reddy, 2008; Demecsová *et al.*, 2020b). To alter the adverse effects of cadmium, presumably, AC29, AC37, and AC38 activated Glutathione reductase (GR) to cope with cadmium stress conditions.

Wu, Sato and Ma (2015) tested 100 barley accessions collected from different regions for the response of these accessions under cadmium stress. They considered BCS318, which is cultivated in Afghanistan as cadmium susceptible, and Haruna Nijo, from Japan, as tolerant. Based on this, the same group of researchers,Lei *et al.* (2020) subjected these two barley cultivars to cadmium stress with a concentration of 0.5 Mm for three days to assess the expression regulation of *HvHMA3*. They revealed the insertion of a Sukkula-like transposable element upstream of this specific gene, which operated as a promoter and improved the expression of *HvHMA3* in translocating the cadmium in the Japanese barley cultivar. Feng, Shen and Shao (2021) also stated in their review the significant role of these transposable elements in crop species under the presence of cadmium. In our study, the data revealed that accessions AC47, AC48, and AC52 are considered cadmium susceptible accessions. Probably these transposable elements were absent in their genome. From this point, introducing this particular gene by backcrossing could result in generating resistance barley for cadmium stress without interfering with the yield product as stated by (Feng, Shen and Shao, 2021).

## 4.4 Field Section

### 4.4.1 The performance of growth traits under the application of (MPE)

An essential agronomic trait for grain yield formation and morphogenesis in cereals is plant height (PH). For achieving the preferred yield product in barley and other cereal breeding programs, remarkable increases in wheat and barley yields have been accomplished during the Green Revolution by the introduction of dwarfing traits into those cereals and reducing their height. A lot of research has been done on the dwarfing gene *sdw1* in order to make new kinds of barley (Xu *et al.*, 2017; Dang *et al.*, 2020). The short cultivars possessed the ability to tolerate the lodging phenomenon and were adapted to mechanical harvest. On the other hand, the other breeding strategy is to produce high vegetative yields of crops to use these parts as biofuel energy.

As a byproduct after harvesting the grains, a significant amount of barley straw is left in the field, which is estimated at fifty-one million tons per year around the world (Qureshi *et al.*, 2010). Therefore, barley leftovers have become an essential resource for bioenergy production. Adapa, Tabil and Schoenau (2009) investigated the chemical composition of straw for four different kinds of cereal, including barley, and they revealed that the straw of barley consists of lignin, hemicellulose, and cellulose with values of 17.13%, 20.36%, and 33.25%, respectively. These compounds have been considered a favorable new version of bioenergy feedstock (Sims *et al.*, 2010; Stolarski *et al.*, 2015). From the above points, cultivating barley for the dual purposes of high yield and biomass would be advantageous.

For assessing the performance of height traits in our investigation, the length measurements were performed one week before harvesting by recording the length of the studied plant from the ground to the base of the spike, and significant effects of moringa were detected on studied barley accessions (Table 4.4.1). The mean pairwise comparison analysis on the experimental data between the accessions showed significant variation between barley accessions (Table 4.4.2). The tallest length was recorded by AC57 with a value of (110.03 cm), followed by AC56 (109.30 cm) and AC55 (104.03 cm). Oppositely, the barley accessions of AC1, AC6, and AC5 were taken into account as the shortest barley accessions, with a value of (52.17, 62.93, and 65.90) cm, respectively. As it is clear from the mean analysis and pairwise comparison form (Appendix 4.1), which represents the interaction between treatment and the accessions, the exogenous application of Moringa plant parts positively affected the plant height. The highest length under presence of moringa was recorded by AC57 (117 cm), followed by the other two accessions (AC29 and AC56) with a value of (113.53 and 110.27) cm, respectively. By contrast, the barley accession AC1, with a length of 51.27 cm under absence of moringa, is considered the shortest accession among 59 barley accessions.

Similar to our investigation, Rehman *et al.* (2017) found a considerable increase in wheat height due to conducting moringa extract in their research. As mentioned previously, the Moringa plant owns a great number of phytohormones, including gibberellin (Ali, Hassan and Elgimabi, 2018). This phytohormone plays a major role in regulating the shape of the leaf, flowering as well as the elongation of steam in barley and other cereal species (Gao and Chu, 2020).

Metabolism and signaling of gibberellin are both crucial for controlling plant height. The presence of this phytohormone enhanced the elongation of the internode, resulting in increased cell division as well as cell elongation (Gao and Chu, 2020). Reducing the catabolism of gibberellin would produce taller plants, and therefore more biomass would be generated. By contrast, dwarf or semi-dwarf phenotypes can be noticed in the case of the present defective mutants for gibberellin biosynthesis (Marzec and Alqudah, 2018). The molecular mechanism underlying the regulation of gibberellin biosynthesis has been widely investigated in rice. The functions of transcription factor *OsMADS57* were well studied in rice by Chu *et al.* (2019), internode elongation, rice height, and gibberellic acid were reduced in the suppression of this particular transcription factor compared to the rice wild type. The results obtained here can provide valuable information for similar research in barley to know the expression level of this phytohormone after applying the foliar application of Moringa extract. In barley, little is known about the molecular mechanism controlling the regulation of these fundamental phytohormones, while very recently regarding the mechanism of barley internode elongation. *HvSS1* on chromosome *6H* regulates internode elongation, according to Pu *et al.* (2021) , and knocking down this candidate gene may be associated with an insufficient amount of endogenous bioactive gibberellin..

The height of the plant is an essential component of plant architecture that is highly associated with the final biomass yield product. To detect quantitative trait loci controlling plant height, Wang *et al.* (2014) conducted research on 182 barley lines within six environmental conditions. They found no significant effects on yield components and other agronomic traits with candidate quantitative trait *loci* belonging to height. However, pleiotropic interaction is one of the most important considerations in manipulating plant height for the production of biomass(Teplyakova *et al.*, 2017). For that reason, the potential for adverse consequences of changes in plant height might affect other important traits. To overcome this issue, high-resolution and large-scale mapping research is required to identify more candidate quantitative trait loci affecting plant height, and nested association mapping is one of them (Gage *et al.*, 2020; Zahn *et al.*, 2021). It is possible to conclude that the foliar application of the Moringa plant not only regulates the yield trait attributes in barley but also produces a high yield for its biomass. Therefore, improving these traits by utilizing this kind of treatment will be the key objective of many kinds of cereal breeding programs in the future.

Some traits, like leaf area (Allel, Ben-Amar and Abdelly, 2018), awn length (Huang, Wu and Hong, 2021), and chlorophyll content (Bahrami, Arzani and Rahimmalek, 2019), have been shown to play a major role in increasing photosynthesis under normal and stressful conditions. The primary organ which takes a huge portion in photosynthesis is the flag leaf. The characteristics of flag leaf are considered essential selection criteria for high grain yields in barley (Xue *et al.*, 2008). For this reason, the lower leaves are mostly covered by the upper plant parts and therefore do not directly take part in absorbing the radiation of solar energy. In our investigation, therefore, the trait of leaf area was measured on flag leaf to observe the effect of Moringa plant extract on the investigated barley accessions, and significant variations were detected (Table 4.4.1).

Pairwise comparison of means between the barley accessions as presented in (Table 4.4.2), demonstrated that AC51, with a value of (16.61 $cm^2$), again recorded the highest measurement of this trait compared to the rest of the investigated barley accessions, followed by AC11 and AC15, with values of (16.18 and 15.36) $cm^2$, respectively. The performance of barley accession of AC36 was the lowest (6.91 $cm^2$), followed by AC54 (7.59 $cm^2$) and AC24 (7.80 $cm^2$). In the presence of foliar application of Moringa, the interaction results demonstrated that the highest measurement (18.48 $cm^2$) was obtained by AC51, followed by AC11 and AC15 with values of (17.43 and 17.40) $cm^2$ under availability of moringa, respectively (Appendix 4.1). whereas barley accession AC36 with lack of moringa application had the lowest measurement (4.85 $cm^2$) for this trait. Similarly, the other two barley accessions (AC26 and AC54) possessed the lowest measurement of leaf area, with a value of 6.97 and 7.37, $cm^2$ respectively.

Several phytohormones with an obvious portion were detected in Moringa leaf extract by Ali, Hassan and Elgimabi (2018), including gibberellins, auxin, and cytokinins. It is well documented that gibberellins improve plant height while auxins improve the elongation of cells and promote the growth of stems, and cytokinins play a critical role in the promotion of cell division and modification of apical dominance (Taiz and Zeiger, 2010). After the application of Moringa plant extract, a significant increase in leaf area was observed, probably due to the presence of these critical phytohormones in their nature. In accordance with our findings, Chattha *et al.* (2018) found similar outcomes in the case of using this type of extract on the wheat plant. Additionally, Ali *et al.* (2010) showed a significant increase in the measurement of this trait on wheat varieties after conducting the same exogenous application of Moringa leaf extract.

Because chlorophyll is required to convert light energy into stored chemical energy, crop growth and yield are directly affected by chlorophyll content (Sid'ko *et al.*, 2017). Correlations between leaf area, chlorophyll content, and yield were shown by many studies for barley cultivars (Klem *et al.*, 2012; Lausch *et al.*, 2013; Begović *et al.*, 2020). This is probably due to capturing more light and including a denser chloroplast. New opportunities to predict total chlorophyll content (TCC) at the various crop growth stages have been provided with the development of remote sensing equipment (SPAD), which is widely accepted by researchers (Donnelly *et al.*, 2020; Shibaeva, Mamaev and Sherudilo, 2020).
In our investigation, therefore, SPAD (CCM-200) was used to measure the Total Chlorophyll Content (TCC) of barley leaves. For this trait, significant outcomes were detected using moringa application (Table 4.4.1). The mean pairwise comparison analysis among barley accessions as expressed in (Table 4.4.2), showed that the (TCC) of barley accession AC28 had the highest value of (20.27 spad unit) compared to the rest of the studied barley accession, whereas barley accession AC27 had the lowest value of (6.67 spad unit). Significant interaction effects were observed between the treatment applications of Moringa and barley accessions as presented in (Appendix 4.1) for the studied trait. This trait gave much more content in the barley accession of AC39 (26.00 spad unit) under moringa application than that in the barley accession of AC20 (5.07 spad unit) when moringa was not available.
Merwad (2018a) revealed that even under saline conditions compared to untreated wheat plants, the total photosynthetic pigments in the leaf increased dramatically under availability of moringa. In the regulation of photosynthesis and many physiological processes, salicylic acid (SA) plays a main role under stress conditions to maintain these regulations within plant cells (Taiz and Zeiger, 2010). After 21 days from the sowing date for barley cultivar (Giza126), foliar application of salicylic acid with a concentration of 0.5 mM was applied under drought stress conditions by Abdelaal *et al.* (2020). They found that the chlorophyll concentration was increased significantly compared to the control treatment. Very recently, on barley genotypes, a foliar application of combination gibberellic acid and (SA) with a concentration of (110 mg/l and 1.5 mM) showed a significant increase in different plant physiological properties, including total chlorophyll content (Askarnejad, Soleymani and Javanmard, 2021).
Many essential developmental processes are modulated by the presence or absence of cytokinins, including the leaf development in the last phase, well-known as senescence, which is associated with the breakdown of chlorophyll and photosynthetic collapse. All of these undesirable changes can be slowed by cytokinins (Hönig *et al.*, 2018).
Regarding the analysis of phytohormones in Moringa leaf extract, a considerable portion of (SA) and cytokinins were detected with a value of (1.87 and 0.63 mg $g^{-1}$ DW), respectively (Ali, Hassan and Elgimabi, 2018). Taking all the phytohormones present in Moringa leaf extract into account, it is possible to conclude that a strong direct correlation is present between the total chlorophyll content and those phytohormones. For all the above reasons, these traits (leaf area, and total chlorophyll content) could be used as growth morphological markers for the selection of barley accessions having higher photosynthetic activity.

### 4.4.2 The performance of some yield contributing traits under the presence of moringa plant extract

The awns of barley are the essential photosynthetic organs after the flag leaf. For the developing grains in spikelets within the spike, this organ is the nearest plant part which acts as a source of assimilation for the formation of grain. The spike organs of barley (including awn) are active for photosynthesis, and their contribution to the accumulation of grain dry weight is more than 75% (Torne, 1963; Abebe, Wise and Skadsen, 2009). It was proven a long time ago that under normal growth conditions in barley, more than 90% of spike photosynthesis can be achieved through the awns organ (Ziegler-Jöns, 1989). Therefore, the proportion of net photosynthesis can significantly be enhanced by these plant parts, resulting in a greater value of grain dry matter.

As indicated in our analysis, the foliar application of Moringa significantly induced the performance of this trait (Table 4.4.1). The mean pairwise comparison analysis among barley accessions presented in (Table 4.4.3) indicated that barley accessions AC21 (14.49 cm), AC15 (14.10 cm), and AC25 (13.94 cm) displayed superior performance in awn length over the rest of the examined accessions. Oppositely, the lowest length measurement was obtained by AC57 (8.95 cm), followed by other barley accessions of AC44 (9.19 cm), AC45, and AC47 with the same value of (9.29 cm). In the same way, mean pairwise comparison and their interaction with moringa between studied barley accessions displayed that the same accessions AC15, AC25, and AC21 with the values of (15.94, 15.76, and 15.02) cm exhibited the maximum length for awn length under moringa application, respectively. While, in the absence of Moringa application, the shortest length was recorded by AC35, followed by AC57, and AC47 with the values of (8.46, 8.71, and 8.81) cm, respectively (Appendix 4.2).

The awn removal in barley genotypes showed a considerable effect on grain yield performance, transpiration rate, and net photosynthetic in which all these measurements were decreased (Jiang *et al.*, 2006). Additionally, under drought stress conditions, Hein *et al.* (2016) showed that the organs of spike, including awn of barley genotypes, respond differently at the grain filling stage compared to vegetative organs in which it preserves more cellular hydration. A notable increase was detected in the case of conducting the foliar application of Moringa in our investigation for this trait. In our investigation, the increase of awn could be explained by revealing the potentails of this trait using moringa applications.

At least two types of tillers are known in cereals (fertile and non-fertile). The first, which is also known as the productive tiller, leads to the formation of spikes, and therefore it is essential for the seed yield. While the second type consumes the plant's mineral resources. This type of tiller often does not survive until the end of the plant's life, so it cannot produce yield (Sreenivasulu and Schnurbusch, 2012). From this point, it is very essential to measure the fertile tiller number per plant to assess the final productivity of studied cereals.

In our experiment regarding the foliar effect of Moringa plant extract, significant results were documented for traits of number of tillers. The mean pairwise comparison among the barley accessions alone, as shown in (Table 4.4.3), revealed that barley accession AC47 had the highest tiller number (30.83). The barley accession AC7, on the other hand, was given the lowest number of tillers (5.67). As shown in (Table 4.4.2), the mean pairwise comparison for its interaction revealed that in the presence of the Moringa application, the barley accession AC47 had the highest tiller number per plant (37.33). In contrast, barley accession AC7 recorded the lowest value (5.33) in the absence of the same treatment conducted in this study.

In agreement with our outcomes, Afzal *et al.* (2020) reported that the application of Moringa leaf extract increased the studied yield traits, including the tiller number in wheat. In addition to these findings, Koprna *et al.* (2021) from Palacký University Olomouc stated the positive role of cytokinin application on the tiller number of barley varieties. The obvious increase in tiller number in our results was probably linked to the presence of a growth regulator of cytokinin in the content of Moringa plant extract (Ali, Hassan and Elgimabi, 2018). The trait of tiller number, therefore, can be carefully chosen for studying the application of Moringa plant extract.

Producing a higher yield of barley and other cereals is the ultimate goal of any breeding approach. Several strategies are being implemented to increase the yield of contributing factors. Crop yield in cereals is mainly determined by measuring the most important traits that are strongly associated with the final yield product, including spike length, spike number per plant, seed weight per spike, 1000 grain weight, seed number per spike, the weight of spike, total yield, and straw yield. Foliar application of Moringa plant extract is well documented as it is important in the enhancement of yield contributing parameters in many plant species (Rady and Mohamed, 2015; Abd El-Mageed, Semida and Rady, 2017; Brockman and Brennan, 2017; Ahmad *et al.,* 2020). In our experiment, exogenous application of Moringa plant part extracts positively impacted these parameters compared with the respective control conditions.

The mean comparison among tested barley accessions revealed that AC30, with a length of 8.50 cm, had the superior length of spike, and then AC32 and AC34 were considered the second and third barley accessions having the highest spike length, with a value of 8.03 cm and 7.96 cm. On the other hand, AC3, AC22, and AC18 are characterized as the shortest barley accessions for a studied trait with a value of (3.79, 4.09, and 4.18) cm, respectively (Table 4.4.3). The mean pairwise comparison for investigating the length of spike between the foliar application of Moringa and accessions showed significant improvement for its spike length (Appendix 4.2). The barley accession AC54 had better performance over other studied accessions, having a value of 8.97 cm, followed by AC30 with and without Moringa, with a length of 8.53 and 8.47 cm, respectively. On the other hand, the barley accession AC3 had the shortest spike length in both cases (with and without treatment), as shown in Appendix 4.2, with values of 3.78 and 3.81cm, respectively. AC5, with a length of 3.94 cm, was next. Similarly, a remarkable increase in spike length was observed on the wheat plant in the field by Khan *et al.* (2020) due to utilizing the same application method. In addition, Zaheer *et al.* (2019) conducted research on wheat cultivars using various foliar applications, including cytokinins with a concentration of 25 $mg{\cdot}L^{-1}$ used under drought stress conditions at three different growth stages (tiller formation, flowering, and grain filling). In the presence of this application, the longevity of spikes in their study was improved significantly. Therefore, it is quite reasonable to conduct the foliar application of Moringa to improve the length of spikes. The higher spike length in the current study may be connected with the presence of various phytohormones and secondary metabolites in Moringa plant parts (Ali, Hassan and Elgimabi, 2018).

The formation of spikelets is the first phase of reproductive growth in cereals. Although this process is short in period, it provides the developmental basis for the variation of the spike (Sreenivasulu and Schnurbusch, 2012). The potential of yield could be increased by having the ability to manipulate this critical phase so the architecture of the spike can be restructured. In addition to this phase, the elongation of the spike is the final stage of reproductive growth. In this particular phase, the fertility and fate of each spikelet can be determined. It is quite clear that a combination of the activity of particular phytohormones as well as the nutritional condition of the reproductive meristem both have significant effects on final grain number (Sreenivasulu and Schnurbusch, 2012).

One of the most critical yield attributes is the number of spikes per plant. The selection of barley genotypes based on the higher number of spike numbers per plant may finally lead to choosing better yielding performance among tested accessions. The yield of barley genotype is significantly influenced by this trait (Araus *et al.*, 2008). In this experiment, the number of spikes per plant ranged between 22.33 and 4.00 for barley accessions AC38 and AC3, respectively, in the case of analysis of the data among barley accessions conducting mean pairwise comparison (Table 4.4.3). On the other hand, to observe the influence of Moringa treatment and know the interaction with the studied barley accessions, the same method of mean pairwise comparison was used (Appendix 4.2). The barley accession AC28 had more spikes per plant (25.33) compared to AC3 as it owned just three spikes per plant. This variation in the number of spikes per plant can be attributed to the genetic potential of barley accessions and their different responses to foliar application of Moringa. The three barley accessions (AC28, AC47, and AC36) with values of (25.33, 25.00, and 23.67) had more potential to produce a large number of spikes per plant in the case of the present foliar treatment (Appendix 4.2). This might be due to the ability of these barley accessions to respond greatly to this management. The results of the present study are in agreement with the findings of another group that studied the effect of Moringa on this specific yield trait. Afzal *et al.* (2020) conducted research on wheat plants to identify the possible effects of three different foliar applications (Moringa leaf, sorghum water extract, and salicylic acid) with a concentration of 3%, 0.075%, and 0.01%, respectively, in the case of present heat stress. At an interval of one month, they applied the foliar application three times, starting from the tillering stage. Among tested foliar applications, moringa extract as well as salicylic acid highly improved the performance of this trait. Similarly, Khan *et al.* (2020) revealed a significant impact of Moringa leaf extract alone and in mixture with other plant growth promoters like ascorbic acid and salicylic acid for this trait on wheat by applying this treatment twice at tillering and flowering stage.

Seed weight per spike (SWS) plays a significant role in the formation of yield. This trait directly reveals the efficient use of nutrients by the plant and the translocation of these substrates into reproductive plant parts (Protich, Todorovich and Protich, 2012).

The value of seed weight per spike in overall barley accession increased significantly under the presence of moringa (Table 4.4.1). The mean comparison among tested barley accessions showed that AC59, as well as AC47, showed the highest and lowest values of this trait (3.49 and 0.66) g, respectively (Table 4.4.3). In terms of having an interaction with Moringa, the value of seed weight per spike for studied barley accessions exhibited the same pattern of response as barley accessions. The values ranged between (3.49 and 0.66) g for the accession of AC59 and AC47, with and without moringa, respectively (Appendix 4.2).

These findings agree with those of Yasmeen *et al.* (2012) and Yasmeen Yasmeen *et al.* (2013a) , who found that under normal and saline conditions, as well as late sowing of wheat, foliar application of moringa leaf extract increased grain weight, kernel yield, as well as grain weight per spike. Similar to our results, a considerable increase in the seed weight per pod in pea plants (Merwad, 2018b), seeds in maize kernel (*Zea mays L.*) (Maswada *et al.*, 2018), and snap bean (*Phaseolus vulgaris* L.) (Elzaawely *et al.*, 2017) was detected due to the treatment of Moringa leaf extract.

For measuring the quality of seed, 1000 grain weight is a very important parameter, which is effective on sprouting, the potential of the seed, and overall plant performance. The embryo size and the reserved nutrient quantity used for sprouting and growth determine the quality of the seed (Sadeghzadeh-Ahari *et al.*, 2010).

Analysis of variance displayed significant effects of barley accessions for its 1000 grain weight (1000 KW) (Table 4.4.1). Among the tested barley accessions, AC25 (63.66) g, AC2 (57.30) g, and AC28 (56.94) g, possessed the highest value for the studied trait compared to the accessions of AC48 (35.01) g, AC47 (36.46) g, and AC7 (36.59) g (Table 4.4.3). With a value of 67.12 g, the interaction between the evaluated barley accessions and the Moringa application showed that AC25 performed better for the studied trait than the other accessions under moringa application. In contrast, the accession of AC48 recorded a minimum value of 33.82 g without moringa (Appendix 4.2).

In the presence of drought stress, the application of Moringa leaf extract improved and recovered the barley genotypes for its 1000 grain yield by 24% compared to untreated conditions (Alghabari, 2020). On sandy soil in Egypt, a field experiment was conducted by (M. A. Merwad and Abdel-Fattah, 2017) to observe the effect of Moringa leaf extract on nutrient uptake and yield of wheat. Utilizing this treatment significantly improved seed yield and straw, biological yield, and 1000 seed weight. The same group of researchers investigated pea plants to see whether the same pattern of response in pea plants could be detected (Merwad, 2018b). They showed that at the level of four percent, foliar spraying of Moringa leaf 100 seed weight improved significantly compared to water spray (control condition).

Positive effects of foliar application of Moringa plant parts were detected among tested barley accessions for the trait of seed number per spike (Table 4.4.1).

The mean pairwise comparison among barley accessions varied significantly, as shown in (Table 4.4.3). AC59, AC21, and AC13 had more seed numbers with values of 65.72, 56.83, and 54.83, respectively, compared to barley accessions AC47, AC52, and AC48, which had less grain per spike with values of 18.22, 19.78, and 19.94, respectively. As shown in Appendix 4.2, significant differences were discovered when the data for this trait were analyzed for observing the interaction between foliar application and barley accessions. The AC59 had the highest seed number per spike in both conditions (with and without Moringa), with values of 72.11 and 59.33, respectively. In addition to AC59, the barley accession AC13 with the foliar treatment showed a high seed number per spike with a value of 57.78. On the other hand, being deprived of the foliar treatment of Moringa caused a significant reduction in seed number per spike, as displayed in (Appendix 4.2) for barley accessions AC53, AC47, and AC48 with a value of 17.22, 17.78, and 19.22, respectively.

An accepted concept is that the grain number and final yield are positively correlated with the dry weight of the spike during the spike growth phase, possibly because of improved photosynthetic capacity under both normal and stress conditions (Van De Velde *et al.*, 2017). Very recently, Zhang *et al.* (2021b) determined the roles of cytokinin oxidase and dehydrogenase in rice among eleven candidate CKXs families for their impacts on grain number, leaf senescence, and regulating the source from leaf and sink from grain using *CRISPR/Cas9* gene-editing techniques. They found that *OsCKX11* knockout significantly increased cellular cytokine levels, causing a delayed leaf senescence phenotype. Additionally, the mutant *OsCKX11* compared to the wild type showed a significant increase in the grain number. The possible conclusion here is that *OsCKX11* regulates both grain number and photosynthesis. Similar attempts in wheat were conducted to observe the accessions between the number of seeds and *CKX* gene families. Downregulation of *TaCKX2.4* expressions on the chromosome (3A) increased the cytokinin endogenous level and increased the seed number per spike. The previous research, as indicated above, showed the positive regulation of cytokinin in increasing the seed number per spike. Therefore, the significant outcomes in our investigation for this trait are possibly linked to owning these fundamental phytohormones in the Moringa plant. Besides, the foliar application of Moringa could change the homeostasis level of cytokinin inside the cell and Knock out the above-mentioned genes which in return the cytokinin level will rise and as a consequence, the seed number per spike will increase. Hence, it is possible to conclude that a higher number of seeds detected in this study were stimulated by higher cytokinin content.

Application of Moringa improved spike weight, which is another significant outcome of the present study (Table 4.4.1). The mean pairwise comparison between studied barley accessions for an investigated trait as presented in (Table 4.4.3) revealed that AC59, with a value of 3.74 g, had the highest spike weight, and AC13 (3.27) g and AC21 (3.23) g, came after. On the other hand, the three barley accessions AC47, AC48, and AC53 showed the least performance for spike weight with a value of (0.77, 0.81, and 0.97) g, respectively. As shown in (Appendix 4.2) for detecting the influence of Moringa plant applications and their interaction with the studied barley accessions, mean pairwise comparison showed significant impacts for the studied trait. The barley accession AC59 (4.03) g with Moringa outperformed compared with other accessions in terms of spike weight, followed by the other two barley accessions AC2 (3.57) g and AC55 (3.45) g. By contrast, the lack of used applications caused a notable reduction in this trait weight. For instance, AC47, AC53, and AC48, with values of 0.77, 0.79, and 0.81g, respectively recorded the lowest value of spike weight.

The higher spike weight of plants sprayed with Moringa was due to the increased length of spike, the number of seeds per spike, and other yield-contributing factors as described previously. A cheap, rich, and natural source of important secondary metabolic products and plant phytohormones plays a key role in the improvement of barley yield. Due to the phenomenon of staying green for a longer period throughout grain filling, the application of the Moringa plant part as a foliar spray significantly increased the studied parameters in our study. This may be linked with the abundance of cytokinin hormone in Moringa, which is the most general coordinator between senescence and staying green traits that eventually improves final yield productivities.

Plant researchers are now focusing on studying bio-stimulants and using them in their research to improve crop yields. Plant stimulants have been shown to improve the health of plants and the quality of their yields by making plants take in more nutrients, changing their physiology, and making them better able to handle stress (Yasmeen *et al.*, 2013a; Khan *et al.*, 2020).

In the present investigation, in response to foliar treatments of Moringa Plant Extract with concentration (1:30) were applied during the growth to indicate the performance of 59 barley accessions for its productivities. High levels of significant variations for all traits of yield components and studied parameters were observed. Significant differences were observed between the foliar treatments and barley accessions in terms of the total yield (TY) and straw weight (ST), with significant differences among all studied barley accessions based on the analysis of variance (Table 4.4.1).

The study of total yield performance and related traits for its production is the function of the genetic character of the cereal crop, the status of nutrients within the soil texture, the exogenous application of growth enhancers, as well as the environmental conditions surrounding the crop plants (Khan *et al.*, 2020).

The total yield of cereal grain and the values of its related components are different from cultivar to cultivar. These differences in the yield are strictly correlated with variation in the number of grains and must thus rely on variation in the shoot number, which last to produce more spikes (Richards *et al.*, 1987; Fageria, Baligar and Clark, 2006). In addition to total yield, the straw weight is an essential trait for plant breeding as it gives details about the capacity of the plant to allocate biomass in the form of reproductive plant parts. It is associated with the yield of grain and biomass following the multiplicative yield component, in which the yield of grain is a product of yield biomass (Wnuk *et al.*, 2013).

Regarding the analysis of total yield performance and harvest index after application of Moringa plant extract, significant results were obtained in response to (MPE). Barley accession AC55 for total yield showed superiority over the rest of barley accessions, followed by AC59 and AC30 with values of (337.47, 291.98, and 255.06) g, respectively. While the lowest value was recorded by the barley accession of AC3 (15.83) g, followed by AC5 (27.29) g, and AC12 (29.66) g (Table 4.4.3). Whereas, in the case of having interaction with the treatment (Moringa application) for the same studied traits,

AC59 showed the highest variability for its mean among the investigated accessions under moringa treatment, with a value of 363.27 g. In contrast, the lowest record was obtained by AC5 with a value of (12.24) g, when moringa was not available (Appendix 4.2). Similarly, AC3 showed the lowest rate (72.51) g in the case of studying the straw weight, followed by AC12 and AC6 with values of (126.24) and (136.54) g, respectively. On the other hand, the highest record was verified by the AC55 with a value of (741.59) g. Subsequently, it was followed by barley accessions of AC25 and AC59 with values of (726.23 and 656.98) g, respectively (Table 4.4.3). However, the mean pairwise comparison and their interaction with Moringa foliar application for straw weight showed a wide range of viability, as shown in (Appendix 4.2). AC3 showed the lowest record of (45.62) g under absence of moringa. Meanwhile, AC25 displayed the highest value (764.32) g compared to the rest of the barley accessions under moringa treatment as stated in appendix 4.2.

In agreement with our outcomes, Khan *et al.* (2020) reported that due to the existence of phenols, antioxidants, essential nutrients, phytohormones, and ascorbates in Moringa leaf extract, it is measured as one of the essential plant biostimulants. In their investigation, as they conducted the combined application of chemical enhancers such as salicylic acid and ascorbic acid as well as foliar application of Moringa leaf extract, they found that both applications improved grain yields, biological yields as well as biochemical parameters as compared to the untreated studied wheat plant. Additionally, the endogenous concentrations of phytohormones might positively be affected in the presence of foliar application of Moringa plant extract in this manner, causing noticeable enhancement in the growth of the plant (Maishanu *et al.*, 2017). In our investigation, significant values for almost all studied yield traits were observed. This is probably due to the role of cytokinins detected in Moringa leaves (Iqbal, 2014; Abd El-Mageed, Semida and Rady, 2017), which motivates the metabolism of carbohydrates. Besides that, this feature produces a new sink source, leading to increased dry matter content. Similarly, a group of researchers led by Brockman and Brennan (2017) found significant outcomes regarding the grain yield and dry biomass utilizing Moringa leaf extract on a wheat cultivar grown under plastic house conditions. However, they conducted their experiment using three different chemical solvents (ethyl-acetate, butanol, and hexane). At the tillering stage, the dry biomass was increased by 37% when the foliar spray of Moringa leaf extract was applied. Subsequently, the grain yield was also raised by 34% compared to no exogenous application (water sprayed).

Unexpected rainfall and heat stress conditions reduce crop yield production. Cytokinin content in the barley plant is greatly decreased by the terminal heat stress that subsequently reduces the quality of grain and yield components (Pospíšilová *et al.*, 2016). Therefore, it is necessary to apply foliar or other forms of application of Moringa plant extract to alter the adverse effect of water limitation and other stress conditions.

Because of all the various compositions of MLE found by other researchers, this treatment can be applied as a bio-stimulant to improve productivity and growth. Improving the growth and productivity characteristics due to the application of Moringa Plant Extract supports the hypothesis of this study that Moringa Plant Extract is an important plant growth enhancer.

**Table 4.4.1 Summary of all pairwise comparison (Duncan) for it is LS means at p value < 0.05.**

| Characters | Status | LS means |
|---|---|---|
| Plant height (PH) | WOM | 86.30 b |
| | WM | 90.95 a |
| Leaf area (LA) | WOM | 10.62 b |
| | WM | 12.31 a |
| Total chlorophyll content (TCC) | WOM | 11.12 b |
| | WM | 12.21 a |
| Awn length (AL) | WOM | 11.15 b |
| | WM | 11.62 a |
| Tiller number/plants (TNP) | WOM | 15.71 b |
| | WM | 17.94 a |
| Spike length (SL) | WOM | 5.88 b |
| | WM | 5.97 a |
| Spike numbrt/plant (SNP) | WOM | 12.29 b |
| | WM | 13.78 a |
| 1000 Grain weight (1000 KW) | WOM | 47.52 a |
| | WM | 45.82 b |
| Seed weight/spike (SWS) | WOM | 1.57 b |
| | WM | 1.68 a |
| Seed number/spike (SNS) | WOM | 33.50 b |
| | WM | 37.06 a |
| Spike weight (SW) | WOM | 1.89 b |
| | WM | 2.01 a |
| Total yield (TY) | WOM | 111.61 b |
| | WM | 138.83 a |
| Straw weight (ST) | WOM | 352.99 b |
| | WM | 412.92 a |

**Table 4.4.2 Mean pairwise comparison between 59 barley accessions for three growth traits after application of Moringa Plant Extract (MPE) based on Multiple Rang Duncan's test at *p* value < 0.05. Any values of means holding common letter are not significant.**

| Accessions | Plant height (PH)-cm | Leaf area (LA)-$cm^2$ | Total chlorophyll content (TCC)-Spad unit |
|---|---|---|---|
| **AC1** | 52.17 a-e | 14.42 cde | 15.17 c-h |
| **AC2** | 78.60 y | 10.78 p-u | 15.47 c-e |
| **AC3** | 74.90 z | 11.15 m-t | 10.58 l-t |
| **AC4** | 81.30 wx | 12.15 i-o | 10.40 l-t |
| **AC5** | 65.90 abc | 12.18 i-n | 10.75 k-t |
| **AC6** | 62.93 ad | 12.30 i-m | 13.48 d-l |
| **AC7** | 93.67 jkl | 15.07 cd | 13.10 d-m |
| **AC8** | 67.83 ab | 12.89 f-k | 12.45 f-o |
| **AC9** | 88.40 rst | 10.90 n-t | 12.02 g-p |
| **AC10** | 80.27 xy | 11.77 k-q | 11.22 j-r |
| **AC11** | 89.13 p-s | 16.18 ab | 11.90 h-q |
| **AC12** | 89.60 p-s | 10.14 s-w | 11.33 i-q |
| **AC13** | 92.63 lmn | 10.86 o-t | 11.13 k-r |
| **AC14** | 92.90 klm | 11.26 m-s | 12.93 e-m |
| **AC15** | 98.43 ef | 15.36 bc | 11.15 k-r |
| **AC16** | 100.10 de | 14.91 cd | 12.87 e-m |
| **AC17** | 95.77 g-j | 11.28 m-s | 15.37 c-g |
| **AC18** | 85.27 v | 10.93 n-t | 7.47 tuv |
| **AC19** | 74.40 z-aa | 14.09 def | 10.27 l-u |
| **AC20** | 75.00 z | 10.50 q-v | 7.62 s-v |
| **AC21** | 102.60 bc | 12.64 h-l | 13.12 d-m |
| **AC22** | 86.60 tuv | 11.86 j-p | 12.68 f-n |
| **AC23** | 82.37 w | 8.22 xy | 7.60 s-v |
| **AC24** | 95.43 g-j | 7.80 yz | 7.02 uv |
| **AC25** | 100.00 de | 11.90 j-p | 11.43 i-q |
| **AC26** | 86.57 tuv | 9.92 t-w | 9.77 m-v |
| **AC27** | 89.20 p-s | 10.78 p-u | 6.67 v |
| **AC28** | 97.30 fg | 12.01 j-p | 20.27 a |
| **AC29** | 101.60 cd | 11.73 k-q | 10.53 l-t |
| **AC30** | 95.17 hij | 13.31 e-i | 18.78 ab |
| **AC31** | 89.13 p-s | 11.55 l-r | 14.67 d-i |
| **AC32** | 94.13 i-l | 10.53 q-v | 12.12 g-p |
| **AC33** | 94.13 jkl | 9.93 t-w | 8.62 q-v |
| **AC34** | 96.23 ghi | 10.02 s-w | 8.88 p-v |
| **AC35** | 90.70 n-q | 10.41 r-v | 17.80 abc |
| **AC36** | 87.57 stu | 6.91 z | 9.25 o-v |
| **AC37** | 89.50 p-s | 8.38 xy | 8.63 q-v |
| **AC38** | 93.03 klm | 11.04 m-t | 7.40 tuv |
| **AC39** | 79.97 xy | 12.30 i-m | 16.23 bcd |
| **AC40** | 89.23 p-s | 9.58 uvw | 12.05 g-p |
| **AC41** | 72.73 a-a | 13.61 e-h | 12.03 g-p |
| **AC42** | 74.40 z-aa | 12.84 g-k | 8.97 p-v |
| **AC43** | 89.57 p-s | 9.57 uvw | 19.80 a |
| **AC44** | 91.13 m-p | 8.34 xy | 10.67 l-t |
| **AC45** | 94.93 ijk | 9.29 vwx | 9.37 n-v |
| **AC46** | 82.83 w | 11.16 m-t | 10.90 k-s |
| **AC47** | 88.50 rst | 10.25 s-w | 7.85 r-v |
| **AC48** | 92.50 l-o | 12.13 i-o | 8.62 q-v |
| **AC49** | 92.17-o | 11.14 m-t | 8.63 q-v |
| **AC50** | 99.43 e | 9.07 wx | 12.13 g-p |
| **AC51** | 97.20 fgh | 16.61 a | 14.12 d-k |
| **AC52** | 92.23 l-o | 11.17 m-t | 12.07 g-p |
| **AC53** | 86.03 uv | 10.54 q-v | 8.85 p-v |
| **AC54** | 88.90 qrs | 7.59 yz | 8.60 q-v |
| **AC55** | 104.03 b | 9.39 vwx | 16.10 b-e |
| **AC56** | 109.30 a | 13.33 e-i | 14.53 d-j |
| **AC57** | 110.03 a | 13.96 d-g | 12.42 f-o |
| **AC58** | 90.47 o-r | 13.53 e-h | 9.78 m-v |
| **AC59** | 92.87 klm | 13.09 f-j | 12.57 f-o |

**Table 4.4.3 Mean pairwise comparison between 59 barley accessions for some yield contributed traits after application of Moringa Plant Extract (MPE) based on Multiple Rang Duncan's test at p value < 0.05. Any values of means holding common letter are not significant.**

| Accession | Awn length (AL)-cm | Tiller number /plant (TNP) | Spike length (SL)-cm | Spike Number/plant (SNP) | Seed weight/spike (SWS)-g | 1000 grain weight (1000 KW)-g | Seed number/spike (SNS) | Spike weight (SW)-g | Total yield (TY)-g | Straw weight (ST)-g |
|---|---|---|---|---|---|---|---|---|---|---|
| **AC1** | 10.86 n-u | 7.67 x-ab | 4.42 v-y | 5.67 r-u | 1.94 jk | 50.60 f-i | 38.28 r | 2.60 ef | 45.73 u-z | 160.26 a-e |
| **AC2** | 12.62 d-g | 6.83 z-ab | 6.85 d-i | 6.00 r-u | 2.30 e | 57.30 b | 40.83 p | 2.75 de | 72.64 q-y | 232.70ab-ac |
| **AC3** | 12.43 d-h | 6.17 aa-ab | 3.79 y | 4.00 u | 1.84 l | 43.17 t-w | 42.72 n | 2.05 k-n | 15.83 z | 72.52 a-g |
| **AC4** | 12.19 e-j | 9.50 w-ab | 6.22 h-n | 6.67 q-u | 2.30 e | 44.30 q-u | 51.89 e | 2.84 d | 55.97 s-z | 191.55 a-d |
| **AC5** | 13.14 bcd | 7.00 z-ab | 4.69 t-x | 5.17 stu | 1.61 rs | 45.13 o-t | 36.06 s | 2.17 jk | 27.30 yz | 136.53 a-f |
| **AC6** | 13.13 bcd | 7.50 y-ab | 4.89 s-w | 5.33 stu | 1.64 qr | 41.44 r-y | 40.06 q | 1.97 l-o | 37.68 v-z | 127.74 a-f |
| **AC7** | 10.19 r-x | 5.67 ab | 5.69 m-r | 4.33 tu | 1.53 t | 36.59 ab-ac | 41.89 o | 1.93 mno | 30.36 w-z | 127.80 a-f |
| **AC8** | 12.37 d-i | 11.33 r-aa | 4.78 t-x | 9.17 n-s | 1.67 pq | 43.40 s-v | 38.89 r | 1.93 mno | 86.25 n-v | 230.76 abc |
| **AC9** | 12.10 e-k | 10.00 u-ab | 4.26 wxy | 7.83 p-u | 2.07 i | 42.84 t-x | 48.33 h | 2.42 ghi | 118.05 j-r | 292.31 yz |
| **AC10** | 11.02 m-r | 10.67 t-ab | 5.85 l-r | 7.83 p-u | 2.15 g | 49.18 g-m | 43.44lm | 2.57 fg | 79.82 o-x | 246.47 ab |
| **AC11** | 12.67 d-g | 9.83 v-ab | 4.90 s-w | 7.33 p-u | 1.83 lm | 47.83 j-n | 38.44 r | 2.22 jk | 97.26 l-t | 274.39 a-a |
| **AC12** | 11.11 l-q | 11.00 s-ab | 5.13 q-v | 8.33 o-t | 1.78 mn | 43.96 r-v | 41.06 p | 2.21 jk | 29.66 xyz | 126.24 a-f |
| **AC13** | 12.28 d-j | 12.17 o-z | 4.72 t-x | 10.50 l-q | 2.63 c | 48.12 i-n | 54.83 c | 3.27 b | 198.89 def | 452.80 lmn |
| **AC14** | 12.43 d-h | 11.33 r-aa | 4.91 s-w | 8.33 o-t | 2.21 f | 48.95 h-m | 46.06 j | 2.50 fgh | 133.30 i-n | 410.97 st |
| **AC15** | 14.10 a | 14.33 l-w | 5.29 p-u | 11.67 h-p | 1.96 jk | 44.61 p-u | 44.00 kl | 2.40 ghi | 129.55 i-o | 367.80 vw |
| **AC16** | 13.87 ab | 12.67 n-y | 5.70 m-r | 11.33 i-p | 2.11 hi | 44.04 r-u | 48.00 hi | 2.45 f-i | 203.69 def | 484.32 j |
| **AC17** | 11.70 h-n | 17.67 h-o | 4.65 t-x | 15.33 c-i | 2.04 j | 46.01 n-r | 44.28 k | 2.49 f-i | 209.53 cde | 464.93 l |
| **AC18** | 9.54 x-ab | 16.33 i-s | 4.18 wxy | 10.50 l-q | 1.85 l | 38.62 z-ab | 47.67 i | 2.10 klm | 51.22 t-z | 227.47 abc |
| **AC19** | 10.99 m-s | 13.00 m-x | 4.58 u-x | 9.83 m-r | 1.98 jk | 37.06a-ac | 53.44 d | 2.20 jk | 94.34 l-u | 300.77 xy |
| **AC20** | 11.81 g-m | 16.67 h-r | 6.63 f-l | 12.50 g-o | 1.07 yz | 45.92 n-s | 23.39 xy | 1.28 uvw | 104.01 l-s | 310.36 x |
| **AC21** | 14.49 a | 11.67 q-z | 7.51 bcd | 10.83 j-q | 2.71 b | 47.67 k-o | 56.83 b | 3.23 b | 202.32 def | 479.92 jk |
| **AC22** | 12.20 e-j | 11.67 q-z | 4.09 xy | 8.83 o-s | 1.94 jk | 47.02 l-p | 41.22 p | 2.19 jk | 140.83g-m | 376.85 uv |
| **AC23** | 9.33 y-ab | 18.17 g-n | 7.28 c-f | 13.17 f-n | 1.03 z | 51.64 efg | 19.98 a-c | 1.34 u | 86.46 n-v | 360.14 w |
| **AC24** | 9.92 w-aa | 16.17 i-s | 6.66 f-k | 12.67 g-o | 1.17 x | 48.96 h-m | 23.83 x | 1.34 u | 104.27 l-s | 313.80 x |
| **AC25** | 13.94 a | 20.83 d-j | 7.35 b-f | 16.00 b-h | 1.44 u | 63.66 de | 22.56 [illegible] | 1.80 op | 229.29 cd | 726.23 b |
| **AC26** | 10.93 n-t | 19.83 d-l | 6.44 g-m | 13.83 d-m | 1.04 yz | 44.04 r-u | 23.67 xy | 1.15 vwx | 129.26 i-o | 377.34 uv |
| **AC27** | 11.66 h-n | 24.83 bcd | 6.34 g-n | 17.67 b-f | 1.21 wx | 52.37 def | 23.06 yz | 1.39 tu | 158.37 f-k | 409.17 st |
| **AC28** | 12.11 e-k | 21.67 c-i | 6.22 h-h | 19.17 abc | 1.43 u | 56.94 b | 25.17 w | 1.72 pq | 188.16 d-g | 537.90 fg |
| **AC29** | 12.79 def | 20.17 d-k | 7.36 b-f | 15.17 c-j | 1.56 st | 56.59 b | 27.50 u | 1.89 no | 184.63 d-g | 539.63 f |
| **AC30** | 13.72 abc | 17.17 h-q | 8.50 a | 15.17 c-j | 1.63 qr | 51.13 e-h | 31.83 t | 2.00 lmn | 255.06 bc | 630.47 d |
| **AC31** | 11.41 j-o | 16.00 j-t | 5.87 k-q | 13.83 d-m | 1.32 v | 56.63 b | 23.44 xy | 1.69 pqr | 189.66 d-g | 509.07 h |
| **AC32** | 10.45 p-w | 17.00 h-q | 8.03 ab | 15.00 c-k | 1.25 w | 49.89 g-k | 24.94 w | 1.61 qrs | 167.30 e-j | 440.33 nop |
| **AC33** | 9.96 v-aa | 19.50 d-l | 6.28 g-n | 15.17 c-j | 0.89 ab | 42.36 u-x | 21.06 ab | 1.05 xy | 117.74 j-r | 458.23 lm |
| **AC34** | 10.07 t-z | 26.33 abc | 7.96 abc | 15.67 c-i | 1.09 y | 46.01 n-r | 23.78 x | 1.26 uvw | 122.67 j-q | 467.58 kl |
| **AC35** | 10.46 p-w | 17.50 h-p | 6.03 j-p | 14.33 d-l | 2.14 g | 48.22 i-n | 44.44 k | 2.38 hi | 134.22 i-n | 499.93 hi |
| **AC36** | 10.40 q-x | 19.67 d-l | 5.79 m-r | 17.50 b-f | 1.44 u | 51.29 e-h | 28.06 u | 1.51 st | 71.85 q-y | 241.77ab-ac |
| **AC37** | 10.71 o-w | 24.00 b-f | 6.35 g-n | 20.17 ab | 1.22 wx | 46.64 m-q | 26.22 v | 1.56 q-t | 128.35 i-p | 369.46 vw |
| **AC38** | 10.01 u-z | 26.17 abc | 6.39 g-n | 22.33 a | 1.10 y | 49.41 g-l | 22.11a-a | 1.32 uv | 119.29 j-r | 415.43 rst |
| **AC39** | 11.95 f-l | 20.00 d-l | 6.13 i-o | 18.00 b-e | 1.22 wx | 55.36 bc | 22.00a-a | 1.52 rst | 180.20 d-h | 485.06 j |
| **AC40** | 12.43 d-h | 15.17 k-v | 6.41 g-n | 13.67 e-m | 1.33 v | 54.07 cd | 24.61 w | 1.68 p-s | 80.56 o-w | 301.92 xy |
| **AC41** | 11.31 k-p | 19.83 d-l | 7.42 b-e | 16.33 b-g | 1.05 yz | 44.41 q-u | 23.61 xy | 1.31 uv | 114.09 k-r | 423.04 qrs |
| **AC42** | 10.85 n-v | 19.33 e-l | 7.03 d-g | 15.83 b-h | 1.32 v | 49.56 g-l | 26.56 v | 1.60 qrs | 101.83 l-t | 314.77 x |

**Continue**

| | | | | | | | | | | |
|---|---|---|---|---|---|---|---|---|---|---|
| **AC43** | 10.17 r-y | 17.00 h-q | 4.20 wxy | 15.67 c-i | 2.18 fg | 43.64 r-z | 49.83 g | 2.45 f-i | 212.27 cde | 525.90 fg |
| **AC44** | 9.19 aa-ab | 28.00 ab | 6.14 h-o | 18.33 a-d | 0.85ab-ac | 40.81 w-z | 21.00 ab | 0.98 xy | 83.43 n-v | 437.39 opq |
| **AC45** | 9.29 z-ab | 18.33 g-m | 7.04 d-g | 13.83 d-m | 1.19 wx | 47.59 k-o | 25.22 w | 1.41 tu | 144.53 g-l | 427.37 pqr |
| **AC46** | 10.73 o-w | 28.17 ab | 6.91 d-h | 19.17 abc | 0.96 a-a | 42.52 u-x | 22.50h- z | 1.13 wxy | 92.38 m-u | 449.60 mno |
| **AC47** | 9.29 z-ab | 28.50 ab | 5.09 r-v | 20.17 ab | 0.66 a-d | 36.46ab-ac | 18.22 ad | 0.77 z | 69.57 r-y | 368.86 vw |
| **AC48** | 10.75 o-w | 30.83 a | 5.62 n-s | 18.33 a-d | 0.70 a-d | 35.01 ac | 19.94 ac | 0.81 z | 77.61 p-x | 402.90 t |
| **AC49** | 10.73 o-w | 15.00 k-v | 4.19 wxy | 10.67 k-q | 2.12 gh | 49.07 h-m | 43.28 mn | 2.40 ghi | 155.21 f-k | 490.05 ij |
| **AC50** | 10.46 p-w | 19.17 e-l | 5.19 q-u | 15.00 c-k | 1.70 op | 40.65 x-z | 41.94 o | 1.95 mno | 213.84 cde | 572.49 e |
| **AC51** | 12.59 d | 12.00 p-z | 6.04 j-p | 8.50 o-t | 1.91 k | 39.21 y-aa | 48.61 h | 2.13 kl | 68.88 r-y | 291.12 yz |
| **AC52** | 9.46 y-ab | 24.33 b-d | 6.73 e-j | 16.50 b-g | 0.91 aa-b | 46.06 n-r | 19.78 ac | 1.06 xy | 70.26 r-y | 367.29 vw |
| **AC53** | 9.92 w-aa | 23.33 b-g | 5.77 m-r | 16.67 b-g | 0.82 a-c | 37.90aa-ab | 21.94 aa | 0.97 y | 61.91 s-z | 371.47 vw |
| **AC54** | 10.12 s-z | 18.83 f-l | 7.86 abc | 14.83 c-l | 1.31 v | 53.31 cde | 24.56 w | 1.56 q-t | 92.70 m-u | 284.98 z-aa |
| **AC55** | 11.50 i-o | 17.00 h-q | 6.08 i-o | 14.33 d-l | 2.45 d | 47.30 k-o | 51.39 e | 3.03 c | 337.47 a | 741.59 a |
| **AC56** | 10.99 m-s | 22.00 c-h | 5.63 n-s | 17.00 b-g | 1.76 no | 43.05 t-x | 40.83 p | 2.21 jk | 84.16 n-v | 378.81 u |
| **AC57** | 8.95 ab | 15.33 j-u | 6.38 g-n | 12.50 g-o | 1.97 jk | 38.67 z-ab | 50.67 f | 2.32 ij | 128.16 i-p | 387.33 u |
| **AC58** | 12.87 de | 16.67 h-r | 5.77 m-r | 14.50 d-l | 2.31 e | 50.32 f-j | 46.11 j | 3.07 c | 176.44 e-i | 523.77 g |
| **AC59** | 12.98 cde | 18.17 g-n | 5.36 o-t | 15.17 c-j | 3.22 a | 49.04 h-m | 65.72 a | 3.74 a | 291.98 b | 656.98 c |

### 4.4.3 Hierarchical clustering arrangement, Principal component analysis among tested barley accessions, and correlation among studied traits

To gain a better understanding of relations between studied barley accessions and studied morphological traits, a heat map of pairwise correlations (two-side dendrogram) based on mean values obtained from all measured traits in the case of present and absence of Moringa plant extract was constructed with the aid of jump software (Fig. 4.4.1). Even though six groups were estimated in both cases, the barley accessions that were studied behaved and grouped in different ways.

As shown in (Fig. 4.4.1-A), under control conditions, the majority of barley accessions in association with studied traits clustered together in (group 5) which comprised seventeen barley accessions, namely (AC20, AC40, AC24, AC36, AC26, AC37, AC33, AC45, AC23, AC54, AC32, AC41, AC42, AC25, AC29, AC27, AC39), indicating that these barley accessions had the same linkage for most of the studied traits. While (Group 3) is considered the smallest group among constructed clades, as only three genotypes were clustered in this particular group, including (AC51, AC56, AC57), demonstrating that these barley accessions share similar associations with investigated traits and they are different from the rest of the studied barley accessions. The rest of the barley accessions were grouped into four other distinct clusters. On the other hand, a different arrangement was observed in the case of the present foliar application of Moringa, showing significant responses to this treatment by the studied barley accessions and its impacts on selected morphological parameters. The largest group included twenty-one barley accessions in the case of the present Moringa treatment shown in (Fig. 4.4.1–B), namely (AC46, AC42, AC41, AC20, AC26, AC37, AC38, AC40, AC27, AC34, AC45, AC24, AC33, AC52, AC44, AC54, AC23, AC53, AC48, AC47). This particular group (Group 2) responded similarly to the examined traits. While the two distinct barley accessions (AC59 and AC55) gathered together and demonstrated a positive relationship with studied traits, they formed a distinct cluster distinct from all other accessions, while the remaining barley accessions fell into other distinct clusters.

Principal component analysis (PCA) was performed on the experimental dataset for multifactorial comparison to display the correlations between the various plant parameters. For this, PCA was performed for all 13 measured morphological traits under both normal and treated conditions. The multifactorial analysis was conducted with the help of XLSTAT software.

PCA revealed that 59 different barley accessions were clustered into four clades in both cases of control (A) and treated conditions (B) (Fig. 4.4.2). The first two factorial axes (F1, F2) collectively represent 62.93% of the variance of the data under normal conditions. whereas in the present foliar application of Moringa, it represented 56.97% of the variance of the data. Under normal conditions, all measured traits except for LA were grouped into 2 main clusters (1 and 2). Cluster 1 included SNP, TNP, SL, PH, ST, and 1000KW; cluster 2 was comprised of TY, TCC, AL, SWS, SW, and SNS. A negative correlation was observed for LA trait in cluster 4 with the traits in cluster 1 (Fig. 4.4.2-A).

Clade 1 comprised eleven barley accessions including (AC25, AC27, AC28, AC29, AC31, AC32, AC34, AC38, AC41, AC50, AC54), which is mostly grown in the south of Iraq, while the opposite clade in which their performance differed from clade 1 included fourteen barley accessions containing (AC2, AC10, AC12, AC8, AC7, AC18, AC5, AC6, AC1, AC3, AC4, AC11, AC9, AC19). In addition to these two clades, sixteen studied barley accessions were distributed over clade 2, which consists of (AC13, AC14, AC16, AC17, AC15, AC35, AC43, AC30, AC21, AC22, AC57, AC58, AC59, AC51, AC55). The studied traits contributed more positively to this clade, suggesting that this component reflected the yield potential of each barley accession in this particular clade. Further, the rest of the eighteen barley accessions, as it is clear from (Fig. 4.4.2-A), are grouped in clade 3. This determines the genetic differences among those groups which can be selected for crossing, especially in the case of AC59 with AC47 in clades 2 and 3 respectively, as well as AC25 with AC1 in clades 1 and 4 respectively for the future breeding program.

Regarding the analysis of PCA in the present, the foliar application of Moringa, different patterns of distribution of the barley accessions as well as studied traits can be observed in comparison with the untreated condition. As it is shown in (Fig. 4.4.2-B), nearly half of the studied barley accessions are separated and distributed over clades 1 and 2 with an equal number of thirteen barley accessions for each. The clade 1 comprised (AC37, AC38, AC46, AC45, AC27, AC32, AC56, AC31, AC29, AC39, AC28, AC25, AC30), while clade 2 involved (AC50, AC35, AC43, AC17, AC49, AC57, AC16, AC15, AC58, AC21, AC55, AC59, AC13). In addition, clade 3 in which, contrary to clade 1, included seventeen barley accessions, namely; AC11, AC2, AC10, AC14, AC22, AC9, AC8, AC18, AC51, AC12, AC19, AC7, AC5, AC6, AC1, AC3, AC4 and finally, the rest of the sixteen barley accessions stayed together in clade 3. The present study discovered that studied traits had a strong correlation with those barley accessions distributed over clades 1 and 2, suggesting the need for more emphasis on these barley accessions to increase the final productivities in the presence of Moringa plant extract. The attributed variations among these accessions might partly reflect their different genetic backgrounds as well as their different responses to the utilized application.

The correlation coefficients are the measure of the degree of similarity and differences between two characters or variables, and the nature of association can be assessed among studied parameters. From this point, Pearson correlations (r) of the studied traits under control and Moringa Foliar application conditions are determined from their mean values and shown in (Fig. 4.4.3).

Under control conditions, in general, thirty-two significant positive and fourteen negative correlations were detected. The positive correlations among the thirteen studied parameters for r value ranged between 0.97 and 0.26, five for each PH * LA. In addition, for TCC, SL, and AL, an equal number of four positive correlations were documented, followed by SW, which showed three positive correlations with studied parameters. Further, two other positive associations were observed for TNP * SNP individually. Finally, the rest of the three positive correlations were verified for each SNS, SWS, and TY. For more details, as indicated in (Fig. 4.3-A), A strong positive significant correlation between SW * SWS traits was observed (r = 0.97***, $p < 0.0001$) followed by TNP * SNP (r = 0.94***, $p < 0.0001$), SNS * SWS (r = 0.93***, $p < 0.0001$) while weak positive associations were noted between AL * TY (r = 0.26*, $p = 0.05$) followed by the same value (r = 0.30*, $p = 0.02$) between (SW * TW) as well as (LA * TCC). On the other hand, three negative correlations with studied traits were recorded for each SL, TNP, and SNP, followed by AL and TCC having two negative correlations separately. Lastly, the only negative relationships were observed between SNS * 1000KW under control conditions.

Regarding the analysis for its correlations between studied parameters under foliar application of Moringa plant part extract, overall twenty-three significant positive and ten negative correlations were obtained. The positive correlations among studied traits for r value ranged between 0.98 and 0.27, for it is the association between (SW*SWS) and (SL*SNP) respectively. Both traits (AL*SW) had three positive associations with the rest of the parameters, followed by seven studied traits, namely; PH, LA, TCC, SL, TNP, SNP, and 1000 kw, which had two positive correlations with the other investigated parameters individually, while the rest of the three studied traits (SNS, SWS, and TY) had only one positive correlation with (SWS, TY, and HI) respectively. As well documented in (Fig. 4.3-B), a very robust positive significant relation between SW and SWS yield-related characters was detected (r = 0.98***, $p < 0.0001$) and then the association among SNS and SWS came after with the value of (r = 0.93***, $p < 0.0001$) followed by TNP*SNP (r = 0.91***, $p < 0.0001$), whereas weak positive linkage was revealed with an almost similar pattern between (SL*SNP, r = 0.28*, $p = 0.03$) and (SL*HI, r = 0.28*, $p = 0.03$). On the other hand, for each trait studied (SL, TNP, and SNP), three negative correlations were found between each trait and other parameters. This was followed by a negative correlation between AL*TNP.

### 4.4.4 Percentage responses of barley accessions for studied growth and yield contributed traits under the foliar application of MPE

Different patterns of responses by the barley accessions under the foliar application of moringa plant parts were detected. As shown in Fig. 4.4.4, our results confirmed that the application used in our investigation increased all growth and yield related studied parameters, especially the yield traits, with the only exception of the trait (1000 KW%). Regarding the analysis for displaying the percentage response by all barley accessions in the case of conducting such a foliar application, the overall view of responses for plant height trait, as shown in (Fig. 4.4.4), indicated the huge impact of such a treatment on the studied barley accession, in which 68% of the accession were increased in their height. Similarly, among tested traits, the highest boost was recorded by leaf area, with a value of 83% (Fig. 4.4.4). Regarding the responses by barley accessions to both total chlorophyll content (TCC) and awn length (AL), almost similar increases with a value of (64 and 63)% were detected, respectively (Fig. 4.4.4). The highest response among all studied traits was observed for leaf area (LA) where 83% of barley accessions responded positively to the application of moringa. In addition, more than half of the barley accessions responded positively to Moringa application for the spike length (SL) parameters (Fig. 4.4.4). In respect to the impact of moringa on the trait of spike weight (SW), 69% of barley accessions increased their weight of spike. A progressive response was observed for the other three traits, namely; seed no/spike (SNS), tiller number per plant (TNP), total yield (TY), and straw weight (ST), in which three quarters of the tested barley accessions responded positively to foliar application of moringa. In addition to the previous parameters, the other two traits, spike number per plant (SNP) and seed weight/spike (SWS), were positively responded by 69% barley accessions. For 1000KW traits, 31% of barley accessions responded positively to foliar application.

# Conclusions and Recommendations

## ➢ Conclusions

- The three different marker systems (SCoT, ISSR, and CDDP) provided a comprehensive pattern of the genetic diversity among 59 collected barley accessions by separating all studied accessions into two distinct groups. This investigation is a contribution to the representation of the core collection of barley accessions in Iraq. In respect to our finding, the ISSR markers further separated the studied materrails.
- Natural stress tolerance is a complex phenomenon involving various physiological and biochemical processes. Use of germination and seedling traits is a cost-effective solution for the rapid identification of tolerant or sensitive barley accessions for a limited period. Germination and seedling growth assessment were conducted under induced osmotic stress mediated by PEG treatments as a reflective basis for drought tolerance of selected accessions. All PEG concentrations significantly minimized germination and seedling early growth. The results of this study showed that the reactions of barley seeds to PEG-induced drought stress initially revealed a reduction in WU and an increase in the content of osmolytes, non-enzymatic, and enzymatic compounds. Based on the results, the AC37 seems to be the most tolerant, while AC47 was the most susceptible to PEG stress. AC37 has expressed the smallest decrease in seedling growth and physiological traits, although there is a significant increase in biochemical traits except for TFC.
- Similar to the seedling experiment, AC36 and AC37 displayed considerable phenotypic variation in drought tolerance under different growth conditions at greengouse, presenting a potential value of these barley accessions for identifying useful parents for drought tolerance programme. Therefore, it is more encouraging to identify the stable and true drought-tolerant barley accessions at early growth stages by creating conditions for drought stress using different osmotic materials, as not many differences were found between PEG simulated drought and water holding conditions in the plastic house by studying barley accessions.
- The mechanism that regulates cadmium uptake, transportation, and accumulation is still not fully understood. Many further studies are required at the molecular level to better understand the genetic functions that reduce the adverse effects of cadmium and to ensure better crop production under cadmium stress conditions. However, it can be concluded that, from our outcomes, a significant genetic distance among 59 barley accessions in the presence of different cadmium exposures (Cd125, Cd-250, and Cd500) µM were exists. The responses by barley accessions to studied traits varied based on the amount of cadmium exposure. As indicated in our results, the intermediate exposure of cadmium (Cd-250) μM stimulated the activities of most biochemical traits tested by most accessions. Favorable behavior under all cadmium exposure was found by barley accessions AC29, AC38, and AC37, respectively. In the short term, the use of physio-chemical traits conducted in our investigation can be considered useful tools for rapid screening in the early growth stage to indicate resistant or sensitive barley accessions in the presence of cadmium heavy metal stress. Based on early screening responses by all barley accessions, six barley accessions under plastic house conditions, AC29, AC37, AC38, AC47, AC48, and AC52 were subjected to cadmium stress with a concentration of 500 μM at tillering (S1) and flowering (S2) and double cadmium stress at both (S1+S2). The selected cadmium-tolerant barley accessions showed much less reduction in total yield and related morphological traits than sensitive accessions under the availability of cadmium at the mentioned stages, which confirmed the reliability of the screening barley accessions at the early stage for indicating cadmium tolerance.

- The modified Oakleaf (*Quercus aegilops*) and Kangar (*Gundelia tournefortii*) residues with NaOH were morphologically and biochemically improved by tested barley accessions in response to the cadmium stress condition at the early growth stage.
- To modulate the productivity and growth of barley crop plants as reflected in our investigation into improving all studied traits, foliar application of Moringa leaf aqueous extract with a concentration of 3% at the critical growth stage can be applied. Furthermore, this type of application may be used as an alternative biostimulant to chemical plant growth hormones, especially when the target is establishing an organic farming system.

## ➢ Recommendations

- Aside from the availability of a wide range of barley accessions in this experiment, it is conceivable to incorporate wild type barley, which is generally accepted for its capacity to survive diverse stress conditions, to see if it is genetically and biochemically connected to resistant accessions.
- Many specific genes for barley have been identified for their ability to give drought (*HvSAP8*, *HvSAP16* and *HvbZIP21* ) and heavy-metal (*HvPCR2)* stress tolerance . As a result, the absence or presence of these genes in any germplasm may be examined for drought resistance through investigating changes in the expression of a certain gene or group of genes by measuring the amount of the gene-specific transcript. Therefore, the insertion of these genes into the susceptible barley accessions may be more advantageous; this may be accomplished by gene editing and RNA editing (specifically CRISPR). CRISPR/Cas9 might aid in the facilitation and manipulation of new genes that may contribute to drought and heavy metal tolerance.

- Based on the outcomes available in this investigation, the resistant and susceptible barley accessions can be crossed to generate moderate barley accessions for both heavy-metal and drought conditions.
- Collaborate with global relevant organizations to provide them with the drought- and heavy-metal-resistant materials we have discovered so that they can be distributed in areas affected by both of these issues.

# References

Abd El-Mageed, T. A., Semida, W. M. and Rady, M. M. (2017) 'Moringa leaf extract as biostimulant improves water use efficiency, physio-biochemical attributes of squash plants under deficit irrigation', Agricultural Water Management, 193, pp. 46-54.

Abdul-Razzak Tahir, N., Ahmad, N., Mustafa, K. and Kareem, D. (2021) 'Diversity maintenance of some barley (Hordeum spp) genetic resources using SSR-based marker', JAPS: Journal of Animal & Plant Sciences, 31(1).

Abebe, T., Wise, R. P. and Skadsen, R. W. (2009) 'Comparative transcriptional profiling established the awn as the major photosynthetic organ of the barley spike while the lemma and the palea primarily protect the seed', The Plant Genome, 2(3).

Abouseadaa, H. H., Atia, M. A., Younis, I. Y., Issa, M. Y., Ashour, H. A., Saleh, I., Osman, G. H., Arif, I. A. and Mohsen, E. (2020) 'Gene-targeted molecular phylogeny, phytochemical profiling, and antioxidant activity of nine species belonging to family Cactaceae', Saudi Journal of Biological Sciences, 27(6), pp. 1649-1658.

Afzal, I., Akram, M., Rehman, H., Rashid, S. and Basra, S. (2020) 'Moringa leaf and sorghum water extracts and salicylic acid to alleviate impacts of heat stress in wheat', South African Journal of Botany, 129, pp. 169-174

Ahmad, J., Ali, A. A., Baig, M. A., Iqbal, M., Haq, I. and Qureshi, M. I. (2019) 'Role of phytochelatins in cadmium stress tolerance in plants', Cadmium toxicity and tolerance in plants, pp. 185-212.

Ahmad, T. A., Ahmad, F. K., Rasul, K. S., Aziz, R. R., Omer, D. A., Tahir, N. A. R., and Mohammed, A. A. (2020) 'Effect of some Plant Extracts and Media Culture on Seed Germination and Seedling Growth of Moringa oleifera'. Journal of Plant Production, 11(7), pp. 669-674.

Ahmed, D. A., Tahir, N. A.-r., Salih, S. H. and Talebi, R. (2021) 'Genome diversity and population structure analysis of Iranian landrace and improved barley (Hordeum vulgare L.) genotypes using arbitrary functional gene-based molecular markers', Genetic Resources and Crop Evolution, 68(3), pp. 1045-1060.

Ahmed, I. M., Nadira, U. A., Bibi, N., Cao, F., He, X., Zhang, G. and Wu, F. (2015) 'Secondary metabolism and antioxidants are involved in the tolerance to drought and salinity, separately and combined, in Tibetan wild barley', Environmental and Experimental Botany, 111, pp. 1-12.

Ahmed, I. M., Nadira, U. A., Qiu, C.-W., Cao, F., Chen, Z.-H., Vincze, E. and Wu, F. (2020) 'The barley S-adenosylmethionine synthetase 3 gene HvSAMS3 positively regulates the tolerance to combined drought and salinity stress in Tibetan wild barley', Cells, 9(6), pp. 1530.

Aimar, D., Calafat, M., Andrade, A., Carassay, L., Abdala, G. and Molas, M. (2011) 'Drought tolerance and stress hormones: From model organisms to forage crops', Plants and environment, 10(6), pp. 137-164.

Ajithan, C., Vasudevan, V., Sathish, D., Sathish, S., Krishnan, V. and Manickavasagam, M. (2019) 'The influential role of polyamines on the in vitro regeneration of pea (Pisum sativum L.) and genetic fidelity assessment by SCoT and RAPD markers', Plant Cell, Tissue and Organ Culture (PCTOC), 139(3), pp. 547-561.

Al-Ajlouni, Z. I., Al-Abdallat, A. M., Al-Ghzawi, A. L. A., Ayad, J. Y., Abu Elenein, J. M., Al-Quraan, N. A. and Baenziger, P. S. (2016) 'Impact of pre-anthesis water deficit on yield and yield components in barley (Hordeum vulgare L.) plants grown under controlled conditions', Agronomy, 6(2), pp. 33.

Alemu, G., Desalegn, T., Debele, T., Adela, A., Taye, G. and Yirga, C. (2017) 'Effect of lime and phosphorus fertilizer on acid soil properties and barley grain yield at Bedi in Western Ethiopia', African journal of agricultural research, 12(40), pp. 3005-3012

Alexander, R. D., Wendelboe-Nelson, C. and Morris, P. C. (2019) 'The barley transcription factor HvMYB1 is a positive regulator of drought tolerance', Plant Physiology and Biochemistry, 142, pp. 246-253.

Alghabari, F. (2020) 'Foliar Applicationof Diannela ensata, Ambrosia dumosa and Moringa oliferia Improved Barley Growth and Yield Traits Under Drought Stress', Applied Ecology and Environmental Research, 18(5), pp. 6041-6052.

Alhasnawi, A. N. (2019) 'Role of proline in plant stress tolerance: A mini review', Research on Crops, 20(1), pp. 223-229.

Ali, E., Hassan, F. and Elgimabi, M. (2018) 'Improving the growth, yield and volatile oil content of Pelargonium graveolens L. Herit by foliar application with moringa leaf extract through motivating physiological and biochemical parameters', South African Journal of Botany, 119, pp. 383-389.

Allel, D., Ben-Amar, A. and Abdelly, C. (2018) 'Leaf photosynthesis, chlorophyll fluorescence and ion content of barley (Hordeum vulgare) in response to salinity', Journal of Plant Nutrition, 41(4), pp. 497-508.

Allel, D., Ben-Amar, A., Lamine, M. and Abdelly, C. (2017) 'Relationships and genetic structure of North African barley (Hordeum vulgare L.) germplasm revealed by morphological and molecular markers: Biogeographical considerations', South African Journal of Botany, 112, pp. 1-10.

Almerekova, S., Genievskaya, Y., Abugalieva, S., Sato, K. and Turuspekov, Y. (2021) 'Population structure and genetic diversity of two-rowed barley accessions from Kazakhstan based on snp genotyping data', Plants, 10(10), pp. 2025.

Amini, R. (2013) 'Drought stress tolerance of barley (Hordeum vulgare L.) affected by priming with PEG', International Journal of Farming and Allied Sciences, 2(20), pp. 803-808.

Amom, T. and Nongdam, P. (2017) 'The use of molecular marker methods in plants: a review', International Journal of Current Research and Review, 9(17), pp. 1-7.

Anjum, S. A., Xie, X.-y., Wang, L.-c., Saleem, M. F., Man, C. and Lei, W. (2011) 'Morphological, physiological and biochemical responses of plants to drought stress', African journal of agricultural research, 6(9), pp. 2026-2032.

Aouadi, M., Guenni, K., Abdallah, D., Louati, M., Chatti, K., Baraket, G. and Salhi Hannachi, A. (2019) 'Conserved DNA-derived polymorphism, new markers for genetic diversity analysis of Tunisian Pistacia vera L', Physiology and Molecular Biology of Plants, 25(5), pp. 1211-1223.

Argun, M. E., Dursun, S., Ozdemir, C. and Karatas, M. (2007) 'Heavy metal adsorption by modified oak sawdust: Thermodynamics and kinetics', Journal of Hazardous Materials, 141(1), pp. 77-85.

Arnao, M. B., Hernández-Ruiz, J., Cano, A. and Reiter, R. J. (2021) 'Melatonin and carbohydrate metabolism in plant cells', Plants, 10(9), pp. 1917.

Asad, S., Fayyaz, M., Majeed, K., Ali, S., Liu, J., Rasheed, A. and Wang, Y. (2022) 'Genetic Variability and Aggressiveness of Tilletia indica Isolates Causing Karnal Bunt in Wheat', Journal of Fungi, 8(3), pp. 219.

Ashfaque, F., Farooq, S., Chopra, P. and Chhillar, H. (2020) 'Improving Heavy Metal Tolerance through Plant Growth Regulators and Osmoprotectants in Plants', Improving Abiotic Stress Tolerance in Plants: CRC Press, pp. 69-98.

Asif, S., Ali, Q. and Malik, A. (2020) 'Evaluation of salt and heavy metal stress for seedling traits in wheat', Biological and Clinical Sciences Research Journal, 2020(1).

Askarnejad, M. R., Soleymani, A. and Javanmard, H. R. (2021) 'Barley (Hordeum vulgare L.) physiology including nutrient uptake affected by plant growth regulators under field drought conditions', Journal of Plant Nutrition, pp. 1-17.

Auesukaree, C., Bussarakum, J., Sirirakphaisarn, S. and Saengwilai, P. (2021) 'Effects of aqueous Moringa oleifera leaf extract on growth performance and accumulation of cadmium in a Thai jasmine rice-Khao Dawk Mali 105 variety'.

Ayachi, I., Ghabriche, R., Kourouma, Y., Naceur, B., Abdelly, C., Thomine, S. and Ghnaya, T. (2021) 'Cd tolerance and accumulation in barley: screening of 36 North African cultivars on Cd-contaminated soil', Environmental Science and Pollution Research, 28(31), pp. 42722-42736.

Baghaie, A. H. and Aghili, F. (2019) 'Health risk assessment of Pb and Cd in soil, wheat, and barley in Shazand County, central of Iran', Journal of Environmental Health Science and Engineering, 17(1), pp. 467-477.

Bali, A. S., Sidhu, G. P. S. and Kumar, V. (2020) 'Root exudates ameliorate cadmium tolerance in plants: a review', Environmental Chemistry Letters, 18(4), pp. 1243-1275

.Bandurska, H., Niedziela, J., Pietrowska-Borek, M., Nuc, K., Chadzinikolau, T. and Radzikowska, D. (2017) 'Regulation of proline biosynthesis and resistance to drought stress in two barley (Hordeum vulgare L.) genotypes of different origin', Plant Physiology and Biochemistry, 118, pp. 427-437.

Barati, M., Majidi, M., Mirlohi, A., Safari, M., Mostafavi, F. and Karami, Z. (2018) 'Potential of Iranian wild barley (Hordeum vulgare ssp. spontaneum) in breeding for drought tolerance', Cereal Research Communications, 46(4), pp. 707-716.

Basal, O., Szabó, A. and Veres, S. (2020) 'PEG-induced drought stress effects on soybean germination parameters', Journal of Plant Nutrition, 43(12), pp. 1768–1779.

Bashir, K., Matsui, A., Rasheed, S. and Seki, M. (2019) 'Recent advances in the characterization of plant transcriptomes in response to drought, salinity, heat, and cold stress', F1000Research, 8.

Basu, S., Prabhakar, A. A., Kumari, S., Kumar, R. R., Shekhar, S., Prakash, K., Singh, J. P., Singh, G. P., Prasad, R. and Kumar, G. (2022) 'Micronutrient and redox homeostasis contribute to Moringa oleifera-regulated drought tolerance in wheat', Plant Growth Regulation, pp. 1-12.

Begović, L., Pospihalj, T., Lončarić, P., Čamagajevac, I. Š., Cesar, V. and Leljak-Levanić, D. (2020) 'Distinct accumulation and remobilization of fructans in barley cultivars contrasting for photosynthetic performance and yield', Theoretical and Experimental Plant Physiology, 32(2), pp. 109-120.

Bello, O. S., Adegoke, K. A. and Akinyunni, O. O. (2017) 'Preparation and characterization of a novel adsorbent from Moringa oleifera leaf', Applied Water Science, 7(3), pp. 1295-1305.

Berni, R., Luyckx, M., Xu, X., Legay, S., Sergeant, K., Hausman, J.-F., Lutts, S., Cai, G. and Guerriero, G. (2019) 'Reactive oxygen species and heavy metal stress in plants: Impact on the cell wall and secondary metabolism', Environmental and Experimental Botany, 161, pp. 98-106.

Berwal, M. K., Kumar, R., Prakash, K., Rai, G. K. and Hebbar, K. (2021) 'Antioxidant Defense System in Plants Against Abiotic Stress', Abiotic Stress Tolerance Mechanisms in Plants: CRC Press, pp. 175-202.

Bhoi, A., Yadu, B., Chandra, J. and Keshavkant, S. (2021) 'Contribution of strigolactone in plant physiology, hormonal interaction and abiotic stresses', Planta, 254(2), pp. 1-21.

Bianchi, L., Sframeli, M., Vantaggiato, L., Vita, G. L., Ciranni, A., Polito, F., Oteri, R., Gitto, E., Di Giuseppe, F. and Angelucci, S. (2021) 'Nusinersen Modulates Proteomics Profiles of Cerebrospinal Fluid in Spinal Muscular Atrophy Type 1 Patients', International journal of molecular sciences, 22(9), pp. 4329.

Bornet, B. and Branchard, M. (2001) 'Nonanchored inter simple sequence repeat (ISSR) markers: reproducible and specific tools for genome fingerprinting', Plant Molecular Biology Reporter, 19(3), pp. 209-215.

Boudiar, R., Casas, A. M., Gioia, T., Fiorani, F., Nagel, K. A. and Igartua, E. (2020) 'Effects of low water availability on root placement and shoot development in landraces and modern barley cultivars', Agronomy, 10(1), pp. 134.

Bowne, J. B., Erwin, T. A., Juttner, J., Schnurbusch, T., Langridge, P., Bacic, A. and Roessner, U. (2012) 'Drought responses of leaf tissues from wheat cultivars of differing drought tolerance at the metabolite level', Molecular plant, 5(2), pp. 418–429.

Brockman, H. G. and Brennan, R. F. (2017) 'The effect of foliar application of Moringa leaf extract on biomass, grain yield of wheat and applied nutrient efficiency', Journal of Plant Nutrition, 40(19), pp. 2728-2736.

Buege, J. A. and Aust, S. D. (1978) ' Microsomal lipid peroxidation', Methods in Enzymology: Elsevier, pp. 302-310.

Cai, K., Chen, X., Han, Z., Wu, X., Zhang, S., Li, Q., Nazir, M. M., Zhang, G. and Zeng, F. (2020) 'Screening of worldwide barley collection for drought tolerance: the assessment of various physiological measures as the selection criteria', Frontiers in Plant Science, pp. 1159.

Carter, A. Y., Hawes, M. C. and Ottman, M. J. (2019) 'Drought-tolerant barley: I. Field observations of growth and development', Agronomy, 9(5), pp. 221.

Carter, A. Y., Ottman, M. J., Curlango-Rivera, G., Huskey, D. A., D'Agostini, B. A. and Hawes, M. C. (2019) 'Drought-tolerant barley: II. Root tip characteristics in emerging roots', Agronomy, 9(5), pp. 220.

Ceccarelli, S., Grando, S., Tutwiler, R., Baha, J., Martini, A., Salahieh, H., Goodchild, A. and Michael, M. (2000) 'A methodological study on participatory barley breeding I. Selection phase', Euphytica, 111(2), pp. 91-104.

Chamon, A., Gerzabek, M., Mondol, M., Ullah, S., Rahman, M. and Blum, W. (2005) 'Influence of cereal varieties and site conditions on heavy metal accumulations in cereal crops on polluted soils of Bangladesh', Communications in soil science and plant analysis, 36(7-8), pp. 889-906.

Chattha, M. U., Khan, I., Hassan, M. U., Chattha, M. B., Nawaz, M., Akhtar, N., Usman, M., Kharal, M. and Ullah, M. A. (2018) 'Efficacy of extraction methods of Moringa oleifera leaf extract for enhanced growth and yield of wheat', Journal of Basic and Applied Sciences, 14, pp. 131-135.

Chehregani, A., Noori, M. and Yazdi, H. L. (2009) 'Phytoremediation of heavy-metal-polluted soils: screening for new accumulator plants in Angouran mine (Iran) and evaluation of removal ability', Ecotoxicology and environmental safety, 72(5), pp. 1349-1353.

Choudhary, K., Choudhary, O. and Shekhawat, N. (2008) 'Marker assisted selection: a novel approach for crop improvement', American-Eurasian Journal of Agronomy, 1(2), pp. 26-30.

Chu, Y., Xu, N., Wu, Q., Yu, B., Li, X., Chen, R. and Huang, J. (2019) 'Rice transcription factor OsMADS57 regulates plant height by modulating gibberellin catabolism', Rice, 12(1), pp. 1-14.

Collard, B. and Mackill, D. (2009) 'Conserved DNA-derived polymorphism (CDDP): a simple and novel method for generating DNA markers in plants', Plant Molecular Biology Reporter, 27(4), pp. 558-562.

Coruh, N., Celep, A. S., Özgökçe, F. and İşcan, M. (2007) 'Antioxidant capacities of Gundelia tournefortii L. extracts and inhibition on glutathione-S-transferase activity', Food Chemistry, 100(3), pp. 1249-1253.

Cruz de Carvalho, M. H. (2008) 'Drought stress and reactive oxygen species: Production, scavenging and signaling', Plant Signaling & Behavior, 3(3), pp. 156–165.

Cui, H., Wang, Y., Yu, T., Chen, S., Chen, Y. and Lu, C. (2020) 'Heterologous expression of three Ammopiptanthus mongolicus dehydrin genes confers abiotic stress tolerance in Arabidopsis thaliana', Plants, 9(2), pp. 193.

Dang, V. H., Hill, C. B., Zhang, X.-Q., Angessa, T. T., McFawn, L.-A. and Li, C. (2020) 'Genetic dissection of the interactions between semi-dwarfing genes sdw1 and ari-e and their effects on agronomic traits in a barley MAGIC population', Molecular breeding, 40(7), pp. 1-14.

Daryanto, S., Wang, L. and Jacinthe, P.-A. (2017) 'Global synthesis of drought effects on cereal, legume, tuber and root crops production: A review', Agricultural Water Management, 179, pp. 18-33.

Dbira, S., Al Hassan, M., Gramazio, P., Ferchichi, A., Vicente, O., Prohens, J. and Boscaiu, M. (2018) 'Variable levels of tolerance to water stress (drought) and associated biochemical markers in Tunisian barley landraces', Molecules, 23(3), pp. 613

Delangiz, N., Khoshru, B., Lajayer, B. A., Ghorbanpour, M. and Kazemalilou, S. (2020) 'Molecular mechanisms of heavy metal tolerance in plants', Cellular and Molecular Phytotoxicity of Heavy Metals, pp. 125-136.

Demecsová, L., Zelinová, V., Liptáková, Ľ. and Tamás, L. (2020) 'Mild cadmium stress induces auxin synthesis and accumulation, while severe cadmium stress causes its rapid depletion in barley root tip', Environmental and Experimental Botany, 175, pp. 104038

Dhanagond, S., Liu, G., Zhao, Y., Chen, D., Grieco, M., Reif, J., Kilian, B., Graner, A. and Neumann, K. (2019) 'Non-invasive phenotyping reveals genomic regions involved in pre-anthesis drought tolerance and recovery in spring barley', Frontiers in Plant Science, pp. 1307.

Dien, D. C., Mochizuki, T. and Yamakawa, T. (2019) 'Effect of various drought stresses and subsequent recovery on proline, total soluble sugar and starch metabolisms in Rice (Oryza sativa L.) varieties', Plant Production Science, 22(4), pp. 530-545.

Djeridane, A., Yousfi, M., Nadjemi, B., Boutassouna, D., Stocker, P. and Vidal, N. (2006) 'Antioxidant activity of some Algerian medicinal plants extracts containing phenolic compounds', Food Chemistry, 97(4), pp. 654-660.

Dodig, D., Kandić, V., Zorić, M., Nikolić-Đorić, E., Živanov, S. T. and Perović, D. (2020) 'Response of kernel growth of barley genotypes with different row type to climatic factors before and after inflection point of grain filling', Field crops research, 255, pp. 107864.

Donnelly, A., Yu, R., Rehberg, C., Meyer, G. and Young, E. B. (2020) 'Leaf chlorophyll estimates of temperate deciduous shrubs during autumn senescence using a SPAD-502 meter and calibration with extracted chlorophyll', Annals of Forest Science, 77(2), pp. 1-12.

Drine, S., Smedley, M., Ferchichi, A. and Harwood, W. (2018) 'Sequence and expression variation in the dehydrin6 gene in barley varieties contrasting in response to drought stress', South African Journal of Botany, 119, pp. 278-285.

Drobek, M., Frąc, M. and Cybulska, J. (2019) 'Plant biostimulants: Importance of the quality and yield of horticultural crops and the improvement of plant tolerance to abiotic stress—A review', Agronomy, 9(6), pp. 335.

Duggan, B., Richards, R., Van Herwaarden, A. and Fettell, N. (2005) 'Agronomic evaluation of a tiller inhibition gene (tin) in wheat. I. Effect on yield, yield components, and grain protein', Australian Journal of Agricultural Research, 56(2), pp. 169-178.

El Rasafi, T., Oukarroum, A., Haddioui, A., Song, H., Kwon, E. E., Bolan, N., Tack, F. M., Sebastian, A., Prasad, M. and Rinklebe, J. (2021) 'Cadmium stress in plants: A critical review of the effects, mechanisms, and tolerance strategies', Critical Reviews in Environmental Science and Technology, pp. 1-52.

Emamverdian, A., Ding, Y. and Xie, Y. (2020) 'The Role of New Members of Phytohormones in Plant Amelioration under Abiotic Stress with an Emphasis on Heavy Metals', Polish Journal of Environmental Studies, 29(2).

Etminan, A., Pour-Aboughadareh, A., Mohammadi, R., Ahmadi-Rad, A., Noori, A., Mahdavian, Z. and Moradi, Z. (2016) 'Applicability of start codon targeted (SCoT) and inter-simple sequence repeat (ISSR) markers for genetic diversity analysis in durum wheat genotypes', Biotechnology & Biotechnological Equipment, 30(6), pp. 1075-1081.

Farooq, S. and Azam, F. (2002) 'Molecular markers in plant breeding-I: Concepts and characterization', Pakistan journal of biological sciences, 5(10), pp. 1135-1140.

Feiziasl, V., Jafarzadeh, J., Sadeghzadeh, B. and Shalmani, M. M. (2022) 'Water deficit index to evaluate water stress status and drought tolerance of rainfed barley genotypes in cold semi-arid area of Iran', Agricultural Water Management, 262, pp. 107395.

Feng, J., Shen, R. F. and Shao, J. F. (2021) 'Transport of cadmium from soil to grain in cereal crops: A review', Pedosphere, 31(1), pp. 3-10.

Fernandez, G. C. 'Effective selection criteria for assessing plant stress tolerance'. Proceeding of the International Symposium on Adaptation of Vegetables and other Food Crops in Temperature and Water Stress, Aug. 13-16, Shanhua, Taiwan, 1992, 257-270.

Flexas, J. and Medrano, H. (2002) 'Energy dissipation in C3 plants under drought', Functional Plant Biology, 29(10), pp. 1209–1215.

Fraser, T. E., Silk, W. K. and Rost, T. L. (1990) 'Effects of low water potential on cortical cell length in growing regions of maize roots', Plant Physiology, 93(2), pp. 648–651.

Fuentes, J. L., Escobar, F., Alvarez, A., Gallego, G., Duque, M. C., Ferrer, M., Deus, J. E. and Tohme, J. M. (1999) 'Analyses of genetic diversity in Cuban rice varieties using isozyme, RAPD and AFLP markers', Euphytica, 109(2), pp. 107-115.

Gage, J. L., Monier, B., Giri, A. and Buckler, E. S. (2020) 'Ten years of the maize nested association mapping population: impact, limitations, and future directions', The plant cell, 32(7), pp. 2083-2093.

Ganeshamurthy, A., Varalakshmi, L. and Sumangala, H. (2016) 'Environmental risks associated with heavy metal contamination in soil, water and plants in urban and periurban agriculture', Journal of Horticultural Sciences, 3(1), pp. 1-29.

Gao, S. and Chu, C. (2020) 'Gibberellin metabolism and signaling: targets for improving agronomic performance of crops', Plant and Cell Physiology, 61(11), pp. 1902-1911.

Geravandi, M., Farshadfar, E. and Kahrizi, D. (2010) 'Evaluation of drought tolerance in bread wheat advanced genotypes in field and laboratory conditions', Seed and Plant Improvement Journal, (2).

Ghomi, K., Rabiei, B., Sabouri, H. and Gholamalipour Alamdari, E. (2021) 'Association analysis, genetic diversity and population structure of barley (Hordeum vulgare L.) under heat stress conditions using SSR and ISSR markers linked to primary and secondary metabolites', Molecular Biology Reports, 48(10), pp. 6673-6694.

Goldsbrough, P. (2020) 'Metal tolerance in plants: the role of phytochelatins and metallothioneins', Phytoremediation of contaminated soil and water: CRC Press, pp. 221-233

Grzesiak, S., Hordyńska, N., Szczyrek, P., Grzesiak, M. T., Noga, A. and Szechyńska-Hebda, M. (2018) 'Variation among wheat ( Triticum easativum L.) genotypes in response to the drought stress: I – selection approaches', Journal of Plant Interactions, 14(1), pp. 30–44.

Guala, S. D., Vega, F. A. and Covelo, E. F. (2010) 'The dynamics of heavy metals in plant–soil interactions', Ecological Modelling, 221(8), pp. 1148-1152.

Guasmi, F., Elfalleh, W., Hannachi, H., Feres, K., Touil, L., Marzougui, N., Triki, T. and Ferchichi, A. (2012) 'The use of ISSR and RAPD markers for genetic diversity among south tunisian barley', International Scholarly Research Notices, 2012.

Guo, F., Ding, C., Zhou, Z., Huang, G. and Wang, X. (2018) 'Effects of combined amendments on crop yield and cadmium uptake in two cadmium contaminated soils under rice-wheat rotation', Ecotoxicology and environmental safety, 148, pp. 303-310.

Gupta, D., Pena, L. B., Romero-Puertas, M. C., Hernández, A., Inouhe, M. and Sandalio, L. M. (2017) 'NADPH oxidases differentially regulate ROS metabolism and nutrient uptake under cadmium toxicity', Plant, Cell & Environment, 40(4), pp. 509-526.

Hagenblad, J., Leino, M. W., Afonso, G. H. and Morales, D. A. (2019) 'Morphological and genetic characterization of barley (Hordeum vulgare L.) landraces in the Canary Islands', Genetic Resources and Crop Evolution, 66(2), pp. 465-480.

Haghi, G., Hatami, A. and Arshi, R. (2011) 'Distribution of caffeic acid derivatives in Gundelia tournefortii L', Food Chemistry, 124(3), pp. 1029-1035.

Hajibarat, Z., Saidi, A., Hajibarat, Z. and Talebi, R. (2015) 'Characterization of genetic diversity in chickpea using SSR markers, start codon targeted polymorphism (SCoT) and conserved DNA-derived polymorphism (CDDP)', Physiology and molecular biology of plants, 21(3), pp. 365-373.

Hamidi, H., Talebi, R. and Keshavarzi, F. (2014) 'Comparative efficiency of functional gene-based markers, start codon targeted polymorphism (SCoT) and conserved DNA-derived polymorphism (CDDP) with ISSR markers for diagnostic fingerprinting in wheat (Triticum aestivum L.)', Cereal Research Communications, 42(4), pp. 558-567.

Hao, Q., Wang, W., Han, X., Wu, J., Lyu, B., Chen, F., Caplan, A., Li, C., Wu, J. and Wang, W. (2018) 'Isochorismate-based salicylic acid biosynthesis confers basal resistance to Fusarium graminearum in barley', Molecular plant pathology, 19(8), pp. 1995-2010.

Harb, A., Awad, D. and Samarah, N. (2015) 'Gene expression and activity of antioxidant enzymes in barley (Hordeum vulgare L.) under controlled severe drought', Journal of Plant Interactions, 10(1), pp. 109-116.

Hasanloo, T., Pazirandeh, M. S., Niknam, V., Shahbazi, M., Ebrahimzadeh Mabood, H. and Ghaffari, A. (2013) 'Effects of drought and methyl jasmonate on antioxidant activities of selected barley genotypes', Journal of Agrobiology, 30(2), pp. 71–82.

Hasanuzzaman, M., Shabala, L., Brodribb, T. J., Zhou, M. and Shabala, S. (2017) 'Assessing the suitability of various screening methods as a proxy for drought tolerance in barley', Functional Plant Biology, 44(2), pp. 253-266.

Hayes, P., Carrijo, D. and Meints, B. (2020) 'Towards low cadmium accumulation in barley', Nature Food, 1(8), pp. 465-465.

He, L., Wang, X., Na, X., Feng, R., He, Q., Wang, S., Liang, C., Yan, L., Zhou, L. and Bi, Y. (2021) 'Alternative Pathway Is Involved in Hydrogen Peroxide-Enhanced Cadmium Tolerance in Hulless Barley Roots', Plants, 10(11), pp. 2329.

Hein, J. A., Sherrard, M. E., Manfredi, K. P. and Abebe, T. (2016) 'The fifth leaf and spike organs of barley (Hordeum vulgare L.) display different physiological and metabolic responses to drought stress', BMC plant biology, 16(1), pp. 1-12.

Hellal, F. A., El-Shabrawi, H. M., Abd El-Hady, M., Khatab, I. A., El-Sayed, S. A. A. and Abdelly, C. (2018) 'Influence of PEG induced drought stress on molecular and biochemical constituents and seedling growth of Egyptian barley cultivars', Journal of Genetic Engineering & Biotechnology, 16(1), pp. 203–212.

Hernandez, J., Meints, B. and Hayes, P. (2020) 'Introgression Breeding in Barley: Perspectives and Case Studies', Frontiers in Plant Science, 11, pp. 761.

Hönig, M., Plíhalová, L., Husičková, A., Nisler, J. and Doležal, K. (2018) 'Role of cytokinins in senescence, antioxidant defence and photosynthesis', International journal of molecular sciences, 19(12), pp. 4045.

Honsdorf, N., March, T. J., Berger, B., Tester, M. and Pillen, K. (2014) 'High-throughput phenotyping to detect drought tolerance QTL in wild barley introgression lines', PloS one, 9(5), pp. e97047.

Hossain, M. M., Khatun, M. A., Haque, M. N., Bari, M. A., Alam, M. F., Mandal, A. and Kabir, A. H. (2018) 'Silicon alleviates arsenic-induced toxicity in wheat through vacuolar sequestration and ROS scavenging', International journal of phytoremediation, 20(8), pp. 796-804.

Hosseini, M., Yassaie, M., Rashed-Mohassel, M. H., Ghorbani, R. and Niazi, A. (2022) 'Genetic diversity of Iranian wild barley (Hordeum spontaneum Koch.) populations', Journal of Crop Science and Biotechnology, pp. 1-11.

Hoyle, G. L., Steadman, K. J., Good, R. B., McIntosh, E. J., Galea, L. M. E. and Nicotra, A. B. (2015) 'Seed germination strategies: an evolutionary trajectory independent of vegetative functional traits', Frontiers in Plant Science, 6(731).

Huang, B., Wu, W. and Hong, Z. (2021) 'Genetic Interactions of Awnness Genes in Barley', Genes, 12(4), pp. 606.

Hussain, H. A., Hussain, S., Khaliq, A., Ashraf, U., Anjum, S. A., Men, S. and Wang, L. (2018) 'Chilling and drought stresses in crop plants: implications, cross talk, and potential management opportunities', Frontiers in Plant Science, 9, pp. 393.

Huybrechts, M., Cuypers, A., Deckers, J., Iven, V., Vandionant, S., Jozefczak, M. and Hendrix, S. (2019) 'Cadmium and plant development: an agony from seed to seed', International journal of molecular sciences, 20(16), pp. 3971.

Igwe, D. O., Ihearahu, O. C., Osano, A. A., Acquaah, G. and Ude, G. N. (2021) 'Genetic diversity and population assessment of Musa L.(Musaceae) employing CDDP markers', Plant Molecular Biology Reporter, pp. 1-20.

Ilyas, M., Nisar, M., Khan, N., Hazrat, A., Khan, A. H., Hayat, K., Fahad, S., Khan, A. and Ullah, A. (2020) 'Drought tolerance strategies in plants: A mechanistic approach', Journal of Plant Growth Regulation, pp. 1-19.

Iqbal, M. A. (2014) 'Role of moringa, brassica and sorghum water extracts in increasing crops growth and yield: A Review', Amer.-Eurasian J. Agric. Environ. Sci, 14(11), pp. 1150-1158.

Iqbal, M. T. (2018) 'Physiological, biochemical and molecular responses of different barley varieties to drought and salinity'.

Ishtiyaq, S., Kumar, H., Varun, M., Ogunkunle, C. O. and Paul, M. S. (2021) 'Role of secondary metabolites in salt and heavy metal stress mitigation by halophytic plants: An overview', Handbook of Bioremediation, pp. 307-327.

Jain, C., Khatana, S. and Vijayvergia, R. (2019) 'Bioactivity of secondary metabolites of various plants: a review', International Journal of Pharmaceutical Sciences and Research, 10(2), pp. 494-498.

Jain, C. K., Malik, D. S. and Yadav, A. K. (2016) 'Applicability of plant based biosorbents in the removal of heavy metals: a review', Environmental Processes, 3(2), pp. 495-523.

Jain, P., Farooq, B., Lamba, S. and Koul, B. (2020) 'Foliar spray of Moringa oleifera Lam. leaf extracts (MLE) enhances the stevioside, zeatin and mineral contents in Stevia rebaudiana Betoni', South African Journal of Botany, 132, pp. 249-257.

Janeczko, A., Gruszka, D., Pociecha, E., Dziurka, M., Filek, M., Jurczyk, B., Kalaji, H. M., Kocurek, M. and Waligórski, P. (2016) 'Physiological and biochemical characterisation of watered and drought-stressed barley mutants in the HvDWARF gene encoding C6-oxidase involved in brassinosteroid biosynthesis', Plant Physiology and Biochemistry, 99, pp. 126-141.

Jiang, Q., Roche, D., Durham, S. and Hole, D. (2006) 'Awn contribution to gas exchanges of barley ears', Photosynthetica, 44(4), pp. 536-541.

Jing, X.-Q., Shalmani, A., Zhou, M.-R., Shi, P.-T., Muhammad, I., Shi, Y., Sharif, R., Li, W.-Q., Liu, W.-T. and Chen, K.-M. (2020) 'Genome-wide identification of malectin/malectin-like domain containing protein family genes in rice and their expression regulation under various hormones, abiotic stresses, and heavy metal treatments', Journal of Plant Growth Regulation, 39(1), pp. 492-506.

Jócsák, I., Malgwi, I., Rabnecz, G., Szegő, A., Varga-Visi, É., Végvári, G. and Pónya, Z. (2020) 'Effect of cadmium stress on certain physiological parameters, antioxidative enzyme activities and biophoton emission of leaves in barley (Hordeum vulgare L.) seedlings', PloS one, 15(11), pp. e0240470.

Juknys, R., Vitkauskaitė, G., Račaitė, M. and Venclovienė, J. (2012) 'The impacts of heavy metals on oxidative stress and growth of spring barley', Central european journal of biology, 7(2), pp. 299-306.

Kalai, T., Bouthour, D., Manai, J., Bettaieb Ben Kaab, L. and Gouia, H. (2016) 'Salicylic acid alleviates the toxicity of cadmium on seedling growth, amylases and phosphatases activity in germinating barley seeds', Archives of Agronomy and Soil Science, 62(6), pp. 892-904.

Kaur, E., Bhardwaj, R. D., Kaur, S. and Grewal, S. K. (2021) 'Drought stress-induced changes in redox metabolism of barley (Hordeum vulgare L.)', Biologia Futura, 72(3), pp. 347-358

Kebede, A., Kang, M. S. and Bekele, E. (2019) 'Advances in mechanisms of drought tolerance in crops, with emphasis on barley', Advances in Agronomy, 156, pp. 265-314.

Keshavarz, H., Hosseini, S., SEDIBE, M. and ACHILONU, M. (2022) 'Arbuscular Mycorrhizal Fungi Used To Support Iranian Barley Cultivated On Cadmium Contaminated Soils (Hordeum vulgare L.)', Applied Ecology and Environmental Research, 20(1), pp. 43-53.

Khalid, M., Afzal, F., Gul, A., Ahanger, M. A. and Ahmad, P. (2016) 'Analysis of novel haplotype variation at TaDREB-D1 and TaCwi-D1 genes influencing drought tolerance in bread/synthetic wheat derivatives: an overview', Water stress and crop plants, pp. 206-226.

Khan, A., Khan, S., Khan, M. A., Qamar, Z. and Waqas, M. (2015) 'The uptake and bioaccumulation of heavy metals by food plants, their effects on plants nutrients, and associated health risk: a review', Environmental Science and Pollution Research, 22(18), pp. 13772-13799.

Khan, S., Basra, S., Nawaz, M., Hussain, I. and Foidl, N. (2020) 'Combined application of moringa leaf extract and chemical growth-promoters enhances the plant growth and productivity of wheat crop (Triticum aestivum L.)', South African Journal of Botany, 129, pp. 74-81.

Khan, S., Basra, S. M. A., Afzal, I., Nawaz, M. and Rehman, H. U. (2017) 'Growth promoting potential of fresh and stored Moringa oleifera leaf extracts in improving seedling vigor, growth and productivity of wheat crop', Environmental Science and Pollution Research, 24(35), pp. 27601-27612.

Khlestkina, E., Shvachko, N., Zavarzin, A. and Börner, A. (2020) 'Vavilov's Series of the "Green Revolution" Genes', Russian Journal of Genetics, 56(11), pp. 1371-1380.

Khodaee, L., Azizinezhad, R., Etminan, A. R. and Khosroshahi, M. (2021) 'Assessment of genetic diversity among Iranian Aegilops triuncialis accessions using ISSR, SCoT, and CBDP markers', Journal of Genetic Engineering and Biotechnology, 19(1), pp. 1-9.

Khwarahm, N. R. (2020) 'Mapping current and potential future distributions of the oak tree (Quercus aegilops) in the Kurdistan Region, Iraq', Ecological Processes, 9(1), pp. 1-16.

Kilian, B., Ozkan, H., Kohl, J., von Haeseler, A., Barale, F., Deusch, O., Brandolini, A., Yucel, C., Martin, W. and Salamini, F. (2006) 'Haplotype structure at seven barley genes: Relevance to gene pool bottlenecks, phylogeny of ear type and site of barley domestication', Molecular genetics and genomics : MGG, 276, pp. 230-41.

Kintlová, M., Blavet, N., Cegan, R. and Hobza, R. (2017) 'Transcriptome of barley under three different heavy metal stress reaction', Genomics data, 13, pp. 15-17.

Kintlová, M., Vrána, J., Hobza, R., Blavet, N. and Hudzieczek, V. (2021) 'Transcriptome Response to Cadmium Exposure in Barley (Hordeum vulgare L.)', Frontiers in Plant Science, 12.

Klem, K., Ač, A., Holub, P., Kováč, D., Špunda, V., Robson, T. M. and Urban, O. (2012) 'Interactive effects of PAR and UV radiation on the physiology, morphology and leaf optical properties of two barley varieties', Environmental and Experimental Botany, 75, pp. 52-64.

Kobayashi, T., Nozoye, T. and Nishizawa, N. K. (2019) 'Iron transport and its regulation in plants', Free Radical Biology and Medicine, 133, pp. 11-20.

Kooyers, N. J. (2015) 'The evolution of drought escape and avoidance in natural herbaceous populations', Plant Science, 234, pp. 155-162.

Koprna, R., Humplík, J. F., Špíšek, Z., Bryksová, M., Zatloukal, M., Mik, V., Novák, O., Nisler, J. and Doležal, K. (2021) 'Improvement of Tillering and Grain Yield by Application of Cytokinin Derivatives in Wheat and Barley', Agronomy, 11(1), pp. 67.

Kosova, K., Vitamvas, P., Urban, M. O., Kholova, J. and Prášil, I. T. (2014) 'Breeding for enhanced drought resistance in barley and wheat-drought-associated traits, genetic resources and their potential utilization in breeding programmes', Czech Journal of Genetics and Plant Breeding, 50(4), pp. 247-261.

Kowalczewski, P. Ł., Radzikowska, D., Ivanišová, E., Szwengiel, A., Kačániová, M. and Sawinska, Z. (2020) 'Influence of Abiotic Stress Factors on the Antioxidant Properties and Polyphenols Profile Composition of Green Barley (Hordeum vulgare L.)', International journal of molecular sciences, 21(2), pp. 397.

Kudo, K., Kudo, H. and Kawai, S. (2007) 'Cadmium uptake in barley affected by iron concentration of the medium: role of phytosiderophores', Soil Science and Plant Nutrition, 53(3), pp. 259-266.

Kumar, A., Bernier, J., Verulkar, S., Lafitte, H. and Atlin, G. (2008) 'Breeding for drought tolerance: direct selection for yield, response to selection and use of drought-tolerant donors in upland and lowland-adapted populations', Field crops research, 107(3), pp. 221-231.

Kumar, S., Stecher, G., Li, M., Knyaz, C. and Tamura, K. (2018) 'MEGA X: molecular evolutionary genetics analysis across computing platforms', Molecular biology and evolution, 35(6), pp. 1547-1549.

Lateef, D., Mustafa, K. and Tahir, N. (2021) 'Screening of Iraqi barley accessions under PEG-induced drought conditions', All Life, 14(1), pp. 308-332.

Latifi, A. M., Nabavi, S. M., Mirzaei, M., Tavalaei, M., Ghafurian, H., Hellio, C. and Nabavi, S. F. (2012) 'Bioremediation of toxic metals mercury and cesium using three types of biosorbent: bacterial exopolymer, gall nut, and oak fruit particles', Toxicological & Environmental Chemistry, 94(9), pp. 1670-1677.

Latutrie, M., Gourcilleau, D. and Pujol, B. (2019) 'Epigenetic variation for agronomic improvement: an opportunity for vegetatively propagated crops', American journal of botany, 106(10), pp. 1281.

Lausch, A., Pause, M., Schmidt, A., Salbach, C., Gwillym-Margianto, S. and Merbach, I. (2013) 'Temporal hyperspectral monitoring of chlorophyll, LAI, and water content of barley during a growing season', Canadian Journal of Remote Sensing, 39(3), pp. 191-207.

Lawlor, D. W. and Cornic, G. (2002) 'Photosynthetic carbon assimilation and associated metabolism in relation to water deficits in higher plants', Plant, Cell & Environment, 25(2), pp. 275–294.

Lei, G. J., Fujii-Kashino, M., Hisano, H., Saisho, D., Deng, F., Yamaji, N., Sato, K., Zhao, F.-J. and Ma, J. F. (2020) 'Breeding for low cadmium barley by introgression of a Sukkula-like transposable element', Nature Food, 1(8), pp. 489-499.

Lentini, M., De Lillo, A., Paradisone, V., Liberti, D., Landi, S. and Esposito, S. (2018) 'Early responses to cadmium exposure in barley plants: effects on biometric and physiological parameters', Acta Physiologiae Plantarum, 40(10), pp. 1-11.

Lin, H., Fang, C., Li, Y., Lin, W., He, J., Lin, R. and Lin, W. (2017) 'Cadmium-stress mitigation through gene expression of rice and silicon addition', Plant Growth Regulation, 81(1), pp. 91-101.

Liu, H., Zang, F., Wu, Q., Ma, Y., Zheng, Y. and Zang, D. (2020) 'Genetic diversity and population structure of the endangered plant Salix taishanensis based on CDDP markers', Global Ecology and Conservation, 24, pp. e01242.

Liu, T., Lawluvy, Y., Shi, Y., Ighalo, J. O., He, Y., Zhang, Y. and Yap, P.-S. (2022) 'Adsorption of cadmium and lead from aqueous solution using modified biochar: a review', Journal of environmental chemical engineering, 10(1), pp. 106502

Lobo, M. C. (2013) 'Growth of four varieties of barley (Hordeum vulgare L.) in soils contaminated with heavy metals and their effects on some physiological traits', American Journal of Plant Sciences, 4(09), pp. 1799.

Luo, L., Xia, H. and Lu, B.-R. (2019) 'Crop breeding for drought resistance', Frontiers in Plant Science, 10, pp. 314.

M. A. Merwad, A.-R. and Abdel-Fattah, M. K. (2017) 'Improving productivity and nutrients uptake of wheat plants using Moringa oleifera leaf extract in sandy soil', Journal of Plant Nutrition, 40(10), pp. 1397-1403.

Mahalingam, R. and Bregitzer, P. (2019) 'Impact on physiology and malting quality of barley exposed to heat, drought and their combination during different growth stages under controlled environment', Physiologia Plantarum, 165(2), pp. 277-289.

Maishanu, H. M., Mainasara, M. M., Yahaya, S. and Yunusa, A. (2017) 'The use of moringa leaves extract as a plant growth hormone on cowpea (Vigna Anguiculata)', Traektoriâ Nauki= Path of Science, 3(12).

Makhtoum, S., Sabouri, H., Gholizadeh, A., Ahangar, L. and Katouzi, M. (2021) 'Quantitative genes controlling chlorophyll fluorescence attributes in barley (Hordeum vulgare L.)', Journal of Genetic Resources, 7(1), pp. 72-86.

Malik, D., Jain, C. and Yadav, A. K. (2017) 'Removal of heavy metals from emerging cellulosic low-cost adsorbents: a review', Applied Water Science, 7(5), pp. 2113-2136

Mansour, E., Desoky, E.-S. M., Ali, M. M., Abdul-Hamid, M. I., Ullah, H., Attia, A. and Datta, A. (2021) 'Identifying drought-tolerant genotypes of faba bean and their agro-physiological responses to different water regimes in an arid Mediterranean environment', Agricultural Water Management, 247, pp. 106754.

Marzec, M. and Alqudah, A. M. (2018) 'Key hormonal components regulate agronomically important traits in barley', International journal of molecular sciences, 19(3), pp. 795.

Marzec, M., Situmorang, A., Brewer, P. B. and Brąszewska, A. (2020) 'Diverse roles of MAX1 homologues in rice', Genes, 11(11), pp. 1348.

Maswada, H. F., Abd El-Razek, U. A., El-Sheshtawy, A.-N. A. and Elzaawely, A. A. (2018) 'Morpho-physiological and yield responses to exogenous moringa leaf extract and salicylic acid in maize (Zea mays L.) under water stress', Archives of Agronomy and Soil Science, 64(7), pp. 994-1010.

Mattina, M. I., Lannucci-Berger, W., Musante, C. and White, J. C. (2003) 'Concurrent plant uptake of heavy metals and persistent organic pollutants from soil', Environmental pollution, 124(3), pp. 375-378.

Merwad, A.-R. M. (2018a) 'Using humic substances and foliar spray with moringa leaf extract to alleviate salinity stress on wheat', Sustainability of Agricultural Environment in Egypt: Part II: Springer, pp. 265-286.

Merwad, A.-R. M. A. (2018b) 'Using Moringa oleifera extract as biostimulant enhancing the growth, yield and nutrients accumulation of pea plants', Journal of Plant Nutrition, 41(4), pp. 425-431.

Moghaddam, M., Ehdaie, B. and Waines, J. G. (2000) 'Genetic diversity in populations of wild diploid wheat Triticum urartu Tum. ex. Gandil. revealed by isozyme markers', Genetic Resources and Crop Evolution, 47(3), pp. 323-334.

Mohamed, A. H., Omar, A. A., Attya, A. M., Elashtokhy, M., Zayed, E. M. and Rizk, R. M. (2021) 'Morphological and Molecular Characterization of Some Egyptian Six-Rowed Barley (Hordeum vulgare L.)', Plants, 10(11), pp. 2527

Mostafavi, A. S., Omidi, M., Azizinezhad, R., Etminan, A. and Badi, H. N. (2021) 'Genetic diversity analysis in a mini core collection of Damask rose (Rosa damascena Mill.) germplasm from Iran using URP and SCoT markers', Journal of Genetic Engineering and Biotechnology, 19(1), pp. 1-14.

Mostofa, M. G., Ha, C. V., Rahman, M., Nguyen, K. H., Keya, S. S., Watanabe, Y., Itouga, M., Hashem, A., Abd_Allah, E. F. and Fujita, M. (2021) 'Strigolactones Modulate Cellular Antioxidant Defense Mechanisms to Mitigate Arsenate Toxicity in Rice Shoots', Antioxidants, 10(11), pp. 1815.

Moualeu-Ngangue, D., Dolch, C., Schneider, M., Leon, J., Uptmoor, R. and Stützel, H. (2020) 'Physiological and morphological responses of different spring barley genotypes to water deficit and associated QTLs', PloS one, 15(8), pp. e0237834.

Moursi, Y. S., Thabet, S. G., Amro, A., Dawood, M. F., Baenziger, P. S. and Sallam, A. (2020) 'Detailed Genetic Analysis for Identifying QTLs Associated with Drought Tolerance at Seed Germination and Seedling Stages in Barley', Plants, 9(11), pp. 1425.

Mrówka, M., Szymiczek, M. and Skonieczna, M. (2021) 'The impact of wood waste on the properties of silicone-based composites', Polymers, 13(1), pp. 7.

Munns, R., James, R. A., Sirault, X. R., Furbank, R. T. and Jones, H. G. (2010) 'New phenotyping methods for screening wheat and barley for beneficial responses to water deficit', Journal of experimental botany, 61(13), pp. 3499-3507.

Nabti, E., Jha, B. and Hartmann, A. (2017) 'Impact of seaweeds on agricultural crop production as biofertilizer', International Journal of Environmental Science and Technology, 14(5), pp. 1119-1134.

Nazar, R., Iqbal, N., Masood, A., Khan, M. I. R., Syeed, S. and Khan, N. A. (2012) 'Cadmium toxicity in plants and role of mineral nutrients in its alleviation'.

Nephali, L., Piater, L. A., Dubery, I. A., Patterson, V., Huyser, J., Burgess, K. and Tugizimana, F. (2020) 'Biostimulants for plant growth and mitigation of abiotic stresses: A metabolomics perspective', Metabolites, 10(12), pp. 505.

Niu, X., Luo, T., Zhao, H., Su, Y., Ji, W. and Li, H. (2020) 'Identification of wheat DREB genes and functional characterization of TaDREB3 in response to abiotic stresses', Gene, pp. 144514.

Nouri, A., Golabadi, M., Etminan, A., Rezaei, A. and Mehrabi, A. A. (2021) 'Comparative assessment of SCoT and ISSR markers for analysis of genetic diversity and population structure in some Aegilops tauschii Coss. accessions', Plant Genetic Resources, 19(5), pp. 375-383.

Nybom, H. and Bartish, I. V. (2000) 'Effects of life history traits and sampling strategies on genetic diversity estimates obtained with RAPD markers in plants', Perspectives in plant ecology, evolution and systematics, 3(2), pp. 93-114.

Paunov, M., Koleva, L., Vassilev, A., Vangronsveld, J. and Goltsev, V. (2018) 'Effects of different metals on photosynthesis: cadmium and zinc affect chlorophyll fluorescence in durum wheat', International journal of molecular sciences, 19(3), pp. 787.

Paux, E., Faure, S., Choulet, F., Roger, D., Gauthier, V., Martinant, J. P., Sourdille, P., Balfourier, F., Le Paslier, M. C. and Chauveau, A. (2010) 'Insertion site-based polymorphism markers open new perspectives for genome saturation and marker-assisted selection in wheat', Plant biotechnology journal, 8(2), pp. 196-210.

Piasecka, A., Sawikowska, A., Kuczyńska, A., Ogrodowicz, P., Mikołajczak, K., Krajewski, P. and Kachlicki, P. (2020) 'Phenolic metabolites from barley in contribution to phenome in soil moisture deficit', International journal of molecular sciences, 21(17), pp. 6032.

Pieczynski, M., Marczewski, W., Hennig, J., Dolata, J., Bielewicz, D., Piontek, P., Wyrzykowska, A., Krusiewicz, D., Strzelczyk-Zyta, D., Konopka-Postupolska, D., Krzeslowska, M., Jarmolowski, A. and Szweykowska-Kulinska, Z. (2013) 'Down-regulation of CBP80 gene expression as a strategy to engineer a drought-tolerant potato', Plant biotechnology journal, 11(4), pp. 459–469.

Pospíšilová, H., Jiskrova, E., Vojta, P., Mrizova, K., Kokáš, F., Čudejková, M. M., Bergougnoux, V., Plíhal, O., Klimešová, J. and Novák, O. (2016) 'Transgenic barley overexpressing a cytokinin dehydrogenase gene shows greater tolerance to drought stress', New biotechnology, 33(5), pp. 692-705.

Pour-Aboughadareh, A., Ahmadi, J., Mehrabi, A. A., Etminan, A. and Moghaddam, M. (2018) 'Insight into the genetic variability analysis and relationships among some Aegilops and Triticum species, as genome progenitors of bread wheat, using SCoT markers', Plant Biosystems-An International Journal Dealing with all Aspects of Plant Biology, 152(4), pp. 694-703.

Pour-Aboughadareh, A., Yousefian, M., Moradkhani, H., Moghaddam Vahed, M., Poczai, P. and Siddique, K. H. M. (2019) 'iPASTIC: An online toolkit to estimate plant abiotic stress indices', Applications in Plant Sciences, 7(7).

Pritchard, J. K., Stephens, M. and Donnelly, P. (2000) 'Inference of population structure using multilocus genotype data', Genetics, 155(2), pp. 945-959.

Protich, R., Todorovich, G. and Protich, N. (2012) 'Grain weight per spike of wheat using different ways of seed protection', Bulgarian Journal of Agricultural Science, 18(2), pp. 185-190.

Pu, X., Tang, Y., Zhang, M., Li, T., Qiu, X., Zhang, J., Wang, J., Li, L., Yang, Z. and Su, Y. (2021) 'Identification and candidate gene mining of HvSS1, a novel qualitative locus on chromosome 6H, regulating the uppermost internode elongation in barley (Hordeum vulgare L.)', Theoretical and applied genetics, pp. 1-14.

Qaim, M. (2020) 'Role of New Plant Breeding Technologies for Food Security and Sustainable Agricultural Development', Applied Economic Perspectives and Policy, 42(2), pp. 129-150.

Qiu, C.-W., Zhang, C., Wang, N.-H., Mao, W. and Wu, F. (2021) 'Strigolactone GR24 improves cadmium tolerance by regulating cadmium uptake, nitric oxide signaling and antioxidant metabolism in barley (Hordeum vulgare L.)', Environmental Pollution, 273, pp. 116486.

Qureshi, N., Saha, B. C., Dien, B., Hector, R. E. and Cotta, M. A. (2010) 'Production of butanol (a biofuel) from agricultural residues: Part I–Use of barley straw hydrolysate', Biomass and bioenergy, 34(4), pp. 559-565.

Rady, M. M. and Mohamed, G. F. (2015) 'Modulation of salt stress effects on the growth, physio-chemical attributes and yields of Phaseolus vulgaris L. plants by the combined application of salicylic acid and Moringa oleifera leaf extract', Scientia Horticulturae, 193, pp. 105-113.

Rahim, D., Kalousek, P., Tahir, N., Vyhnánek, T., Tarkowski, P., Trojan, V., Abdulkhaleq, D., Ameen, A. H. and Havel, L. (2020) 'In vitro assessment of Kurdish rice genotypes in response to PEG-induced drought stress', Applied Sciences, 10(13), pp. 4471.

Rahimi, M., Nazari, L., Kordrostami, M. and Safari, P. (2018) 'SCoT marker diversity among Iranian Plantago ecotypes and their possible association with agronomic traits', Scientia Horticulturae, 233, pp. 302-309.

Rahimpour, F., Shojaeimehr, T. and Sadeghi, M. (2017) 'Biosorption of Pb (II) using Gundelia tournefortii: Kinetics, equilibrium, and thermodynamics', Separation Science and Technology, 52(4), pp. 596-607.

Ramesh, P., Mallikarjuna, G., Sameena, S., Kumar, A., Gurulakshmi, K., Reddy, B. V., Reddy, P. C. O. and Sekhar, A. C. (2020) 'Advancements in molecular marker technologies and their applications in diversity studies', Journal of Biosciences, 45(1), pp. 1-15.

Rasul, K. S., Grundler, F. M., and Abdul-razzak Tahir, N. (2022) 'Genetic diversity and population structure assessment of Iraqi tomato accessions using fruit characteristics and molecular markers'. Horticulture, Environment, and Biotechnology, 63(4), pp. 523-538.

Raza, A., Razzaq, A., Mehmood, S. S., Zou, X., Zhang, X., Lv, Y. and Xu, J. (2019) 'Impact of climate change on crops adaptation and strategies to tackle its outcome: A review', Plants, 8(2), pp. 34.

Rehman, A., Jingdong, L., Shahzad, B., Chandio, A. A., Hussain, I., Nabi, G. and Iqbal, M. S. (2015) 'Economic perspectives of major field crops of Pakistan: An empirical study', Pacific science review b: humanities and social sciences, 1(3), pp. 145-158.

Rehman, H., Basra, S., Rady, M., Ghoneim, A. and Wang, Q. (2017) 'Moringa Leaf Extract Improves Wheat Growth and Productivity by Affecting Senescence and Source-sink Relationship', International Journal of Agriculture and Biology, 19, pp. 479-484.

Rigane, G., Ghazghazi, H., Aouadhi, C., Ben Salem, R. and Nasr, Z. (2017) 'Phenolic content, antioxidant capacity and antimicrobial activity of leaf extracts from Pistacia atlantica', Natural product research, 31(6), pp. 696–699.

Riyazuddin, R., Nisha, N., Singh, K., Verma, R. and Gupta, R. (2021) 'Involvement of dehydrin proteins in mitigating the negative effects of drought stress in plants', Plant cell reports, pp. 1-15.

Rohman, M. M., Alam, S. S., Akhi, A. H., Begum, F. and Amiruzzaman, M. (2020) 'Response of catalase to drought in barley (Hordeum vulgare L.) seedlings and its purification', African Journal of Biotechnology, 19(7), pp. 478-486.

Romdhane, L., Dal Ferro, N., Slama, A. and Radhouane, L. (2020) 'Optimizing irrigation and determining the most sensitive development stage to drought in barley (Hordeum vulgare L.) in a semi-arid environment', Acta Botanica Croatica, 79(1), pp. 0-0.

Roy, D., Basu, N., Bhunia, A. and Banerjee, S. K. (1993) 'Counteraction of exogenous L-proline with NaCl in salt-sensitive cultivar of rice', Biologia plantarum, 35(1), pp. 69.

Sabagh, A. E., Hossain, A., Islam, M. S., Barutcular, C., Hussain, S., Hasanuzzaman, M., Akram, T., Mubeen, M., Nasim, W. and Fahad, S. (2019) 'Drought and salinity stresses in barley: consequences and mitigation strategies', Australian Journal of Crop Science, 13(6), pp. 810-820.

Sadeghzadeh-Ahari, D., Hass, M., Kashi, A., Amri, A. and Alizadeh, K. (2010) 'Genetic variability of some agronomic traits in the Iranian fenugreek landraces under drought stress and non-stress conditions', African Journal of Plant Science, 4(2), pp. 012-020.

Sadok, W. and Tamang, B. G. (2019) 'Diversity in daytime and night-time transpiration dynamics in barley indicates adaptation to drought regimes across the Middle-East', Journal of Agronomy and Crop Science, 205(4), pp. 372-384.

Saidi, A., Jabalameli, Z. and Ghalamboran, M. (2018) 'Evaluation of genetic diversity of carnation cultivars using CDDP and DAMD markers and morphological traits', The Nucleus, 61(2), pp. 129-135.

Salam, O. E. A., Reiad, N. A. and ElShafei, M. M. (2011) 'A study of the removal characteristics of heavy metals from wastewater by low-cost adsorbents', Journal of Advanced Research, 2(4), pp. 297-303.

Samarah, N., Alqudah, A., Amayreh, J. and McAndrews, G. (2009) 'The effect of late-terminal drought stress on yield components of four barley cultivars', Journal of Agronomy and Crop Science, 195(6), pp. 427-441.

Saroei, E., Cheghamirza, K. and Zarei, L. (2017) 'Genetic diversity of characteristics in barley cultivars', Genetika, 49(2), pp. 495-510.

Sarshad, A., Talei, D., Torabi, M., Rafiei, F. and Nejatkhah, P. (2021) 'Morphological and biochemical responses of Sorghum bicolor (L.) Moench under drought stress', SN Applied Sciences, 3(1), pp. 1-12.

Scheibe, R. (2019) 'Maintaining homeostasis by controlled alternatives for energy distribution in plant cells under changing conditions of supply and demand', Photosynthesis research, 139(1), pp. 81-91.

Schutzendubel, A. and Polle, A. (2002) 'Plant responses to abiotic stresses: heavy metal-induced oxidative stress and protection by mycorrhization', Journal of experimental botany, 53(372), pp. 1351-1365.

Sehar, S., Adil, M. F., Zeeshan, M., Holford, P., Cao, F., Wu, F. and Wang, Y. (2021) 'Mechanistic Insights into Potassium-Conferred Drought Stress Tolerance in Cultivated and Tibetan Wild Barley: Differential Osmoregulation, Nutrient Retention, Secondary Metabolism and Antioxidative Defense Capacity', International journal of molecular sciences, 22(23), pp. 13100.

Serrote, C. M. L., Reiniger, L. R. S., Silva, K. B., dos Santos Rabaiolli, S. M. and Stefanel, C. M. (2020) 'Determining the Polymorphism Information Content of a molecular marker', Gene, 726, pp. 144175.

Shakirova, F., Allagulova, C. R., Maslennikova, D., Klyuchnikova, E., Avalbaev, A. and Bezrukova, M. (2016) 'Salicylic acid-induced protection against cadmium toxicity in wheat plants', Environmental and Experimental Botany, 122, pp. 19-28.

Sharif, R., Mujtaba, M., Ur Rahman, M., Shalmani, A., Ahmad, H., Anwar, T., Tianchan, D. and Wang, X. (2018) 'The multifunctional role of chitosan in horticultural crops; a review', Molecules, 23(4), pp. 872.

Shibaeva, T., Mamaev, A. and Sherudilo, E. (2020) 'Evaluation of a SPAD-502 Plus Chlorophyll Meter to Estimate Chlorophyll Content in Leaves with Interveinal Chlorosis', Russian Journal of Plant Physiology, 67(4), pp. 690-696

Shinozaki, K., Yamaguchi-Shinozaki, K. and Seki, M. (2003) 'Regulatory network of gene expression in the drought and cold stress responses', Current opinion in plant biology, 6(5), pp. 410-417.

Shiyu, Q., Hongen, L., Zhaojun, N., Rengel, Z., Wei, G., Chang, L. and Peng, Z. (2020) 'Toxicity of cadmium and its competition with mineral nutrients for uptake by plants: A review', Pedosphere, 30(2), pp. 168-180.

Sid'ko, A., Botvich, I. Y., Pis'man, T. and Shevyrnogov, A. (2017) 'Estimation of the chlorophyll content and yield of grain crops via their chlorophyll potential', Biophysics, 62(3), pp. 456-459.

Sims, R. E., Mabee, W., Saddler, J. N. and Taylor, M. (2010) 'An overview of second generation biofuel technologies', Bioresource technology, 101(6), pp. 1570-1580.

Singh, S., Kumar, V., Datta, S., Dhanjal, D. S., Sharma, K., Samuel, J. and Singh, J. (2020) 'Current advancement and future prospect of biosorbents for bioremediation', Science of the Total Environment, 709, pp. 135895

Sirijan, M., Drapal, M., Chaiprasart, P. and Fraser, P. D. (2020) 'Characterisation of Thai strawberry (Fragaria× ananassa Duch.) cultivars with RAPD markers and metabolite profiling techniques', Phytochemistry, 180, pp. 112522.

Soon, C. Y., Tee, Y. B., Tan, C. H., Rosnita, A. T. and Khalina, A. (2018) 'Extraction and physicochemical characterization of chitin and chitosan from Zophobas morio larvae in varying sodium hydroxide concentration', International Journal of Biological Macromolecules, 108, pp. 135-142.

Souahi, H., Chebout, A., Akrout, K., Massaoud, N. and Gacem, R. (2021) 'Physiological responses to lead exposure in wheat, barley and oat', Environmental Challenges, 4, pp. 100079.

Souri, Z., Cardoso, A. A., da-Silva, C. J., de Oliveira, L. M., Dari, B., Sihi, D. and Karimi, N. (2019) 'Heavy metals and photosynthesis: recent developments', Photosynthesis, Productivity and Environmental Stress, pp. 107-134.

Sreenivasulu, N. and Schnurbusch, T. (2012) 'A genetic playground for enhancing grain number in cereals', Trends in plant science, 17(2), pp. 91-101.

Stevens, J., Jones, M. A. and Lawson, T. (2021) 'Diverse Physiological and Physical Responses among Wild, Landrace and Elite Barley Varieties Point to Novel Breeding Opportunities', Agronomy, 11(5), pp. 921.

Stolarski, M. J., Krzyżaniak, M., Łuczyński, M., Załuski, D., Szczukowski, S., Tworkowski, J. and Gołaszewski, J. (2015) 'Lignocellulosic biomass from short rotation woody crops as a feedstock for second-generation bioethanol production', Industrial Crops and Products, 75, pp. 66-75.

Štorchová, H., Hrdličková, R., Chrtek Jr, J., Tetera, M., Fitze, D. and Fehrer, J. (2000) 'An improved method of DNA isolation from plants collected in the field and conserved in saturated NaCl/CTAB solution', Taxon, 49(1), pp. 79-84.

Su, T., Fu, L., Kuang, L., Chen, D., Zhang, G., Shen, Q. and Wu, D. (2022) 'Transcriptome-wide m6A methylation profile reveals regulatory networks in roots of barley under cadmium stress', Journal of Hazardous Materials, 423, pp. 127140.

Syeda, H. I., Sultan, I., Razavi, K. S. and Yap, P.-S. (2022) 'Biosorption of heavy metals from aqueous solution by various chemically modified agricultural wastes: A review', Journal of Water Process Engineering, 46, pp. 102446.

Szalai, G., Tajti, J., Hamow, K. Á., Ildikó, D., Khalil, R., Vanková, R., Dobrev, P., Misheva, S. P., Janda, T. and Pál, M. (2020) 'Molecular background of cadmium tolerance in Rht dwarf wheat mutant is related to a metabolic shift from proline and polyamine to phytochelatin synthesis', Environmental Science and Pollution Research, 27(19), pp. 23664-23676.

Tahir, N. A. (2014). Genetic variability evaluation among Iraqi rice (Oryza sativa L.) varieties using RAPD markers and protein profiling. Jordan Journal of Biological Sciences (JJBS), 7(1), 13-18.

Tahir, N. A. R., Rasul, K. S., & Lateef, D. D. (2023). Effect of mixing oak leaf biomass with soil on cadmium toxicity and translocation in tomato genotypes. Heliyon, 9(8).

Tahir, N. A. R., Lateef, D. D., Mustafa, K. M., Rasul, K. S., & Khurshid, F. F. (2023). Determination of Physiochemical Characteristics Associated with Various Degrees of Cadmium Tolerance in Barley Accessions. Agronomy, 13(6), 1502.

Tahir, N., Lateef, D., Rasul, K., Rahim, D., Mustafa, K., Sleman, S., ... & Aziz, R. Assessment of genetic variation and population structure in Iraqi barley accessions using ISSR, CDDP, and SCoT markers.

Tahir, N. A. R., Lateef, D. D., Mustafa, K. M., and Rasul, K. S. (2022) 'Under natural field conditions, exogenous application of moringa organ water extract enhanced the growth-and yield-related traits of barley accessions'. Agriculture, 12(9), p. 1502.

Tahir, N. A., Ahmed, J. O., Azeez, H. A., Palani, W. R. M. and Omer, D. A. (2019) 'Phytochemical, antibacterial, antioxidant and phytotoxicity screening of the extracts collected from the fruit and root of wild Mt. Atlas Mastic tree (Pistacia atlantica Subsp. Kurdica)', Applied Ecology and Environmental Research, 17(2), pp. 4417–4429.

Taiz, L. and Zeiger, E. (2010) 'Plant physiology 5th Ed', Sunderland, MA: Sinauer Associates, 464.

Takahashi, R., Bashir, K., Ishimaru, Y., Nishizawa, N. K. and Nakanishi, H. (2012) 'The role of heavy-metal ATPases, HMAs, in zinc and cadmium transport in rice', Plant Signaling & Behavior, 7(12), pp. 1605-1607.

Talakayala, A., Ankanagari, S. and Garladinne, M. (2020) 'Role of Silicon Transportation Through Aquaporin Genes for Abiotic Stress Tolerance in Plants', Protective Chemical Agents in the Amelioration of Plant Abiotic Stress: Biochemical and Molecular Perspectives, pp. 622-634.

Talebi, R., Nosrati, S., Etminan, A. and Naji, A. M. (2018) 'Genetic diversity and population structure analysis of landrace and improved safflower (Cartamus tinctorious L.) germplasm using arbitrary functional gene-based molecular markers', Biotechnology & Biotechnological Equipment, 32(5), pp. 1183-1194.

Tamás, L., Dudíková, J., Ďurčeková, K., Halušková, Ľ., Huttová, J. and Mistrík, I. (2009) 'Effect of cadmium and temperature on the lipoxygenase activity in barley root tip', Protoplasma, 235(1), pp. 17-25.

Tamás, L., Dudíková, J., Ďurčeková, K., Halušková, L. u., Huttová, J., Mistrík, I. and Ollé, M. (2008) 'Alterations of the gene expression, lipid peroxidation, proline and thiol content along the barley root exposed to cadmium', Journal of plant physiology, 165(11), pp. 1193-1203.

Tan, C., Chapman, B., Wang, P., Zhang, Q., Zhou, G., Zhang, X.-q., Barrero, R. A., Bellgard, M. I. and Li, C. (2020) 'BarleyVarDB: a database of barley genomic variation', Database, 2020.

Tani, E., Chronopoulou, E., Labrou, N., Sarri, E., Goufa, M., Vaharidi, X., Tornesaki, A., Psychogiou, M., Bebeli, P. and Abraham, E. (2019) 'Growth, physiological, biochemical, and transcriptional responses to drought stress in seedlings of Medicago sativa L., Medicago arborea L. and their hybrid (Alborea)', Agronomy, 9(1), pp. 38.

Tarawneh, R. A., Alqudah, A. M., Nagel, M. and Börner, A. (2020) 'Genome-wide association mapping reveals putative candidate genes for drought tolerance in barley', Environmental and Experimental Botany, 180, pp. 104237.

Teplyakova, S., Lebedeva, M., Ivanova, N., Horeva, V., Voytsutskaya, N., Kovaleva, O. and Potokina, E. (2017) 'Impact of the 7-bp deletion in HvGA20ox2 gene on agronomic important traits in barley (Hordeum vulgare L.)', BMC plant biology, 17(1), pp. 1-10.

Thabet, S. G., Moursi, Y. S., Karam, M. A., Börner, A. and Alqudah, A. M. (2020) 'Natural variation uncovers candidate genes for barley spikelet number and grain yield under drought stress', Genes, 11(5), pp. 533.

Thabet, S. G., Moursi, Y. S., Karam, M. A., Graner, A. and Alqudah, A. M. (2018) 'Genetic basis of drought tolerance during seed germination in barley', PloS one, 13(11), pp. e0206682.

Thakare, M., Sarma, H., Datar, S., Roy, A., Pawar, P., Gupta, K., Pandit, S. and Prasad, R. (2021) 'Understanding the holistic approach to plant-microbe remediation technologies for removing heavy metals and radionuclides from soil', Current Research in Biotechnology, 3, pp. 84-98.

Tiwari, G., Singh, R., Singh, N., Choudhury, D. R., Paliwal, R., Kumar, A. and Gupta, V. (2016) 'Study of arbitrarily amplified (RAPD and ISSR) and gene targeted (SCoT and CBDP) markers for genetic diversity and population structure in Kalmegh [Andrographis paniculata (Burm. f.) Nees]', Industrial Crops and Products, 86, pp. 1-11.

Tong, H., Wang, J., Chen, H., Wang, Z., Fan, H. and Ni, Z. (2017) 'Transcriptomic analysis of gene expression profiles of stomach carcinoma reveal abnormal expression of mitotic components', Life sciences, 170, pp. 41-49.

Torres, E. (2020) 'Biosorption: A review of the latest advances', Processes, 8(12), pp. 1584.

Torun, H. (2019) 'Time-course analysis of salicylic acid effects on ROS regulation and antioxidant defense in roots of hulled and hulless barley under combined stress of drought, heat and salinity', Physiologia Plantarum, 165(2), pp. 169-182.

Tripathi, P., Subedi, S., Khan, A. L., Chung, Y.-S. and Kim, Y. (2021) 'Silicon effects on the root system of diverse crop species using root phenotyping technology', Plants, 10(5), pp. 885.

Tubaña, B. S. and Heckman, J. R. (2015) 'Silicon in soils and plants',  Silicon and plant diseases: Springer, pp. 7-51.

Vaezi, B., Bavei, V. and Shiran, B. (2010) 'Screening of barley genotypes for drought tolerance by agro-physiological traits in field condition', African journal of agricultural research, 5(9), pp. 881-892.

Van De Velde, K., Ruelens, P., Geuten, K., Rohde, A. and Van Der Straeten, D. (2017) 'Exploiting DELLA signaling in cereals', Trends in plant science, 22(10), pp. 880-893.

Vasanthi, N., Saleena, L. M. and Raj, S. A. (2014) 'Silicon in crop production and crop protection-A review', Agricultural Reviews, 35(1), pp. 14-23.

Vats, S. (ed.) (2018) Biotic and Abiotic Stress Tolerance in Plants. Singapore: Springer Singapore.

Vega, I., Nikolic, M., Pontigo, S., Godoy, K., Mora, M. d. L. L. and Cartes, P. (2019) 'Silicon improves the production of high antioxidant or structural phenolic compounds in barley cultivars under aluminum stress', Agronomy, 9(7), pp. 388.

Verma, S., Yashveer, S., Rehman, S., Gyawali, S., Kumar, Y., Chao, S., Sarker, A. and Verma, R. P. S. (2021) 'Genetic and Agro-morphological diversity in global barley (Hordeum vulgare L.) collection at ICARDA', Genetic Resources and Crop Evolution, 68(4), pp. 1315-1330.

Verret, F., Gravot, A., Auroy, P., Leonhardt, N., David, P., Nussaume, L., Vavasseur, A. and Richaud, P. (2004) 'Overexpression of AtHMA4 enhances root-to-shoot translocation of zinc and cadmium and plant metal tolerance', FEBS letters, 576(3), pp. 306-312.

Vieira, E. A., Carvalho, F. I. F. d., Bertan, I., Kopp, M. M., Zimmer, P. D., Benin, G., Silva, J. A. G. d., Hartwig, I., Malone, G. and Oliveira, A. C. d. (2007) 'Association between genetic distances in wheat (Triticum aestivum L.) as estimated by AFLP and morphological markers', Genetics and Molecular Biology, 30(2), pp. 392-399.

Vinceti, B., Loo, J., Gaisberger, H., van Zonneveld, M. J., Schueler, S., Konrad, H., Kadu, C. A. and Geburek, T. (2013) 'Conservation priorities for Prunus africana defined with the aid of spatial analysis of genetic data and climatic variables', PloS one, 8(3), pp. e59987.

Vítámvás, P., Kosová, K., Musilová, J., Holková, L., Mařík, P., Smutná, P., Klíma, M. and Prášil, I. T. (2019) 'Relationship between dehydrin accumulation and winter survival in winter wheat and barley grown in the field', Frontiers in Plant Science, 10, pp. 7

Vladimirovich, D. A., Grigorievich, D. V. and Sergeevna, D. N. (2021) 'The assessment of cadmium nitrate effect on morphological and cytogenetic indices of spring barley (Hordeum vulgare) seedlings', Brazilian Journal of Botany, 44(1), pp. 43-56.

Wang, A., Yu, Z. and Ding, Y. (2009) 'Genetic diversity analysis of wild close relatives of barley from Tibet and the Middle East by ISSR and SSR markers', Comptes rendus biologies, 332(4), pp. 393-403.

Wang, J., Yang, J., Jia, Q., Zhu, J., Shang, Y., Hua, W. and Zhou, M. (2014) 'A new QTL for plant height in barley (Hordeum vulgare L.) showing no negative effects on grain yield', PloS one, 9(2), pp. e90144.

Wang, W., Zhang, G., Yang, S., Zhang, J., Deng, Y., Qi, J., Wu, J., Fu, D., Wang, W. and Hao, Q. (2021) 'Overexpression of isochorismate synthase enhances drought tolerance in barley', Journal of plant physiology, 260, pp. 153404.

Wang, X.-K., Gong, X., Cao, F., Wang, Y., Zhang, G. and Wu, F. (2019) 'HvPAA1 encodes a P-Type ATPase, a novel gene for cadmium accumulation and tolerance in barley (Hordeum vulgare L.)', International journal of molecular sciences, 20(7), pp. 1732.

Wang, X., Ma, R., Cui, D., Cao, Q., Shan, Z. and Jiao, Z. (2017) 'Physio-biochemical and molecular mechanism underlying the enhanced heavy metal tolerance in highland barley seedlings pre-treated with low-dose gamma irradiation', Scientific reports, 7(1), pp. 1-14.

Wang, Z., Guo, H., Shen, F., Yang, G., Zhang, Y., Zeng, Y., Wang, L., Xiao, H. and Deng, S. (2015) 'Biochar produced from oak sawdust by Lanthanum (La)-involved pyrolysis for adsorption of ammonium (NH4+), nitrate (NO3−), and phosphate (PO43−)', Chemosphere, 119, pp. 646-653.

Wehner, G. G., Balko, C. C., Enders, M. M., Humbeck, K. K. and Ordon, F. F. (2015) 'Identification of genomic regions involved in tolerance to drought stress and drought stress induced leaf senescence in juvenile barley', BMC plant biology, 15(1), pp. 125.

Weiss, E., Zohary, D. and Hopf, M. (2012) Domestication of Plants in the Old World - The Origin and Spread of Domesticated Plants in South-west Asia, Europe, and the Mediterranean Basin.

Wildermuth, M. C., Dewdney, J., Wu, G. and Ausubel, F. M. (2001) 'Isochorismate synthase is required to synthesize salicylic acid for plant defence', Nature, 414(6863), pp. 562-565.

Wójcik-Jagła, M., Fiust, A., Kościelniak, J. and Rapacz, M. (2018) 'Association mapping of drought tolerance-related traits in barley to complement a traditional biparental QTL mapping study', Theoretical and applied genetics, 131(1), pp. 167-181.

Wu, L.-M., Fang, Y., Yang, H.-N. and Bai, L.-Y. (2019) 'Effects of drought-stress on seed germination and growth physiology of quinclorac-resistant Echinochloa crusgalli', PloS one, 14(4), pp. e0214480.

Xie, W., Xiong, W., Pan, J., Ali, T., Cui, Q., Guan, D., Meng, J., Mueller, N. D., Lin, E. and Davis, S. J. (2018) 'Decreases in global beer supply due to extreme drought and heat', Nature plants, 4(11), pp. 964-973.

Xu, Y., Jia, Q., Zhou, G., Zhang, X.-Q., Angessa, T., Broughton, S., Yan, G., Zhang, W. and Li, C. (2017) 'Characterization of the sdw1 semi-dwarf gene in barley', BMC plant biology, 17(1), pp. 1-10.

Xue, D.-w., Chen, M.-c., Zhou, M.-x., Chen, S., Mao, Y. and Zhang, G.-p. (2008) 'QTL analysis of flag leaf in barley (Hordeum vulgare L.) for morphological traits and chlorophyll content', Journal of Zhejiang University Science B, 9(12), pp. 938-943.

Xue, W., Yan, J., Jiang, Y., Zhan, Z., Zhao, G., Tondelli, A., Cattivelli, L. and Cheng, J. (2019) 'Genetic dissection of winter barley seedling response to salt and osmotic stress', Molecular breeding, 39(9), pp. 137.

Yadav, S. (2010) 'Heavy metals toxicity in plants: an overview on the role of glutathione and phytochelatins in heavy metal stress tolerance of plants', South African Journal of Botany, 76(2), pp. 167-179.

Yang, L., Liu, B., Zhai, H., Wang, S., Liu, H., Zhou, Y., Meng, F., Yang, J., Zhu, G. and Chui, S. (2009) 'Dwarf male-sterile wheat: a revolutionary breeding approach to wheat', Induced plant mutations in the genomics era.'(Ed. QY Shu) pp, pp. 370-372.

Yasmeen, A., Basra, S., Farooq, M., ur Rehman, H. and Hussain, N. (2013a) 'Exogenous application of moringa leaf extract modulates the antioxidant enzyme system to improve wheat performance under saline conditions', Plant Growth Regulation, 69(3), pp. 225-233

Yasmeen, A., Basra, S. M. A., Ahmad, R. and Wahid, A. (2012) 'Performance of late sown wheat in response to foliar application of Moringa oleifera Lam. leaf extract', Chilean Journal of Agricultural Research, 72(1), pp. 92.

Yasmeen, A., Basra, S. M. A., Wahid, A., Nouman, W. and REHMAN, H. U. (2013b) 'Exploring the potential of Moringa oleifera leaf extract (MLE) as a seed priming agent in improving wheat performance', Turkish Journal of Botany, 37(3), pp. 512-520

Yemm, E. W. and Willis, A. J. (1954) 'The estimation of carbohydrates in plant extracts by anthrone', The Biochemical journal, 57(3), pp. 508–514.

Yongcui, H., Zehong, Y. and Xiujin, L. (2005) 'Genetic diversity among barley germplasm with known origins based on the RAMP and ISSR markers', Scientia Agricultura Sinica.

Yu, G., Ma, J., Jiang, P., Li, J., Gao, J., Qiao, S. and Zhao, Z. 'The mechanism of plant resistance to heavy metal'. IOP Conference Series: Earth and Environmental Science: IOP Publishing, 052004.

Yu, Z., Wang, X., Mu, X. and Zhang, L. (2019b) 'RNAi mediated silencing of dehydrin gene WZY2 confers osmotic stress intolerance in transgenic wheat', Functional Plant Biology, 46(10), pp. 877-884.

Yuan, W., Suo, J., Shi, B., Zhou, C., Bai, B., Bian, H., Zhu, M. and Han, N. (2019) 'The barley miR393 has multiple roles in regulation of seedling growth, stomatal density, and drought stress tolerance', Plant Physiology and Biochemistry, 142, pp. 303-311.

Zaheer, M. S., Raza, M. A. S., Saleem, M. F., Erinle, K. O., Iqbal, R. and Ahmad, S. (2019) 'Effect of rhizobacteria and cytokinins application on wheat growth and yield under normal vs drought conditions', Communications in Soil Science and Plant Analysis, 50(20), pp. 2521-2533.

Zahn, S., Schmutzer, T., Pillen, K. and Maurer, A. (2021) 'Genomic Dissection of Peduncle Morphology in Barley through Nested Association Mapping', Plants, 10(1), pp. 10.

Zakaria, Z., Zulkafflee, N. S., Mohd Redzuan, N. A., Selamat, J., Ismail, M. R., Praveena, S. M., Tóth, G. and Abdull Razis, A. F. (2021) 'Understanding potential heavy metal contamination, absorption, translocation and accumulation in rice and human health risks', Plants, 10(6), pp. 1070.

Zare, S., Mirlohi, A., Saeidi, G., Sabzalian, M. R. and Ataii, E. (2021) 'Water stress intensified the relation of seed color with lignan content and seed yield components in flax (Linum usitatissimum L.)', Scientific reports, 11(1), pp. 1-15.

Zeraatkar, A. K., Ahmadzadeh, H., Talebi, A. F., Moheimani, N. R. and McHenry, M. P. (2016) 'Potential use of algae for heavy metal bioremediation, a critical review', Journal of environmental management, 181, pp. 817-831.

Zhang, C., Yang, Q., Zhang, X., Zhang, X., Yu, T., Wu, Y., Fang, Y. and Xue, D. (2021a) 'Genome-Wide Identification of the HMA Gene Family and Expression Analysis under Cd Stress in Barley', Plants, 10(9), pp. 1849.

Zhang, L. (2011) 'Strong plant-soil associations in a heterogeneous subtropical broad-leaved forest', Plant and Soil, 347(1-2), pp. 211–220.

Zhang, M., Jin, Z.-Q., Zhao, J., Zhang, G. and Wu, F. (2015) 'Physiological and biochemical responses to drought stress in cultivated and Tibetan wild barley', Plant Growth Regulation, 75(2), pp. 567-574.

Zhang, W., Peng, K., Cui, F., Wang, D., Zhao, J., Zhang, Y., Yu, N., Wang, Y., Zeng, D. and Wang, Y. (2021b) 'Cytokinin oxidase/dehydrogenase OsCKX11 coordinates source and sink relationship in rice by simultaneous regulation of leaf senescence and grain number', Plant biotechnology journal, 19(2), pp. 335-350.

Zhao, H., Ni, S., Cai, S. and Zhang, G. (2021) 'Comprehensive dissection of primary metabolites in response to diverse abiotic stress in barley at seedling stage', Plant Physiology and Biochemistry, 161, pp. 54-64.

Zheng, B., Zhong, S., Tang, Y. and Chen, L. (2020) 'Understanding the nutritional functions of thermally-processed whole grain highland barley in vitro and in vivo', Food Chemistry, 310, pp. 125979.

Zheng, Y., Jia, A., Ning, T., Xu, J., Li, Z. and Jiang, G. (2008) 'Potassium nitrate application alleviates sodium chloride stress in winter wheat cultivars differing in salt tolerance', Journal of plant physiology, 165(14), pp. 1455–1465.

Ziegler-Jöns, A. (1989) 'Gas-exchange of ears of cereals in response to carbon dioxide and light', Planta, 178(2), pp. 164-175.